\documentclass[a4paper,twoside,12pt,titlepage]{mybook}
\usepackage{a4}

\usepackage{amsmath,amsfonts,amssymb,amsthm}
\usepackage{url}
\usepackage{float}
\usepackage{graphicx}
\usepackage{subfigure}
\usepackage[sectionbib]{bibunits}
\usepackage{algorithmic}
\usepackage{setspace}
\usepackage{fancyhdr}
\usepackage{hyperref}
  \renewcommand{\chaptermark}[1]{\markboth{\chaptername \ \thechapter \ \ #1}{}}
  
\usepackage{adfatitlepage}
\usepackage{rotating}
\usepackage[none]{hyphenat}
\usepackage[intoc]{nomen}
\usepackage{pstricks}
\usepackage{array}
\usepackage{substr}
\usepackage[chapter]{algorithm}
\usepackage{url}
\usepackage{verbatim}
\usepackage{cite}
\usepackage{mathrsfs}
\usepackage{multirow}
\usepackage{bm}
\usepackage{lineno}
\usepackage{array}
\usepackage{color}
\usepackage{latexsym}
\usepackage{amsxtra}
\usepackage{acronym}
\usepackage{longtable}
\usepackage{nomencl}
\usepackage{titletoc}
\usepackage{stfloats}
\usepackage{hhline}
\usepackage{tabularx}
\usepackage{makecell}
\usepackage{enumerate}
\usepackage{epstopdf}
\usepackage{epsfig}
\usepackage[latin1]{inputenc}
\usepackage{fontenc}

\newtheorem{theo}{Theorem}[chapter]
\newtheorem{lemm}{Lemma}[chapter]
\newtheorem{coro}{Corollary}[chapter]

\newcommand*{\QEDA}{\hfill\ensuremath{\blacksquare}}

\makenomenclature

\usepackage{geometry}
\begin{document}

\begin{titlepage}
\begin{center}
\vspace*{1cm}
\Huge \hspace{-15mm}\textbf{Random Access for Massive Machine-Type Communications} \\
\vspace{1.5cm}
\hspace{-15mm}\normalsize\textbf{Zhuo Sun}
\\
\vspace{2cm}
\normalsize
{\hspace{-15mm}A thesis submitted to the Graduate Research School of\\
\hspace{-15mm}The University of New South Wales\\
\hspace{-15mm}in partial fulfillment of the requirements for the degree of\\
\text{ \ }\\
\hspace{-15mm}\textbf{Doctor of Philosophy}}\\
\vspace{2cm}
\hspace{-15mm}\includegraphics[width=0.4\textwidth]{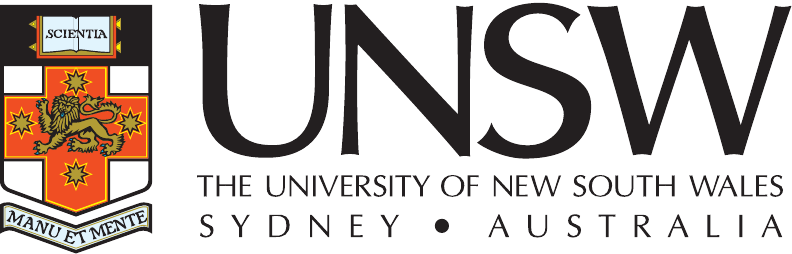}\\
\vspace{2cm}
\textbf{\hspace{-15mm}School of Electrical Engineering and Telecommunications\\
\hspace{-15mm}Faculty of Engineering\\
\hspace{-15mm}The University of New South Wales}\\
\vspace{2cm}
\hspace{-15mm}{March 2019}
\end{center}
\end{titlepage}

\frontmatter
\onehalfspacing
%\pagestyle{empty}
%\include{misc/library}
%\clearpage{\thispagestyle{empty}\cleardoublepage}
%\include{misc/declaration}
%\clearpage{\thispagestyle{empty}\cleardoublepage}

\pagenumbering{roman}
\pagestyle{fancy}
\fancyhf{}
\setlength{\headheight}{15pt}
\fancyhead[LE,RO]{\footnotesize \textbf \thepage}
%\begin{center}
%    \large \emph{Dedicated to my parents, my husband, and my daughter.}
%\end{center}
\doublespacing
\chapter*{Abstract}
\addcontentsline{toc}{chapter}{\protect\numberline{}{Abstract}}

As a key enabler for the Internet-of-Things (IoT), machine-type communications (MTC) has emerged to be an essential part for future communications.
In MTC, a number of machines (called users or devices) ubiquitously communicate to a base station or among themselves with no or minimal human interventions.
It is envisioned that the number of machines to be connected will reach tens of billions in the near future.
In order to accommodate such a massive connectivity, random access schemes are deemed as a natural and efficient way.
Different from scheduled multiple access schemes in existing cellular networks, users with transmission demands will access the channel in an uncoordinated manner via random access schemes, which can substantially reduce the signalling overhead.
However, the reduction in signalling overhead may sacrifice the system reliability and efficiency, due to the unknown user activity and the inevitable interference from contending users.
This seriously hinders the application of random access schemes in MTC.
Therefore, this thesis is dedicated to studying methods to improve the efficiency of random access schemes and to facilitate their deployment in MTC.

In the first part of this thesis, we design a joint user activity identification and channel estimation scheme for grant-free random access systems.
We first propose a decentralized transmission control scheme by exploiting the channel state information (CSI).
With the proposed transmission control scheme, we design a compressed sensing (CS) based user activity identification and channel estimation scheme.
We analyze the packet delay and throughput of the proposed scheme and optimize the transmission control scheme to maximize the system throughput.

The second part of this thesis focuses on the design and analysis of a random access scheme, i.e., the coded slotted ALOHA (CSA) scheme, in the presence of channel erasures, to improve the system throughput.
First, we design the code probability distributions for CSA schemes with repetition codes and maximum distance separable (MDS) codes to maximize the expected traffic load, under both
packet erasure channels and slot erasure channels.
In particular, we derive the extrinsic information transfer (EXIT) functions of CSA schemes over the two erasure channels.
By optimizing the convergence behavior of the derived EXIT functions, we obtain the code probability distributions
to maximize the expected traffic load.
Then, we derive the asymptotic throughput of CSA schemes over erasure channels for an infinite frame length, which is verified to well approximate the throughput for CSA schemes with a practical frame length.
Numerical results demonstrate that the designed code distributions can maximize the expected traffic load and improve the throughput for CSA schemes over erasure channels.

In the third part of this thesis, we concentrate on designing efficient data decoding algorithms for the CSA scheme, to further improve the system efficiency.
First, we present a low-complexity physical-layer network coding (PNC) method to obtain linear combinations of collided packets.
Then, we design an enhanced message-level successive interference cancellation (SIC) algorithm to exploit the obtained linear combinations to recover more users' packets.
In addition, we propose an analytical framework for the PNC-based decoding scheme and derive a tight approximation of the system throughput for the proposed scheme.
Furthermore, we optimize the CSA scheme to maximize the system throughput and energy efficiency, respectively.

\doublespacing
\chapter*{Acknowledgments}
This work would not have been done without the encouragement and the support from many people I have met during my Ph.D. journey.

First and foremost, I would like to thank my supervisor Prof. Jinhong Yuan, who gave me the opportunity to pursue this degree in the first place, and has been always supporting me all these years.
He guided me into good research topics, helped me formulate research problems and was generous to share his time on my doubts and confusions.
His rigor and conciseness shaped my style and improved my skills in presenting complex technical ideas in papers. It is my pleasure and honor to be his student.

Second, I would like to thank my co-supervisor Dr. Lei Yang, for his guidance, constructive suggestions, and help on my research.
Lei is very kind that each time he proposed to help me before I even asked.
His hard working and dedication also impressed me a lot.

I am also grateful to work with Dr. Derrick Wing Kwan Ng. Derrick is so gifted and so hard working at the same time. His remarkable perception on research and his efficient working style have been invaluable.

I also would like to thank Dr. Tao Yang, who had supported me in my first year of Ph.D study. Special thanks to Dr. Yixuan Xie for helping me get a better understanding of coding theory. Many thanks to Prof. Marco Chiani at the University of Bologna, Italy, for his valuable contributions on my second research work.

I would also like to thank all my friends and colleagues at the Wireless Communication Lab at UNSW, for many discussions we had leading to a better understanding of wireless communication theories, and the great pleasure we
have working together.

Finally, I would like to give my deepest appreciation to my beloved parents for everything they have done for me.
I cannot thank enough to my husband for his understanding and unconditional support.
I would also like to express the special thanks to my daughter, who brings me many happy times.
None of this would happen without them.
I would like to dedicate this thesis to my parents, my husband, and my daughter.

\singlespacing
%\fancyhead[CE,CO]{\MakeUppercase{List of Publications}}
\chapter*{List of Publications} \label{listofpublications}
\addcontentsline{toc}{chapter}{\protect\numberline{}{List of Publications}}
%\nomenclature{$a$}{The number of angels per unit area}%
%\nomenclature{$N$}{The number of angels per needle point}%
%\nomenclature{$A$}{The area of the needle point}%
%

\ifpdf
    \graphicspath{{1_introduction/figures/PNG/}{1_introduction/figures/PDF/}{1_introduction/figures/}}
\else
    \graphicspath{{1_introduction/figures/EPS/}{1_introduction/figures/}}
\fi

\noindent{\large\textbf{Journal Articles:}} \vspace{0.1in}
\begin{enumerate}
\item \textbf{Z. Sun}, Y. Xie, J. Yuan and T. Yang, ``Coded Slotted {ALOHA} for Erasure Channels: {D}esign and Throughput Analysis,'' {\bf\em IEEE Trans. Commun.}, vol. 65, no. 11, pp. 4817-4830, Nov. 2017.
\item \textbf{Z. Sun}, L. Yang, J. Yuan and D. W. K. Ng, ``Physical-Layer Network Coding Based Decoding Scheme for Random Access,'' in \textit{IEEE Trans. Veh. Technol.}, vol. 68, no. 4, pp. 3550-3564, Apr. 2019.
\item \textbf{Z. Sun}, Z. Wei, L. Yang, J. Yuan, X. Cheng, W. Lei, ``Exploiting Transmission Control for Joint User Identification and Channel Estimation in Massive Connectivity,'' \textit{IEEE Trans. Commun.}, accepted, May 2019.
\end{enumerate}
\vspace{0.1in} \noindent{\large\textbf{Conference Articles:}}
\vspace{0.1in}
\begin{enumerate}
\item \textbf{Z. Sun}, Y. Xie, J. Yuan, and T. Yang, ``Coded slotted ALOHA schemes for erasure channels,'' in {\bf\em Proc. IEEE Intern. Commun. Conf. (ICC)}, Kuala Lumpur, May 2016, pp. 1-6.
\item \textbf{Z. Sun}, L. Yang, J. Yuan, and M. Chiani, ``A novel detection algorithm for random multiple access based on physical-layer network coding,'' in {\bf\em Proc. IEEE Intern. Commun. Conf. Workshops (ICC)}, Kuala Lumpur, May 2016, pp. 608-613.
\item \textbf{Z. Sun}, Z. Wei, L. Yang, J. Yuan, X. Cheng, and L. Wan, ``Joint User Identification and Channel Estimation in Massive Connectivity with Transmission Control,'' in {{\bf\em IEEE Intern. Sympos. on Turbo Codes and Iterative Inf. Process. (ISTC)}}, Hong Kong, Dec. 2018, pp. 1-5.
\end{enumerate}
\chapter*{Abbreviations} \label{abbreviations}
\addcontentsline{toc}{chapter}{\protect\numberline{}{Abbreviations}}
\markboth{ABBREVIATIONS}{}

\begin{longtable}[t]{ll}
\textbf{AMP} \quad\quad&\mbox{Approximate message passing} \vspace{0.1in}\\
\textbf{APP} \quad\quad&\mbox{A posteriori probability}\vspace{0.1in}\\
\textbf{AWGN} \quad\quad&\mbox{Additive white Gaussian noise}\vspace{0.1in}\\
\textbf{BEC} \quad\quad&\mbox{Binary erasure channel}\vspace{0.1in}\\
\textbf{BER} \quad\quad&\mbox{Bit error rate}\vspace{0.1in}\\
\textbf{BP} \quad\quad&\mbox{Basis pursuit}\vspace{0.1in}\\
\textbf{BPDN} \quad\quad&\mbox{Basis pursuit denoising}\vspace{0.1in}\\
\textbf{BPSK} \quad\quad&\mbox{Binary phase-shift keying}\vspace{0.1in}\\
\textbf{BS} \quad\quad&\mbox{Base station} \vspace{0.1in}\\
\textbf{CDF} \quad\quad&\mbox{Cumulative distribution function}\vspace{0.1in}\\
\textbf{CRDSA} \quad\quad&\mbox{Contention resolution diversity slotted ALOHA} \vspace{0.1in}\\
\textbf{CS} \quad\quad&\mbox{Compressed sensing} \vspace{0.1in}\\
\textbf{CSA} \quad\quad&\mbox{Coded slotted ALOHA} \vspace{0.1in}\\
\textbf{CSI}  \quad\quad&\mbox{Channel state information} \vspace{0.1in}\\
\textbf{CSMA} \quad\quad&\mbox{Carrier-sense multiple access}\vspace{0.1in}\\
\textbf{dB} \quad\quad&\mbox{Decibel} \vspace{0.1in}\\
\textbf{DE} \quad\quad&\mbox{Density evolution}\vspace{0.1in}\\
\textbf{DMC} \quad\quad&\mbox{Discrete memoryless channel}\vspace{0.1in}\\
\textbf{DV} \quad\quad&\mbox{Difference vector}\vspace{0.1in}\\
\textbf{ECC} \quad\quad&\mbox{Error control coding}\vspace{0.1in}\\
\textbf{EXIT} \quad\quad&\mbox{Extrinsic information transfer} \vspace{0.1in}\\
\textbf{FDMA} \quad\quad&\mbox{Frequency division multiple access}\vspace{0.1in}\\
\textbf{IHT} \quad\quad&\mbox{Iterative hard thresholding}\vspace{0.1in}\\
\textbf{i.i.d.} \quad\quad&\mbox{Independent and identically distributed}\vspace{0.1in}\\
\textbf{IoT} \quad\quad&\mbox{Internet-of-Things} \vspace{0.1in}\\
\textbf{IRSA} \quad\quad&\mbox{Irregular repetition slotted ALOHA} \vspace{0.1in}\\
\textbf{IST} \quad\quad&\mbox{Iterative soft thresholding}\vspace{0.1in}\\
\textbf{JUICE}  \quad\quad&\mbox{Joint user identification and channel estimation} \vspace{0.1in}\\
\textbf{LS} \quad\quad&\mbox{Least squares}\vspace{0.1in}\\
\textbf{LASSO} \quad\quad&\mbox{Least absolute shrinkage and selection operator}\vspace{0.1in}\\
\textbf{LDPC} \quad\quad&\mbox{Low-density parity-check}\vspace{0.1in}\\
\textbf{LLR} \quad\quad&\mbox{Log-likelihood ratio}\vspace{0.1in}\\
\textbf{MAC} \quad\quad&\mbox{Media access control}\vspace{0.1in}\\
\textbf{MAP} \quad\quad&\mbox{Maximum-a-posterior}\vspace{0.1in}\\
\textbf{MDS} \quad\quad&\mbox{Maximum distance separable}\vspace{0.1in}\\
\textbf{MIMO} \quad\quad&\mbox{Multiple-input multiple output} \vspace{0.1in}\\
\textbf{ML} \quad\quad&\mbox{Maximum likelihood}\vspace{0.1in}\\
\textbf{MMSE} \quad\quad&\mbox{Minimum mean square error}\vspace{0.1in}\\
\textbf{MSE} \quad\quad&\mbox{Mean squared error}\vspace{0.1in}\\
\textbf{MTC} \quad\quad&\mbox{Machine-type communications} \vspace{0.1in}\\
\textbf{mMTC} \quad\quad&\mbox{Massive machine type communication} \vspace{0.1in}\\
\textbf{MUD} \quad\quad&\mbox{Multiuser decoding}\vspace{0.1in}\\
\textbf{NC} \quad\quad&\mbox{Network-coded}\vspace{0.1in}\\
\textbf{NCN} \quad\quad&\mbox{Network-coded node}\vspace{0.1in}\\
\textbf{NMSE} \quad\quad&\mbox{Normalized mean square error}\vspace{0.1in}\\
\textbf{NOMA} \quad\quad&\mbox{Non-orthogonal multiple access}\vspace{0.1in}\\
\textbf{OMA} \quad\quad&\mbox{Orthogonal multiple access}\vspace{0.1in}\\
\textbf{OMP} \quad\quad&\mbox{Orthogonal matching pursuit}\vspace{0.1in}\\
\textbf{PAM} \quad\quad&\mbox{Pulse amplitude modulation}\vspace{0.1in}\\
\textbf{PDF} \quad\quad&\mbox{Probability density function}\vspace{0.1in}\\
\textbf{PEC} \quad\quad&\mbox{Packet erasure channel}\vspace{0.1in}\\
\textbf{PLR} \quad\quad&\mbox{Packet loss rate}\vspace{0.1in}\\
\textbf{PNC}  \quad\quad&\mbox{Physical-layer network coding} \vspace{0.1in}\\
\textbf{QPSK} \quad\quad&\mbox{Quadrature phase-shift keying}\vspace{0.1in}\\
\textbf{SA}  \quad\quad&\mbox{Slotted ALOHA} \vspace{0.1in}\\
\textbf{SBL} \quad\quad&\mbox{Sparse bayesian learning}\vspace{0.1in}\\
\textbf{SIC}  \quad\quad&\mbox{Successive interference cancellation} \vspace{0.1in}\\
\textbf{SNR} \quad\quad&\mbox{Signal-to-noise ratio}\vspace{0.1in}\\
\textbf{SPA} \quad\quad&\mbox{Sum-product algorithm}\vspace{0.1in}\\
\textbf{SN} \quad\quad&\mbox{Slot node}\vspace{0.1in}\\
\textbf{TDD} \quad\quad&\mbox{Time division duplex} \vspace{0.1in}\\
\textbf{UN} \quad\quad&\mbox{User node}\vspace{0.1in}\\
\end{longtable}

\chapter*{List of Notations} \label{listofnotations}
\addcontentsline{toc}{chapter}{\protect\numberline{}{List of Notations}}
\markboth{LIST OF NOTATIONS}{}
Lowercase letters denote scalars, boldface lowercase letters denote vectors, and boldface uppercase letters denote matrices.
\begin{longtable}[t]{ll}
$\mathbf{x}$ \quad\quad&\mbox{A column vector} \vspace{0.1in}\\
$\mathbf{X}$ \quad\quad&\mbox{A matrix} \vspace{0.1in}\\
$\mathcal{X}$ \quad\quad&\mbox{A set} \vspace{0.1in}\\
$\mathbf{X}^T$ \quad\quad&\mbox{Transpose of $\mathbf{X}$} \vspace{0.1in}\\
$\mathbf{X}^{\ast}$ \quad\quad&\mbox{Conjugate transpose of $\mathbf{X}$} \vspace{0.1in}\\
$\mathbf{X}^{-1}$ \quad\quad&\mbox{Inverse of $\mathbf{X}$ } \vspace{0.1in}\\
$(\cdot)^{'}$ \quad\quad&\mbox{Derivation} \vspace{0.1in}\\
$\mathbf{X}_{i,j}$ \quad\quad&\mbox{The element in the row $i$ and the column $j$ of $\mathbf{X}$ } \vspace{0.1in}\\
$|\mathcal{X}|$ \quad\quad&\mbox{Cardinality of $\mathcal{X}$} \vspace{0.1in}\\
$|x|$ \quad\quad&\mbox{Absolute value (modulus) of the scalar $x$} \vspace{0.1in}\\
%$[x]^{+}$ \quad\quad&\mbox{$\max\left(0,x\right)$.} \vspace{0.1in}\\
$\mathbb{R}^{M\times N}$ \quad\quad&\mbox{The set of all $M\times N$ matrices with real-valued entries} \vspace{0.1in}\\
$\mathbb{C}^{M\times N}$ \quad\quad&\mbox{The set of all $M\times N$ matrices with complex-valued entries} \vspace{0.1in}\\
%$\doteq$ quad\quad&\mbox{Asymptotically equivalent.} \vspace{0.1in}\\
$\|\cdot\|_{p}$ \quad\quad&\mbox{$\ell_{p}$-norm of a vector or a matrix, $p=0,1,2$} \vspace{0.1in}\\
$\mathbf{0}$ \quad\quad&\mbox{A zero matrix} \vspace{0.1in}\\
%$\mathbf{I}_{\mathrm{N}}$ \quad\quad&\mbox{$N$ dimension identity matrix} \vspace{0.1in}\\
$\mathrm{Re}(\cdot)$ \quad\quad&\mbox{Real part of a complex number} \vspace{0.1in}\\
$\mathrm{Im}(\cdot)$ \quad\quad&\mbox{Imaginary part of a complex number} \vspace{0.1in}\\
$E_{x}\left[\cdot\right]$ \quad\quad&\mbox{Statistical expectation with respect to random variable $x$} \vspace{0.1in}\\
$\mathcal{CN}(\mu,\sigma^2)$ \quad\quad&\mbox{A complex Gaussian random variable with mean $\mu$ and variance $\sigma^2$} \vspace{0.1in}\\
$\max\left\{\cdot\right\}$ \quad\quad&\mbox{Maximization} \vspace{0.1in}\\
$\min\left\{\cdot\right\}$ \quad\quad&\mbox{Minimization} \vspace{0.1in}\\
$O(n)$ \quad\quad&\mbox{The computation complexity is order $n$ operations}
\vspace{0.1in}\\
$\text{sign}(x)$ \quad\quad&\mbox{The sign of $x$} \vspace{0.1in}\\
$(x)_{+}$ \quad\quad&\mbox{A function equaling $x$ if $x > 0$ and otherwise equaling zero} \vspace{0.1in}\\
$\ln(\cdot)$ \quad\quad&\mbox{Natural logarithm} \vspace{0.1in}\\
$\mathrm{s.t.}$ \quad\quad&\mbox{Subject to} \vspace{0.1in}\\
$\otimes$ \quad\quad&\mbox{Multiplication in Galois field $\mathrm{GF}(2)$} \vspace{0.1in}\\
$\text{Rank}(\cdot)$ \quad\quad&\mbox{Rank of a matrix} \vspace{0.1in}\\
$\bmod(x,y)$ \quad\quad&\mbox{Modulo operation of $x$ and $y$} \vspace{0.1in}\\
%$||\cdot||$ \quad\quad&\mbox{$\ell_{1}$-norm of a vector} \vspace{0.1in}\\
$\lim$ \quad\quad&\mbox{Limit} \vspace{0.1in}\\

\end{longtable}

\fancyhead[CE]{\footnotesize \leftmark}
\fancyhead[CO]{\footnotesize \rightmark}

% -- Header modification --
\renewcommand{\chaptermark}[1]{%
\markboth{\MakeUppercase{%
\thechapter.%
\ #1}}{}}
%--------------

% Table of Contents
\tableofcontents

% List of figures
\listoffigures
%\addcontentsline{toc}{chapter}{\protect\numberline{}{List of Figures}}

\listoftables
%\listoftables
%\addcontentsline{toc}{chapter}{\protect\numberline{}{List of Tables}}

% List of algorithms
\listofalgorithms
\addcontentsline{toc}{chapter}{\protect\numberline{}{List of Algorithms}}

\mainmatter
\doublespacing

\chapter{Overview}\label{C1:chapter1}

\section{Introduction and Challenges}
Driven by the proliferation of new applications in the paradigm of Internet-of-Things (IoT), e.g. smart home, autonomous driving, smart industry, etc., machine-type communications (MTC) has emerged as an essential part for future communications \cite{Mukhopadhyay14,Andrews14,Wong17}.
% and the number of connected machine-type devices is expected to reach tens of billions by 2020.
For MTC, a number of machine type devices or users need to communicate to a base station (BS) or among themselves with no or minimal human interventions \cite{Atzori10,Gubbi13,Xu14,Zanella14,Guo183,Guo182,Boshkovska15,Jiajstsp}, which will undoubtedly improve our life quality and bring great business opportunities.
%supporting such a massive connectivity, i.e., a large number of users communicating to an access point,

One important type of MTC is massive MTC (mMTC), where a massive number of users sporadically transmit short packets to the BS.
Compared to the mature human-centric communication, mMTC possesses many distinctive features \cite{Lien11,Mukhopadhyay14,Stankovic14,Suntwcnoma}, which include the massive connectivity requirement, the sporadic traffic pattern, and the small size of transmitted packets.
These features of mMTC yield existing protocols designed for human-centric communications significantly inefficient and call for radical changes in the communication protocol.
For example, in human-centric cellular systems, the commonly adopted user access approach is the grant-based communication protocols \cite{Hasan13,Bursalioglu16,Bjornson17}.
In particular, each user with the accessing demand selects and transmits a pilot sequence to request an access grant from the BS, prior to the data transmission.
After being granted, the BS allocates resource blocks to the accessed users for their following data transmission.
Due to the small size of transmitted packets for mMTC, such a signalling overhead from requesting access makes the grant-based protocols very inefficient.
In addition, as the users select pilot sequences without coordination, two or more users may select the same sequence simultaneously and thus a collision occurs.
In this case, the collided users cannot get the access grant and reattempt to send the grant request after waiting for a random duration.
With the increasing number of users, the collision probability becomes high and a large number of users need to request the access grant twice or more times.
%
%while extensive efforts [] have been devoted to enhance the performance of this grant-based random access process.
This results in an intolerant access delay for the massive connectivity system \cite{Dai15,Liu18m,Chen19noma,Chen18noma}.
Therefore, designing the efficient user access approach is very desirable for mMTC.
%Inspired by these observations, providing efficient user access and reliable communications for mMTC is a very challenging task.

A new communication protocol, called grant-free random access scheme, was proposed and has achieved the industrial and academic consensus on its applicability for mMTC \cite{Dai15, Sun18, Liu18m, Abbas18, Sun18j}.
By contrast to the grant-based schemes, each user directly transmits its pilot and payload data in one shot, once it has a transmission demand \cite{Liu18m}.
It implies that no access requesting procedure is required in the grant-free random access schemes.
As a result, the signalling overhead from requesting access is eliminated and the access latency is significantly reduced, which stimulates the application of grant-free random access schemes to mMTC.
%As a result, both the access latency and the signalling overhead are significantly reduced, which ensures the efficiency of grant-free communications for mMTC.

While attractive features of the grant-free communication exist, some key issues still remain to be addressed.
%
%The first one is to identify active users.
Firstly, without the access grant procedure, the BS needs to identify the user activity from received pilot sequences.
Unfortunately, due to the massive number of users in the system and the limited channel coherence time, it is impossible to allocate orthogonal pilot sequences to all users \cite{Liu18m,Chen18massive}.
Hence, the user activity cannot be simply identified by exploiting the orthogonality among pilot sequences, which imposes a challenge for mMTC.
In addition, the utilization of non-orthogonal pilots causes severe inter-user interference for channel estimation and the conventional pilot-based channel estimation techniques \cite{Marzetta10, Larsson14} are not applicable to mMTC with invoking non-orthogonal pilots.
Therefore, the accurate channel estimation is another difficulty for mMTC.
%This causes the channel estimation very difficult for mMTC.
%This is because the conventional pilot-based channel estimation techniques mainly rely on the assumption of orthogonal pilot sequences \cite{Marzetta10,Larsson14}, and are not applicable to mMTC with invoking non-orthogonal pilots.

Furthermore, in the grant-free random access system, the users transmit their payload data in an uncoordinated way, which results in the inevitable data collisions and dramatically deteriorates the system efficiency.
For the conventional random access systems, e.g. ALOHA \cite{Abramson1970}, the collided data packets are directly discarded and the corresponding users keep retransmitting until their packets can be successfully recovered by the receiver.
However, the ultra high connectivity density of mMTC significantly increases the collision probability, which causes the frequent retransmissions and then an intolerant delay.
Therefore, instead of disregarding the collided packets, it desires to wisely exploit the packet collisions to retrieve more packets and to improve the system efficiency in mMTC \cite{zhuo17, Liva2011, Sun16icc, Paolini2014, Zhuo19tvt, Casini2007, Sun16iccw}.
Unfortunately, the tremendous number of users causes a large collision size, which makes the efficient collision resolution and data decoding very challenging.

%%The probability of data collisions increases with the number of users.
%Then, the efficient recovery of data from collisions is crucial to increase the reliability of grant-free random access systems.
%However, as the number of users explosively increases, the probability of data collisions challenging for mMTC.
%
%With using the grant-free random access scheme as an efficient user access approach, providing reliable communications for mMTC is another essential issue.
%Moreover, this issue becomes more
%
%Even with the perfect user activity identification and channel estimation, the efficient data recovery is another challenging task for the grant-free random access scheme.
%In the grant-free random access scheme, the users transmit data packets in an uncoordinated way.
%The collisions
%the design of efficient random access scheme
%This is because that users randomly occupy resource blocks to transmit their data and a high data collision probability is generated in the grant-free communication protocol.
%Such a high collision probability results in a large number of retransmissions and even a significant loss of data detection performance for conventional random access schemes \cite{Abramson1970,Choudhury1983,Gitman1975,Yang03}, e.g. ALOHA and slotted ALOHA (SA).

This thesis is devoted to tackling the aforementioned challenges for mMTC in future wireless networks, including the user activity identification, the channel estimation, and the design of efficient data detection schemes.

In the first part of this thesis, we focus on the design of joint user activity identification and channel estimation scheme for the grant-free random access system.
In particular, we propose a decentralized transmission control scheme by exploiting the channel state information (CSI) at the user side.
By characterizing the impact of the proposed transmission control scheme on the distribution of received signals, we design a compressed sensing (CS) based joint user activity identification and channel estimation scheme.
We analyze the user activity identification performance via employing a state evolution technique \cite{Donoho06}.
Additionally, the system performance in terms of the packet delay and the network throughput is analyzed for the proposed scheme.
Based on the analysis, we optimize the introduced transmission control strategy to maximize the network throughput.

In the second part of this thesis, the focus is on designing the transmission scheme for the random access system, particularly the coded slotted ALOHA (CSA) scheme \cite{Paolini2014}, in order to facilitate the collision resolution and date decoding in mMTC.
In particular, we design the code probability distributions for CSA schemes with repetition codes and maximum distance separable (MDS) codes over packet erasure channels and slot erasure channels.
In order to characterize the impact of channel erasures on the design of code probability distributions, we first derive the extrinsic information transfer (EXIT) functions for the CSA schemes over the two erasure channels, respectively.
By optimizing the convergence behavior of derived EXIT functions, we optimize the code probability distributions to achieve the maximum expected traffic load.
Finally, we derive the asymptotic throughput of the CSA scheme over erasure channels by considering an infinite frame length, to theoretically evaluate the system performance of CSA schemes with designed code probability distributions.

The third part of this thesis is mainly dedicated to proposing an efficient and low-complexity data decoding scheme to further improve the system efficiency of CSA systems.
We first propose an enhanced low-complexity physical-layer network coding (PNC)-based data decoding scheme to obtain linear combinations of collided packets.
Then, we design an enhanced message-level successive interference cancellation (SIC) algorithm to wisely exploit the obtained linear combinations and to improve the system throughput.
Moreover, we propose an analytical framework for the PNC-based decoding scheme in the CSA system and derive an accurate approximation for the system throughput of the proposed scheme.
With employing the proposed data decoding scheme, we optimize the transmission scheme to further improve the system throughput and energy efficiency of the CSA system, respectively.
%Thus, it calls for new MUD methods to more efficiently resolve collisions and recover data in the grant-free transmission schemes.
%In fact, the improvement of collision resolution schemes can be considered from both the transmission side and the receiver side.
%In particular, the design of transmission schemes can facilitate the collision resolution and the receiver design can provide an effective method to recover more information from the collisions.

\section{Literature Review}
In this section, we provide an intensive review of the existing works in dealing with the challenging issues of mMTC mentioned in the last section, which will be briefly summarized in the following.

\subsection{User Identification and Channel Estimation}\label{juice_background}
In the grant-free random access system, the employment of non-orthogonal pilot sequences among users causes the accurate user identification very difficult \cite{Ding18, Bockelmann16}.
Fortunately, the sporadic transmission of mMTC, i.e., the fact that the number of active users in a specific time is much smaller than the number of potential users, provides a possibility to deal with this difficulty by exploiting the compressed sensing (CS) techniques \cite{Xue17, Eldar12, Donoho06}.
In particular, if all users' signals are collected as a signal vector and the inactive users' signals are considered as zeros, the identification of user activity is equivalent to detecting the support set of the received signal vector.
Due to the sporadic transmission, the received signal vector is sparse and its support set can be detected by adopting the CS techniques.
In the literature, the CS techniques have been intensively exploited to identify the active users for the grant-free random access systems.
In \cite{Zhu11,Schepker11,Schepker15,Shim12,Wang16}, the CS algorithms were employed to jointly identify users' activity and detect their data by assuming the perfect users' CSI at the receiver, which have demonstrated the great potentials of CS techniques for the sparse user activity identification.
However, the perfect receiver-side CSI is an impractical assumption in mMTC.
In practice, the users' CSI has to be estimated by the receiver.

In conjunction with the user activity identification, the channel estimation for grant-free random access systems has been studied in \cite{Yu17,Schepker13,Xu15,Wunder15,Chen18,Liu18,Liu18ma,Senel17globecom}.
In \cite{Yu17}, the author derived an upper bound on the overall transmission rate for the multiple access channel with massive connectivity and presented a practical two-phase scheme to approach the upper bound.
In this proposed scheme \cite{Yu17}, the user identification and channel estimation are jointly performed by using CS techniques in the first phase.
The data detection is executed in the second phase by using conventional multiuser detection (MUD) techniques \cite{David04}.
Note that, the data detection performance highly relies on the accuracy of the user identification and channel estimation in the first phase.
Therefore, improving the joint user identification and channel estimation (JUICE) performance is essential to increase the efficiency of grant-free random access systems.
To improve the JUICE performance, several algorithms were proposed in \cite{Schepker13,Xu15,Wunder15,Chen18,Liu18}.
In \cite{Schepker13}, the authors designed a greedy CS algorithm based on the orthogonal matching pursuit.
By exploiting the statistical CSI, the Bayesian CS method was modified and applied to JUICE in cloud radio access networks \cite{Xu15}.
The authors in \cite{Wunder15} introduced a one-shot random access procedure and analyzed the achievable rate by using the standard basis pursuit denoising.
The computationally efficient approximate message passing (AMP) algorithm was employed to identify the user activity and to estimate channels in \cite{Chen18,Liu18}.
The paper \cite{Liu18ma} demonstrated the great benefits of the AMP algorithm combined with massive multiple-input multiple output (MIMO) technique for enhancing the JUICE performance in the mMTC.
In \cite{Senel17globecom}, a novel transmission scheme for MTC was introduced, where the information bits are embedded into the pilot sequences.
As a result, when performing the JUICE, the data detection can also be achieved.
These CS algorithms developed in \cite{Schepker13,Xu15,Wunder15,Chen18,Liu18,Liu18ma,Senel17globecom} demonstrated that the JUICE problem can be effectively solved by CS methods.
Note that, both the sparsity of user activity and the strength of received signals can affect the performance of CS algorithms \cite{Donoho06}.
These two aforementioned factors can be controlled in designing novel transmission schemes.
However, most works mainly focus on the design of CS algorithms at the receiver side to improve the system performance of mMTC in the literature.
Designing transmission schemes at the user side for the CS-based user activity identification and channel estimation has not been addressed, but it could potentially provide a significant improvement on the system performance of mMTC.
%
%From the compressed sensing theory\cite{Donoho09}, it is known that enhancing the sparsity of user activity, i.e., decreasing the number of users transmitting simultaneously, can significantly improve the performance of CS algorithms, for a given pilot length \cite{Donoho09}.
%Moreover, given the sparsity of user activity, the probability that an active user is correctly identified by CS algorithms is proportional to its received signal strength.
%%
%In fact, the weaker that an active user's signal, the smaller probability that it is correctly identified, and vice versa.

\subsection{ALOHA based Random Access Schemes}\label{aloha_background}
For the classical ALOHA random access schemes \cite{Abramson1970,Gitman1975,Choudhury1983}, the collided packets are directly abandoned and the corresponding users keep retransmitting until their packets are successfully recovered.
It is obvious that such an access mechanism seriously limits the system performance improvement and results in an intolerant delay, particular for the massive connectivity of mMTC.
Therefore, the collision resolution mechanism is very desirable to enhance the system efficiency.

From a collision resolution point of view, several variants of ALOHA have been proposed over last few years. Among them, the contention resolution diversity slotted ALOHA (CRDSA) scheme introduces the SIC technique to resolve packet collisions \cite{Casini2007}. In other words, each packet is transmitted twice at two random slots within one frame. The two replicas know the location of their respective copy by using a pointer. When one copy is received in a collision-free slot and recovered successfully, the pointer is extracted and the interference generated by its twin replica can be removed from the corresponding slot. The process of recovering packets is performed iteratively, until no more collision-free slots exist or all packets are recovered successively. This iterative process results in an improved throughput, compared to the ALOHA and slotted ALOHA (SA) schemes \cite{Casini2007}.

Recently, the CRDSA scheme has been further enhanced by transmitting a variable number of replicas of a packet in one frame, named as irregular repetition slotted ALOHA (IRSA)\cite{Liva2011}. For the IRSA scheme, the SIC process of CRDSA scheme is represented by a bipartite graph and the threshold behavior of iterative packet recovery process can be analyzed. In particular, given a IRSA scheme, there exists a traffic load threshold, which is the largest traffic load such that all but a vanishing small fraction of users' packets can be recovered successfully for large frame sizes. Moreover, compared to the CRDSA scheme, a higher throughput is achieved by designing a repetition code probability distribution in the IRSA scheme. Based on that, the CSA scheme was proposed as a further generalization of the IRSA scheme\cite{Paolini2011}. Before the transmission, the packet from each user is partitioned and encoded into multiple packets via local packet-oriented codes \cite{Paolini2014} at the media access control (MAC) layer. At the receiver side, the SIC process is combined with the decoding of packet-oriented codes to recover collided packets \cite{Paolini2014a}. Compared to the IRSA scheme, the CSA scheme achieves a much higher peak throughput for medium code rates\cite{Paolini2014}.
%From the literature, it is known that the probability distribution of packet-oriented codes selected by users could be designed to maximize the traffic load threshold of the CSA scheme for asymptotically long frames in \cite{Paolini2011a,Chiani2012,Wu2013}.
Most of the existing CSA designs are based on the assumption of collision channels without erasures or noise in the literature \cite{Paolini2011,Paolini2014}, where only the effect of collisions is considered for transmitted packets.
This assumption is impractical, since in practice both the channel fading and the external interference exist and they can corrupt the transmitted packets.
Therefore, it is crucial to design the CSA scheme over more practical channels.
For packet erasure channels, the error floor of packet loss rate was analyzed for the IRSA scheme with finite frame lengths in \cite{IvanovBAP2015,IvanovBAP2015All}.
However, the design of CSA schemes over other practical channels, such as fading channels and slot erasure channels, to improve the throughput remains as an open yet important research problem, which needs to be addressed.
%However, the analysis is less accurate in the high traffic load region and the design of CSA scheme for erasure channels has not been sufficiently considered.

\subsection{Data Detection for Random Access}\label{pnc_background}
Besides the design of CSA-based transmission schemes in \cite{Liva2011,Paolini2014}, the efficient data detection scheme also plays an essential role to improve the throughput of the random access systems.

For ALOHA-based random access schemes, a \emph{signal-level} SIC \emph{in each time slot} is adopted to explore the capture effect in \cite{Roberts1975,Adireddy05,Xu13,Zhang15,Wu12,Wu13}, so that more users' packets are recovered from the collisions.
Here, the capture effect means that a packet with the higher power can be successfully recovered from the received superimposition of multiple packets, when the received power of different packets are imbalanced.
The scheme can efficiently improve the network throughput, when the power imbalance is significant.
In particular, the paper \cite{Roberts1975} adopted the signal-level SIC technique to resolve collision slots in ALOHA, and \cite{Adireddy05} proposed a channel-aware SA scheme by designing the transmission control.
With employing the signal-level SIC at the receiver, a decentralized random power transmission strategy was proposed to maximize the system throughput in \cite{Xu13}, and this strategy was further extended to multiple time slots in \cite{Zhang15}.
Moreover, a modified message passing algorithm was exploited to design the cross-layer random access scheme in \cite{Wu12,Wu13}.

In \cite{Cocco14}, the authors proposed to obtain multiple linear combinations of collided packets in each collided time slot through an exhaustive decoding of all possible linear combinations, where the exhaustive decoding is achieved by employing PNC techniques \cite{Zhang06,Liew2013,Yang16}.
The users' packets are then recovered from all the decoded linear combinations in a MAC frame via matrix manipulations.
In \cite{Lu2013,You2015}, the joint utilization of MUD and PNC decoding was proposed to decode individual native packets and network-coded packets in each time slot. The decoded packets in all time slots are then exploited by a MAC-layer bridging and decoding scheme to recover users' packets.
%In \cite{Lu2013,You2015}, the joint utilization of MUD and PNC decoding was proposed to recover multiple packets in a time slot, and a \emph{message-level} SIC scheme was presented to recover users' packets.
%
Although the schemes in \cite{Cocco14} and \cite{Lu2013} provide excellent throughput performance, they suffer from a very high decoding complexity, which is less favorable for some applications which require simple machine-type devices.
Therefore, the low-complexity and efficient data detection scheme needs to be proposed for random access systems in mMTC.

\vspace{-5mm}
\section{Thesis Outline and Contributions}
Chapter 1 presents the motivation of this thesis. Chapter 2 provides an overview of some basic concepts that will be used extensively in this thesis. Chapters 3 - 5 present my novel research results on the random access for mMTC, which will be detailed in the following. In Chapter 6, the conclusion and future research topics are presented.

\subsection{Contributions of Chapter 3}
The work on the user activity identification and channel estimation for grant-free communications in mMTC is presented in Chapter 3.
As discussed in Section \ref{juice_background}, most works in the literature \cite{Schepker13,Xu15,Wunder15,Chen18,Liu18,Liu18ma,Senel17globecom} focused on the design of CS algorithms for the user activity identification and channel estimation at the receiver side.
In fact, the design of transmission schemes can adjust the sparsity of received signal vector and the strength of received signals to achieve an enhanced performance for mMTC.
%
%
%
%the CS techniques can be exploited to identify the active users and estimate their channels for mMTC, due to its sporadic transmission.
%For the CS-based user activity identification and channel estimation method, most works focus on designing the CS algorithms to improve the system performance.
%However, the design of transmission schemes is not considered in these works.
%In fact, both the sparsity of received signal vector and the strength of received signals can affect the performance of CS algorithms, and the two factors can be controlled by designing transmission schemes.
Therefore, this thesis studies how to design the transmission scheme to improve the performance of CS algorithms and the system performance.
In particular, we propose a transmission control scheme for the AMP based joint user identification and channel estimation in massive connectivity networks.
In the proposed transmission control scheme, a transmission control function is designed to determine a user's transmission probability, when it has a transmission demand.
By employing a step transmission control function for the proposed scheme, we derive the channel distribution experienced by the receiver to describe the effect of transmission control on the design of AMP algorithm.
Based on that, we modify the AMP algorithm by designing a minimum mean squared error (MMSE) denoiser, to jointly identify the user activity and estimate their channels.
%Based
%In order to design an minimum mean squared error (MMSE) denoiser for AMP, we derive the channel distribution experienced by the receiver .
%%
%Then, a JUICE algorithm is developed.
%
We further derive the false alarm and missed detection probabilities to characterize the user identification performance of the proposed scheme.
Closed-form expressions of the average packet delay and the network throughput are obtained.
Furthermore, we optimize the transmission control function to maximize the network throughput.
%
%It is verified that the analytical results match well with simulation results.
%
We demonstrate that the proposed scheme can significantly improve the user identification and channel estimation performance, reduce the packet delay, and boost the throughput, compared to the conventional scheme without transmission control.

These results have been published in one conference paper and one journal
paper.

\begin{itemize}
	\item \textbf{Z. Sun}, Z. Wei, L. Yang, J. Yuan, X. Cheng, W. Lei, ``Joint User Identification and Channel Estimation in Massive Connectivity with Transmission Control,'' in \textit{Proc. IEEE Intern. Symposium on Turbo Codes and Iterative Inform. Processing (ISTC)}, Hong Kong, Dec. 2018, pp. 1-5.
	\item \textbf{Z. Sun}, Z. Wei, L. Yang, J. Yuan, X. Cheng, W. Lei, ``Exploiting Transmission Control for Joint User Identification and Channel Estimation in Massive Connectivity,'' \textit{IEEE Trans. Commun.}, accepted, May 2019.
\end{itemize}
	
\subsection{Contributions of Chapter 4}
The work on the design of transmission schemes for the CSA system over erasure channels is presented in Chapter 4.
In the literature \cite{Casini2007, Liva2011, Paolini2011}, most of existing works focused on designing CSA schemes for the ideal collision channel without channel fading or noise, in order to improve the system throughput.
To capture the effects of practical channels, this thesis proposes a new transmission design for the CSA scheme over erasure channels.
In particular, we consider both packet erasure channels and slot erasure channels.
We first design the code probability distributions for CSA schemes with repetition codes and MDS codes to maximize the expected traffic load.
We then derive the extrinsic information transfer (EXIT) functions of CSA schemes over erasure channels. By optimizing the convergence behavior of derived EXIT functions, the code probability distributions to achieve the maximum expected traffic load are obtained. Then, we derive the asymptotic throughput of CSA schemes over erasure channels. In addition, we validate that the asymptotic throughput can give a good approximation to the throughput of CSA schemes over erasure channels.

These results have been published in one conference paper and one journal paper.

\begin{itemize}
	\item \textbf{Z. Sun}, Y. Xie, J. Yuan and T. Yang, ``Coded slotted ALOHA schemes for erasure channels,'' in \textit{Proc. IEEE Intern. Commun. Conf. (ICC)}, Kuala Lumpur, May 2016, pp. 1-6.
	\item \textbf{Z. Sun}, Y. Xie, J. Yuan and T. Yang, ``Coded Slotted ALOHA for Erasure Channels: Design and Throughput Analysis,'' in \textit{IEEE Trans. on Commun.}, vol. 65, no. 11, pp. 4817-4830, Nov. 2017.
\end{itemize}

\subsection{Contributions of Chapter 5}
The work on the design of data decoding algorithm for the CSA system is presented in Chapter 5.
As discussed in Section \ref{pnc_background}, the efficient data decoding algorithm is essential to enhance the system throughput of random access systems.
By exploring the characteristic of packet transmission in the CSA scheme, this thesis proposes an effective and low-complexity data decoding algorithm for the CSA.
In particular, we first propose an enhanced low-complexity binary PNC-based decoding scheme for random access systems with binary phase-shift keying (BPSK) modulation to improve the system throughput.
In the proposed scheme, the \textit{linear combinations} of users' packets in each time slot are first obtained by exploiting a low-complexity PNC decoding scheme.
Based on the decoded linear combinations within a MAC frame, we then propose an enhanced message-level SIC algorithm to recover more users' packets.
An analytical framework for the PNC-based decoding scheme is proposed and a tight approximation of the system throughput is derived for the proposed scheme.
Subsequently, we optimize the transmission schemes of CSA systems, i.e., the number of replicas transmitted by each user, to further improve the system throughput and energy efficiency, respectively.
Interestingly, the optimization results show that the optimal number of replicas for maximizing the energy efficiency is a constant for all offered loads.
On the other hand, the optimal number of replicas that maximizes the system throughput decreases as the offered load increases.
Numerical results show that the derived analytical results closely match with the simulation results.
Furthermore, the proposed scheme achieves a considerable throughput improvement, compared to the CRDSA scheme with more than two replicas.

These results have been published in one conference paper and one journal paper.

\begin{itemize}
	\item \textbf{Z. Sun}, L. Yang, J. Yuan and M. Chiani, ``A novel detection algorithm for random multiple access based on physical-layer network coding,'' in \textit{Proc. IEEE Intern. Commun. Conf. (ICC) Workshops}, Kuala Lumpur, May 2016, pp. 608-613.
	\item \textbf{Z. Sun}, L. Yang, J. Yuan and D. W. K. Ng, ``Physical-Layer Network Coding Based Decoding Scheme for Random Access,'' in \textit{IEEE Trans. Veh. Technol.}, vol. 68, no. 4, pp. 3550-3564, Apr. 2019.
\end{itemize}

\chapter{Background}\label{C2:chapter2}
This chapter presents some essential background knowledge required to understand materials presented in the subsequent chapters.
In particular, the first three sections present the basic knowledge for Chapter 3, Chapter 4, and Chapter 5, respectively, which include Bayes' theorem and compressed sensing techniques, modern coding techniques, and physical-layer network coding (PNC).
The forth section introduces some fundamentals of wireless communications.

\section{Bayes' Theorem and Compressed Sensing}
In this section, we present Bayes' theorem and an overview of classical compressed sensing algorithms, which provide the basis for understanding the proposed joint user identification and channel estimation scheme in Chapter 3.

\subsection{Bayes' Theorem}
Bayes' theorem provides a mathematical framework for performing inference and reasoning from a probability point of view \cite{bayes63}.
In particular, Bayes' theorem describes the posterior probability of an event, based on the prior knowledge that is related to the event \cite{Gray04}.
%updates the distribution over parameters from the prior to the posterior distribution in light of an observed event.
Mathematically, Bayes' theorem can be stated as
\begin{align}
p(A|B)=\frac{p(B|A)p(A)}{p(B)},
\label{bayes_theory}
\end{align}
where $A$ and $B$ are events and $p(B) \neq 0$.
The term $p(A|B)$ is the likelihood of event $A$ occurring given that $B$ is true which is called the posterior probability of event $A$.
The term $p(B|A)$ is the likelihood of event $B$ occurring given that $A$ is true, and $p(A)$ and $p(B)$ are the probabilities of observing $A$ and $B$ independently of each other, respectively. Here, $p(A)$ usually refers to the prior probability of event $A$, which reflects the original knowledge of $A$ before having any knowledge of $B$, and $p(B)$ can be regarded as a normalizing parameter, due to its independence of $A$.
The four basic terms constitute the Bayes' theorem, which will be exploited for the design of CS algorithms.
%
%
%In theory, the posterior distribution can capture all information inferred from the data about the parameters.
%For Bayes' theorem \cite{Gray04}, the relation from the prior distribution to the posterior distribution is given by
%\begin{align}
%p(x|y)=\frac{p(y|x)p(x)}{p(y)}.
%\label{bayes_theory}
%\end{align}
%In Eq. \eqref{bayes_theory}, the term $p(y|x)$ is called the likelihood function and it captures the stochastic mapping from the parameter $x$ to the observed data $y$.
%The likelihood function $p(y|x)$ is usually known, since it expresses one's knowledge of how one expects the data to look given the parameter.
%The term $p(x)$ is called the prior and it reflects the original knowledge of $x$ before observing the data.
%The term $p(y)$ is the evidence or marginal likelihood.
%It is obtained by integrating $p(y|x)p(x)$ over all $x$, and usually plays the role of an ignorable normalizing constant.
%Finally, the term $p(x|y)$ is the posterior distribution and it represents the updated knowledge of $x$ after the data is considered.
%The four basic blocks that are mentioned above constitute the Bayesian model, which will be exploited for the design of CS algorithm.

\subsection{Compressed Sensing Algorithms}
Compressed sensing (CS, also known as compressive sensing or compressive sampling), is an extensively developed signal processing technique for efficiently reconstructing a signal from under-sampled measurements \cite{Eldar12,Donoho06,Donoho09}.
For the conventional signal reconstruction, Nyquist sampling theorem provides a lower bound of the sampling rate to completely recover a signal \cite{Shannon98}.
On the other hand, in large amount of practical applications, e.g. the imaging and video processing, the reconstructed signals are sparse and the CS is able to provide accurate recovery of high-dimensional signals from a much smaller number of sampling measurements.
It implies that exploiting the sparsity of signals can indeed reduce the number of sampling measurements and ensure an exact recovery of a signal.
In view of this, the CS algorithms, in particular efficient sparse signal reconstruction algorithms, have drawn much attention from the academic society \cite{Donoho06,Aeron10}.

The sparse signal reconstruction can be expressed as the recovery of a sparse signal $\mathbf{x} \in \mathbb{C}^{n \times 1}$ from $m$ linear combinations of its elements, given by
\begin{align}
\mathbf{y}=\mathbf{Ax},
\label{cs_model}
\end{align}
where $\mathbf{y} \in \mathbb{C}^{m \times 1}$ is the measurement vector, $m \leq n$, and $\mathbf{A} \in \mathbb{C}^{m \times n}$ is the measurement matrix.
When the signal vector $\mathbf{x}$ has $s$ or fewer non-zero elements such that $s \ll n$, the vector $\mathbf{x}$ is said to be $s$-sparse.

One of the theoretically best approaches to recover such a signal vector $\mathbf{x}$ from the measurement vector $\mathbf{y}$ is to solve the $\ell_{0}$-minimization problem \cite{Ge2011}:
\begin{align}
\min_{\mathbf{x}}\lVert\mathbf{x}\rVert_{0} \hspace{3mm} \text{subject to} \hspace{3mm} \mathbf{Ax}=\mathbf{y}.
\label{norm_0}
\end{align}
Here, $\lVert\cdot\rVert_{0}$ is the $\ell_{0}$-pseudo norm of a vector, which counts the number of non-zero elements in the input vector.
%For $1 \leq p \leq \infty$, we denote $\lVert\cdot\rVert_{p}$ as the usual $p$-norm, given by
%\begin{align}
%\lVert\mathbf{x}\rVert_{p}=\left(\sum_{i=1}^{n}|x_{i}|^{p}\right)^{\frac{1}{p}},
%\label{p_norm}
%\end{align}
%and $\lVert\mathbf{x}\rVert_{\infty}=\max|x_{i}|$.
The sufficient and necessary condition of existing a unique solution in Eq. \eqref{norm_0} is that the measurement matrix $\mathbf{A}$ has a rank larger than $2s$ \cite{Needell09}.
A simple proof is as follows.
Assume that $\mathbf{x_{1}}$ and $\mathbf{x_{2}}$ are both solutions of Eq. \eqref{norm_0}.
Due to $\mathbf{Ax_{1}}=\mathbf{y}$ and $\mathbf{Ax_{2}}=\mathbf{y}$, we have $\mathbf{A(x_{1}-x_{2})}=\mathbf{0}$.
Since $\mathbf{x_{1}-x_{2}}$ is $2s$-sparse and $\mathbf{A}$ has the rank larger than $2s$, we can obtain $\mathbf{x_{1}=x_{2}}$.
In fact, the $\ell_{0}$-minimization problem is generally intractable.
In particular, this problem has been proved as a NP-complete problem \cite{Muthukrishnan03} and finding its optimal solution relies on the combinatorial search.
%
%Thus, the $\ell_{0}$-minimization problem can work perfectly in theory.
%However, since the $\ell_{0}$-norm is non-convex and only relies on the combinatorial search to obtain the solution, this problem in Eq. \eqref{norm_0} is computationally known as an NP-complete problem \cite{Muthukrishnan03}.

Fortunately, several numerically feasible suboptimal alternatives \cite{Donoho06, Candes08, Chen01, Tropp07,Wipf04,Bayati11, Donoho09} to this NP-complete problem have been developed as the pioneering work on CS algorithms in the past few years.
Among them, the cornerstone algorithms include $\ell_{1}$-norm minimization (also called Basis Pursuit (BP) algorithm) \cite{Donoho06, Candes08, Chen01}, greedy algorithm \cite{Tropp07}, statistical sparse recovery technique \cite{Wipf04}, and iterative algorithm \cite{Bayati11, Donoho09}.

Starting from the $\ell_{1}$-norm minimization, we will briefly introduce all the cornerstone algorithms in the following.
Note that, the presentations of all algorithms are based on the model in Eq. \eqref{cs_model}.

\subsubsection{$\ell_{1}$-norm Minimization:}
The $\ell_{1}$-norm minimization (BP) was proposed by \cite{Donoho06, Candes08, Chen01}, which relaxes the reconstruction of $\ell_{0}$-minimization problem.
In particular, the non-convex $\ell_{0}$-norm is replaced by a convex $\ell_{1}$-norm and the problem can be reformulated as
\begin{align}
\min_{\mathbf{x}}\lVert\mathbf{x}\rVert_{1} \hspace{3mm} \text{subject to} \hspace{3mm} \mathbf{Ax}=\mathbf{y}.
\label{norm_1}
\end{align}
By using the standard linear programming \cite{Donoho06} to solve Eq. \eqref{norm_1}, the signal $\mathbf{x}$ can be successfully reconstructed \cite{Donoho06,Candes08}.
In addition, when the measurement is corrupted by the noise, the system model is written as
\begin{align}
\mathbf{y}=\mathbf{Ax}+\mathbf{w}.
\label{cs_noise_model}
\end{align}
where $\mathbf{w} \in \mathbb{R}^{m \times 1}$ is the additive measurement noise.
Then, the $\ell_{1}$-minimization problem can be formulated as
\begin{align}
\min_{\mathbf{x}}\lVert\mathbf{x}\rVert_{1} \hspace{3mm} \text{subject to} \hspace{3mm} \lVert\mathbf{Ax}-\mathbf{y}\rVert_{2}^{2} \leq \epsilon,
\label{norm_1_noise}
\end{align}
where $\epsilon$ is a pre-determined noise level of the system.
This type of problem, called basis pursuit denoising (BPDN), has been well studied in the convex optimization field \cite{Chen01,Candes06} and can be solved by many effective approaches, e.g. the interior-point method.
When the noise information is not known, the Lagrangian unconstrained form is exploited to obtain an alternative problem formulation as \cite{Tibshirani96}
\begin{align}
\min_{\mathbf{x}}\lVert\mathbf{Ax}-\mathbf{y}\rVert_{2}^{2}+\lambda\lVert\mathbf{x}\rVert_{1}.
\label{lasso_noise}
\end{align}
This is known as the least absolute shrinkage and selection operator (LASSO) problem \cite{Tibshirani96}.
The fixed regularization parameter $\lambda > 0$ is used to control the sparsity level of the solution, via tuning the weight between the least squared error term and the sparsity term.
Since the solution of Eq. \eqref{lasso_noise} is sensitive to the value of $\lambda$, the least angle regression stage wise (LARS) algorithm \cite{Drori06} is used to simultaneously optimize the parameter $\lambda$ and find the solution of Eq. \eqref{lasso_noise}.

\subsubsection{Greedy Algorithm:}
While the $\ell_{1}$-minimization (BP) algorithm can effectively reconstruct the sparse signal vector via the linear programming technique, it requires substantial computational cost, in particular for large-scale applications.
For example, the solver based on the interior point method has an associated computational complexity order of $O(n^{3})$ \cite{Donoho06}, where $n$ is the signal dimension.
Such a computational cost is burdensome for some real-time systems, e.g. wireless communication systems.

Faced with this, the greedy algorithm was proposed and drew much attention, due to its lower computational overhead than the BP algorithm \cite{Blanchard15}.
In the greedy algorithm, the subset of signal support, i.e., the index set of non-zero entries, is iteratively updated until a good estimation is obtained.
When the support is accurately estimated, the underdetermined system can be converted into the overdetermined one by removing columns of measurement matrix corresponding to zero elements. Then, the elements of support can be estimated by using conventional estimation techniques, e.g. least squares (LS) estimator.
%
%by finding a local optimal selection and reconstructs it by obtaining the global optimum solution in the end, which provides a lower computational complexity than the BP algorithm.
The most popular greedy algorithm is the orthogonal matching pursuit (OMP) \cite{Tropp07}.
It iteratively updates the estimate of signal support by choosing the column of measurement matrix that has the largest correlation with the residual.
Consider the model $\mathbf{y}=\mathbf{Ax}$, where each column of $\mathbf{A}$ is normalized and the sparsity of signal $\mathbf{x}$, i.e., $s$, is known.
The OMP algorithm starts from the initial estimation $\mathbf{\hat{x}}^{0}=\mathbf{0}$ and the residual $\mathbf{r}^{0}=\mathbf{y}$.
The support set of the initial estimate is $\Lambda^{0}=\varnothing$.
In the $t$-th iteration, for $\mathbf{A}$ the column that has largest correlation with the residual $\mathbf{r}^{t-1}$ is chosen, i.e.,
\begin{align}
i=\arg \max_{i} |\mathbf{A }^{T}_{:,i}\mathbf{r}^{t-1}|,
\label{omp_1}
\end{align}
and its index is added into the support set $\Lambda^{t}=\Lambda^{t-1}\cup\{i\}$.
Then, the estimate and the residual of this iteration are updated via
\begin{align}
\mathbf{\hat{x}}^{t}=\mathbf{A}_{:,\Lambda^{t}}^{\ast}\mathbf{y},\\
\mathbf{r}^{t}=\mathbf{y}-\mathbf{A}\mathbf{\hat{x}}^{t}.
\label{omp_2}
\end{align}
The iteration continues until the size of estimated support set $\Lambda^{t}$ reaches $s$.
Based on the OMP algorithm, where only one column is selected in each iteration, many variants of OMP are proposed, e.g. generalized OMP (gOMP) \cite{Wang12}, compressive sampling matching pursuit (CoSaMP) \cite{Needell09}, subspace pursuit (SP) \cite{Dai09}, and multipath matching pursuit (MMP) \cite{Kwon14}.
For these variants, multiple promising columns are selected in each iteration and the support set is then refined by adding the indices of selected columns, which can outperform the OMP algorithm at the cost of a higher computational complexity.

\subsubsection{Statistical Sparse Recovery:}
For the model in Eq. \eqref{cs_model}, the signal vector $\mathbf{x}$ can be treated as a random vector and inferred by using the Bayesian framework in statistical sparse recovery algorithms.
For example, in the maximum-a-posterior (MAP) approach, an estimate of $\mathbf{x}$ can be expressed as
\begin{align}
\mathbf{\hat{x}}=\arg \max_{\mathbf{x}} \ln f(\mathbf{x}|\mathbf{y})=\arg \max_{\mathbf{x}} \ln f(\mathbf{y}|\mathbf{x})+\ln f(\mathbf{x}),
\label{map}
\end{align}
where $f(\mathbf{x})$ is the prior distribution of signal $\mathbf{x}$.
In order to model the sparsity of the signal vector $\mathbf{x}$, $f(\mathbf{x})$ is designed in such a way that it decreases with increasing the magnitude of $\mathbf{x}$.
Well-known examples include independent and identically distributed (i.i.d.) Gaussian and Laplacian distribution.
In addition, the other widely used statistical sparse recovery algorithm is sparse Bayesian learning (SBL) \cite{Ji08}.
In the SBL, the prior distribution of signal vector $\mathbf{x}$ is modeled as zero-mean Gaussian with the variance parameterized by a hyper-parameter. Then, the hyper-parameter and the signal vector are estimated simultaneously.
It is noteworthy that the hyper-parameter can control the sparsity and the distribution of signal vector $\mathbf{x}$.
With approximately choosing the hyper-parameter, the SBL algorithm can outperform the $\ell_{1}$-minimization algorithm \cite{Wipf04}.

\subsubsection{Iterative Thresholding Algorithm:}
For iterative thresholding algorithms, the signal vector is estimated in an iterative way, which particularly include iterative hard thresholding (IHT) algorithm \cite{Blumensath14}, iterative soft thresholding (IST) algorithm \cite{Maleki10}, and approximate message passing (AMP) algorithm \cite{Bayati11}.
Based on the model $\mathbf{y}=\mathbf{Ax}$, the three algorithms will be briefly presented in the next.
For the IHT, it can be expressed by
\begin{align}
\mathbf{\hat{x}}^{t+1}=\eta^{H}(\mathbf{\hat{x}}^{t}+\mathbf{A}^{\ast}(\mathbf{y}-\mathbf{A}\mathbf{\hat{x}}^{t})),
\end{align}
where $\eta^{H}(\cdot)$, called the hard thresholding function, is a non-linear operator that sets all but the largest (in magnitude) $s$ elements of input vector to zero and $s$ is the known sparsity of estimated signal vector $\mathbf{x}$. In the $t$-th iteration, the estimate of $\mathbf{x}$ is denoted as $\mathbf{\hat{x}}^{t}$.
%
%We consider that a hard threshold function $\eta^{H}(x,\mu)$ is applied to vectors in an elementwise way, given by $\eta^{H}(x,\mu)=x\mathbf{1}(|x|>\mu)$, where $\mathbf{1}$ is an indicator function.
%The IHT algorithm for the model $\mathbf{y}=\mathbf{Ax}$ is defined by the following iteration
%\begin{align}
%\mathbf{x}^{t+1}=\eta^{H}(\mathbf{x}^{t}+\mathbf{A}^{\ast}(\mathbf{y}-\mathbf{Ax^{t}}),\lambda^{t}),
%\end{align}
%where $\lambda^{t}$ is the threshold value depending on the iteration $t$ and $\mathbf{x}^{t}$ is the estimate of sparse signal vector $\mathbf{x}$ at the $t$-th iteration.
Intuitively, the algorithm makes progress by moving in the direction of the gradient of $\lVert\mathbf{y}-\mathbf{A\hat{x}^{t}}\rVert_{2}$ and then promotes sparsity by applying the hard thresholding function $\eta^{H}(\cdot)$ \cite{Blumensath14}.

The IST algorithm is another iterative thresholding algorithm, which uses a soft thresholding function instead of a hard thresholding function.
Similarly to the hard thresholding function, the soft thresholding function has input and output of vectors and it operates in an element-wise way.
Then, the soft thresholding function associated with the $i$-th element of input vector $\mathbf{a}$ is given by
\begin{align}
\eta^{S}_{i}(a_{i},\mu)=\text{sign}(a_{i})(|a_{i}|-\mu)_{+},
\label{ist}
\end{align}
where $a_{i}$ is the $i$-th element of input vector $\mathbf{a}$, $\mu$ is a threshold control parameter, and $(|a_{i}|-\mu)_{+}$ equals itself if $|a_{i}|>\mu$ and equals zero otherwise.
Based on the soft thresholding function, the IST algorithm proceeds with the iteration
\begin{align}
\mathbf{\hat{x}}^{t+1}=\eta^{S}(\mathbf{\hat{x}}^{t}+\mathbf{A}^{\ast}(\mathbf{y}-\mathbf{A}\mathbf{\hat{x}}^{t}),\mu).
\label{ist_algo}
\end{align}
Since the soft thresholding function is proved to be the proximity operator of the $\ell_{1}$-norm \cite{Donoho10}, the IST algorithm with a determined threshold control parameter can be equivalent to the $\ell_{1}$-minimization problem.
For the family of iterative thresholding algorithms, including the IHT algorithm and the IST algorithm, as only the multiplication of a vector by a measurement matrix is required in each iteration, the computational complexity is very small and the storage requirement is low, particularly compared to the $\ell_{1}$-minimization algorithm and the greedy algorithm.
Then, the iterative thresholding algorithms are efficient for large-scale systems.
However, these algorithms fall short of the sparsity-undersampling tradeoff, compared to that from the $\ell_{1}$-minimization algorithm.
Therefore, from the theory of belief propagation in graphical models, the AMP algorithm is proposed to achieve a satisfactory sparsity-undersampling tradeoff that can match the theoretical tradeoff for $\ell_{1}$-minimization algorithm \cite{Donoho10itw}.
Besides, the AMP algorithm poses a much lower computational cost than the $\ell_{1}$-minimization algorithm.
As the AMP algorithm acts as a building block for the work in Chapter 3, the algorithm and its analysis will be briefly presented in the next part.

\vspace{-0.5cm}
\subsection{Approximate Message Passing Algorithm}\label{amp}
The AMP algorithm was first proposed in \cite{Donoho09} and further developed by exploiting the distribution of unknown vector $\mathbf{x}$ as a prior information in \cite{Donoho10,Chen17}.
The gist of AMP algorithm is that it exploits an iterative refining process to recover the sparse unknown vector via using the Gaussian approximation during message passing.
It enjoys a dramatically low computational complexity while achieving the identical performance with linear programming in terms of the sparsity-undersampling tradeoff \cite{Donoho09}.
Based on the system model $\mathbf{y}=\mathbf{Ax}$ with $\mathbf{y} \in \mathbb{C}^{m \times 1}$, $\mathbf{A} \in \mathbb{C}^{m\times n}$, and $\mathbf{x} \in \mathbb{C}^{n \times 1}$, the AMP algorithm proceeds with
\begin{align}
{\mathbf{\hat{x}}^{t + 1}} &= \eta \left({{\mathbf{A}^*}}{{\mathbf{z}}^t} + {\mathbf{\hat{x}}^t}\right),
\label{amp_1}\\
{{\bf{z}}^{t + 1}}&= {\bf{y}} - {\mathbf{A}}{{\bf{\hat{x}}}^{t + 1}} + \frac{n}{m}{{\bf{z}}^{t}}\langle{{{\eta ^{'}}\left({\mathbf{A}^*}{{\bf{z}}^t} + {\mathbf{\hat{x}}^t}\right)}}\rangle,
\label{amp_2}
\end{align}
where $t=0,1,\ldots$ is the index of iteration,  vector $\mathbf{\hat{x}}^{t}=[\hat{x}^{t}_{1},\hat{x}^{t}_{2},\ldots,\hat{x}^{t}_{n}]^{T} \in \mathbb{C}^{n \times 1}$ is the estimate of $\mathbf{x}$ in the $t$-th iteration, $\mathbf{z}^{t} \in \mathbb{C}^{m \times 1}$ is the residual of received signal corresponding to the estimate $\mathbf{\hat{x}}^{t}$, $\eta(\cdot)=[\eta_{1}(\cdot),\ldots,\eta_{n}(\cdot)]$ with $\eta_{i}(\cdot)$ being a designed denoiser function for the $i$-th element of input vector, $\eta'(\cdot)$ is the first-order derivative of $\eta(\cdot)$, and $\langle\cdot\rangle$ is the average of all entries of the input vector.
%
%
%
%is the index of iteration, $\mathbf{\hat{x}}^{t}$ is the estimate of $x_{n}$ in the $t$-th iteration, vector $\hat{\mathbf{x}}^{t+1}=[\hat{x}^{t+1}_{1},\hat{x}^{t+1}_{2},\ldots,\hat{x}^{t+1}_{N}]^{T} \in \mathbb{C}^{N \times 1}$ collects all the estimates $\hat{x}^{t+1}_{n}$, $\mathbf{z}^{t+1} \in \mathbb{C}^{M \times 1}$ is the residual of received signal corresponding to the estimate $\hat{\mathbf{x}}^{t+1}$, $\varpi=\frac{M}{N}$, $\beta_{n}$ is the parameter of prior distribution of $x_{n}$, $\eta(\cdot)$ is a designed denoiser function, and $\eta'(\cdot)$ is the first-order derivative of $\eta(\cdot)$ with respect to (w.r.t.) its first argument.
%
Note that, the third term in the right hand side of Eq. \eqref{amp_2} is the correction term, which is known as the ``Onsager term'' from the statistical physics \cite{Onsager53}.
From the point $\mathbf{\hat{x}}^{0}=\mathbf{0}$ and $\mathbf{z}^{0}=\mathbf{y}$, the AMP algorithm starts to proceed.
%Note that, the starting point of this iterative process is $\mathbf{\hat{x}}^{0}=\mathbf{0}$ and $\mathbf{z}^{0}=\mathbf{y}$.
%
It can be seen from Eq. \eqref{amp_1} that by exploiting the measurement matrix $\mathbf{A}$, a matched filter is first performed on the residual $\mathbf{z}^{t}$ to obtain the variable $\mathbf{\tilde{x}}^{t}={{\mathbf{A}^*}}{{\mathbf{z}}^t} + \mathbf{\hat{x}}^t$.
The denoiser function receives the vector $\mathbf{\tilde{x}}^{t}$ as the input and outputs the estimate $\mathbf{\hat{x}}^{t+1}$ in the $(t+1)$-th iteration.
Here, the denoiser function input $\mathbf{\tilde{x}}^{t}$ can be modeled \cite{Donoho10} as
\begin{align}
\mathbf{\tilde{x}}^{t}=\mathbf{x}+\tau_{t}\mathbf{v},
\label{GaussianAppro}
\end{align}
where each entry of $\mathbf{v}$ follows the standard Gaussian distribution due to the correction term, and $\tau_{t}$ denotes a state variable that will be analyzed in the following.
Eq. \eqref{amp_2} is used to compute the residual corresponding to the estimate $\hat{\mathbf{x}}^{t+1}$ in the $(t+1)$-th iteration, and then the AMP algorithm proceeds to the next iteration.

For the iterative process in the AMP algorithm, a state variable, denoted by $\tau_t$, $t=0,1,\ldots$, and its evolution are introduced to characterize the performance of AMP in each iteration \cite{Donoho10itw}.
In particular, for a large-scale system, where the measurement length $m$, the length of estimated signal vector $n$, and the sparsity of estimated signal vector $s$ are infinite but with fixed ratios $\frac{m}{n}$ and $\frac{s}{n}$, the state evolution is given by \cite{Donoho10itw}
\begin{align}
\tau_{t+1}^{2}&=\frac{1}{m}E\left[\lVert\mathbf{\hat{x}}^{t+1}-\mathbf{x}\rVert_{2}^{2}\right]=\frac{1}{m}E\left[\lVert\eta \left({{A^*}}{{\mathbf{z}}^t} + \mathbf{\hat{x}}^t\right)-\mathbf{x}\rVert_{2}^{2}\right] \notag \\
&\overset{(a)}{=}\frac{1}{m}E\left[\lVert\eta\left(\mathbf{x}+\tau_{t}\mathbf{v}\right)-\mathbf{x}\rVert_{2}^{2}\right],
\label{state_evolo}
\end{align}
where $\tau_{t+1}$ is the state variable in the $(t+1)$-th iteration.
The equality $(a)$ in Eq. \eqref{state_evolo} is obtained from the Gaussian approximation in Eq. \eqref{GaussianAppro} in the $t$-th iteration.
The expectation in Eq. \eqref{state_evolo} is taken over the random vectors $\mathbf{x}$ and $\mathbf{v}$.
%
%Note that, the expectation should also be taken over $\beta$, when users have different large-scale fading $\beta$ \cite{Donoho10itw,Chen17}.
%
The state evolution begins with \cite{Donoho10itw}
\begin{align}
\tau_{0}^{2}=\frac{1}{m}E\left[\lVert\mathbf{x}\rVert_{2}^{2}\right].
\label{state_evolo_initial}
\end{align}
It can be observed from Eq. \eqref{state_evolo} that the squared state variable $\tau^2_{t+1}$ characterizes the mean squared error (MSE) of each entry of the estimate $\mathbf{\hat{x}}^{t+1}$ in the $(t+1)$-th iteration.
It implies that from iteration to iteration, the evolution of the performance of AMP algorithm in terms of MSE, can be tracked by exploiting the state variable $\tau_{t}$ \cite{Donoho09}.
%
%As the AMP algorithm proceeds, the state variable $\tau_{t}$ will decrease and converge to a fixed value, denoted by $\tau_{\infty}$, where the convergence is guaranteed for the system with the sparsity level below a sparsity threshold \cite{Donoho09}.
%%
%Therefore, the evolution of $\tau_{t}$ can predict the dynamic behavior of the estimation performance in terms of MSE, with the AMP iterations evolving.
%
Note that, the state variable $\tau_{t}$ is also involved in the AMP algorithm through the Gaussian approximation of $\mathbf{\tilde{x}}^{t}$ in Eq. \eqref{GaussianAppro}.
However, obtaining $\tau_{t}$ via the state evolution in Eq. \eqref{state_evolo} requires a high computational complexity.
Therefore, in the literature \cite{Montanari12}, an empirical estimate of $\tau_{t}$, i.e.,
\begin{align}
\tau_{t}=\frac{1}{\sqrt{m}}||\mathbf{z}^{t}||_{2},
\label{EmpiricalState}
\end{align}
is usually adopted during the implementation of AMP algorithm.
\section{Modern Coding Techniques}
Some modern coding techniques are presented in this section, which will provide the theoretical tool for our work in Chapter \ref{C4:chapter4}.
\subsection{Distance Metrics and Channel Codes}
During the data transmission, original transmitted signals are likely to be corrupted by the channel and the noise at the receiver.
This results in the received signal with errors, which affects the reliability of reconstructing the original data from the received signal.
In order to deal with this problem, error control coding (ECC) is developed \cite{Ryan09}.
In the ECC, some redundant bits are added to the transmitted data, so that the received errors can be corrected and the original data can be retrieved.
Using an ECC can help achieve the same bit error rate (BER) at a lower signal-to-noise ratio (SNR) in a coded system than in a comparable uncoded system \cite{Burr01}.
The reduction of the required SNR to achieve the same BER is called the coding gain.
For an ECC in digital systems with hard-decision decoding, its error detection and correction capacity can be determined by the Hamming distance of this ECC.
Other the hand, for digital systems with soft-decision decoding, the error performance of an ECC is guided by its Euclidean distance.
Therefore, we first introduce two commonly used distance metrics in the coding theory and the error detection and correction capacity of codes in the following.
% exploits many key techniques of codes on the graph, which are widely used for LDPC codes, and

\subsubsection{Distance Metrics:}
There are two important distance metrics widely used for the channel coding \cite{Ryan09}.
One is the Hamming distance, which is defined as the number of different bits between two codewords.
The other is Euclidean distance, which refers to the straight-line distance between two points in Euclidean space.
For a linear block code, the Hamming distance can indicate its error detection and correction capability, particularly when the source emits binary strings over a binary channel.
On the other hand, if the source emits the codewords in $R^{n}$ over a Gaussian channel and the soft decoder is exploited at the receiver, Euclidean distance determines the code error performance.
The minimum Hamming (or Euclidean) distance of a set $\mathbf{C}$ of codes is given by
\begin{align}
d_\text{min}=\min_{\mathbf{v},\mathbf{v}' \in \mathbf{C}: \mathbf{v}\neq \mathbf{v}'}d(\mathbf{v},\mathbf{v}'),
\label{min_dis}
\end{align}
where $d(\mathbf{v},\mathbf{v}')$ denotes the Hamming (or Euclidean) distance of the two codewords $\mathbf{v}, \mathbf{v}' \in \mathbf{C}$.
For a code set $\mathbf{C}$ with the dimension $k$ and the length $n$, its minimum Hamming distance should satisfy
\begin{align}
d_\text{min} \leq n-k+1,
\label{singleton}
\end{align}
which is the Singleton bound \cite{Richardson08}.

\subsubsection{Error Detection and Correction:}
In the coding theory, the error detection and correction are a key enabler for the reliable delivery of digital data over unreliable communication channels, where communication channels are subject to channel noise and errors can be introduced during the transmission \cite{Ryan09}.
In particular, the error detection technique allows detecting errors, and the error correction enables the reconstruction of original data.
For practical communication systems, the ECC is an efficient way to perform the error detection and correction.
%In practice, some redundant bits are added into the information message to construct the ECC, so that some errors can be detected and corrected.
For an ECC in digital systems with hard-decision decoding, its error detection and correction capability is determined by its minimum Hamming distance.
In particular, the ECC with the minimum Hamming distance $d_\text{min}$, it can detect up to $(d_\text{min}-1)$ error bits and correct up to $\lfloor\frac{d_\text{min}-1}{2}\rfloor$ error bits \cite{Ryan09}.

With the knowledge on distance metrics and the error detection and correction capacity of codes, we briefly introduce some widely used channel codes.
Generally, there are two structurally different types of channel codes for the error control in communication and storage systems, i.e., block codes and convolutional codes.
Block codes can be further divided into two categories, i.e., linear and nonlinear block codes.
Nonlinear block codes are not widely used in practical applications and have not been widely investigated \cite{Ryan09}.
Therefore, we mainly focus on the linear block codes here.

\subsubsection{Linear Block Codes:}
A code is linear if the sum of any two codewords, i.e., $\mathbf{v}+\mathbf{v}'$ for $\mathbf{v}, \mathbf{v}' \in \mathbf{C}$, is still a codeword in the code set $\mathbf{C}$ \cite{MacKay03}.
We assume that an information source is a sequence of binary symbols over Galois field of two, i.e., GF(2).
%The binary symbols of this information sequence are information bits, where a bit refers to a binary digit.
In a block coding system, the information sequence is segmented into message blocks of $k$ information bits and there are $2^{k}$ distinct messages.
For the channel encoder, each input message $\mathbf{u}=(u_{0},u_{1},\ldots,u_{k-1})$ of $k$ information bits is encoded into a longer sequence $\mathbf{v}=(v_{0},v_{1},\ldots,v_{n-1})$ of $n$ binary digits according to certain encoding rules, where $k$ and $n$ are called the dimension and length of a codeword, respectively, and they satisfy $n>k$.
The binary sequence $\mathbf{v}$ is called the codeword of the message $\mathbf{u}$.
The classical linear encoding rule can be expressed as \cite{MacKay03}
\begin{align}
\mathbf{v}=\mathbf{u}\mathbf{G},
\label{channel_ec}
\end{align}
where the binary matrix $\mathbf{G}$ is called the generator matrix of dimension $k \times n$. A column of $\mathbf{G}$ corresponds to an encoded bit of a codeword and a row corresponds to an information bit of the message. If the message $\mathbf{u}$ is contained into the codeword $\mathbf{v}$ in an unaltered way, the encoding mapping is called systematic.
Since $2^{k}$ distinct information messages exist, there are $2^{k}$ distinct codewords accordingly.
This set of $2^{k}$ codewords is said to form an $(n,k)$ block code set, and each codeword $\mathbf{v}$ satisfies \cite{Ryan09}
\begin{align}
\mathbf{v}\mathbf{H}^{T}=\mathbf{0},
\label{channel_pc}
\end{align}
where the matrix $\mathbf{H}$ is called the parity-check matrix. The columns of $\mathbf{H}$ correspond to the bits of a codeword and the rows correspond to the parity check equations fulfilled by a valid codeword.
It implies that if a codeword is valid in the code set, it should satisfy Eq. \eqref{channel_pc}.
The code rate is defined as $R = \frac{k}{n}$, which can be interpreted as the average number of information bits carried by each code bit.
%For a block code to be useful, the $2^{k}$ codewords for the $2^{k}$ distinct messages must be distinct.
For an $(n,k)$ block code, the $(n-k)$ bits added to each input message by the channel encoder are called redundant bits.
These redundant bits carry no new information and their main function is to provide the code with the capability of detecting and correcting transmission errors caused by the channel noise or interferences.

Classical linear block codes include repetition codes and maximum distance separable (MDS) codes \cite{Singleton64,Liu18}.
The repetition code is one of the most basic linear block codes, which repeats the message several times.
If the channel corrupts some repetitions, the receiver can detect the occurrence of transmission errors, according to the difference of received messages.
Moreover, the receiver can recover the original message by choosing the received one that occurs most often.
The implementation of a repetition code is extremely simple, while it has a very low code rate.
As a result, the repetition code can be concatenated into other codes, e.g. repeat accumulate (RA) codes \cite{Qiu16,Qiu18j,Qiu17} and turbo like codes \cite{Richardson08}, to achieve an excellent error correction performance.

The MDS code is a kind of linear block codes which meet the Singleton bound.
Since the error detecting and correcting capability is determined by the minimum Hamming distance, it can be seen that given the code dimension $k$ and code length $n$, the MDS code has the largest minimum Hamming distance and thus the largest error detecting and correcting capacity.
According to the Singleton bound in Eq. \eqref{singleton}, the $(n,k)$ MDS code has the minimum Hamming distance $d$ that is equal to $(n-k+1)$.
Then, it can detect up to $(n-k)$ error bits.
It implies that as long as any $k$ bits in an MDS codeword are correctly received, this codeword can be successfully decoded.

\subsubsection{Convolutional Codes:}
\begin{figure}[!t]
	\par
	\begin{center}
		{\includegraphics[width=3.5in]{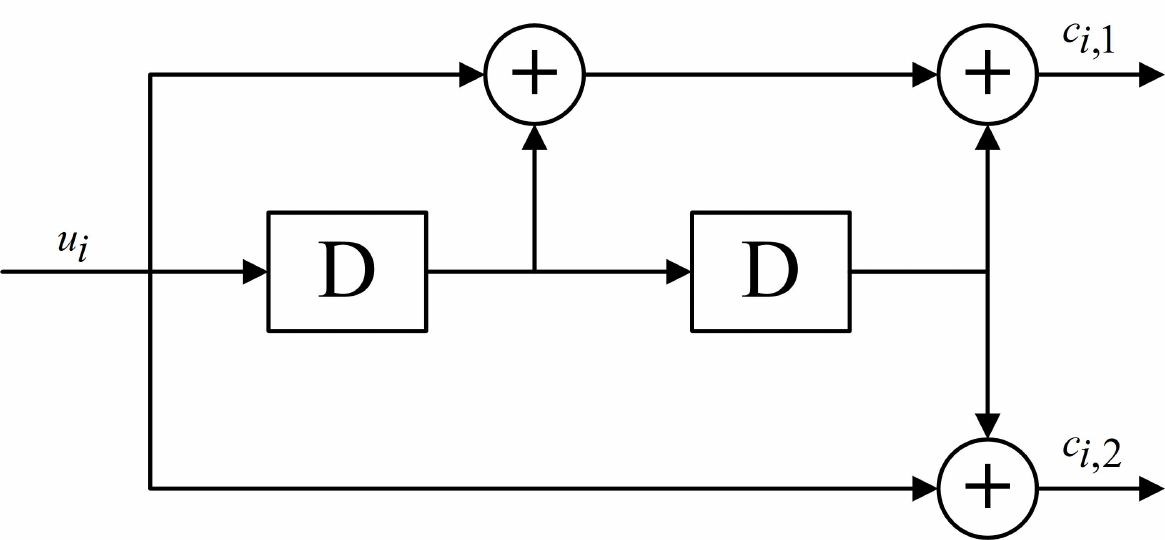}}
	\end{center}\vspace{-5mm}
	\caption{Convolutional encoder with rate $1/2$.}
	\label{fig_cc}
\end{figure}

Convolutional codes, introduced by \cite{Gray11}, refer to codes in which the encoder maps streams of data into more streams of data.
These codes are highly structured to allow a simple implementation and a good performance with the short block length.
The encoding is realized by sending the input streams over linear filters.
An example of a convolutional code with rate $1/2$ is shown in Fig. \ref{fig_cc}.
The information bits are fed into the linear encoder circuit and this circuit outputs the corresponding codeword.
This code construction sets additional constraints on the characteristics of the corresponding matrices $\mathbf{G}$ and $\mathbf{H}$.
Note that, the filtering operation can be expressed as a convolution, which leads to the name, i.e., convolutional codes.
The most popular decoding algorithm for convolutional codes is the Viterbi algorithm \cite{Viterbi67}, which is an efficient implementation of the optimal maximum likelihood decoder.
The gist of the Viterbi algorithm is the sequential computation of the metric and the tracking of survivor paths in the code trellis.
This algorithm was extended in \cite{Hagenauer89} for generating soft-outputs, called soft-output Viterbi algorithm (SOVA) algorithm.
Alternatively, one can also use the Bahl-Cocke-Jelinek-Raviv (BCJR) algorithm as proposed in \cite{Bahl74} to generate the soft-outputs for the iterative decoding.

Based on these classical channel codes, several modern codes have been developed over the past years, which include the polar codes \cite{Arikan09, Wu18, Wu18acc}, turbo codes \cite{Berrou93,Yuan99,Vucetic00}, and low-density parity check (LDPC) codes \cite{Gallager63,Xie16,Kang18}.
Polar codes are a class of linear block codes, whose encoding construction is based on a multiple recursive concatenation of a short kernel code to transform the physical channel into virtual outer channels.
When the number of recursions becomes large, the virtual channels tend to either have high or low reliability (i.e., they polarize), and the data bits are allocated to the most reliable channels.
For the turbo codes, their encoding is a concatenation of two (or more) convolutional encoders separated by interleavers, and its decoding consists of two (or more) soft-in/soft-out convolutional decoders, which iteratively feed probabilistic information back and forth to each other.
LDPC codes are another class of linear block codes on the graph \cite{Ryan09}, which is constructed by using a sparse Tanner graph \cite{Richardson08} and can provide the near-capacity performance with implementable message-passing decoders.
Since we will exploit some key techniques of codes on the graph, e.g. LDPC codes, for the work in Chapter 4, we present more information on LDPC codes in the next.

%
%The turbo code is one important class of  convolutional codes, which was first presented in [] and widely used in 3G/4G mobile communications.

%
%The convolution code, which was invented by Elias in 1955 \cite{Ungerboeck82}, has been widely used for wireless and space communications.

\subsection{Codes on Graph and Tanner Graph}
\subsubsection{LDPC Codes:}
The LDPC code is one important class of linear block codes with reasonably low complexity and implementable decoders \cite{Ungerboeck82, Richardson01, Richardson012}, which can provide near-capacity performance on a large set of data transmissions and are proposed as the standard code for 5G.
%LDPC codes were first invented by Gallager in 1960 \cite{Ungerboeck82} and resurrected in the mid 1990s \cite{Wachsmann99, Zehavi92}.
An LDPC code can be given by the null space of an $m\times n$ parity-check matrix $\mathbf{H}$ that has a low density.
A regular LDPC code is a linear code whose parity-check matrix $\mathbf{H}$ has constant column weight $g$ and row weight $r$, where $r=g(n/m)$ and $g \ll m$.
If $\mathbf{H}$ is a low density matrix but with variable $g$ and $r$, the code is called an irregular LDPC code \cite{Ryan09}.
It is noteworthy that the density of LDPC codes needs to be sufficiently low to permit an effective iterative decoding, which is the key innovation behind the invention of LDPC codes.
Moreover, LDPC decoding is verifiable in the sense that decoding to a correct codeword is a detectable event.

\subsubsection{Tanner Graph:}
\begin{figure}[!t]
	\par
	\begin{center}
		{\includegraphics[width=1.8in]{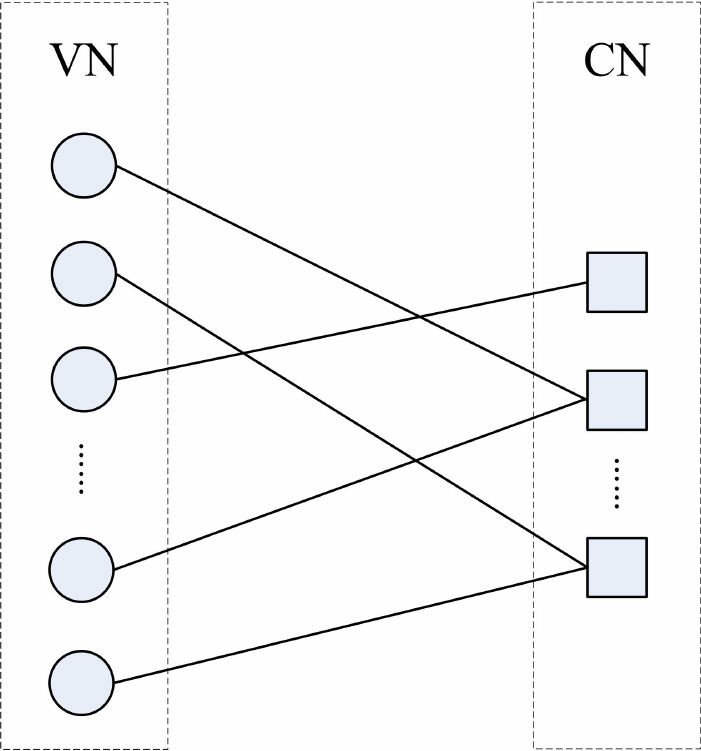}}
	\end{center}\vspace{-5mm}
	\caption{The Tanner graph.}\vspace{-5mm}
	\label{fig_bipartite_graph_model}
\end{figure}

A graphical representation of an LDPC code, which is called a Tanner graph, can provide a complete representation of the code and aid in the description of its decoding algorithm \cite{Ryan09}.
A Tanner graph is a bipartite graph, that is, a graph whose nodes can be divided into two disjoint and independent sets, with edges connecting only nodes of different sets.
The two sets of nodes in a Tanner graph are called the variable nodes (VNs) and the check nodes (CNs).
For a code with the parity-check matrix $\mathbf{H}$ of dimension $m \times n$, its Tanner graph can be drawn as follows: CN $j$ is connected to VN $i$ when $\mathbf{H}_{j,i}=1$.
Then, according to this rule, there are $m$ CNs and $n$ VNs in the Tanner graph, which correspond to $m$ check equations and $n$ code bits, respectively, as shown in Fig. \ref{fig_bipartite_graph_model}.
%
%
% $\mathbf{U}$ and $\mathbf{V}$, so that every edge connects a vertex in $\mathbf{U}$ to a vertex in $\mathbf{V}$ \cite{Richardson08}.
%In coding theory, the bipartite graph, called Tanner graph \cite{Richardson08}, is used to state constraints or equations that specify error control codes.
%Let $\mathbf{C}$ be a binary linear block code and $\mathbf{H}$ be a parity-check matrix of $\mathbf{C}$ with dimension $m \times n$.
%For the Tanner graph associated with $\mathbf{H}$, it has $n$ variable nodes (VN) that correspond to components of a codeword, and $m$ check nodes (CN) that correspond to the set of parity-check constraints, as shown in Fig. \ref{fig_bipartite_graph_model}.
%The check node $j$ is connected to the variable node $i$, if the variable $i$ participates in the $j$-th parity-check constraint, i.e., $\mathbf{H}_{ji}=1$.
The number of edges connected to a VN (or a CN) is called the degree of this VN (or CN).
Denote the number of VNs of degree $i$ and the number of CNs of degree $j$ as $\Lambda_{i}$ and $\Psi_{j}$, respectively.
Since the edge counts must match up, we have $\sum_{i}i\Lambda_{i}=\sum_{j}j\Psi_{j}$.
It is convenient to introduce the following compact notation \cite{Richardson08}
\begin{align}
\Lambda(x)=\sum_{i=1}^{l_{\max}}\Lambda_{i}x^{i} \text{        and       } \Psi(x)=\sum_{j=1}^{r_{\max}}\Psi_{j}x^{j},
\label{degree_poly_node}
\end{align}
where $l_{\max}$ and $r_{\max}$ are the maximum degrees of the VN and the CN, respectively.
Here, $\Lambda(x)$ and $\Psi(x)$ are the polynomial representations of the VN degree distribution and the CN degree distribution from a node perspective, respectively.
Moreover, the polynomials $\Lambda(x)$ and $\Psi(x)$ are non-negative expansions around zero whose integral coefficients are equal to the number of nodes of various degrees.
For the asymptotic analysis, it is more convenient to introduce the VN and CN degree distributions from an edge perspective, given by \cite{Richardson08}
\begin{align}
\lambda(x)=\sum_{i}\lambda_{i}x^{i-1}=\frac{\Lambda'(x)}{\Lambda'(1)} \text{        and       } \rho(x)=\sum_{j}\rho_{j}x^{j-1}=\frac{\Psi'(x)}{\Psi'(1)}.
\label{degree_poly_edge}
\end{align}
Note that, $\lambda(x)$ and $\rho(x)$ are also polynomials with non-negative expansions around zero.
In addition, $\lambda_{i}(\rho_{j})$ is equal to the fraction of edges that connect to VNs of degree $i$ (CNs of degree $j$).
In other words, $\lambda_{i}(\rho_{j})$ is the probability that an edge chosen uniformly at random from the graph is connected to a VN of degree $i$ (a CN of degree $j$).

\subsection{Iterative Decoding Algorithm on Code Graph}
Iterative decoding is a generic term to refer to decoding algorithms that proceed in iterations \cite{Yuan99,Kang18,Wu18acc}.
An important subclass of iterative algorithms are message-passing algorithms, which obey the rule that an outgoing message along an edge only depends on the incoming messages along all edges other than this edge itself \cite{Kschischang01}.
When the messages are probabilities, or called ``belief'', the algorithm is known as sum-product algorithm (SPA) \cite{Kschischang01} and also called the belief propagation algorithm (BPA) \cite{Pearl82}.
In the SPA, the passing probability refers to the log-likelihood ratio (LLR) of a bit, given by \cite{Ryan09}
\begin{align}
L(v_{j}|\mathbf{y})=\ln \left(\frac{\mathrm{Pr}(v_{j}=0|\mathbf{y})}{\mathrm{Pr}(v_{j}=1|\mathbf{y})}\right),
\end{align}
where $\mathrm{Pr}(v_{j}=0|\mathbf{y})$ and $\mathrm{Pr}(v_{j}=1|\mathbf{y})$ are a posteriori probability (APP) that given the received codeword $\mathbf{y}=[y_{0},y_{1},\ldots,y_{n-1}]$, the bit $v_{j}$ equals $0$ and $1$, respectively.
Denote the passing messages from VN $i$ to CN $a$ and from CN $a$ to VN $i$ as $L_{i\rightarrow a}$ and $L_{a\rightarrow i}$, respectively, which are given by \cite{Kschischang01}
\begin{align}
L_{i\rightarrow a}&=\prod_{b \in N(i)\backslash \{a\}}L_{b\rightarrow i},\\
L_{a\rightarrow i}&=\sum_{\sim \{i\}}f(\mathbf{I})\prod_{j \in N(a)\backslash \{i\}}L_{j\rightarrow a},
\label{cn2vn}
\end{align}
where $\sim \{i\}$ denotes the indices of all variable nodes except $i$, $\mathbf{I}$ is the vector of  variable nodes neighbouring to the check node $a$, and $f(\mathbf{I})$ is the joint mass function of all elements in $\mathbf{I}$.
It can be seen that when the VN $i$ computes the information transmitted to CN $a$, i.e., $L_{i\rightarrow a}$, the multiplication of LLR information from all its neighboring CNs $N(i)$ except the recipient CN $a$ is obtained.
Similarly, when computing the information from CN $a$ to VN $i$, i.e., $L_{a\rightarrow i}$, the joint mass function of all elements in $\mathbf{I}$ is multiplied by the LLR information from all its neighboring VNs $N(a)$ except the recipient VN $i$.
Since the information transmitted to a certain CN or VN does not contain the information from itself, the transmitted information is called the extrinsic information \cite{Richardson08}.
\begin{figure}[t]
	\centering
	\subfigure[A VN decoder.]
	{\label{fig_cn2vn} %% label for first subfigure
		\includegraphics[width=1.67in]{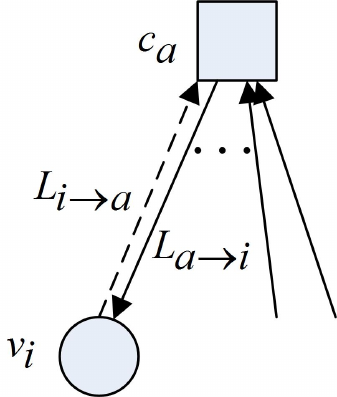}}\vspace{-1mm}
	\hspace{20mm}
	\subfigure[A CN decoder.]
	{\label{fig_pdf_d2} %% label for second subfigure
		\includegraphics[width=1.67in]{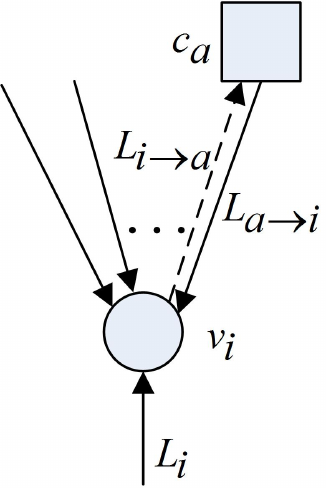}}\vspace{-1mm}
	\caption{VN and CN decoders.}
	\label{cdf_vn2cn}
\end{figure}
At each iteration, all VNs process their inputs and pass extrinsic information up to their neighboring CNs.
All CNs then process their inputs and pass extrinsic information down to their neighboring VNs.
The procedure repeats, starting from the variable nodes.
After a preset maximum number of repetitions (or iterations) of this VN/CN decoding round, or after some stopping criterion has been met, the decoder estimates the LLRs from which decisions on the bits are made.
When the code graph has no cycles or the lengths of cycles are large, the estimates will be very accurate and the decoder will have near-optimal MAP performance.
It is noteworthy the development of SPA relies on an assumption that the LLR quantities received at each node from its neighbors are independent.
However, this assumption difficultly holds when the number of iterations exceeds half of the Tanner graph's girth \cite{Richardson08}.

\subsection{Density Evolution and EXIT Chart}
For an iterative decoder with a finite message alphabet, the distribution of messages passed along edges in each iteration can be expressed by a system of coupled recursive functions.
The procedure of using the system of coupled recursive functions to track the evolution of message distributions is termed as density evolution (DE) \cite{Richardson08}.
The DE provides an analytical performance tracking of the iterative decoder in each iteration, which allows to design the code structure to improve the decoding performance.
For example, the performance of an irregular LDPC code can be improved via the design of its optimal and near-optimal degree distribution.
And, the design of degree distributions can be obtained using the DE.
Moreover, one is interested in the error probability of messages that are passed along edges and its evolution as a function of the iteration number.
Based on that, the decoding threshold, which is defined as the lowest SNR to ensure that the decoder error probability can asymptotically converge to zero for an infinite number of iterations, can be predicted through the DE.
In the following, we adopt the binary erasure channel (BEC) to illustrate the derivation of DE.
Consider the degree distributions of variable nodes and check nodes for an LDPC code from an edge perspective as $\lambda$ and $\rho$, which are given in Eq. \eqref{degree_poly_edge}.
Let $\epsilon$ be the probability that a packet is erased, $\epsilon \in [0,1]$.
From the definition of SPA, the initial VN-to-CN message is equal to the received message, which is an erasure message with the probability $\epsilon$.
%We start with the check-to-variable messages in the $(l+1)$-th iteration.
%Recall that by definition of the algorithm, a
The CN-to-VN message that is emitted by a CN of degree $a$ is an erasure message, if any one of the $a-1$ incoming messages is an erasure.
Denote $x_{l}$ as the probability that an incoming message is an erasure in the $l$-th iteration.
The probability that the outgoing message is an erasure is equal to $1-(1-x_{l})^{a-1}$, where all incoming messages are independent.
As an edge has a probability $\rho_{a}$ to be connected to a CN of degree $a$, the erasure probability of a CN-to-VN message in the $(l+1)$-th iteration is equal to \cite{Richardson08}
\begin{align}
\sum_{a}\rho_{a}(1-(1-x_{l}))^{a-1}=1-\rho(1-x_{l}).
\label{degree_poly_edge_1}
\end{align}
Consider an edge is connected to a VN of degree $i$.
For the VN-to-CN message along this edge in the $(l+1)$-th iteration, it is an erasure if the received value of the associated VN is an erasure and all $i-1$ incoming messages are erasures.
This happens with probability $\epsilon(1-\rho(1-x_{l}))^{i-1}$.
By averaging this probability over the edge degree distribution $\lambda$, we obtain \cite{Richardson08}
\begin{align}
x_{l+1}=\epsilon\lambda(1-\rho(1-x_{l})).
\label{de}
\end{align}
Eq. \eqref{de} is an recursive function of the probability $x_{l}$.
Its evolution characterizes the decoder performance of the LDPC code in each iteration.

As an alternative to DE, the extrinsic-information-transfer (EXIT) chart technique, which is introduced by \cite{Clark1999}, provides a graphical tool for estimating the convergence behaviour of an iterative decoder.
The basic idea behind EXIT chart is based on the fact that the VN processors and CN processors work cooperatively and iteratively to make each bit decision in the iterative decoder, with the metric of interest improving with each half-iteration \cite{Hagenauer04}.
For both the VN processors and the CN processors, the relations of the input metric versus the output metric can be obtained by using the transfer curves, and the output metric of one processor is the input metric of its companion processor.
Therefore, both transfer curves can be plotted on the same axes, but with the abscissa and ordinate reserved for one processor, which generates the EXIT chart.
Furthermore, the EXIT chart provides a graphical method to predict the decoding threshold of the ensemble of codes that are characterized by given VN and CN degree distributions.
In particular, when the VN processor transfer curve just touches the CN processor transfer curve, the SNR is obtained as the decoding threshold.
%
%the decoding threshold is the signal-to-noise ratio (SNR), at which the transfer curve of VN processors just touches that of CN processors and the decoder error probability can asymptotically converge to zero for an infinite number of iterations.
%A iterative decoder actually constitutes two component decoders which exchange the extrinsic information between each other.
%Each constituent decoder can be represented by a mutual information transfer function, which describes the flow of extrinsic information through the soft in/soft out decoder.
%The exchange of extrinsic information between constituent decoders is then plotted in an EXIT chart.
%During the iterative decoding procedure, the extrinsic output of one decoder becomes the a-prior input of the other decoder, which is indicated by the line between the transfer functions.
%Each line indicates a single decoding step of the
%iterative decoding procedure.
%Then, the convergence of iterative decoding procedure is only possible if the two transfer characteristics do not intersect.
%In this case,  the average number of required decoding steps can be estimated.
\section{Physical-layer Network Coding}
\begin{figure}[!t]
\par
\begin{center}
{\includegraphics[width=3in]{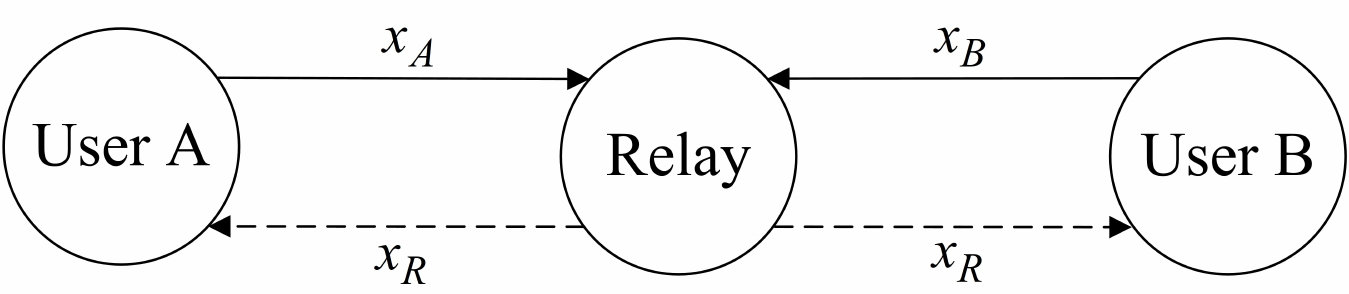}}
\end{center}\vspace{-5mm}
\caption{A PNC strategy for two-way relay channels.}
\label{fig_pnc_model}
\end{figure}

Physical-layer network coding (PNC) was firstly proposed for a two-way relay channel by \cite{Zhang06} in 2006, which exploits the network coding operation \cite{Ahlswede00, Li03, Sadeghi09, Sadeghi10} in the superimposed electromagnetic (EM) waves to embrace the interference.
It has been shown to be able to boost the throughput and significantly improve the reliability in a multi-way relay channel \cite{Akino09, Louie10,Yang13,Huang13,Huang14,Guo18}.
The basic idea of PNC can be summarized as follows, where the two-way relay channel is considered for simplicity.
Consider the two-way relay channel, where users A and B want to exchange packets via a relay node as shown in Fig. \ref{fig_pnc_model}.
Each round of packet exchange consists of two equal-duration time slots, which are called the uplink phase and the downlink phase, respectively.
In the uplink phase, two users transmit their own signal simultaneously to the relay, where the signals are denoted by $x_{A}$ and $x_{B}$, respectively.
In the downlink phase, the relay broadcasts a signal $x_{R}$, which can be represented as a function of the two signals, i.e.,
\begin{align}
x_{R}=f(x_{A},x_{B}).
\label{pnc}
\end{align}
Upon receiving the signal $x_{R}$, each user extracts the other user's information by exploiting its own signal, which finishes the round of packet exchange.
Obviously, the function $f(\cdot)$, called the PNC mapping function \cite{Zhang06}, acts as a pivotal role for the system performance.
Since the mapping function refers to the mapping from a superimposed EM signal to a desired network-coded signal, it should be designed according to the employed modulation constellation.
The original work on PNC suggests using a XOR function of the two users' packets \cite{Zhang06,Liew2013} for the two-way relay channel with binary phase-shift keying (BPSK) and quadrature phase-shift keying (QPSK) modulations, which is a well-known function for PNC operations.
In fact, $f(\cdot)$ can be a linear function \cite{Fang14,Yang14 } or a non-linear function \cite{Akino09,Muralidharan13}.
For the linear mapping function, $f(\cdot)$ is the linear combination of the two users' packets in $\text{GF}(2^{n})$.
It offers low computational complexity and scalability, compared to the non-linear mapping function.
In \cite{Yang12}, a linear PNC scheme is proposed for real Rayleigh fading two-way relay channels with pulse amplitude modulation (PAM).
%The linear PNC scheme is extended to Rayleigh fading two-way relay channels with QPSK modulation in [huang:14].
Furthermore, the linear PNC is extended to complex Rayleigh fading two-way relay channels in \cite{Yang14}, where a design criterion of linear PNC, namely minimum set-distance maximization, is proposed to achieve the optimal error performance at high SNRs.

In addition, the PNC technique can be adopted to multiple access networks \cite{Lu2013,You2015,Cocco11con}.
A cross-layer scheme design is proposed to improve the throughput of wireless network in \cite{Lu2013,You2015}.
In particular, the PNC decoding and the MUD are jointly used to obtain multi-reception results at the physical layer, and the multi-reception results are exploited at the MAC layer to recover users' packets.

\section{Introduction to Wireless Communicaiton}
This section provides a brief overview of wireless communications. The presentation is not intended to be exhaustive and does not provide new results, but it is intended to provide the necessary background to understand Chapters 3-5.

\subsection{Channel Models}
A wireless channel is one of the essential elements in the wireless transmission system.
By sufficiently understanding the wireless channel, we can mathematically model its physical properties to facilitate the design of communication systems \cite{David04}.

Wireless channels operate through electromagnetic radiation from the transmitter to the receiver \cite{David04}.
The transmitted signal propagates through the physical medium which contains obstacles and surfaces for reflection.
This causes multiple reflected signals of the same source to arrive at the receiver in different time slots.
In order to model these effects from the physical medium, the concept of channel model is proposed.
The effect of multiple wavefronts is represented as multiple paths of a channel.
The fluctuation in the envelope of a transmitted radio signal is represented as the channel fading \cite{David04}.
The process to estimate some information about the channel refers to the channel estimation, which is essential to recover the transmitted signal at the receiver.

In principle, with the transmitted signal, one could solve the electromagnetic field equations to find the electromagnetic field impinging on the receiver antenna \cite{David04}. However, this task is non-trivial in practice, since it requires to know the physical
properties of obstructions in a more accurate manner.
Instead, a mathematical model of the physical channel is used, which is simpler and tractable.
%Although it is only an approximation of the real physical environment, it can give a good performance.
In the following, two mathematical models of channel are presented, which are widely used in the communication system designs and in this thesis.

\subsubsection{Fading Channels:}
As mentioned above, the strength change of transmitted signals through the channel is represented by the channel fading.
In particular, the channel fading can be divided into two types \cite{David04}, i.e., the large-scale fading and the small-scale fading.
The large-scale fading mainly refers to the path loss, which is a function of distance and shadowing by large objects, e.g. building and hills.
In this case, the variation of signal strength is over distances of the order of cell sizes.
The small-scale fading is caused by the constructive and destructive interference of the multiple signal paths between the transmitter and the receiver.
This occurs at the spatial scale of the order of the carrier wavelength.

The Rayleigh fading model is the simplest model for wireless channels \cite{David04}.
It is based on the assumption that there are a large number of statistically independent reflected and scattered paths with random amplitudes in the delay window corresponding to a single tap of the tapped delay line model.
Each tap gain $h_{l}$ is the sum of many independent random variables.
According to the Central Limit Theorem, the net effect can be modeled as a zero-mean complex Gaussian random variable, given by
\begin{align}
h_{l}\sim \mathcal{CN}(0,\sigma_{l}^{2}),
\label{rayleigh_1}
\end{align}
where $\sigma_{l}^{2}$ is the variance of tap $h_{l}$, $l \in \{0,1,\ldots,L\}$, and $L$ is the number of taps.
The magnitude $|h_{l}|$ of the $l$-th tap is a Rayleigh random variable with density
\begin{align}
p(x)=\frac{x}{\sigma_{l}^{2}}\exp\left(\frac{-x^2}{2\sigma_{l}^{2}}\right), x \leq 0,
\label{rayleigh_2}
\end{align}
and the squared magnitude $|h_{l}|^2$ is exponentially distributed with density
\begin{align}
p(x)=\frac{1}{\sigma_{l}^{2}}\exp\left(\frac{-x}{2\sigma_{l}^{2}}\right), x \leq 0.
\label{rayleigh_3}
\end{align}
The Rayleigh fading model is quite reasonable for the scattering mechanisms, where many small reflectors and no line of sight exist.
The other widely used fading model is the Rician fading model, in which the line of sight path is dominating.
\subsubsection{Erasure Channels:}
\begin{figure}[!t]
\par
\begin{center}
{\includegraphics[width=1.8in]{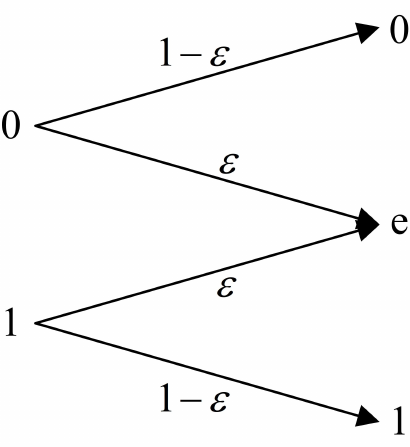}}
\end{center}\vspace{-5mm}
\caption{Binary erasure channel.}
\label{fig_bec}
\end{figure}

Although the transmitted and received signals are continuous-valued for most of the channels, many crucial processes of practical communication systems, e.g. the coding/decoding and modulation/demodulation, are based on the discrete signals in nature.
%In particular, the transmitter sends one out of a finite number of codewords and the receiver would like to figure out which codeword is transmitted.
Therefore, for the performance or channel capacity analysis, we usually model channels with the discrete input/output, called discrete memoryless channels (DMCs) \cite{Richardson08}.
One important DMC is the BEC, which is shown in Fig. \ref{fig_bec}.
It can be seen from the figure that the BEC has binary input and ternary output.
The input symbols cannot be flipped but can be erased with a probability $p(e|0)=p(e|1)=\epsilon$.
This BEC channel is widely used in the information theory, since it is one of simplest channels to analyze.
In addition, the packet erasure channel (PEC) is proposed as a generalization of the BEC.
For the PEC, the transmitted packets are either received or lost.
It is noteworthy that the erasure of PEC can be seen as the result of deep-fading in the practical fading channel.
Then, the PEC can be seen as a simplified version of the fading channel, and it is commonly used for the system design \cite{Wangtit12}.
%\subsection{Channel Estimation}

\subsection{Multiple Access Techniques}
\subsubsection{Multiple Access System:}
The idea of using a communication channel to enable several transmitters to send information data simultaneously starts from Thomas A. Edison's 1873 invention of the diplex \cite{Verdu1998}.
For this revolutionary system, two telegraphic packets are simultaneously transmitted in the same direction through the same wire, which is the embryonic form of multiple access systems.

\begin{figure}[tph]
\par
\begin{center}
{\includegraphics[width=3in]{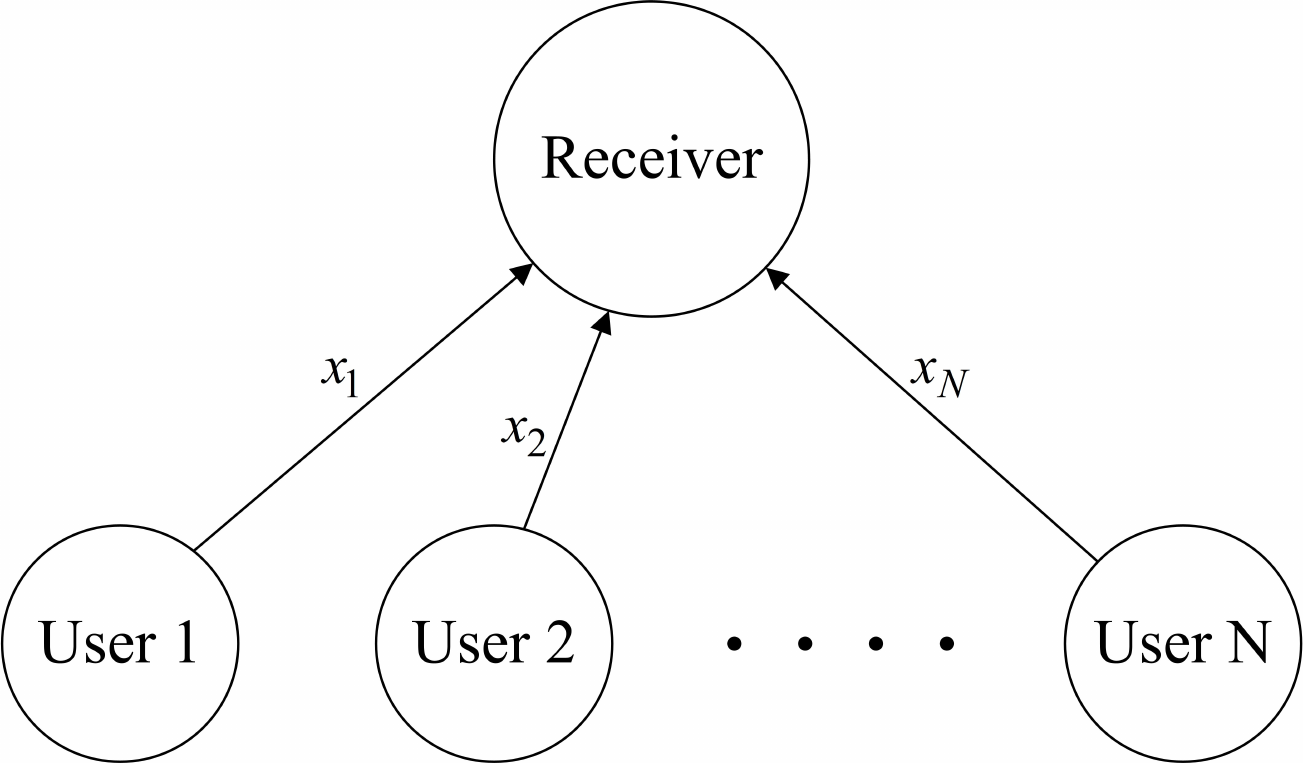}}
\end{center}\vspace{-5mm}
\caption{Multiple access system with $N$ users and a common receiver.}
\label{fig_ma_uplink}
\end{figure}

Nowadays, multiple access systems have been intensively developed and widely used in many areas, e.g. multiple cellular users transmitting to a base station and local area networks.
A common feature of those communication systems \cite{Verdu1998} is that multiple transmitters simultaneously send signals to a common receiver, and the transmitted signals are superimposed at the receiver, as depicted in Fig. \ref{fig_ma_uplink}.
%
%Multiple access lies at the heart of wireless communication systems.
%It refers to a technique that allows multiple users to transmit data to a common receiver via a shared channel, which is depicted in Fig. \ref{fig_ma_uplink}.

The multiple access communication lies at the heart of wireless communication systems.
The first-generation (1G) to the forth-generation (4G) of cellular networks have adopted radically different multiple access schemes, which include time division multiple access (TDMA), frequency division multiple access (FDMA), code division multiple access (CDMA), and orthogonal frequency division multiple access (OFDMA) \cite{Ng12}. In addition, space division multiple access (SDMA) \cite{Miao16, David04} is also used in other practical systems.
The common gist of these schemes is to allocate orthogonal resources for different transmitters to send their packets.
Therefore, all these schemes belong to the orthogonal multiple access (OMA).
%Consider each channel use is of duration $T$ seconds.
For example, in FDMA that has widely been used in the 1G mobile wireless communication network, the whole bandwidth is divided into several non-overlapping frequency subchannels and each transmitter employs a subchannel to send its voice.
In other words, one orthogonal spectral resource is allocated to only one user.
As a result, the signals transmitted by different users can be easily separated and recovered.
While the data detection process is simple for OMA schemes, the spectrum occupation is inefficient and cannot satisfy the requirements of high throughput, high traffic load, and low latency for current communication systems.
Therefore, it triggers the proposal of non-orthogonal multiple access (NOMA) schemes \cite{Ding17,WeiSurvey2016,Liu17,Xiangjstsp,Weijstsp,Yangnmamag,Weinomaletter,Henomajsac}.
By allowing multiple users to share a same resource block, NOMA schemes can increase the spectral efficiency and the user-fairness.
In particular, NOMA schemes include the power-domain NOMA and the code-domain NOMA.
Power-domain NOMA \cite{Ding17,WeiTCOM2017,wei2018multibeam,WeiCOML,Qiu182,Qiu19,Wei2016NOMA,wei2018multiICC,wei2017fairness,wei2017performance,Wei2018PerformanceGain,Wei2018BeamWidthControl,Sun17} allocates different power levels to different users and exploits the successive interference cancellation (SIC) to retrieve their packets.
For the code-domain NOMA \cite{Nikopour13,Di18}, data streams are mapped to multi-dimensional sparse codewords, where each codeword represents a spread transmission layer.
As a result, more users can share the same time-frequency resource block and the system efficiency is improved.

\subsubsection{Random Multiple Access:}
Among multiple access communications, random multiple access is one of the approaches to dynamic channel sharing for dealing with the burst traffic.
In a random multiple access system, when a user has the packet to transmit, it randomly occupies a resource block for its transmission.
It implies that the random multiple access system has a flexible and dynamic resource occupation.
However, it always exists an inevitable probability that two users occupy the same resource block and their packets collide with each other, which refers to the collision probability.
In this case, the collided packets cannot be reliably decoded by the receiver and the users need to retransmit their packets after a duration.
%
%However, due to the uncoordinated transmission of users, it always exists the probability that one transmitted message will interfere with another in a same resource block, which is called a collision.
%In this case, it is typically assumed that the receiver cannot reliably demodulate these collided messages in random multiple access systems.
%Instead, the receiver can notify the collision to all collided transmitters, and their messages will be retransmitted later.
In order to reduce the collision probability for retransmission, the collided transmitters usually wait a random period of time before retransmitting \cite{Lien11}.
The algorithm used by the transmitter to determine the retransmission delay plays an essential role in the random access schemes.

\begin{figure}[!t]
	\par
	\begin{center}
		{\includegraphics[width=3.55in]{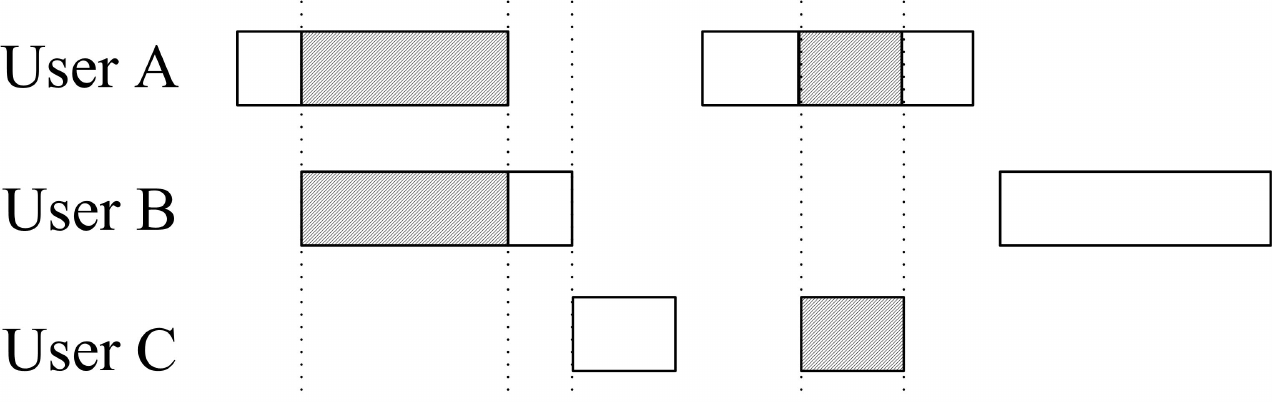}}
	\end{center}\vspace{-5mm}
	\caption{ALOHA scheme with three transmitters.}
	\label{fig_aloha}\vspace{-5mm}
\end{figure}

The first random multiple access system is the ALOHA system, which was proposed for a wireless connection between the computer resources of different islands of the state of Hawaii in \cite{Abramson1970}.
For the ALOHA, when a user is ready to transmit its packet, it simply transmits the packet and shares the channel with other users in an uncoordinated way, as shown in Fig. \ref{fig_aloha}.
If there is only one user transmitting, the user's packet can be successfully received over a noiseless channel and it receives an acknowledgment via the feedback channel.
Otherwise, the packet collision occurs and the user cannot receive the acknowledgment.
In Fig. \ref{fig_aloha}, the collided parts of packets are highlighted by shaded areas.
%In this case, the user's message is said to suffer from the interference and it retransmits a copy of its message at another randomly chosen time, where the interference is caused by other messages transmitting at the same time.
The starting times of packets can be modeled as a Poisson point process with parameter $\lambda$ packets/second \cite{Abramson1970}. If each packet lasts $\tau$ seconds, the normalized channel traffic can be given by
\begin{align}
G=\lambda\tau.
\label{traffic_aloha}
\end{align}
The normalized throughput is equal to \cite{Abramson1970}
\begin{align}
T=Ge^{-2G}.
\label{thpt_aloha}
\end{align}
When the traffic $G$ equals $0.5$, the maximum value of this normalized throughput is obtain as $\frac{1}{2e}=0.184$.

\begin{figure}[!t]
	\par
	\begin{center}
		{\includegraphics[width=2.8in]{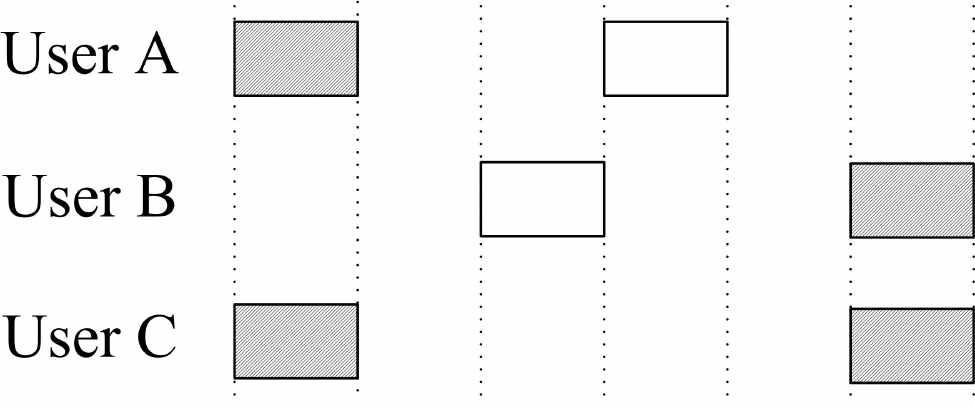}}
	\end{center}\vspace{-5mm}
	\caption{Slotted ALOHA scheme with three transmitters.}
	\label{fig_sa}
\end{figure}

Based on the ALOHA scheme, slotted ALOHA (SA) is proposed in \cite{Gitman1975}.
By defining a set of contiguous equal-duration time slots, the users align the start of their packet transmissions to the start of a time slot and transmit packets within the time slot, which is depicted in Fig. \ref{fig_sa}.
The use of synchronous transmission reduces the number of collisions and improves the throughput, compared to the ALOHA scheme.
In particular, the maximum achievable throughput is doubled, i.e., $\frac{1}{e}=0.37$.
Note that, this improvement is based on the assumption that the length of transmitted packets equals the duration of a time slot.
%If this assumption cannot hold, some portions of time slots are wasted and the throughput has a large loss.
%Moreover, the SA scheme requires the perfect synchronization, which means more overhead than the ALOHA scheme.

Carrier-sense multiple access (CSMA) scheme was proposed as another important random access system in \cite{Kleinrock75} and has been widely used in many practical wireless networks \cite{Kleinrock75,Bianchi00}, such as WiFi.
In the CSMA, users sense the channel to see if it is occupied or not before transmitting, in order to avoid as many collisions as possible.
If no occupation is sensed, the user transmits.
Otherwise, it retries after a while.
The sensing can be done by measuring the received power and comparing it to a predetermined threshold, which increases the hardware and software complexities.
Furthermore, the CSMA with Collision Avoidance (CSMA-CA) scheme is proposed to further reduce the collision probability by wisely using the back-off.
The intuition behind this is that the more collisions occur, the more congested the network is.
Thus, retransmissions need to be less frequent (more spaced apart in time).
It is noteworthy that compared to the ALOHA-based schemes, the CSMA scheme may suffer from a large sensing overhead.
In this thesis, we mainly focus on the ALOHA-based random access scheme to develop our proposed random access schemes in Chapter 4.

\subsubsection{Data Detection:}
For random access systems, particularly the conventional ALOHA and SA schemes, collided packets are directly discarded and the corresponding users retransmit their packets after a while.
This mechanism dramatically degrades the system performance and increases the delay.
Then, instead of directly discarding collided packets, exploiting efficient data detection techniques to recover collided packets is very essential to improve the system performance and decrease the delay \cite{Hong08}.
%
%The data detection technique provide a efficient way to recover signals from the superimposed signal.
%Therefore, the data detection acts as an important role for the ALOHA based random access system to enhance its performance.
In the following, we overview two classical data detection methods.

In general wireless networks, the data detection can be divided into the single-user detection and the MUD.
For the single-user detection, a signal is detected by treating other collided signals as the interference and each signal is detected separately.
While the single-user detection has a low complexity, its performance is not good, in particular when the number of collided signals is very large or the power difference of collided signals is significant.
Then, the MUD was proposed and summarized by \cite{Verdu1998}, where collided signals are detected simultaneously.

We consider a multiple access system, where $N$ single-antenna users simultaneously transmit data to a common single-antenna receiver and each user's data is one symbol from a finite alphabet set $\mathcal{X}$.
The received signal $y$ is the sum of the received signals for all users plus noise, given by
\begin{align}
y=\mathbf{h}^{T}\mathbf{x}+z,
\label{mud}
\end{align}
where $\mathbf{h} \in \mathbb{C}^{N\times1}$ is the channels from $N$ users to the receiver, $\mathbf{x}\in \mathcal{X}^{N\times1}$ is the transmitted data from $N$ users, and $z$ is AWGN.
Based on the model in Eq. \eqref{mud}, the optimal detector of MUD, i.e., the maximum likelihood (ML) detector \cite{Verdu89}, is given by
\begin{align}
\hat{\mathbf{x}}_{ML}=\arg \max_{\hat{\mathbf{x}}\in \mathcal{X}^{N}}f_{y|\mathbf{x},\mathbf{h}}(y|\mathbf{x}=\hat{\mathbf{x}},\mathbf{h}).
\label{ml}
\end{align}
%Here, $\hat{\mathbf{s}}$ is the detected signal, $\mathbf{H}$ is the given channel, and $\mathbf{y}$ is the received signal.
It can be seen from Eq. \eqref{ml} that given the channel $\mathbf{h}$, the ML detector chooses the signal $\hat{\mathbf{x}}$ that maximizes the probability of received signal as the detected signal.
When the prior distribution of detected signal $\mathbf{x}$, i.e., $P(\mathbf{x}=\hat{\mathbf{x}})$, is exploited, the MAP detection \cite{Bassett18,Choi06} can be obtained, given by
\begin{align}
\hat{\mathbf{x}}_{MAP}&=\arg \max_{\hat{\mathbf{x}}\in \mathcal{X}^{N}}\frac{P(\mathbf{x}=\hat{\mathbf{x}})f_{y|\mathbf{x},\mathbf{h}}(y|\mathbf{x}=\hat{\mathbf{x}},\mathbf{h})}{f_{y|\mathbf{h}}(y|\mathbf{h})} \notag \\
&\propto\arg \max_{\hat{\mathbf{x}}\in \mathcal{X}^{N}}P(\mathbf{x}=\hat{\mathbf{x}})f_{y|\mathbf{x},\mathbf{h}}(y|\mathbf{x}=\hat{\mathbf{x}},\mathbf{h}).
\label{map}
\end{align}
We ignore the marginal likelihood $f_{y|\mathbf{h}}(y|\mathbf{h})$, since it does not depend on $\mathbf{x}$ and has no effect on the optimization.
Comparing Eq. \eqref{ml} with \eqref{map}, it can be observed that the MAP detection is equivalent to the ML detection, when the prior distribution of detected signal i.e., $P(\mathbf{x}=\hat{\mathbf{x}})$, is uniform.

For both the ML detector and the MAP detector, they can jointly detect all collided signals at a time.
Although the joint detection methods can provide excellent performance, they suffer from a high computational complexity.
Therefore, many suboptimal approaches are proposed and the SIC is a popular one among them.
Instead of jointly detecting all collided signals, the SIC algorithm iteratively detects the collided signals.
In particular, when a collided signal is successively detected, it will be removed from the superimposed signal and the next signal is detected from the remaining superimposed signal.
Due to removing the previously successfully detected signals, the following signals can be detected with less interference and a higher successful probability for the SIC algorithm.
Moreover, in order to further improve the performance of SIC algorithm, an optimal detection ordering is proposed, that is the stronger signal is detected earlier \cite{Verdu1998}.

\section{Chapter Summary}
In this chapter, we have provided the background knowledge on CS techniques, modern coding techniques, and the PNC technique, which are essential to understand the works given in Chapters 3-5, respectively.
Besides, we have presented a brief overview of fundamental concepts of wireless communications, which include the widely used channel models, multiple access systems, random multiple access systems, and the data detection techniques.

%\subsection{Detection Techniques in Uplink Multiple Access Systems}
%\subsubsection{Single User Detection}
%\subsubsection{Multi-User Detection}
%
%\subsection{Channel Estimation in Uplink Multiple Access Systems}
%\subsubsection{Pilot Aided Channel Estimation}
%\subsubsection{Blind Channel Estimation}

\chapter{Joint User Activity Identification and Channel Estimation in Random Access Systems for mMTC}\label{C3:chapter3}

%\nomenclature{$a$}{The number of angels per unit area}%
%\nomenclature{$N$}{The number of angels per needle point}%
%\nomenclature{$A$}{The area of the needle point}%
%
%\ifpdf
%    \graphicspath{{1_introduction/figures/PNG/}{1_introduction/figures/PDF/}{1_introduction/figures/}}
%\else
%    \graphicspath{{1_introduction/figures/EPS/}{1_introduction/figures/}}
%\fi
\section{Introduction}
%While the conventional scheme works well in general, its performance is not enough considered for the case with a small pilot length and a high reliability requirement.
%Moreover, the small pilot length is favorable in mMTC and the high reliability on user identification is essential to improve the following data detection.
%From the compressed sensing theory\cite{Donoho09}, it is known that enhancing the sparsity of user activity, i.e., decreasing the number of users transmitting simultaneously, can significantly improve the performance of CS algorithms, for a given pilot length \cite{Donoho09}.
%Moreover, given the sparsity of user activity, the probability that an active user is correctly identified by CS algorithms is proportional to its received signal strength.
%%
%In fact, the weaker that an active user's signal, the smaller probability that it is correctly identified, and vice versa.
%%
%Therefore, in this chapter we propose a simple transmission control scheme and modify an AMP algorithm at the receiver to address the user identification and channel estimation problem for massive random multiple-access networks, which is different from the methods developed from the receiver point of view in \cite{Schepker13,Xu15,Wunder15,Hannak15,Chen18,Liu18}.
%
In this work, we focus on the user activity identification and channel estimation in random access systems for mMTC.
In particular, we first propose a transmission control scheme to enhance the system sparsity and to achieve an improved performance for the user activity identification and channel estimation, especially when a small pilot length and a high reliability on the user identification are required.
By employing a step function for the proposed transmission control scheme, we then design an MMSE denoiser and modify the AMP algorithm to jointly identify the active users and estimate their channels.
%
%An MMSE denoiser was designed for the modified AMP algorithm, with a step transmission control function employed at the transmitter.
%
Additionally, we derive the false alarm probability and the missed detection probability to characterize the user identification performance for the proposed scheme.
We also analyze the packet delay and the network throughput.
Based on the analysis, we optimize the transmission control threshold of the step function to maximize the network throughput.
%
%It was verified that the analytical results matched well with simulation results.
%
%Simulation results demonstrated that compared to the conventional scheme without transmission control, the proposed scheme can significantly improve the user identification and channel estimation performance, reduce the average delay, and enhance the network throughput.
%
%
%we first propose a transmission control scheme to enhance the system sparsity and then to achieve an improved performance for the user activity identification and channel estimation, especially when a small pilot length and a high reliability are required.
The main contributions of this work are summarized below:

1. We propose a simple transmission control scheme at the transmitter to improve the performance of joint user identification and channel estimation (JUICE).
In the scheme, each user decides to transmit a packet or postpone its transmission based on a transmission control function of its instantaneous local CSI, when there is a transmission demand.
We design the function, so that the users with better channel gains have a higher probability to transmit their packets, and vice versa.
This effectively postpones the transmissions of users with small channel gains and therefore enhances the sparsity of user activity.
2. We modify an AMP algorithm to jointly identify the users' activity and estimate their channels at the receiver, for the system with a step transmission control function.
The channel distribution experienced by the receiver is first derived based on the adopted step transmission control function.
Then, we design an MMSE denoiser to modify the AMP algorithm at the receiver.

3. We derive the false alarm and the missed detection probabilities of user identification for the proposed scheme, by using the state evolution \cite{Donoho10itw}.
Based on the user identification performance, we obtain closed-form expressions of the average packet delay and the network throughput.
Moreover, we optimize the transmission control function to maximize the network throughput.

4. We verify that our analytical results match well with simulation results.
We demonstrate that compared to the conventional scheme without transmission control in \cite{Chen18}, the proposed scheme can significantly improve the system performance for mMTC, in terms of the missed detection probability and the normalized mean squared error (NMSE) of channel estimation.
In addition, we show that the average packet delay is reduced and the network throughput is enhanced for the proposed scheme.

\section{System Model}
Consider $N$ potential users who may transmit packets to a receiver through a common channel.
Both the receiver and users are equipped with a single antenna\footnote{Note that, it is the first work to propose the transmission control scheme for the joint user identification and channel estimation with compressed sensing. Therefore, we use a simple case where the base station has a single antenna, to facilitate the presentation and highlight insights of the proposed scheme for practical implementations.
	For the case where the base station has multiple antennas \cite{Liu18ma, Senel17globecom}, we will consider it in the further work.}.
We assume that in a time slot each user has a transmission demand with a probability $\epsilon$, $0\leq \epsilon\leq 1$, in an i.i.d. manner.
When user $n$, $n \in \{1,2,\ldots,N\}$, has a transmission demand, it transmits its packet with an average transmission probability $\lambda_{n}$, which is determined by our proposed transmission control scheme.
%
%This is different from the existing work in \cite{Chen17,liang17}, where a user will transmit the packet once it has a transmission demand, i.e., $\lambda_{n}=1$, $\forall n$.
%
In this work, we define an active user as a user who has a transmission demand and transmits its packet to the receiver.
Let $a_{n}\in \{0,1\}$ indicate the activity of user $n$.
In particular, if $a_{n}=1$, user $n$ is active. Otherwise, user $n$ is inactive.
As a result, we have $\mathrm{Pr}(a_{n}=1)=\epsilon\lambda_{n}$ and $\mathrm{Pr}(a_{n}=0)=1-\epsilon\lambda_{n}$, where $\epsilon\lambda_{n}$ is called the active probability of user $n$. The set of active users is given by
\begin{align}
\mathcal{A}=\{n: a_{n}=1, n=1,2,\ldots,N\},
\label{set_users_withpackets}
\end{align}
and the number of active users is $K=|\mathcal{A}|$.

If user $n$ is active, it will transmit a packet, which includes an $M$-length pilot sequence $\mathbf{s}_{n}\in \mathbb{C}^{M\times 1}$ and the information data, to the receiver within a time slot.
Otherwise, it will keep silent.
%
%Note that, in this paper, we only focus on the JUICE problem based on the pilot transmission, while the packet detection problem will be considered in our future work.
%
Then, in a specific time slot, the received signal during pilot transmission is written as
\begin{align}
\mathbf{y}&=\sum_{n=1}^{N}\mathbf{s}_{n}a_{n}h_{n}+\mathbf{w}=\sum_{n=1}^{N}\mathbf{s}_{n}x_{n}+\mathbf{w}
=\mathbf{S}\mathbf{x}+\mathbf{w},
\label{yp_model}
\end{align}
where $h_{n} \in \mathbb{C}$ denotes the channel fading coefficient from user $n$ to the receiver in the time slot\footnote{We assume that the channel fading coefficient $h_{n}$ remains constant during a time slot in this work.} and $x_{n} = a_{n}h_{n}$ captures the joint effect of user activity and channel fading of user $n$.
Vector $\mathbf{y}\in \mathbb{C}^{M\times 1}$ denotes the received signal during $M$ pilot symbols and $\mathbf{w}\in \mathbb{C}^{M\times 1}$ is the AWGN with each entry following the distribution $w_{n}\sim\mathcal{CN}(0,\sigma^{2}_{w})$, where $\sigma^{2}_{w}$ denotes the noise variance.
The pilot matrix $\mathbf{S}=[\mathbf{s}_{1},\mathbf{s}_{2},\ldots,\mathbf{s}_{N}] \in \mathbb{C}^{M\times N}$ collects all users' pilots and the vector $\mathbf{x}=[x_{1},x_{2},\ldots,x_{N}]^{T} \in \mathbb{C}^{N\times 1}$ collects all unknown variables $x_{n}$.

The channel from user $n$ to the receiver is modeled by $h_{n}=\sqrt{\beta}g_{n}$, where $g_{n}\sim\mathcal{CN}(0,1)$ is the Rayleigh block fading component and $\beta$ is the large-scale fading component.
We assume that all users have the same channel distribution, i.e., the same large-scale fading coefficient $\beta$, and the coefficient $\beta$ is known by the receiver as the prior information\footnote{In this work, we consider symmetric users with the identical large-scale fading to simplify the AMP algorithm design, which can be achieved by using the long term power control to compensate the different large-scale fading among users.
	The proposed scheme can be easily extended to the scenario with different $\beta$ among users.}.
In addition, it is assumed that each user knows its own instantaneous channel\footnote{Note that, the time division duplex (TDD) system is considered and the channel reciprocity is assumed to be held in this work. In particular, each user's uplink channel coefficient can be obtained by exploiting broadcast pilot from the base station, like the narrowband reference signal (NRS) of narrowband machine-type devices \cite{Schwarz16,Liberg17,Wang17}. Moreover, we assume that the perfect CSI is used for the user's transmission control in this work. It can provide a performance upper bound and reveal more insights for exploiting the transmission control in the user identification and channel estimation scheme. For the case with imperfect CSI, the proposed transmission control scheme is still applicable and can improve the user activity identification and channel estimation performance, which will be discussed in the future work.} $h_{n}$ and the receiver has the pilot matrix $\mathbf{S}$.
Based on the received signal $\mathbf{y}$ and the pilot matrix $\mathbf{S}$, the receiver jointly identifies active users and estimates their channels, i.e., recovers the vector $\mathbf{x}$ in Eq. \eqref{yp_model}.
%
%This problem is fundamentally different from the channel estimation problem in conventional multiple access scenarios \cite{Wei17,Yang16}, where the user activity is known in advance and only channel coefficients need to be estimated by the receiver. %
%
This is a challenging problem, since the length of pilot sequences $M$ is much smaller than the number of potential users $N$ in a massive connectivity system.
%
%As a result, Eq. \eqref{yp_model} is a highly underdetermined system.
%
However, due to the sporadic transmission in mMTC \cite{Bockelmann16,Monsees15}, the number of active users at a time is much smaller than the number of potential users, i.e., $K \ll N$, thereby leading to a sparse vector $\mathbf{x}$.
Therefore, recovering $\mathbf{x}$ can be formulated as a compressed sensing problem \cite{Donoho06,Candes06}.
%
%Due to the large-scale feature of mMTC, a computationally efficient AMP algorithm \cite{Donoho09,Donoho10} is employed to recover the vector $\mathbf{x}$ in this paper. \color{black}
%
Note that, due to $M \ll N$, pilot sequences of all potential users are mutually non-orthogonal but it is assumed that each user has an unique pilot.
Moreover, we assume that the pilot sequence $\mathbf{s}_{n}$, $\forall n$, is constructed with each entry $s_{n,m}\sim\mathcal{CN}(0,\frac{1}{M})$, $m\in\{1,2,\ldots,M\}$, according to the CS theory\cite{Choi17,Eldar12}.
%

%However, it is known that when the pilot length decreases or the number of active users increases, the performance of AMP algorithm will degrade, thereby resulting in a decreased performance of JUICE \cite{Donoho09}.
%%
%Moreover, it is motivated by the fact that a signal with the higher signal-to-noise ratio (SNR) can be successfully detected and recovered by the AMP algorithm with a higher probability.
%%
%On the other hand, the signal with a lower SNR can be successfully recovered with a lower probability. Even worse, its existence also reduces the sparsity of unknown signal vector and negatively affects the recovery of other signals.
%%
%Therefore, in this paper, we propose a transmission control scheme to limit the number of active users, and thus to improve the performance of JUICE.
%%
%In particular, when a user has the transmission demand, whether it transmits or not is determined by its local channel state information (CSI) for the proposed transmission control scheme.
%%
%The better channel condition that a user has, the larger probability that it transmits its packet.
%%
%In this way, it is expected to achieve a better performance of JUICE, due to the enhanced sparsity of unknown signal vector and the high SNR of each received signal.
%%
We also note that the considered system model with transmission control ($0\leq\lambda_{n}\leq 1$) is general.
It includes the conventional system model without transmission control, i.e., $\lambda_{n}= 1$, as its special case.
It implies that the designed AMP algorithm and the presented performance analysis in this work can be applied to the systems without transmission control.

\vspace{-5mm}
\section{The Proposed Transmission Control Scheme}
In this section, we present our proposed transmission control scheme.
To characterize the effect of the introduced transmission control strategy on the design of AMP algorithm, we derive the channel distribution experienced by the receiver.
It serves as a building block for the design of MMSE denoiser in the AMP algorithm, as shown in Section \ref{amp}.
\subsection{Transmission Control Scheme}
\begin{figure}[!t]
	\par
	\begin{center}
		{\includegraphics[width=5in]{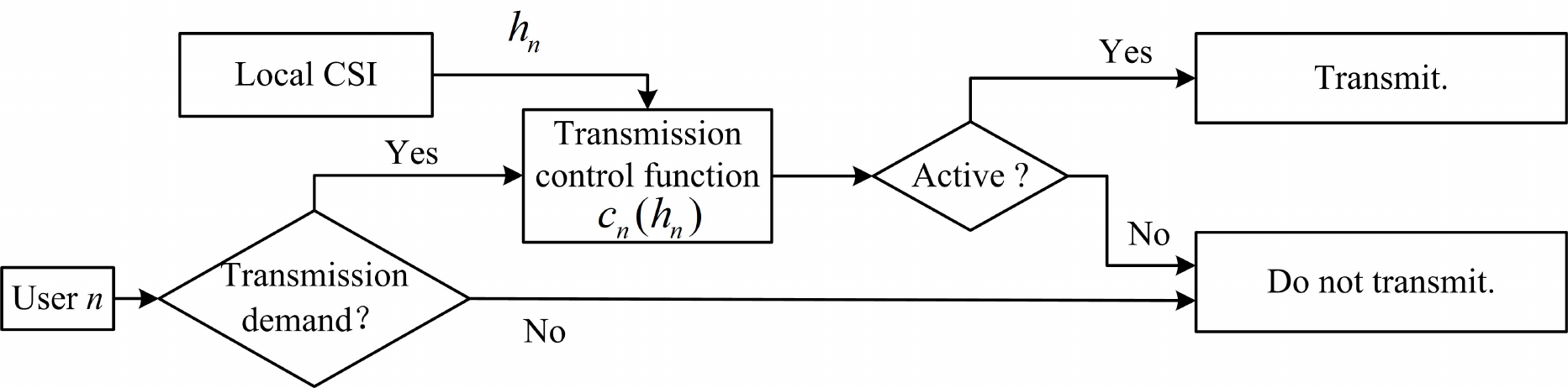}}
	\end{center}\vspace{-2mm}
	\caption{The proposed transmission control scheme for user $n$, $\forall n$.}
	\label{TransmissionControlStrategy}
\end{figure}

Fig. \ref{TransmissionControlStrategy} illustrates our proposed transmission control scheme at user $n$, $\forall n \in\{1,2,\ldots,N\}$.
It can be seen from the figure that user $n$ computes its instant transmission probability $c_n(h_{n})$ through a designed transmission control function $c_n(\cdot)$ and its local CSI $h_{n}$, $\forall n$, when it has a transmission demand.
Then, the user transmits its packet with the probability $c_n(h_{n})$.
As a result, the average transmission probability $\lambda_{n}$ can be given by
\begin{align}
\lambda_{n}=\int c_n(u)P_{h_n}(u)du,
\label{prob_transmission}
\end{align}
where $P_{h_{n}}(u)$ denotes the probability density function (PDF) of the channel coefficient of user $n$. Since $h_{n}$ is a complex channel coefficient, $u$ refers to a complex number in Eq. \eqref{prob_transmission}.
%
%Note that, $\lambda_{n}$ is given by averaging $c_n(h_{n})$ over the distribution of fading channel, and it is very important for the JUICE algorithm design in the following.
%

The transmission control function $c_n(\cdot)$ can be designed in a way such that users with better channel gains have higher transmission probabilities and vice versa.
This leads to the enhanced sparsity of received signals, as users with small channel gains may postpone their transmissions.
%
%As a result, the JUICE performance can be improved for the proposed scheme.
%
Since users have i.i.d. channels, we consider that all the users have the same transmission control function $c(\cdot)$ in this work.
Note that, for the general case where users have different channel distributions, i.e., different $\beta_{n}$ for $n \in \{1,\ldots,N\}$, users can exploit their own instantaneous channel coefficient normalized by $\beta_{n}$ for the transmission control function, i.e., $c\left(\frac{h_{n}}{\sqrt{\beta_{n}}}\right)$.
As a result, each user's control threshold in Eq. \eqref{step_cntrl_function} is proportional to the large-scale fading coefficient of its channel, and all users can achieve an identical average transmission probability for the proposed transmission control scheme.
Moreover, the following design and analysis that are based on $c(\cdot)$ can be directly applied to this general case\footnote{Note that, designing the transmission control scheme that is only associated with the users' channel gains is insufficient, in particular when considering the fairness among users. In fact, the users' quality of service (QoS) requirement is also an essential factor to be considered for designing the transmission control scheme. As a first work to introduce the transmission control into the JUICE, and we mainly focus on investigating the effects of transmission control on the JUICE performance. Therefore, we consider the simple transmission control scheme that is only determined by the users' channel gains in this work. For the future work, we will propose a more practical transmission control scheme by taking into account both the users' channels and their QoS requirements.}.
Furthermore, to emphasize the effectiveness of the proposed transmission control scheme, we consider the simplest case with $c(\cdot)$ as a step function\footnote{It is noteworthy that there are many transmission control function candidates to be chosen and the employed threshold-based transmission control function may not be the optimal one in general. However, selecting an optimal transmission control function to minimize the MSE of estimated vectors is challenging, since it is difficult to quantize the effect of selecting different transmission control functions on the AMP algorithm performance, i.e., the MSE of estimation. Therefore, we employ the threshold-based function that is commonly used in \cite{Bhaskaran09,Davis11}, to shed light on how a transmission control can be used to improve the performance of AMP based joint user activity identification and channel estimation in this work.}, given by
\begin{align}
c(u) = \left\{ {\begin{array}{*{20}{c}}
	{\begin{array}{*{20}{c}}
		1&{|u| > \varsigma }
		\end{array}}\\
	{\begin{array}{*{20}{c}}
		0&{|u| \le \varsigma }
		\end{array}}
	\end{array}} \right.,
\label{step_cntrl_function}
\end{align}
where $\varsigma$ is the control threshold.
Eq. \eqref{step_cntrl_function} means that if the instant channel gain of user $n$, $|h_n|$, is larger than the control threshold $\varsigma$, it will be active once it has a transmission demand. Otherwise, user $n$ will be inactive, even though it has a transmission demand.
Based on the step transmission control function in Eq. \eqref{step_cntrl_function}, the transmission control design is equivalent to the threshold design, which will be given in Section \ref{OptimalThresholdDesign}.
Note that, each user can control its activity distributedly based on its own instant channel realization.
We also note that the transmission control function $c(\cdot)$ can be generalized to any other more complicated functions.
\subsection{Channel Distribution Experienced by the Receiver}
According to Eqs. \eqref{prob_transmission} and \eqref{step_cntrl_function}, the average transmission probability of user $n$ can be given by
\begin{align}
\lambda&=\int_{|u|>\varsigma}P_{h_n}(u)du
%&\overset{(a)}{=}\int_{|h|>\varsigma}\frac{1}{\pi \beta}\exp\left(-\frac{|h|^2}{\beta}\right)dh \notag \\
\overset{(a)}{=}\exp\left(-\frac{{\varsigma}^{2}}{\beta}\right), \forall n,
\label{step_tx_prob}
\end{align}
where the equality $(a)$ follows the complementary cumulative distribution function (CCDF) of Rayleigh distribution.
Due to the fact that $P_{h_1}(u) = P_{h_2}(u), \ldots, =P_{h_N}(u)$, all users have the same average transmission probability $\lambda$ and the subscript $n$ of $\lambda_{n}$ is omitted in Eq. \eqref{step_tx_prob}.
It can be seen from Eq. \eqref{step_tx_prob} that the average transmission probability $\lambda$ decreases with the increasing control threshold $\varsigma$ for a given $\beta$.
Then, adjusting the threshold $\varsigma$ can effectively control each user's active probability $\epsilon\lambda$ and thus change the sparsity of $\mathbf{x}$.

Now, we define a new random variable $\hat {h}_n$ to characterize the channel distribution of user $n$ experienced by the receiver.
Since the receiver can experience a realization of $\hat {h}_n$ only if user $n$ is active, i.e., $a_{n}=1$, the cumulative distribution function (CDF) of $\hat {h}_n$ can be given by the following conditional distribution:
\begin{align}
{F}_{\hat{h}_n}(u)&=F_{h_n|a_{n}=1}\left(u\right)
= \frac{\mathrm{Pr}\{h_n\le u,a_{n}=1\}}{\mathrm{Pr}\{ a_{n}=1\}} = \frac{\int_{z \le u}c(z)P_{h_{n}}(z)dz}{\int c(z)P_{h_{n}}(z)dz}, \forall n.
\label{CDFhHAT}
\end{align}
% Details 12-02 2017
%\begin{align}
%\begin{array}{l}
%f(x) = \int_{z \le x} {c\left( z \right)} {P_{{h_n}}}\left( z \right)dz\mathop  = \limits^{z = a + bj} \\
%\frac{{\partial f(x)}}{{\partial {\mathop{\rm Re}\nolimits} \left\{ x \right\}}} = \int\limits_{b \le {\mathop{\rm Im}\nolimits} \left\{ x \right\}} {c\left( {{\mathop{\rm Re}\nolimits} \left\{ x \right\},b} \right){P_{{h_n}}}\left( {{\mathop{\rm Re}\nolimits} \left\{ x \right\},b} \right)db} \\
%\frac{{\partial f(x)}}{{\partial {\mathop{\rm Im}\nolimits} \left\{ x \right\}}} = \int\limits_{a \le {\mathop{\rm Re}\nolimits} \left\{ x \right\}} {c\left( {a,{\mathop{\rm Im}\nolimits} \left\{ x \right\}} \right){P_{{h_n}}}\left( {a,{\mathop{\rm Im}\nolimits} \left\{ x \right\}} \right)da}
%\end{array}
%\end{align}
Note that, since $u$ refers to a complex number, $\mathrm{Pr}\{h_n\le u,a_{n}=1\}$ represents the probability $\mathrm{Pr}\{\mathrm{Re}(h_n)\le \mathrm{Re}(u), \mathrm{Im}(h_n)\le \mathrm{Im}(u), a_{n}=1\}$ in Eq. \eqref{CDFhHAT}, for simplification.
As a result, the PDF of $\hat {h}_n$ can be given by
\begin{align}
{P}_{\hat{h}_n}(u)&=\frac{c(u)P_{h_n}(u)}{\lambda}\overset{(a)}{=} \left\{ {\begin{array}{*{20}{c}}
	{\frac{1}{\lambda }\frac{1}{{\pi \beta }}\exp \left( { - \frac{{|u{|^2}}}{\beta }} \right)}&{\left| u \right| > \varsigma }\\
	0&{\left| u \right| \le \varsigma }
	\end{array}} \right. ,
\label{step_post_distri_h}
\end{align}
where the equality $(a)$ is obtained for $c(\cdot)$ given as a step function in Eq. \eqref{step_cntrl_function}.
It can be seen from Eq. \eqref{step_post_distri_h} that all the $\hat{h}_n$, $n \in \{1,\ldots,N\}$, possess the same truncated complex Gaussian distribution \cite{Ryan11} with a truncating threshold at $\varsigma$, i.e., the PDF of $\hat{h}_n$ is a scaled Gaussian function when $|u| > \varsigma$ and is zero otherwise.
Note that, when $\varsigma = 0$, the proposed scheme degenerates to the conventional scheme without transmission control and $\hat{h}_n = {h_n}$.
%
%\color{black}Moreover, the design of MMSE denoiser in the AMP algorithm is based on the channel distribution experienced by the receiver.
%%
%Then, exploiting the proposed transmission control scheme can affect the design of AMP algorithm, through manipulating the channel distribution experienced by the receiver,
%%
%%the channel distribution experienced by the receiver is employed to design the MMSE denoiser of the AMP.
%%
%Therefore, for the proposed transmission control scheme we will design the AMP algorithm to jointly identify users' activity and estimate their channels in the next section.\color{black}
%
%Then, we can observe that the channel distribution experienced by the receiver is manipulated by the designed transmission control function.
%%
%Therefore, it opens the opportunity for us to design the transmission control function to adjust the channel distribution experienced by the receiver, such that the performance of JUICE can be improved.

Based on the distribution of $\hat{h}_n$, we derive the distribution of unknown variable $x_{n}=a_{n}h_{n}$, which serves as a prior information for the MMSE denoiser design in the following\footnote{Note that, the distributions of $a_{n}$, $h_{n}$, and $x_{n}$ are considered to be known in this work. In practice, these distributions may be unknown and need to be learned by using the machine learning methods, which will be considered in the future work.}.
In particular, the CDF of $x_{n}$ can be given by
\begin{align}
{F}_{x_n}(u)&= \mathrm{Pr}\{{a_n}{h_n}\le u\}={\Pr \left\{ {{a_n} = 1,{h_n} \le u} \right\}} + {\Pr \left\{ {{a_n} = 0,0 \le u} \right\}},
\label{CDFx}
\end{align}
where
\begin{align}
{\Pr \left\{ {{a_n} = 1,{h_n} \le u} \right\}} &= {\Pr \left\{ {{a_n} = 1} \right\}}{\Pr \left\{ {h_n} \le u| {{a_n} = 1} \right\}} \notag \\
&= {\Pr \left\{ {{a_n} = 1} \right\}}{F}_{\hat{h}_n}(u).
\label{Probability2}
\end{align}
Since $\mathrm{Pr}(a_{n}=1)=\epsilon\lambda$ and $\mathrm{Pr}(a_{n}=0)=1-\epsilon\lambda$, the PDF of $x_{n}$ can be given by
\begin{align}
P_{x_{n}}(u)=\left(1-\epsilon\lambda\right)\delta_{0}(u)+\epsilon\lambda P_{\hat{h}_{n}}(u),
\label{distri_x}
\end{align}
where $\delta_{0}(u)$ denotes the point mass measure at zero and it is given by
\begin{align}
{\delta _0}(u) = \left\{ {\begin{array}{*{20}{c}}
	1&{u = 0},\\
	0&{u \ne 0}.
	\end{array}} \right.
\label{point_mass_measure_function}
\end{align}
It can be observed from Eq. \eqref{distri_x} that the distribution of $x_{n}$ is a mixture of a Bernoulli distribution for $a_{n}$ and a truncated complex Gaussian distribution for $\hat{h}_{n}$.
It can also be seen from Eqs. \eqref{step_post_distri_h} and \eqref{distri_x} that the distribution of unknown variable $x_{n}$ in the proposed transmission control scheme depends on the transmission control function $c(\cdot)$ through the PDF of $\hat{h}_{n}$.
This dependance allows us to steer the distribution of $x_{n}$ by designing the transmission control function, thereby affecting the MMSE denoiser design for the AMP algorithm, which will be presented in the next section.
\section{AMP based User Identification and Channel Estimation}
In this section, we first present an overview of the general AMP framework, including the iterative algorithm and the state evolution.
Then, for the proposed transmission control scheme we design the AMP algorithm to jointly identify the user activity and estimate their channels.
\subsection{Design of MMSE Denoiser for AMP Algorithm}
From the iterative functions of the AMP algorithm, given by Eqs. \eqref{amp_1} and \eqref{amp_2}, and the state evolution in Eq. \eqref{state_evolo}, it is known that the denoiser function $\eta(\cdot)$ acts as an important component in the AMP algorithm and its design significantly affects the performance for the users' activity detection and channel estimation.
Generally, the denoiser function receives the variable ${\bf{\tilde{x}}}^{t}$, modeled as the unknown variable $\bf{x}$ plus noise, and outputs an estimate of the unknown variable, i.e., ${\bf{\hat{x}}}^{t+1}$.
The denoiser function is typically designed to minimize the MSE of the output estimate at each iteration by effectively removing the noise.
It implies that the small value $\tilde{x}^{t}_{n}$ can be decreased to approach zero by the denoiser function, thereby obtaining a sparse output vector $\hat{\mathbf{x}}^{t+1}$ as an estimate of the unknown vector $\mathbf{x}$ for the next iteration.
In the literature, a soft thresholding denoiser function was proposed by employing a minimax framework in \cite{Donoho09}, where the prior information of unknown vector $\mathbf{x}$ is not exploited.
When utilizing the distribution of unknown vector $\mathbf{x}$, the MMSE denoiser is obtained by using the Bayesian framework in \cite{Donoho10,Chen17}.
In this work, we aim to employ the MMSE denoiser for the designed AMP algorithm.
However, the design of MMSE denoiser depends on the distribution of $\mathbf{x}$, which is in turn associated with the proposed transmission control function $c(\cdot)$.
Therefore, the conventional MMSE denoiser proposed by \cite{Donoho10,Chen17} is not applicable to our proposed transmission control scheme and we need to propose an MMSE denoiser in this work.

In order to facilitate the derivation of the denoiser function, we define a random variable $\bar{x}_{n}^{t}$ as $\bar{x}_{n}^{t}=\hat{h}_{n}+\tau_{t}v$, where $\hat{h}_{n}$ follows the distribution in Eq. \eqref{step_post_distri_h}.
It can be seen that there is $\bar{x}_{n}^{t}=\tilde{x}_{n}^{t}$ when $a_n = 1$.
In Lemma \ref{lemma_dis_x_bar}, we first derive the distribution of $\bar{x}_{n}^{t}$, and then we design the MMSE denoiser for the proposed transmission control scheme in Theorem \ref{theo_denoiser_tx_cntrl}.
\begin{lemm} Given the average transmission probability $\lambda$, the large-scale fading coefficient $\beta$, and the control threshold $\varsigma$, the PDF of the random variable $\bar{x}_{n}^{t}$ is given by
\vspace{-5mm}	
\begin{align}
	{P_{\bar{x}_{n}^{t}}}\left(u\right)
	=\frac{{{\mathrm{Q}_1}\left(\frac{{|{\mu_{t}u}|}}{{\sqrt {{{\sigma_t^2} \mathord{\left/
								{\vphantom {{\sigma_t^2} 2}} \right.
								\kern-\nulldelimiterspace} 2}} }},\frac{\varsigma }{{\sqrt {{{\sigma_t^2} \mathord{\left/
								{\vphantom {{\sigma_t^2} 2}} \right.
								\kern-\nulldelimiterspace} 2}} }}\right)}}{{\lambda \pi \left(\beta  + \tau_{t}^2\right)}}\exp \left( - \frac{{\left|u\right|^2}}{{\beta  + \tau_{t}^2}}\right),
	\label{dis_x_bar}
	\end{align}\vspace{-5mm}
	where $\mathrm{Q}_1(\cdot)$ is the Marcum-Q-function with order one, $\mu_{t}=\frac{\beta}{\beta+\tau_{t}^2}$, and $\sigma_t^2=\frac{\beta\tau_{t}^2}{\beta+\tau_{t}^2}$.
	\label{lemma_dis_x_bar}
\end{lemm}

\emph{\quad Proof: } Please refer to Appendix \ref{AppendixA3}. \QEDA
%\begin{proof}
%	See Appendix A.
%\end{proof}

%
From Lemma \ref{lemma_dis_x_bar}, we can observe that for the proposed transmission control scheme, the PDF of $\bar{x}_{n}^{t}$ is equal to the PDF of a complex Gaussian distribution, i.e., $\mathcal{CN}(0,\beta+\tau_{t}^2)$, multiplied by a scalar $\frac{{{\mathrm{Q}_1}\left(\frac{{|{\mu_{t}u}|}}{{\sqrt {{{\sigma_t^2} \mathord{\left/
							{\vphantom {{\sigma_t^2} 2}} \right.
							\kern-\nulldelimiterspace} 2}} }},\frac{\varsigma }{{\sqrt {{{\sigma_t^2} \mathord{\left/
							{\vphantom {{\sigma_t^2} 2}} \right.
							\kern-\nulldelimiterspace} 2}} }}\right)}}{\lambda}$. %
This is different from the case without transmission control, where $\bar{x}_{n}^{t}$ follows the distribution $\mathcal{CN}(0,\beta+\tau_{t}^2)$.
In addition, the scalar is an increasing function of $|u|$.
It means that compared to the standard complex Gaussian variable, $\bar{x}_{n}^{t}$ possesses a higher probability density when it is large and vice versa.
This is consistent with the employed transmission control scheme, i.e., the better channel condition that a user has, the larger probability that it transmits its packet, and vice versa.
%
%Note that, in Eq. \eqref{dis_x_bar}, $\mu_{t}x$ and $\sigma_t^2$ can be interpreted as the posterior mean and variance of channel coefficient for the case without transmission control, respectively.

%
%$\tau_{t}$ can be obtained from the empirical estimate $\hat{\tau}_{t}=\frac{1}{\sqrt{M}}||\mathbf{z}^{t}||_{2}$ in practice, where $||\cdot||_{2}$ denotes the $\ell_{2}$ norm of a vector.

%Based on the distribution of $\bar{x}_{n}^{t}$ in Lemma \ref{lemma_dis_x_bar}, we design the MMSE denoiser for the proposed threshold based transmission control scheme in Theorem \ref{theo_denoiser_tx_cntrl}.

\begin{theo}
	For the proposed transmission control scheme with a control threshold $\varsigma$, the MMSE denoiser of the AMP algorithm is given by
	\begin{align}
	\eta \left(u,\beta,t\right)=\frac{\mu_{t}u}{{\alpha_{1}\left(u,t\right) + \alpha_{2}\left(u,t\right)\frac{{1 - \epsilon \lambda}}{{\epsilon}}\frac{{\beta  + \tau_{t}^2}}{{\tau_{t}^2}}\exp \left( { - \frac{{\beta |u|^2}}{{\tau_{t}^2(\beta  + \tau_{t}^2)}}} \right)}},
	\label{denoiser_tx_cntrl}
	\end{align}
	where $\alpha_{1}\left(u,t\right)=\frac{{{\mathrm{Q}_1}\left(\frac{|\mu_{t}u|}{{\sqrt {{{\sigma_{t}^2} \mathord{\left/
								{\vphantom {{\sigma_t^2} 2}} \right.
								\kern-\nulldelimiterspace} 2}} }},\frac{\varsigma }{{\sqrt {{{\sigma_{t}^2} \mathord{\left/
								{\vphantom {{\sigma_{t}^2} 2}} \right.
								\kern-\nulldelimiterspace} 2}} }}\right)}}{{{\mathrm{Q}_2}\left(\frac{|\mu_{t}u|}{{\sqrt {{{\sigma_{t}^2} \mathord{\left/
								{\vphantom {{\sigma_t^2} 2}} \right.
								\kern-\nulldelimiterspace} 2}} }},\frac{\varsigma }{{\sqrt {{{\sigma_{t}^2} \mathord{\left/
								{\vphantom {{\sigma_{t}^2} 2}} \right.
								\kern-\nulldelimiterspace} 2}} }}\right)}}$, $\alpha_{2}\left(u,t\right)=\frac{1}{{{\mathrm{Q}_2}\left(\frac{|\mu_{t}u|}{{\sqrt {{{\sigma_{t}^2} \mathord{\left/
								{\vphantom {{\sigma_t^2} 2}} \right.
								\kern-\nulldelimiterspace} 2}} }},\frac{\varsigma }{{\sqrt {{{\sigma_{t}^2} \mathord{\left/
								{\vphantom {{\sigma_{t}^2} 2}} \right.
								\kern-\nulldelimiterspace} 2}} }}\right)}}$, and $\mathrm{Q}_2(\cdot)$ is the Marcum-Q function with order two.
	
	% Note that, since $x$ is a realization of $\tilde{x}^{t}_{n}$, $\mu_{t}x$ depends on the user index $n$ and $\mu_{t}x$ also depends on the iteration index $t$ through $\tau_{t}$.
	\label{theo_denoiser_tx_cntrl}
\end{theo}

\emph{\quad Proof: } Please refer to Appendix \ref{AppendixB3}. \QEDA
%\begin{proof}
%	See Appendix B.
%\end{proof}

\begin{figure}[!t]
	\par
	\begin{center}
		{\includegraphics[width=3.4in]{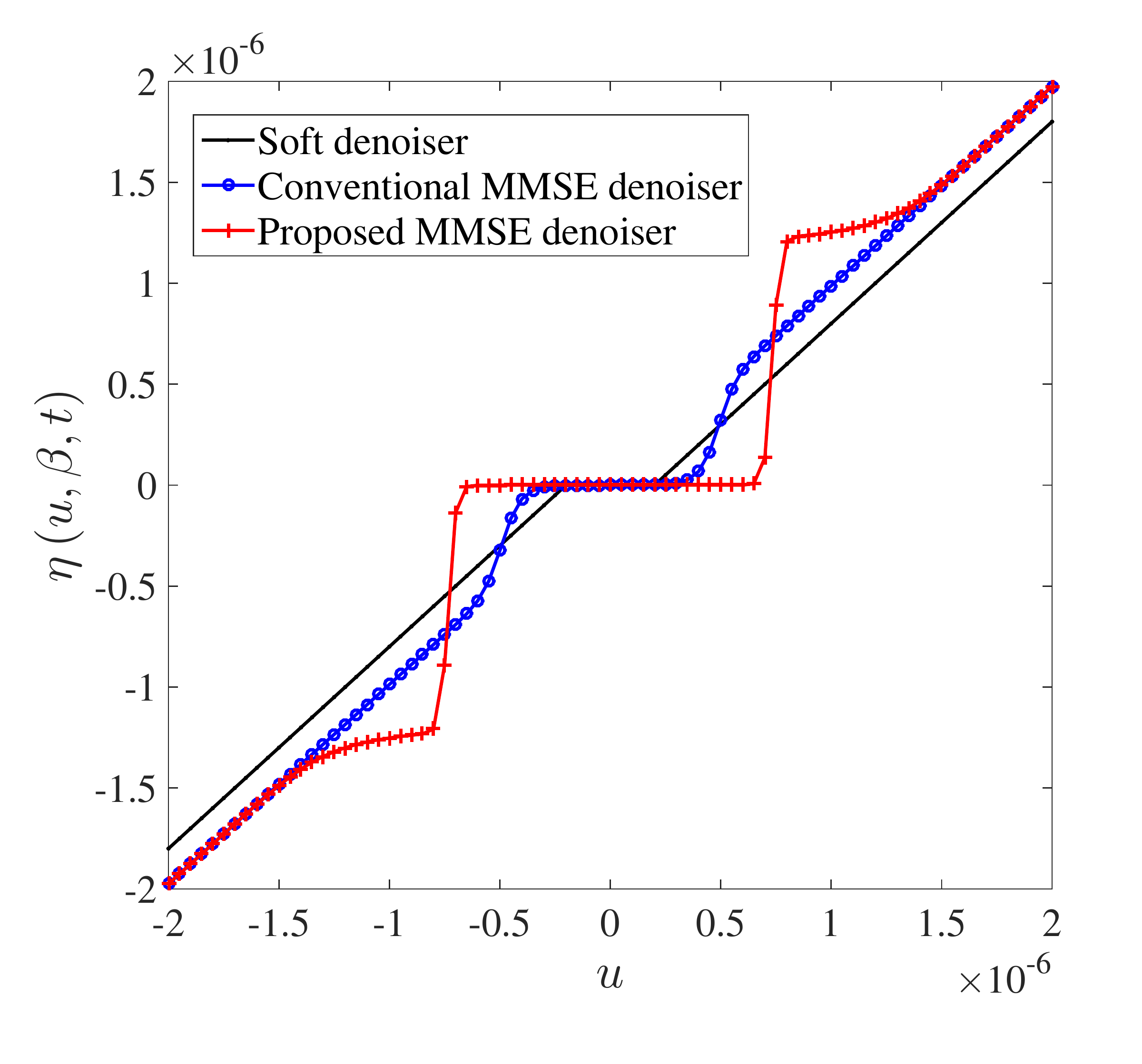}}
	\end{center}\vspace{-10mm}
	\caption{Comparison of the proposed denoiser, the conventional MMSE denoiser, and the soft denoiser, where $\tau_{t}=2\times 10^{-7}$, $\beta=-115$ dB, $\epsilon=0.24$, $\lambda=0.63$, and $\varsigma=1.2\times 10^{-6}$. The soft denoiser is given by $\eta^\mathrm{soft}(u)=\left(u-\frac{\theta u}{|u|}\right)\mathbf{1}(|u|>\theta)$, where $\mathbf{1}(\cdot)$ is the indicator function.}
	\label{fig_denoiser_soft_conventionalmmse_designedmmse_v2}
\end{figure}

Compared to the conventional MMSE denoiser without transmission control \cite{Chen17}, the main difference that our proposed MMSE denoiser presents is the two functions $\alpha_{1}\left(u,t\right)$ and $\alpha_{2}\left(u,t\right)$, as shown in Theorem \ref{theo_denoiser_tx_cntrl}.
For the MMSE denoiser without transmission control, the two functions are constant ones, i.e., $\alpha_{1}\left(u,t\right)=1$ and $\alpha_{2}\left(u,t\right)=1$.
In contrast, for our proposed MMSE denoiser, the two functions $0<\alpha_{1}\left(u,t\right)\leq 1$ and $\alpha_{2}\left(u,t\right)\geq 1$ vary with $|u|$, and $\alpha_{2}\left(u,t\right)$ is a decreasing function of $|u|$.
In particular, in Eq. \eqref{denoiser_tx_cntrl}, for a small input $|u|$, i.e., $|u| \ll {{\varsigma} \mathord{\left/
		{\vphantom {{\varsigma} \mu_{t}}} \right.
		\kern-\nulldelimiterspace} \mu_{t}}$, we have $\alpha_{1}\left(u,t\right) \to 1$ and $\alpha_{2}\left(u,t\right)\to \infty$.
As a result, the output of the proposed MMSE denoiser $\eta \left(u,\beta,t\right)$ approaches zero.
On the other hand, for a large input $|u|$, i.e., $|u| \gg {{\varsigma} \mathord{\left/
		{\vphantom {{\varsigma} \mu_{t}}} \right.
		\kern-\nulldelimiterspace} \mu_{t}}$, both functions $\alpha_{1}\left(u,t\right)$ and $\alpha_{2}\left(u,t\right)$ approach one, and the proposed denoiser degenerates to the conventional MMSE denoiser without transmission control, which will be verified by Fig. \ref{fig_denoiser_soft_conventionalmmse_designedmmse_v2}.
In other words, in the high channel gain regime, the proposed transmission control function has a limited effect on the design of MMSE denoiser.
It is caused by the fact that the proposed transmission control scheme almost does not affect users' transmissions when they have good channel conditions.

In Fig. \ref{fig_denoiser_soft_conventionalmmse_designedmmse_v2}, we compare the proposed MMSE denoiser, the conventional MMSE denoiser \cite{Donoho10,Chen17}, and the soft denoiser \cite{Donoho09}, by showing the relation between $\eta \left(u,\beta,t\right)$ and its input $u$.
We can observe that similar to the conventional MMSE denoiser and the soft denoiser, the proposed MMSE denoiser can shrink the small input towards zero, thereby enhancing the sparsity of the estimate $\hat{\mathbf{x}}^{t+1}$.
However, for the proposed MMSE denoiser the shrinkage occurs in a wider range of small inputs, compared to the conventional MMSE denoiser and the soft denoiser.
This is because $\alpha_{2}\left(u,t\right)$ acts as a decreasing function of $|u|$ in the denominator of the proposed MMSE denoiser $\eta \left(u,\beta,t\right)$, and accelerates the decrease of $\eta \left(u,\beta,t\right)$ with decreasing $|u|$.
%
%Then, it means that the proposed MMSE denoiser function can further enhance the sparsity of the estimate $\hat{\mathbf{x}}^{t+1}$ and thus improve the algorithm performance, compared to the other two denoiser functions.
%
%Note that, it can be seen from Fig. \ref{fig_denoiser_soft_conventionalmmse_designedmmse_v2} that the output of the designed MMSE denoiser $\eta \left(x,\beta,t\right)$, i.e., the estimate of $x_{n}$ in the $(t+1)$-th iteration, $n \in \{1,\ldots,N\}$, jumps from zero to the control threshold and the negative control threshold as $x>0$ increases and $x<0$ decreases, respectively, since the unknown variable $x_{n}$ follows the mixture distribution of a Bernoulli distribution and a complex Gaussian distribution truncated at the control threshold.
%

\subsection{AMP based User Activity Identification and Channel Estimation}
In this part, we propose the AMP based user identification and channel estimation, which is summarized in \textbf{Algorithm} \ref{juice}.
\begin{table}[t]
	\begin{algorithm} [H]  % enter the algorithm environment
		\caption{JUICE in Random Access Systems for mMTC}     % give the algorithm a caption
		\label{juice}                             % and a label for \ref{} commands later in the document
		\begin{algorithmic} [1]
			% enter the algorithmic environment
			\STATE \textbf{INPUT}: The measurement $\mathbf{y}$, the pilot matrix $\mathbf{S}$, the large-scale fading coefficient $\beta$, the transmission demand probability $\epsilon$, the average transmission probability $\lambda$, and the control threshold $\varsigma$
			\STATE \textbf{OUTPUT}: The active user set $\mathcal{A}$ and the channel estimates ${\mathbf{h}}_{\mathcal{A}}$
			\STATE \textbf{Initialization}: $\mathbf{\hat{x}}^{0}=\mathbf{0}$, $\mathbf{z}^{0}=\mathbf{y}$, $t=0$, $\mathcal{A} = \emptyset$, ${\mathbf{h}}_{\mathcal{A}} = \mathbf{0}$, the tolerance error $o = 10^{-3}$, and the maximum number of iteration $t_{\rm{max}} = 50$
			\REPEAT
			\STATE Compute state variable in the $t$-th iteration, by using Eq. \eqref{EmpiricalState}
			\STATE Calculate the denoiser function input $\tilde{\mathbf{x}}^{t}$ by exploiting a matched filter on $\mathbf{z}^{t}$ and the current estimate $\hat{\mathbf{x}}^{t}$
			\STATE Calculate the estimate $\hat{\mathbf{x}}^{t+1}$ in the $(t+1)$-th iteration based on $\tilde{\mathbf{x}}^{t}$ by using the proposed MMSE denoiser in Eq. \eqref{denoiser_tx_cntrl}
			\STATE Calculate the residual $\mathbf{z}^{t+1}$ corresponding to the estimate $\hat{\mathbf{x}}^{t+1}$ based on Eq. \eqref{amp_2} and Eq. \eqref{derivative_denoiser}
			\STATE $t = t+1$
			\UNTIL
			$\frac{ \left\| \hat{\mathbf{x}}^{t} - \hat{\mathbf{x}}^{t-1}\right\| }{\left\| \hat{\mathbf{x}}^{t-1}\right\|} \le o$ or $t \ge t_{\rm{max}}$
			\FOR{$n = 1$, $n{+}{+}$, while $n \leq N$}
			\IF{$|\tilde{x}_{n}^{t}|> l_{n}$}
			\STATE User $n$ is detected as an active user and its channel estimate is $\hat{x}_{n}^{t}$, i.e., $\mathcal{A} = \mathcal{A} \bigcup \{n\}$ and ${\mathbf{h}}_{\mathcal{A}}\left(n\right) = \hat{x}_{n}^{t}$
			\ELSE
			\STATE User $n$ is detected as an inactive user and ${\mathbf{h}}_{\mathcal{A}}\left(n\right) = 0$
			\ENDIF
			\ENDFOR
		\end{algorithmic}
	\end{algorithm}
\end{table}
Firstly, the unknown vector $\mathbf{x}$ is recovered by exploiting the AMP algorithm, as shown in step $3-10$.
Based on the recovered vector, the user identification is then performed, as shown in step $11-17$.

According to the proposed MMSE denoiser in Theorem \ref{theo_denoiser_tx_cntrl}, we derive the first-order derivative of the denoiser function w.r.t. $u$, i.e., $\eta'(u,\beta,t)$ so as to design the AMP algorithm, given by
\begin{align}
&\eta'(u,\beta,t)= \frac{\mu_t}{\left( {{\alpha _1} + {\alpha _2}\frac{{1 - \epsilon\lambda }}{\epsilon}\frac{\beta }{{\sigma _t^2}}\exp \left( { - \frac{{{\mu _t}|u{|^2}}}{{\tau _t^2}}} \right)} \right)} - \frac{{{\mu _t}u}}{{{{\left( {{\alpha _1} + {\alpha _2}\frac{{1 - \epsilon\lambda }}{\epsilon}\frac{\beta }{{\sigma _t^2}}\exp \left( { - \frac{{{\mu _t}|u{|^2}}}{{\tau _t^2}}} \right)} \right)}^2}}} \notag \\
&\left[ {\alpha'_{1} - {\alpha _2}\frac{{1 - \epsilon\lambda }}{\epsilon}\frac{{\beta {u^ * }}}{{\tau _t^4}}\exp \left( { - \frac{{{\mu _t}|u{|^2}}}{{\tau _t^2}}} \right) + \alpha'_{2}\frac{{1 - \epsilon\lambda }}{\epsilon}\frac{\beta }{{\sigma _t^2}}\exp \left( { - \frac{{{\mu _t}|u{|^2}}}{{\tau _t^2}}} \right)} \right],
\label{derivative_denoiser}
\end{align}
where the functions $\alpha_{1}\left(u,t\right)$ and $\alpha_{2}\left(u,t\right)$ are denoted by $\alpha_{1}$ and $\alpha_{2}$ for simplicity, respectively.
%
%The fact $\frac{d\left(|\hat{x}^{t}|^2\right)}{d(\hat{x}^{t})}=\left(\hat{x}^{t}\right)^\ast$ is exploited.
%
%
Functions $\alpha'_{1}$ and $\alpha'_{2}$ are the first-order derivatives of $\alpha_{1}$ and $\alpha_{2}$ w.r.t. $u$, which are given by
\begin{align}
\alpha'_{1}&=\frac{a'b}{\mathrm{Q}^2_{2}}\exp\left(-\frac{a^2+b^2}{2}\right)\left(\mathrm{Q}_{2}\mathbf{I}_{1}\left(ab\right)-\frac{b}{a}\mathrm{Q}_{1}\mathbf{I}_{2}\left(ab\right)\right) \; \text{and}
\label{derivative_alpha1} \\
\alpha'_{2}&=\frac{a'b}{\mathrm{Q}^2_{2}}\exp\left(-\frac{a^2+b^2}{2}\right)\left(-\frac{b}{a}\mathbf{I}_{2}\left(ab\right)\right),
\label{derivative_alpha2}
\end{align}
respectively.
Functions $\mathrm{Q}_{1}$ and $\mathrm{Q}_{2}$ are given by
\begin{align}
{\mathrm{Q}_1}\left(\frac{|\mu_{t}u|}{{\sqrt {{{\sigma_t^2} \mathord{\left/
					{\vphantom {{\sigma_t^2} 2}} \right.
					\kern-\nulldelimiterspace} 2}} }},\frac{\varsigma }{{\sqrt {{{\sigma_t^2} \mathord{\left/
					{\vphantom {{\sigma_t^2} 2}} \right.
					\kern-\nulldelimiterspace} 2}} }}\right) \text{and   }
{{\mathrm{Q}_2}\left(\frac{|\mu_{t}u|}{{\sqrt {{{\sigma_t^2} \mathord{\left/
						{\vphantom {{\sigma_t^2} 2}} \right.
						\kern-\nulldelimiterspace} 2}} }},\frac{\varsigma }{{\sqrt {{{\sigma_t^2} \mathord{\left/
						{\vphantom {{\sigma_t^2} 2}} \right.
						\kern-\nulldelimiterspace} 2}} }}\right)},
\end{align}
respectively,
$a=\frac{ {\mu _t} |u|}{\sqrt {{{\sigma_t^2} \mathord{\left/
				{\vphantom {{\sigma_t^2} 2}} \right.
				\kern-\nulldelimiterspace} 2}} }$, $b=\frac{\varsigma}{\sqrt {{{\sigma_t^2} \mathord{\left/
				{\vphantom {{\sigma_t^2} 2}} \right.
				\kern-\nulldelimiterspace} 2}} }$, and the first-order derivative of $a$ w.r.t. $u$ is $a'=\frac{{\mu _t} u}{2|u|{\sqrt {{{\sigma_t^2} \mathord{\left/
					{\vphantom {{\sigma_t^2} 2}} \right.
					\kern-\nulldelimiterspace} 2}} }}$.

Based on the proposed MMSE denoiser and its first-order derivative, given by Eq. \eqref{denoiser_tx_cntrl} and Eq. \eqref{derivative_denoiser}, the AMP algorithm proceeds iteratively between Eqs. \eqref{amp_1} and \eqref{amp_2}, until the algorithm converges or the number of iterations exceeds the given maximum iteration number $t_{\rm{max}}$.
The estimate in the last iteration is used as the recovered value of $\mathbf{x}$.
Here, we assume that when the normalized difference of the estimates in two consecutive iterations is no more than the given tolerance error $o$, the AMP algorithm converges.
%
%, the identification of user activity and their channel estimation are jointly performed, which is summarized in \textbf{Algorithm} \ref{alg2}.
%%
%It can be seen that during the user identification and channel estimation process, the unknown vector $\mathbf{x}$ is first recovered by exploiting the designed AMP algorithm.
%%
%Define a tolerance error $o$ as the threshold on the normalized difference of the estimated $\mathbf{x}$ in two consecutive iterations.
%%
%When the normalized difference is no more than the given tolerance error $o$, the iteration stops and the estimate in this iteration is used as the recovered value of $\mathbf{x}$.

\begin{figure}[t]
	\par
	\begin{center}
		{\includegraphics[width=3.5in]{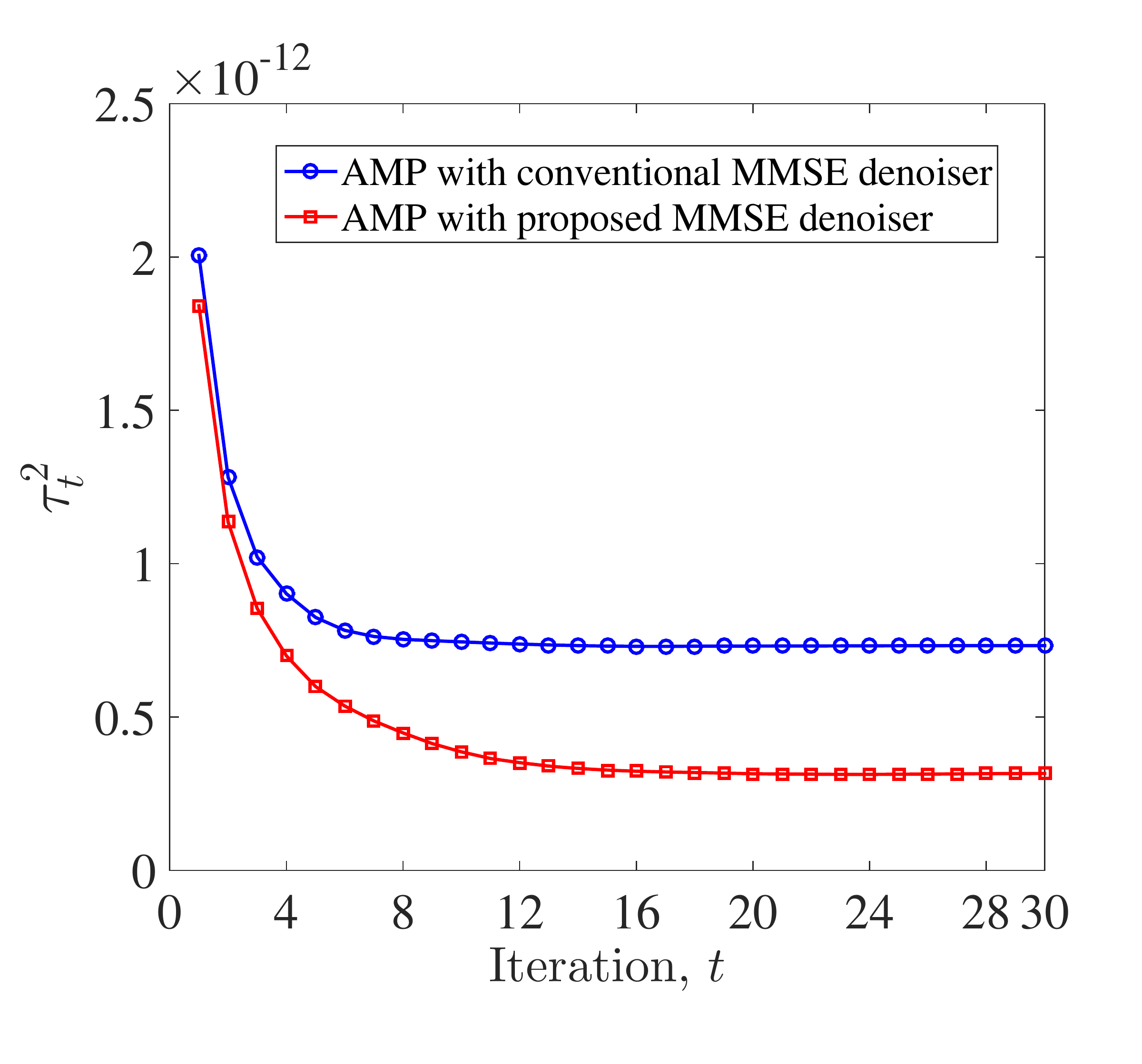}}
	\end{center}\vspace{-10mm}
	\caption{Simulated state evolution for conventional MMSE denoiser and proposed MMSE denoiser with $N=10000$, $\varpi=0.4$, $\epsilon=0.24$, $\varsigma=1.2\times10^{-6}$, and $\beta = -115$ dB.}
	\label{fig_simu_tau_w_wo_0406}
\end{figure}

%
According to the concept of phase transition in the AMP algorithm \cite{Donoho10itw}, it is known that the convergence of AMP algorithm can be characterized by the convergence of the state evolution $\tau_{t}$, and the convergency value of $\tau_{t}$, denoted by $\tau_{\infty}$, indicates the performance of AMP algorithm in terms of MSE of the recovered vector \cite{Donoho10,Chen17}.
Then, we rely on simulations to show the evolution of the squared state variable $\tau_{t}^2$ for the designed AMP algorithm and the AMP algorithm with the conventional MMSE denoiser in Fig. \ref{fig_simu_tau_w_wo_0406}, in order to evaluate the performance of the designed AMP algorithm.
We can observe that the state variable $\tau_{t}$ converges for both cases.
Moreover, compared to the AMP algorithm with the conventional MMSE denoiser, the convergency value of $\tau_{t}^2$, denoted by $\tau_{\infty}^2$, is much smaller for the designed AMP algorithm.
The smaller $\tau_{\infty}^2$ predicts a smaller MSE of the recovered vector for the designed AMP algorithm, which will be verified in Section \ref{numerical_result}.

%
%In fact, our proposed transmission control strategy and the designed MMSE denoiser can effectively exploit the time diversity and reduce the sparsity level to improve the performance of JUICE.

After obtaining the estimate of $\bf{x}$ via the AMP algorithm, the detection of user activity is performed.
According to Eq. \eqref{denoiser_tx_cntrl}, it is known that the output of the proposed MMSE denoiser $\eta \left(\tilde{x}_{n}^{t},\beta,t\right)$ approaches zero or $\mu_{t} \tilde{x}_{n}^{t}$, when the input $\tilde{x}_{n}^{t}$ is small or large, respectively.
Therefore, we can exploit $\tilde{x}_{n}^{t}$ to determine whether the output of the denoiser approaches zero or not and thus to detect the user activity.
In particular, we compare $|\tilde{x}_{n}^{\infty}|$ to a given detection threshold $l_{n}$, $\forall n$, where $\tilde{x}_{n}^{\infty}$ is the convergency value of $\tilde{x}_{n}^{t}$ when the AMP algorithm ends.
If $|\tilde{x}_{n}^{\infty}|> l_{n}$, user $n$ is active and it will be included into the active user set $\mathcal{A}$ and ${\mathbf{h}}_{\mathcal{A}}\left(n\right) = \hat{x}_{n}^{\infty}$, where $\hat{x}_{n}^{\infty}$ is the convergency value of $\hat{x}_{n}^{t}$.
Otherwise, user $n$ is inactive and ${\mathbf{h}}_{\mathcal{A}}\left(n\right) = 0$.
Note that, the detection threshold $l_{n}$, $n \in \{1,\ldots,N\}$, significantly affects the detection performance of user activity and needs to be appropriately selected.
The selection of $l_{n}$ will be presented in the next section.
\section{Performance Analysis and Optimal Control Threshold Design}
In this section, we first derive the state evolution to predict the performance of the designed AMP algorithm.
Then, we derive the false alarm probability and missed detection probability to characterize the user identification performance for the proposed scheme.
Based on the derived missed detection probability, we obtain closed-form expressions of the average packet delay and the network throughput.
Furthermore, the optimal control threshold design is proposed to maximize the network throughput.

\subsection{State Evolution for AMP Algorithm with the Proposed MMSE Denoiser}
From the state evolution for general AMP algorithm in Eq. \eqref{state_evolo}, it is known that the state evolution depends on the denoiser function of AMP algorithm.
For the AMP algorithm with proposed MMSE denoiser, we derive the state evolution in the following theorem, which provides the basis for the user identification performance analysis.
%
%In this part, we derive the evolution of state variable $\tau_{t}$ to characterize the performance of designed AMP algorithm. Recall the overview of general AMP algorithm in Section II.B, it is known that the state variable $\tau_{t}$ can characterize the performance of AMP algorithm at the $t$th iteration and the evolution of $\tau_{t}$ can predict the dynamical convergence behavior of iterative process. In addition, the evolution of $\tau_{t}$ is dependent on the denoiser function of AMP algorithm, as seen in Eq. \eqref{state_evolo}. Based on the designed AMP algorithm for the threshold based transmission control scheme, we derive the evolution of $\tau_{t}$ in Theorem \ref{theo_state_evolo_tx_contrl}.

\begin{theo}
	Consider the proposed transmission control scheme with a control threshold $\varsigma$ in the asymptotic regime, where both the number of potential users $N$ and the pilot length $M$ go to infinity, while the ratio $\varpi=\frac{M}{N}$ is a fixed positive value.
	The evolution from the state variable $\tau_{t}$ in the $t$-th iteration to $\tau_{t+1}$ in the $(t+1)$-th iteration is given by
	\begin{align}
	\tau_{t+1}^{2}&=\sigma^{2}_{w}+\frac{1}{\varpi}E_{\tilde{x}_{n}^{t}}\left[\frac{\phi \left(\tilde{x}_{n}^{t}\right)}{\alpha_{1}\left(\tilde{x}_{n}^{t},t\right)}{\sigma^2_{t}}+\left(\frac{\phi \left(\tilde{x}_{n}^{t}\right)}{\alpha_{3}\left(\tilde{x}_{n}^{t},t\right)}-\frac{\phi^{2}\left(\tilde{x}_{n}^{t}\right)}{\alpha^2_{1}\left(\tilde{x}_{n}^{t},t\right)}\right)\mu_{t}^2 |\tilde{x}_{n}^{t}|^2\right],
	\label{state_evolo_tx_contrl}
	\end{align}
	where $\sigma^{2}_{w}$ is the noise variance and the expectation is taken over $\tilde{x}_{n}^{t}$.
	%
	%The functions $\alpha_{1}\left(x,\varsigma,t\right)$ and $\alpha_{3}\left(x,\varsigma,t\right)$ are denoted by $\alpha_{1}$ and $\alpha_{3}$ for simplicity in Eq. \eqref{state_evolo_tx_contrl}, respectively.
	%
	The function $\alpha_{1}\left(u,t\right)$ has been defined in Theorem \ref{theo_denoiser_tx_cntrl} and $\alpha_{3}\left(u,t\right)$ is given by $\alpha_{3} \left(u,t\right)=\frac{{{\mathrm{Q}_1}\left(\frac{|\mu_{t}u|}{{\sqrt {{{\sigma_{t}^2} \mathord{\left/
								{\vphantom {{\sigma_t^2} 2}} \right.
								\kern-\nulldelimiterspace} 2}} }},\frac{\varsigma }{{\sqrt {{{\sigma_{t}^2} \mathord{\left/
								{\vphantom {{\sigma_{t}^2} 2}} \right.
								\kern-\nulldelimiterspace} 2}} }}\right)}}{{{\mathrm{Q}_3}\left(\frac{|\mu_{t}u|}{{\sqrt {{{\sigma_{t}^2} \mathord{\left/
								{\vphantom {{\sigma_t^2} 2}} \right.
								\kern-\nulldelimiterspace} 2}} }},\frac{\varsigma }{{\sqrt {{{\sigma_{t}^2} \mathord{\left/
								{\vphantom {{\sigma_{t}^2} 2}} \right.
								\kern-\nulldelimiterspace} 2}} }}\right)}}$, with $\mathrm{Q}_3(\cdot)$ denoting the Marcum-Q functions with order three.
	The function $\phi\left(u\right)$ is given by $\phi\left(u\right)=\frac{{\epsilon\lambda{P_{{\bar{x}_{n}^{t}}}}\left(u\right)}}{{{P_{\tilde{x}^{t}_{n}}}\left(u\right)}}$.
	\label{theo_state_evolo_tx_contrl}
\end{theo}

\emph{\quad Proof: } Please refer to Appendix \ref{AppendixC3}. \QEDA
%\begin{IEEEproof}
%	See Appendix C.
%\end{IEEEproof}

From Theorem \ref{theo_state_evolo_tx_contrl}, we can observe that for the proposed scheme, the evolution of $\tau_{t}$ depends on the control threshold $\varsigma$.
It allows us to design the threshold $\varsigma$ to adjust the convergency value $\tau_{\infty}$ and then to improve the AMP algorithm, thereby achieving a better performance.
Note that, the derived state evolution in Theorem \ref{theo_state_evolo_tx_contrl} is a generalized case for that of AMP algorithm with the conventional MMSE denoiser.
In particular, if $\varsigma=0$ and $\alpha_{1}=\alpha_{3}=1$, the derived state evolution degenerates to that of AMP algorithm with the conventional MMSE denoiser.

\subsection{False Alarm and Missed Detection Probabilities}
In \textbf{Algorithm} \ref{juice}, the user activity detection is conducted by comparing $\tilde{x}_{n}^{t}$ and $l_{n}$, $\forall n$, after the AMP algorithm terminates.
In order to characterize the performance of this activity detection, the false alarm probability and missed detection probability will be analyzed.
In particular, the false alarm probability is defined as the probability that an inactive user is detected as being active.
The missed detection probability is defined as the probability that an active user is detected as being inactive.
Note that, both the missed detection probability and the false alarm probability only represent the user identification performance. The effect of transmission delay caused by the transmission control will be considered in the following defined packet delay and throughput.
For the proposed scheme, we derive the analytical false alarm probability and missed detection probability in Theorem \ref{theo_pf_pm_tx_contrl}.

\begin{theo}
	Consider the proposed transmission control scheme with a control threshold $\varsigma$.
	Given the convergency value of state variable $\tau_{\infty}$ and the detection threshold $l_{n}$ for user $n$, the false alarm probability and missed detection probability for user $n$ are given by
	\begin{align}
	P_{f,n}=&\exp\left(-\frac{l_{n}^2}{\tau_{\infty}^2}\right),
	\; \text{and}
	\label{pf_t_iteration} \\
	P_{m,n} =& \frac{1}{{\lambda}}\left[ {\exp \left( { - \frac{\varsigma ^2}{\mu_{\infty} \beta  + \sigma _{\infty}^2}} \right){{\mathop{\rm Q}\nolimits} _1}\left({l_n} \sqrt{2 \left(\frac{\mu_{\infty}^2}{\sigma_{\infty}^2}+\frac{\mu_{\infty}}{\beta}\right)}, \frac{\sqrt{2\beta \mu_{\infty}}\varsigma}{\sigma_{\infty}\sqrt{\beta\mu_{\infty}+\sigma_{\infty}^2}}\right)\notag }\right.\\
	&\left.{-\exp \left( { - \frac{{l_n^2}}{{\beta  + \tau_{\infty}^2}}}\right){{\mathop{\rm Q}\nolimits} _1}\left( {\frac{\mu_{\infty} {l_n}}{\sqrt {{{\sigma _{\infty}^2} \mathord{\left/
							{\vphantom {{\sigma _{\infty}^2} 2}} \right.
							\kern-\nulldelimiterspace} 2}}},\frac{\varsigma }{{\sqrt {{{\sigma _{\infty}^2} \mathord{\left/
								{\vphantom {{\sigma _{\infty}^2} 2}} \right.
								\kern-\nulldelimiterspace} 2}} }}} \right)}\right],
	\forall n,
	%P_{m,n} =& \frac{1}{{\lambda}}\left[ {\exp \left( { - \frac{{(\beta  + \tau _{\infty}^2){\varsigma ^2}}}{{{\beta ^2} + \sigma _{\infty}^2(\beta  + \tau _{\infty}^2)}}} \right){{\mathop{\rm Q}\nolimits} _1}\left( {\frac{{{l_n}\sqrt {2({\beta ^2} + \sigma _{\infty}^2(\beta  + \tau _{\infty}^2))} }}{{{\sigma _{\infty}}(\beta  + \tau _{\infty}^2)}},\frac{{2\beta \varsigma }}{{{\sigma _{\infty}}\sqrt {2({\beta ^2} + \sigma _{\infty}^2(\beta  + \tau _{\infty}^2))} }}} \right)} \right. \notag \\
	%&\left. { - \exp \left( { - \frac{{l_n^2}}{{\beta  + \tau _{\infty}^2}}} \right){{\mathop{\rm Q}\nolimits} _1}\left( {\frac{{\beta {l_n}}}{{(\beta  + \tau _{\infty}^2)\sqrt {{{\sigma _{\infty}^2} \mathord{\left/
	% {\vphantom {{\sigma _{\infty}^2} 2}} \right.
	% \kern-\nulldelimiterspace} 2}} }},\frac{\varsigma }{{\sqrt {{{\sigma _{\infty}^2} \mathord{\left/
	% {\vphantom {{\sigma _{\infty}^2} 2}} \right.
	% \kern-\nulldelimiterspace} 2}} }}} \right)} \right], \forall n,
	\label{pm_t_iteration}
	\end{align}
	respectively. Here, $\mu_{\infty}$ and $\sigma^2_{\infty}$ are the convergency values of $\mu_{t}$ and $\sigma^2_{t}$, given by $\mu_{\infty}=\frac{\beta}{\beta+\tau_{\infty}^2}$ and $\sigma^2_{\infty}=\frac{\beta\tau_{\infty}^2}{\beta+\tau_{\infty}^2}$, respectively.
	\label{theo_pf_pm_tx_contrl}
\end{theo}

\emph{\quad Proof: } Please refer to Appendix \ref{AppendixD3}. \QEDA
%\begin{IEEEproof}
%	See Appendix D.
%\end{IEEEproof}

According to the signal detection theory \cite{Wickens01}, there is a tradeoff between the false alarm probability $P_{f,n}$ and missed detection probability $P_{m,n}$.
In particular, a small $P_{f,n}$ results in a large $P_{m,n}$, and vice versa.
From Eq. \eqref{pf_t_iteration}, we can observe that a small $P_{f,n}$ can be obtained from a large detection threshold $l_{n}$.
In other words, when employing a larger detection threshold, less users are detected as active users and then the false alarm probability is smaller.
However, it will result in a larger missed detection probability.
Therefore, the tradeoff between the false alarm probability $P_{f,n}$ and the missed detection probability $P_{m,n}$ can be controlled by the detection threshold $l_{n}$, which will be verified by the numerical results in Section \ref{numerical_result}.
%
%It provides us a way to select the detection threshold $l_{n}$ to achieve a trade-off between the false alarm probability and missed detection probability.
%
In this work, we select the detection threshold $l_{n}$ for a given false alarm probability since the false alarm probability of user $n$ only depends on its detection threshold $l_{n}$ for a given $\tau_{\infty}$.
According to the selected $l_{n}$ and Eq. \eqref{pm_t_iteration}, we then obtain the missed detection probability $P_{m,n}$.
To ease the presentation in the following, we consider the case that all users have the same given false alarm probability $P_{f}$ and then the same detection threshold $l$.
It leads to the same missed detection probability $P_{m}$ for all users.
Therefore, the subscript $n$ of the detection threshold $l_n$, the false alarm probability $P_{f,n}$, and the missed detection probability $P_{m,n}$ will be omitted for simplicity in the following.

In addition, we can observe from Eq. \eqref{pf_t_iteration} that for a given $P_{f}$, $l$ decreases with $\tau_{\infty}^2$.
As shown in Fig. \ref{fig_simu_tau_w_wo_0406}, it is known that for the AMP algorithm with the proposed MMSE denoiser, $\tau_{\infty}^2$ is smaller than that with the conventional MMSE denoiser.
%According to the comparison of $\tau_{\infty}^2$ for the AMP algorithm with the proposed MMSE denoiser and the conventional MMSE denoiser
It implies that the detection threshold in the proposed scheme is smaller than that for the conventional scheme without transmission control, given the same false alarm probability.
As we mentioned before, a smaller detection threshold leads to a smaller missed detection probability.
Therefore, compared to the conventional scheme without transmission control, the missed detection probability $P_{m}$ is decreased by exploiting the proposed scheme for the same false alarm probability $P_{f}$, which will be verified by numerical results in Section \ref{numerical_result}.

\subsection{Delay and Throughput}
Apart from the performance of user identification and channel estimation, exploiting the transmission control may also affect the user's transmission probability.
Since the network performance is associated with both the performance of user identification and channel estimation and the user's transmission probability, it cannot be easily determined that how using the transmission control affects the network performance.
Therefore, we introduce the packet delay \cite{Yang03} and the network throughput \cite{Abramson77} to characterize the effect from exploiting the transmission control on the network performance, which are given by Theorem \ref{theo_pdf_cdf_delay} and Theorem \ref{theo_anal_thpt}, respectively.

The packet delay is defined as the number of time slots required from a user initially attempting to transmit a packet to the successful reception of this packet\footnote{
	It is noteworthy that the imperfect user identification and channel estimation can result in the data detection error and degrade the network performance, for which the analysis is non-trivial and is not our main concern in this work.
	Then, in order to emphasize the insights of exploiting the transmission control for the network performance, we assume the perfect data detection for defining the packet delay and the throughput.
	The data detection error and its impact on the network performance will be considered in the future work.}.
Here, the packet delay includes a transmission delay caused by the transmission control and a retransmission delay due to the missed detection.
As the users randomly access the channel with a certain probability, the packet delay is a random variable, whose probability mass function (PMF) and cumulative distribution function (CDF) are derived in Theorem \ref{theo_pdf_cdf_delay}.
On the other hand, the network throughput is defined as the average number of successfully received packets in a time slot.
Note that, the backoff time \cite{Yang03} of users and the time consumption for delivering the feedback information of packet reception from the receiver are not taken into account for the delay analysis in this work.
In other words, once an active user is miss-detected, the user immediately knows the unsuccessful reception of its packet and restarts the attempt to retransmit the packet in the next time slot.
Moreover, it is assumed that if a user's transmission attempt is prevented by the proposed transmission control function, the user will keep attempting until it is allowed to transmit the packet.

\begin{theo}
	The probability mass function (PMF) and cumulative distribution function (CDF) of the packet delay for the proposed scheme are given by
	\begin{align}
	f_{D}(d)=\lambda(1-P_{m})(1-\lambda(1-P_{m}))^{d-1}
	\label{pdf_delay}
	\end{align}
	and
	\begin{align}
	F_{D}(d)=1-(1-\lambda(1-P_{m}))^{d},
	\label{cdf_delay}
	\end{align}
	respectively, where the variable $D$ denotes the packet delay, the average transmission probability $\lambda$ and the missed detection probability $P_{m}$ are obtained by Eq. \eqref{step_tx_prob} and \eqref{pm_t_iteration}, respectively.
	\label{theo_pdf_cdf_delay}
\end{theo}
%It can be seen from Theorem 4 that the packet delay captures the effects from both the transmission control and the missed detection through $\lambda$ and $P_{m}$, respectively. \\

\emph{\quad Proof: }In the proposed transmission control scheme, when a user has a transmission attempt, it actually transmits its packet with an average probability $\lambda$ and its transmission activity may be missed detected with a probability $P_{m}$.
	Then, we consider that a user consumes $d$ time slots from it initially attempting to transmit a packet to the successful reception of its packet.	
	Among the first $(d-1)$ time slots when the packet is not successfully received, there can be $(i-1)$, $i\in \{1,2,\ldots,d\}$, time slots when the user transmits but is missed detected, and the remaining $(d-i)$ time slots when the user does not transmit.
	At the $d$-th time slot, the packet must be successfully received and the probability is $\lambda(1-P_{m})$.
	%Denote the random variable $d$ as the packet delay.
	Thus, the probability that $d$ time slots need to be consumed to successfully receive a packet can be given by
	\begin{align}
	f_{D}(d)&=\lambda(1-P_{m})\sum_{i=1}^{d}\binom{d-1}{i-1}\lambda^{i-1}(1-\lambda)^{d-i}P_{m}^{i-1} \notag \\
	&=\lambda(1-P_{m})\left(1-\lambda(1- P_{m})\right)^{d-1},
	\label{pdf_delay_pf}
	\end{align}
	where the random variable $D$ denotes the packet delay and the second equality is based on the binomial theorem.
	Eq. \eqref{pdf_delay_pf} is the PMF of the packet delay $D$.
	
	Based on that, the CDF of the packet delay $D$ can be obtained as
	\begin{align}
	F_{D}(d)&=\sum_{i=1}^{d}f_{D}(i)=\sum_{i=1}^{d}\lambda(1-P_{m})(1-\lambda(1-P_{m}))^{i-1} \notag \\
	&=1-(1-\lambda(1-P_{m}))^{d}.
	\label{cdf_delay_pf}
	\end{align}
	The proof ends. \QEDA
	
It can be seen from Theorem \ref{theo_pdf_cdf_delay} that the packet delay captures the effects from both the transmission control and the missed detection through $\lambda$ and $P_{m}$, respectively.
\begin{coro}
	The average packet delay for the proposed scheme is given by
	\begin{align}
	\overline{D}=\frac{1}{\lambda\left(1-P_{m}\right)},
	\label{average_delay}
	\end{align}
	where the average transmission probability $\lambda$ and the missed detection probability $P_{m}$ are obtained by Eq. \eqref{step_tx_prob} and \eqref{pm_t_iteration}, respectively.
	\label{coro_anal_delay}
\end{coro}

\emph{\quad Proof: } From the PMF of the packet delay $D$, we have the average packet delay as
	\begin{align}
	\overline{D}=\sum_{d=1}^{\infty}df_{D}(d)=\frac{1}{\lambda\left(1-P_{m}\right)},
	\label{average_delay_pf}
	\end{align}
	where the second equality is obtained by incorporating Eq. \eqref{pdf_delay} into the equation.
	The proof ends. \QEDA

\begin{theo}
	Consider the proposed transmission control scheme with a control threshold $\varsigma$.
	Given the number of potential users $N$ and the transmission demand probability $\epsilon$, the network throughput is given by
	\begin{align}
	T=&N\epsilon \left(\lambda-\left[ {\exp \left( { - \frac{\varsigma ^2}{\mu_{\infty} \beta  + \sigma _{\infty}^2}} \right){{\mathop{\rm Q}\nolimits} _1}\left({l} \sqrt{2 \left(\frac{\mu_{\infty}^2}{\sigma_{\infty}^2}+\frac{\mu_{\infty}}{\beta}\right)}, \frac{\sqrt{2\beta \mu_{\infty}}\varsigma}{\sigma_{\infty}\sqrt{\beta\mu_{\infty}+\sigma_{\infty}^2}}\right)\notag }\right.\right. \notag \\
	&\left.\left.{-\exp \left( { - \frac{{l^2}}{{\beta  + \tau_{\infty}^2}}}\right){{\mathop{\rm Q}\nolimits} _1}\left( {\frac{\mu_{\infty} {l}}{\sqrt {{{\sigma _{\infty}^2} \mathord{\left/
							{\vphantom {{\sigma _{\infty}^2} 2}} \right.
							\kern-\nulldelimiterspace} 2}}},\frac{\varsigma }{{\sqrt {{{\sigma _{\infty}^2} \mathord{\left/
								{\vphantom {{\sigma _{\infty}^2} 2}} \right.
								\kern-\nulldelimiterspace} 2}} }}} \right)}\right]\right),
	\label{anal_thpt_3}
	\end{align}
	where the average transmission probability $\lambda$ and the convergency value of state variable $\tau_{\infty}$ are obtained from Eqs. \eqref{step_tx_prob} and \eqref{state_evolo_tx_contrl}, respectively. The detection threshold $l$ is obtained from $l=\sqrt{-\tau_{\infty}^2\ln P_{f}}$, given a false alarm probability $P_{f}$.
	\label{theo_anal_thpt}
\end{theo}

\emph{\quad Proof: }For the proposed scheme, among $N$ potential users, each user is active with a probability $\epsilon\lambda$ in a time slot.
	It implies that on average, there are $N\epsilon\lambda$ transmitted packets in a time slot.
	As each packet can be successfully received with a probability $(1-P_{m})$, the average number of successfully received packets in a time slot is obtained by
	\begin{align}
	T=N\epsilon\lambda\left(1-P_{m}\right)\overset{(a)}{=}\frac{N\epsilon}{\overline{D}},
	\label{thpt}
	\end{align}
	where the equality $(a)$ is obtained by substituting Eq. \eqref{average_delay} into \eqref{thpt}.
	%
	%From the average delay of a user, we have the average probability that an active user can be correctly detected in a time slot as $\frac{1}{D}$.
	%%
	%Then, the average number of correctly detected active users in a time slot is given by
	%\begin{align}
	%T=\frac{N\epsilon}{D}=N\epsilon\lambda\left(1-P_{m}\right).
	%\label{thpt}
	%\end{align}
	%where $K=N\epsilon\lambda$ is the number of active users.
	%
	By incorporating Eq. \eqref{pm_t_iteration} into \eqref{thpt}, we derive the network throughput in Eq. \eqref{anal_thpt_3}, which ends the proof.  \QEDA
	
\subsection{Design of Optimal Control Threshold}\label{OptimalThresholdDesign}
From the proof of Theorem \ref{theo_anal_thpt}, it is known that the network throughput is affected by the proposed transmission control scheme via the average transmission probability $\lambda$ and the missed detection probability $P_{m}$.
The stricter transmission control leads to the smaller average transmission probability as well as the lower missed detection probability, and vice versa.
Then, there is an optimal transmission control scheme to balance the two effects and then to achieve a maximum network throughput.
From Eq. \eqref{anal_thpt_3}, it can be seen that with the step transmission control function, the effect of the proposed transmission control scheme on the throughput is characterized by the control threshold $\varsigma$.
In this case, the design of optimal transmission control is equivalent to the design of optimal threshold to maximize the throughput, i.e., $\varsigma^\ast=\arg \max\limits_{\varsigma}T$.
However, the throughput is a complicated function of $\varsigma$, since the throughput is not only directly associated with $\varsigma$ but also depends on $\varsigma$ through $\lambda$ and $\tau_{\infty}$, as shown in Eq. \eqref{step_tx_prob} and \eqref{state_evolo_tx_contrl}, respectively.
Then, it makes the design of optimal $\varsigma$ very difficult.
Instead, we exploit the one-dimension search to obtain the optimal $\varsigma$.
Note that, the one-dimension search of the optimal $\varsigma$ is performed for a given false alarm probability, since the throughput is also affected by the false alarm probability.
We also note that maximizing the throughput is equivalent to minimizing the average packet delay, since the throughput is inversely proportional to the average delay, as shown in Eq. \eqref{thpt}.

\begin{figure}[!t]
	\par
	\begin{center}
		{\includegraphics[width=3.5in]{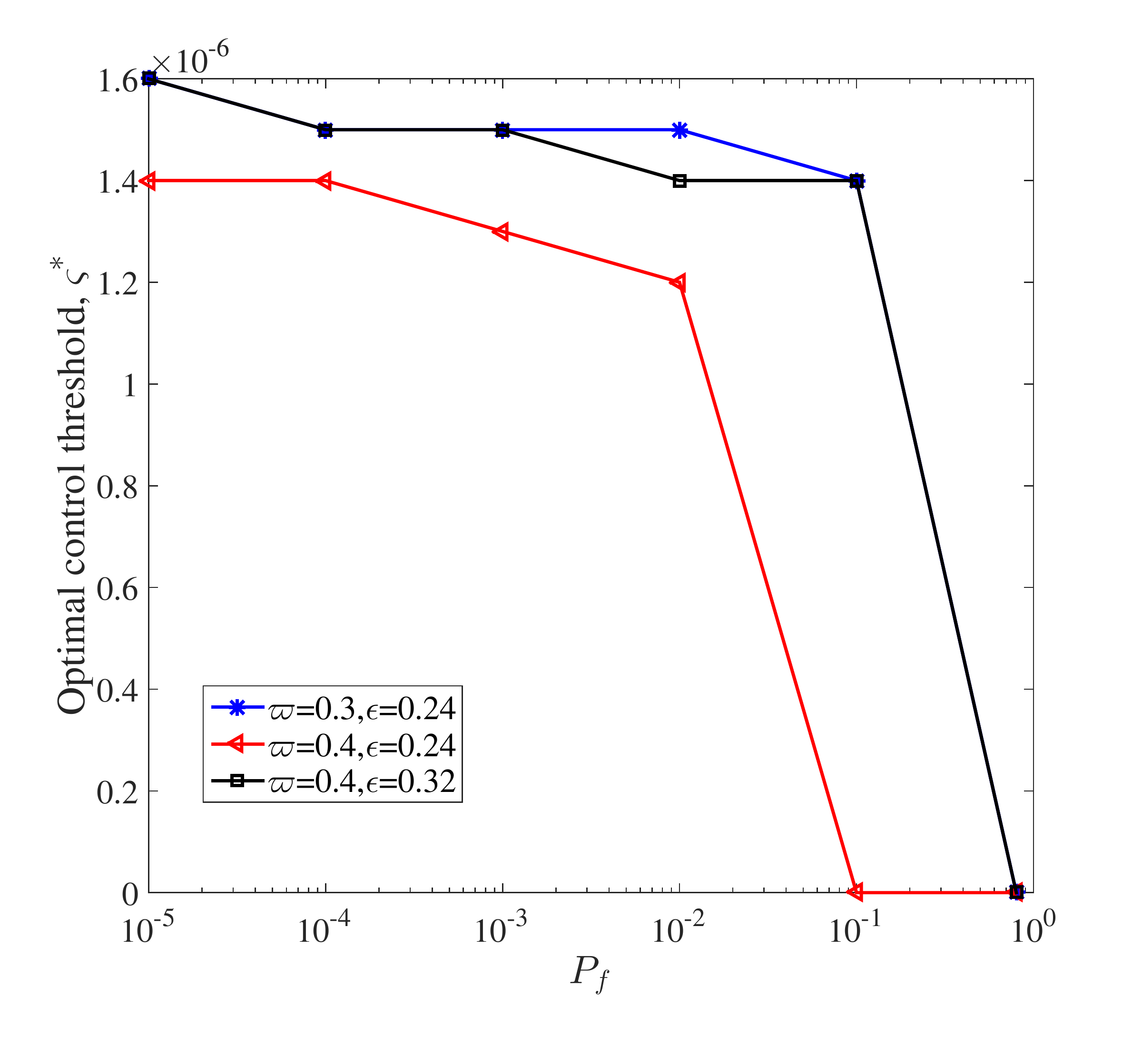}}
	\end{center}\vspace{-10mm}
	\caption{Optimal control threshold $\varsigma^\ast$ for $N=10000$ and $\beta=-115$ dB.}
	\label{fig_opt_ctrl_thehd_pf_v3}
\end{figure}
In Fig. \ref{fig_opt_ctrl_thehd_pf_v3}, we show the optimal control threshold $\varsigma^\ast$ for different false alarm probabilities $P_{f}$.
From the figure, we can observe that the optimal control threshold $\varsigma^{\ast}$ increases with decreasing the false alarm probability $P_{f}$.
For example, when $P_{f}$ decreases from $10^{-2}$ to $10^{-4}$, $\varsigma^{\ast}$ increases from $1.2\times10^{-6}$ to $1.4\times10^{-6}$ for $\varpi=0.4$ and $\epsilon=0.24$.
It means that the transmission control should be stricter for a smaller false alarm probability.
This is because a smaller false alarm probability $P_{f}$ leads to a larger detection threshold $l$, for a given squared state variable $\tau_{\infty}^2$.
Moreover, a larger $l$ results in a larger missed detection probability $P_{m}$ and then a lower network throughput $T$.
Therefore, in order to satisfy the smaller false alarm probability $P_{f}$ as well as to improve the network throughput, it is necessary to employ a larger control threshold $\varsigma$ to decrease the squared state variable $\tau_{\infty}^2$, thereby achieving a smaller detection threshold $l$ and a smaller missed detection probability $P_{m}$.
%
%Moreover, the transmission control threshold $\varsigma$ increases with decreasing the targeted false alarm probability $P_{f}$.
%
Note that, while increasing the transmission control threshold decreases the missed detection probability and positively affects the network throughput, it decreases the average transmission probability and then negatively affects the network throughput.
Thus, the optimal control threshold $\varsigma^\ast$ remains almost unchanged in a certain range of the false alarm probability $P_{f}$, such as $10^{-4}\leq P_{f}\leq 10^{-2}$ for $\varpi=0.3$ and $\epsilon=0.24$ in Fig. \ref{fig_opt_ctrl_thehd_pf_v3}.
In addition, we evaluate the effect of pilot lengths and the number of users with transmission demand on the optimal control threshold, by employing different $\varpi$ and $\epsilon$ in Fig. \ref{fig_opt_ctrl_thehd_pf_v3}.
From the figure, we can observe that for a shorter pilot sequence, i.e., a smaller $\varpi$, the optimal control threshold $\varsigma^\ast$ is larger.
This is because the sparsity threshold that guarantees the convergence of the AMP algorithm decreases with a shorter pilot, according to the concept of phase transition in the AMP \cite{Donoho10itw}.
Then, it needs to employ a larger control threshold to enhance the sparsity.
We can also observe from the figure that the optimal control threshold $\varsigma^\ast$ is larger, when more users have the transmission demand, i.e., a larger $\epsilon$.
This is because a larger $\epsilon$ results in a lower sparsity.
Then, a larger control threshold is utilized to enhance the sparsity.
\section{Numerical Results}\label{numerical_result}
\begin{figure}[t]
	\par
	\begin{center}
		{\includegraphics[width=3.5in]{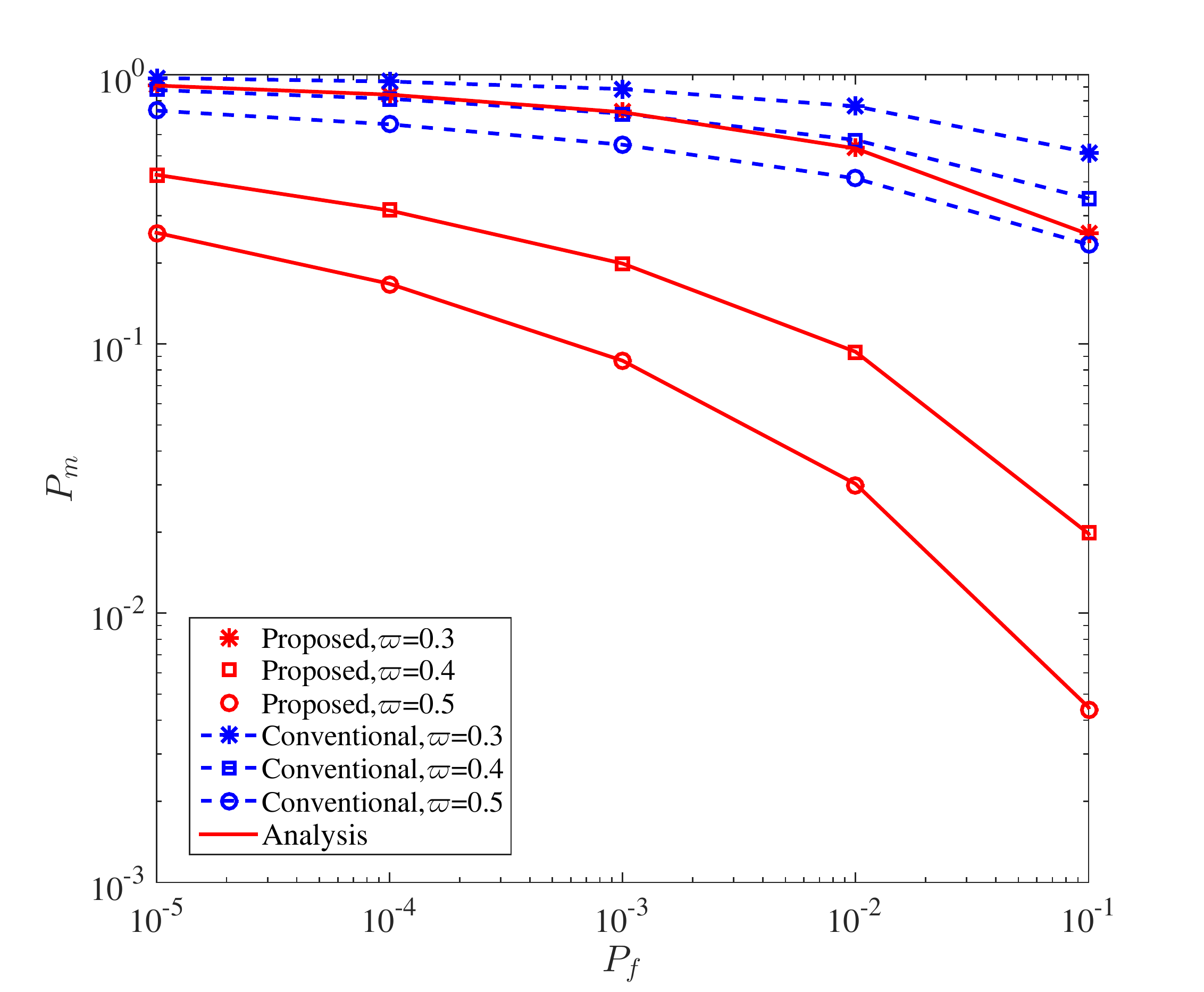}}
	\end{center}\vspace{-7mm}
	\caption{Simulated and analytical missed detection probability for $\epsilon=0.24$ and $\varsigma=1.2\times10^{-6}$.}
	\label{fig_analysis_pm_w_wo_control_v4}
\end{figure}
In this section, we evaluate the performance of the proposed scheme via Monte-Carlo simulations.
%
%In particular, we investigate the JUICE performance in terms of the missed detection probability, the network throughput, and the NMSE.
%
Consider an uplink system, where the number of potential users is $N=10000$, the pilot length is $M=4000$, i.e., $\varpi=0.4$, and the transmission demand probability for each user is $\epsilon=0.24$.
The users' channels follow the i.i.d. Rayleigh distribution with zero mean and variance of $\beta$, where $\beta=-115$ dB \cite{TR36814}.
The power spectral density of AWGN at the receiver is $-169$ dBm/Hz and the bandwidth is $1$ MHz.
%
%For transmission control scheme with control threshold $\varsigma=1.2\times10^{-6}$ is employed.
%
%In order to illustrate the performance improvement of the proposed scheme, we employ the conventional scheme without transmission control for comparison in the following simulations.
%%
All numerical results are obtained by averaging over $1000$ channel realizations.

\begin{figure}[t]
	\par
	\begin{center}
		{\includegraphics[width=3.5in]{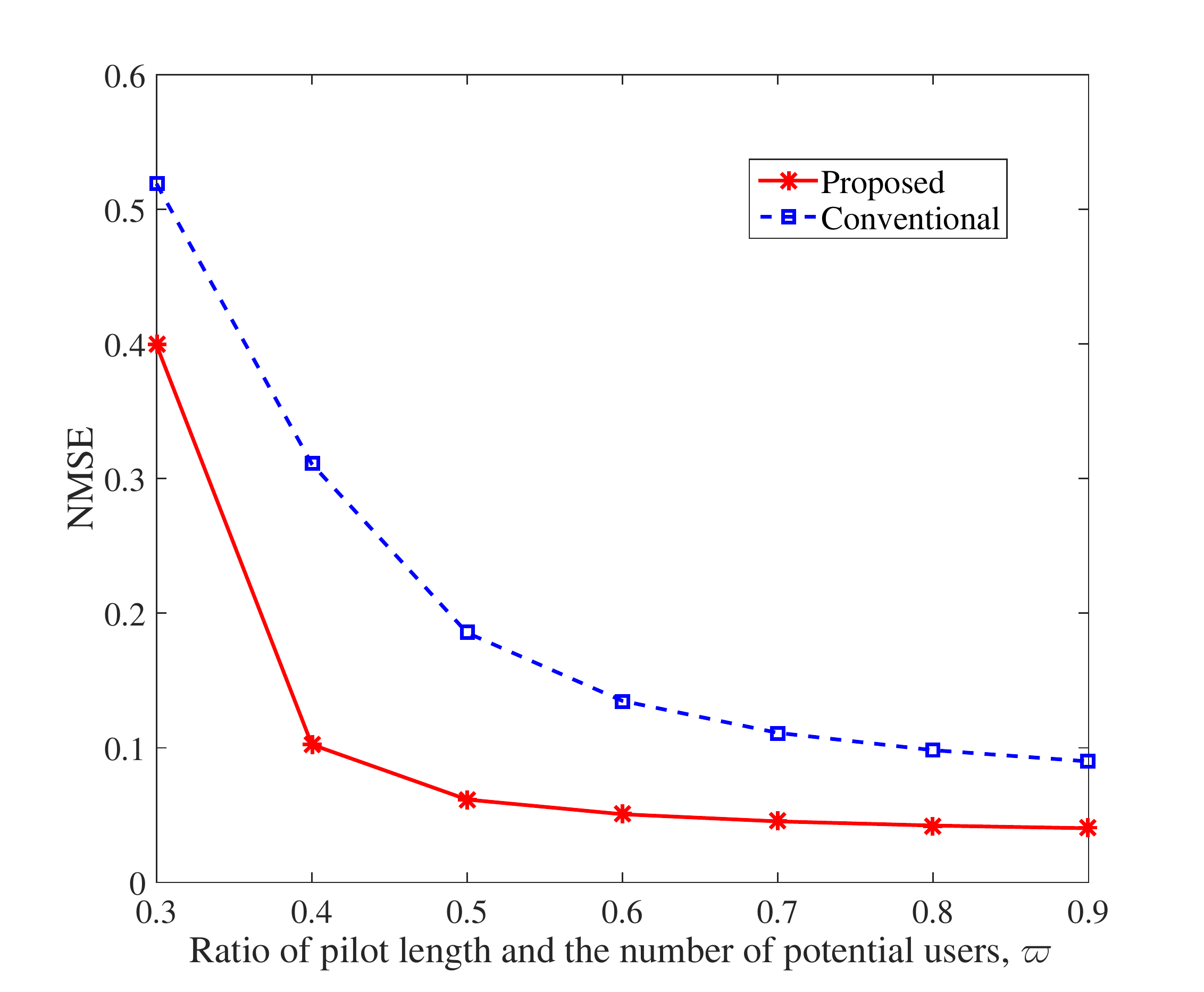}}
	\end{center}\vspace{-8mm}
	\caption{Comparison of NMSE for the conventional scheme without transmission control and the proposed scheme with $P_{f}=10^{-3}$, $\epsilon=0.24$, and $\varsigma=1.2\times10^{-6}$.}
	\label{fig_simu_nmse}
\end{figure}
We first present the missed detection probability $P_{m}$ for the proposed scheme in Fig. \ref{fig_analysis_pm_w_wo_control_v4}.
From the figure, it can be verified that there is a tradeoff between the false alarm probability $P_{f}$ and the missed detection probability $P_{m}$, i.e., an increase of $P_{f}$ leads to a decrease of $P_{m}$, and vice versa.
%
%In addition, the tradeoff is achieved via the detection threshold $l$, since each value of $P_{f}$ corresponds to a detection threshold $l$ and the detection threshold $l$ affects $P_{m}$ for a given $\varpi$.
%
We also show the missed detection probability $P_{m}$ for different pilot lengths in this figure, i.e., different $\varpi$.
It can be seen that $P_{m}$ decreases by increasing $\varpi$ and this trend is more obvious for a smaller $\varpi$.
This is because that increasing the pilot length, i.e., $\varpi$, can introduce more measurements for the receiver whereby the AMP performance can be improved, i.e., $P_{m}$ is reduced.
In addition, analytical results are presented in the figure.
It is shown that analytical results match with simulation results tightly.
Moreover, we can observe that compared to the conventional scheme without the transmission control, $P_{m}$ is significantly decreased by exploiting the proposed scheme, for all three considered pilot lengths.
The proposed scheme with $\varpi=0.3$ achieves a similar or even smaller missed detection probability $P_{m}$, compared to the conventional scheme with $\varpi=0.4$.
It implies that the pilot length can be decreased by exploiting the proposed scheme, in order to achieve the same miss detection performance.

\begin{figure}[t]
	\begin{center}
		{\includegraphics[height=2.8in,width=3.5in]{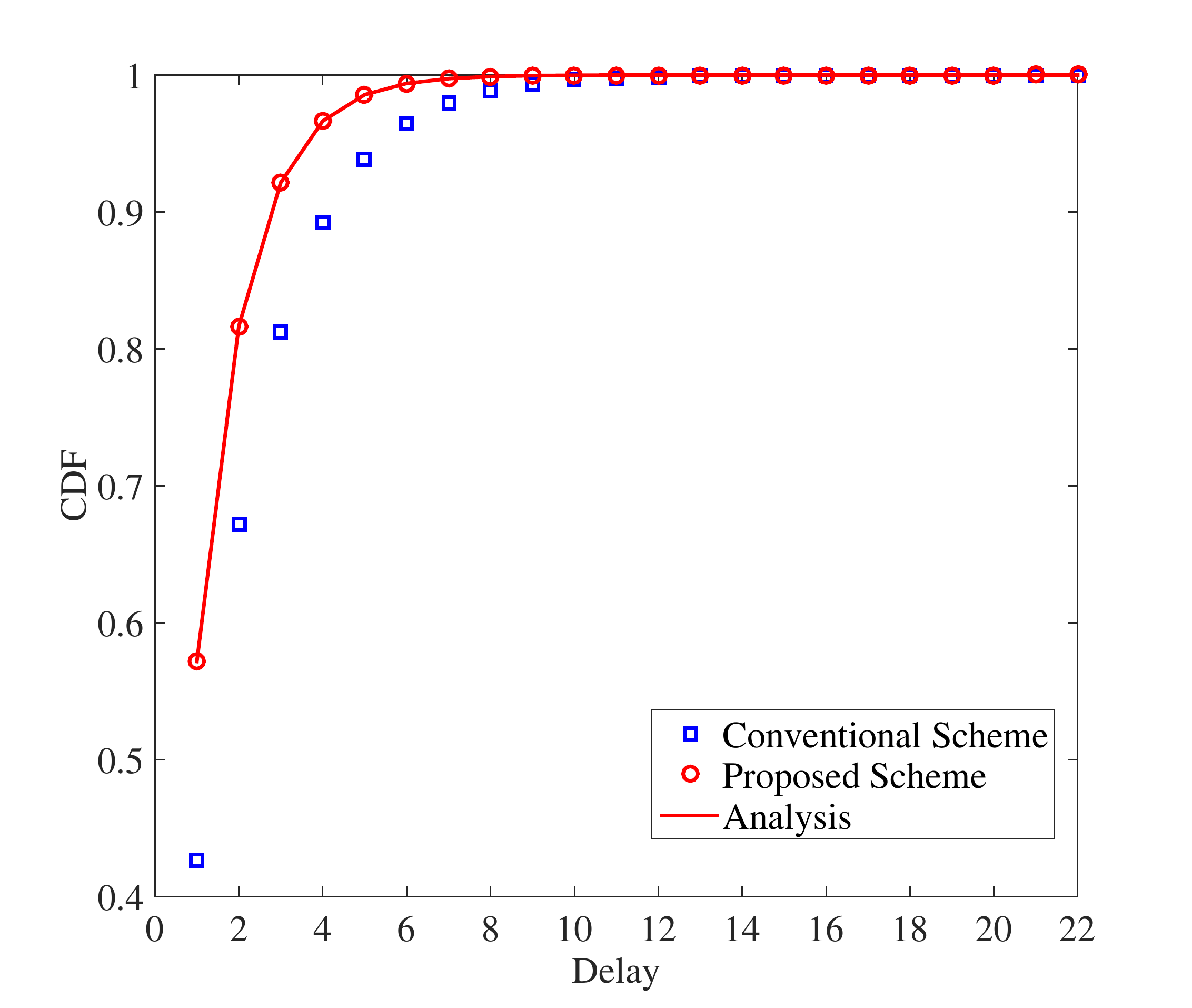}}
	\end{center}\vspace{-8mm}
	\caption{Simulated and analytical CDFs of packet delay for the proposed scheme and the conventional scheme with  $N=10000$, $M=4000$, $\epsilon=0.24$, $\varsigma=1.2\times 10^{-6}$, and $P_{f}=10^{-5}$.}
	\label{cdf_delay2}	
\end{figure}
In addition, we compare the NMSE of the proposed scheme and the conventional scheme in Fig. \ref{fig_simu_nmse}, where the NMSE is defined as the MSE of the estimate $\hat{\mathbf{x}}^{\infty}$ and the actual vector $\mathbf{x}$ normalized by $||\mathbf{x}||_2^2$, given by $\frac{||\hat{\mathbf{x}}^{\infty}-\mathbf{x}||_2^2}{||\mathbf{x}||_2^2}$.
From the figure, we can observe that the NMSE is significantly reduced by employing the proposed scheme.
For example, the NMSE is decreased by $66\%$ when $\varpi=0.4$.
Moreover, when achieving the same NMSE performance, the pilot length required by the proposed scheme is much smaller than that for the conventional scheme.
For example, when NMSE equals $0.1$, we have $\varpi=0.4$ for the proposed scheme and $\varpi=0.8$ for the conventional scheme.
This means the pilot length can be reduced by $50\%$ in the proposed scheme for the same NMSE.
In addition, we can see from the figure that for both the conventional scheme and the proposed scheme, the NMSE first decreases significantly and then gets flat with increasing the pilot length $\varpi$.
It implies that when the pilot length increases to a certain value, increasing the pilot length will not significantly improve the NMSE performance.

We also show the analytical and simulated CDF of packet delay for the proposed scheme in Fig. \ref{cdf_delay2}, where the CDF of packet delay for the conventional scheme is also provided for a comparison.
It can be seen from the figure that the analysis matches well with the simulation, and the CDF of packet delay for the proposed scheme increases faster than that for the conventional scheme.
It implies that compared to the conventional scheme, the packet delay can be effectively decreased by using the proposed transmission scheme.

\begin{figure}[!t]
	\par
	\begin{center}
		{\includegraphics[width=3.5in]{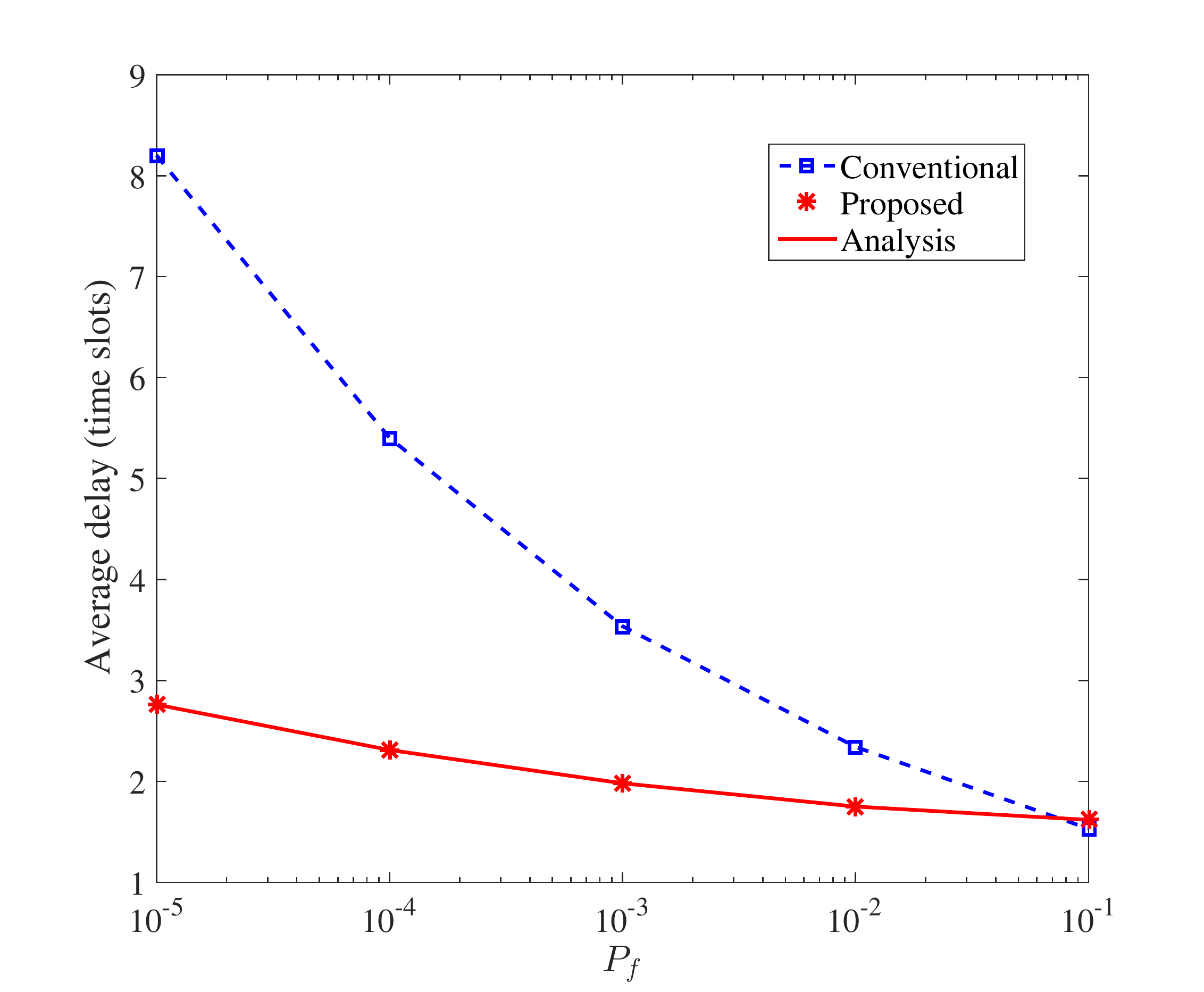}}
	\end{center}\vspace{-8mm}
	\caption{Simulated and analytical average packet delay for $\epsilon=0.24$, $\varpi=0.4$, and $\varsigma=1.2\times 10^{-6}$.}
	\label{fig_delay_w_wo}
\end{figure}
In Fig. \ref{fig_delay_w_wo}, we present the average packet delay with different false alarm probabilities for the proposed scheme.
It can be observed that compared to the conventional scheme without transmission control, the average delay is significantly reduced for the proposed scheme, particularly in the small to medium false alarm probability regime.
For example, when $P_{f}=10^{-3}$, the average delay decreases by $42.8\%$ for the proposed scheme.
This is because that the significant decrease of missed detection probability is obtained by employing the proposed scheme, as shown in Fig. \ref{fig_analysis_pm_w_wo_control_v4}, thereby reducing the probability of retransmissions and the average delay.
Moreover, it is seen from the figure that the average delay decreases with increasing the false alarm probability, due to the tradeoff between the false alarm probability $P_{f}$ and the missed detection probability $P_{m}$.
In addition, the analytical average delay is shown to match well with the simulation.

\begin{figure}[!t]
	\par
	\begin{center}
		{\includegraphics[width=3.5in]{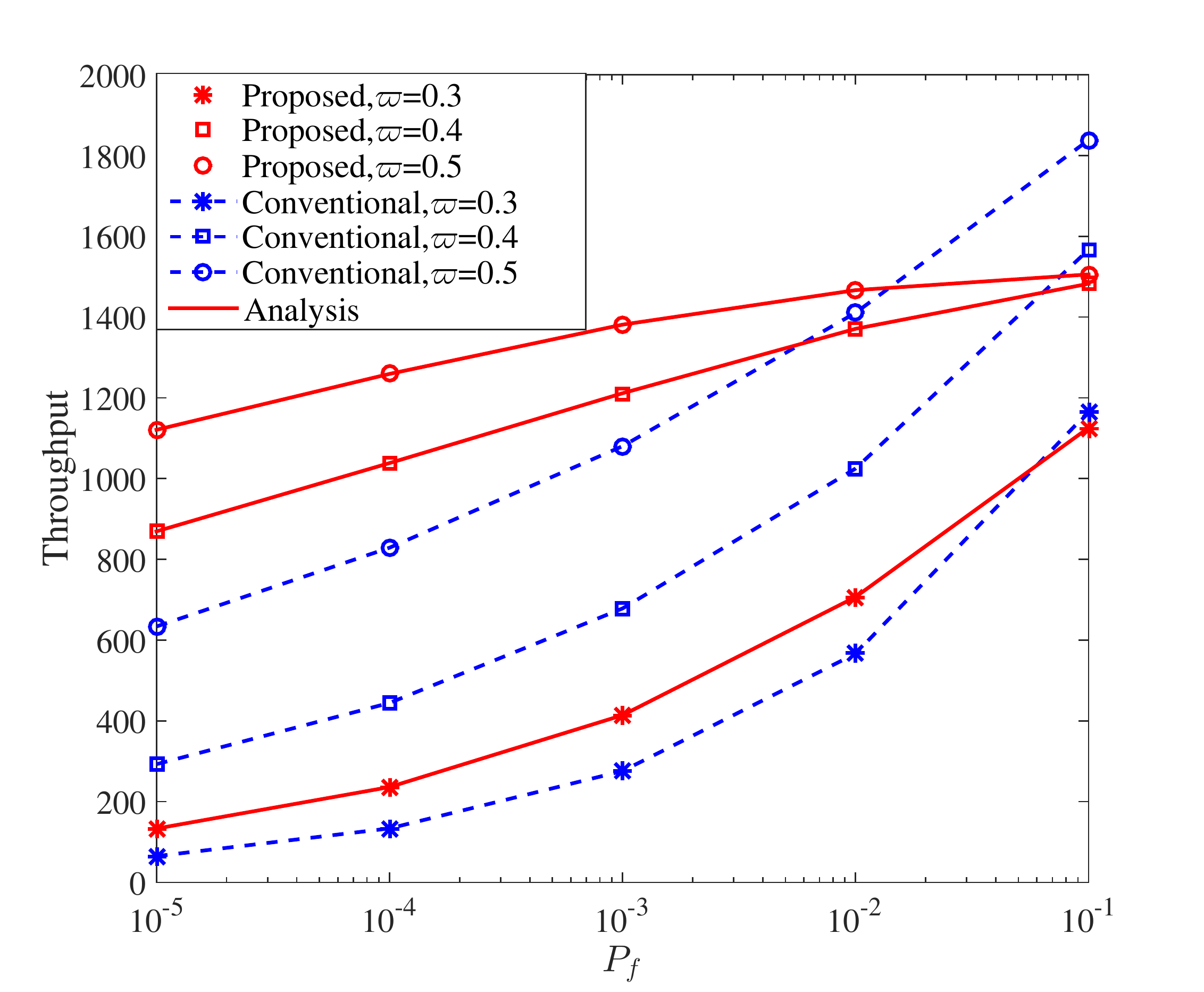}}
	\end{center}\vspace{-8mm}
	\caption{Simulated and analytical network throughput for $\epsilon=0.24$ and $\varsigma=1.2 \times 10^{-6}$.}
	\label{fig_thpt_pm_w_wo_control_v4}
\end{figure}
Furthermore, we show the network throughput for the proposed scheme in Fig. \ref{fig_thpt_pm_w_wo_control_v4}.
From the figure, we can observe that the network throughput increases with increasing the false alarm probability $P_{f}$.
This is because that the increase of false alarm probability $P_{f}$ leads to the decrease of missed detection probability $P_{m}$, thereby enhancing the throughput.
For the same $P_{f}$, the network throughput can be improved by increasing the pilot length, and the improvement is more obvious for a smaller pilot length.
This is due to the fact that the increase of pilot length decreases the missed detection probability, in particular when the pilot length is small.
The analytical result is also presented in the figure.
It is shown that analytical results match well with simulation results.
In addition, we compare the network throughput of the proposed scheme and that of the conventional scheme in Fig. \ref{fig_thpt_pm_w_wo_control_v4}.
It can be seen that the network throughput is substantially improved by exploiting the proposed scheme, in particular in the small-to-medium false alarm probability regime.
In particular, the proposed scheme achieves a $2.5$-fold network throughput improvement compared to the conventional scheme for $P_{f}=10^{-4}$ and $\varpi=0.4$.
Note that, the throughput of the proposed scheme is a bit smaller than that of the conventional scheme for the large false alarm probability, such as $P_{f}=0.1$.
This is because for a large false alarm probability, the missed detection probability is small and the network throughput is mainly determined by the transmission probability.
The proposed scheme decreases the transmission probability and then results in a smaller throughput, compared to the conventional scheme.
However, such a high false alarm probability is impractical for random access systems, which will deteriorate the multiuser detection performance and reduce the system performance in terms of the sum-rate.
Therefore, in a practical region of small-to-medium false alarm probability, the proposed scheme can effectively improve the network throughput.
\section{Chapter Summary}
To summarize, this chapter proposed a transmission control scheme and designed an AMP algorithm to improve the user identification and channel estimation performance.
An MMSE denoiser was proposed for the AMP algorithm design, with a step transmission control function employed at the transmitter.
The false alarm probability and missed detection probability were derived for the proposed scheme.
Based on that, closed-form expressions of the average packet delay and the network throughput were obtained.
The transmission control threshold was also optimized to maximize the network throughput.
%
%It was verified that the analytical results matched well with simulation results.
%
Simulation results demonstrated that compared to the conventional scheme without transmission control, the proposed scheme can significantly improve the user identification and channel estimation performance, reduce the packet delay, and enhance the network throughput.
\newpage
\section{Appendix}
\subsection{Proof of Lemma \ref{lemma_dis_x_bar}} \label{AppendixA3}
	According to the definition $\bar{x}_{n}^{t}=\hat{h}_{n}+\tau_{t}v$, we obtain the posterior distribution of the random variable $\hat{h}_{n}$ conditioned on $\bar{x}_{n}^{t}$, given by
	\begin{align}
	&{P_{\hat{h}_{n}|\bar{x}_{n}^{t}}}\left(z|\bar{x}_{n}^{t} = u\right) = \frac{1}{{{P_{\bar{x}_{n}^{t}}}\left(u\right)}}{P_{\bar{x}_{n}^{t}|\hat{h}_{n}}}\left(u|\hat{h}_{n} = z\right)P_{\hat{h}_{n}}\left(z\right) \notag \\
	%&= \frac{1}{{{P_{\bar{x}_{n}^{t}}}\left(u\right)}}\frac{1}{{\pi \tau_{t}^2}}\exp \left( - \frac{{|u- z|^2}}{{\tau_{t}^2}}\right)\frac{1}{{\lambda}}\frac{1}{{\pi \beta }}\exp \left( - \frac{{|z|^2}}{\beta }\right)c\left(z\right) \notag \\
	&\overset{(a)}{=}\frac{1}{{{P_{\bar{x}_{n}^{t}}}\left(u\right)}}\frac{1}{{\lambda \pi \left(\beta  + \tau_{t}^2\right)}}\exp \left( - \frac{{|u|^2}}{{\beta  + \tau_{t}^2}}\right) \frac{1}{{\pi \left(\frac{{\beta \tau_{t}^2}}{{\beta  + \tau_{t}^2}}\right)}}\exp \left( - \frac{{|z - \frac{{\beta u}}{{\beta  + \tau_{t}^2}}|^2}}{{\frac{{\beta \tau_{t}^2}}{{\beta  + \tau_{t}^2}}}}\right)c\left(z\right) \notag \\
	&=\frac{1}{{{P_{\bar{x}_{n}^{t}}}\left(u\right)}} \frac{1}{{\lambda \pi \left(\beta  + \tau_{t}^2\right)}}\exp \left( - \frac{{|u|^2}}{{\beta  + \tau_{t}^2}}\right) \frac{1}{{\pi \sigma_{t}^{2}}}\exp \left( - \frac{{|z - \mu_{t}u|^2}}{\sigma_{t}^{2}}\right)c\left(z\right),
	\label{post_dis_hn}
	\end{align}
	where $\mu_{t}=\frac{\beta}{\beta+\tau_{t}^2}$, $\sigma_t^2=\frac{\beta\tau_{t}^2}{\beta+\tau_{t}^2}$, $c\left(z\right)$ is the transmission control function defined in Eq. \eqref{step_cntrl_function}, and the equality $(a)$ is based on the derivation in \cite{Bromiley03}.
	Due to the fact $\int {P_{\hat{h}_{n}|\bar{x}_{n}^{t}}}\left(z|\bar{x}_{n}^{t} = u\right)dz=1$, we have
	\begin{align}
	&\int \frac{1}{{{P_{\bar{x}_{n}^{t}}}\left(u\right)}}\frac{1}{{\lambda \pi \left(\beta  + \tau_{t}^2\right)}}\exp \left( - \frac{{|u|^2}}{{\beta  + \tau_{t}^2}}\right) \frac{1}{{\pi \sigma_{t}^{2}}}\exp \left( - \frac{{|z - \mu_{t}u|^2}}{\sigma_{t}^{2}}\right)c\left(z\right)dz \notag \\
	&=\frac{1}{{{P_{\bar{x}_{n}^{t}}}\left(u\right)}}\frac{1}{{\lambda \pi \left(\beta  + \tau_{t}^2\right)}}\exp \left( - \frac{{|u|^2}}{{\beta  + \tau_{t}^2}}\right) \int \frac{1}{{\pi \sigma_{t}^{2}}}\exp \left( - \frac{{|z - \mu_{t}u|^2}}{\sigma_{t}^{2}}\right)c\left(z\right)dz \notag \\
	&=\frac{1}{{{P_{\bar{x}_{n}^{t}}}\left(u\right)}}\frac{1}{{\lambda \pi \left(\beta  + \tau_{t}^2\right)}}\exp \left( - \frac{{|u|^2}}{{\beta  + \tau_{t}^2}}\right) \mathrm{Q}_{1}\left(\frac{|\mu_{t}u|}{\sqrt{\frac{\sigma_{t}^{2}}{2}}},\frac{\varsigma}{\sqrt{\frac{\sigma_{t}^{2}}{2}}}\right)=1.
	\label{int_post_dis_hn}
	\end{align}
	As a result, the distribution of $\bar{x}_{n}^{t}$ can be obtained as
	\begin{align}
	{P_{\bar{x}_{n}^{t}}}\left(u\right)=\frac{\mathrm{Q}_{1}\left(\frac{|\mu_{t}u|}{\sqrt{\frac{\sigma_{t}^{2}}{2}}},\frac{\varsigma}{\sqrt{\frac{\sigma_{t}^{2}}{2}}}\right)}{\lambda \pi \left(\beta  + \tau_{t}^2\right)}\exp\left(-\frac{|u|^2}{\beta+\tau_{t}^2}\right).
	\label{dis_overline_x}
	\end{align}
	
	\subsection{Proof of Theorem \ref{theo_denoiser_tx_cntrl}} \label{AppendixB3}
	From the analysis in \cite{Donoho10,Chen17}, it is known that the MMSE denoiser for user $n$ is given by the posterior mean of $x_{n}$ conditioned on the given $\tilde{x}^{t}_{n}=x_{n}+\tau_{t}v$, where $\tau_{t}$ is the state variable in the $t$-th iteration and $v\sim\mathcal{CN}(0,1)$.
	In particular, it can be expressed as
	\begin{align}
	&\eta \left(u,\beta,t\right)=E_{x_{n}|\tilde{x}^{t}_{n} = u}\left[x_{n}\right] \overset{(a)}{=} \frac{{\epsilon\lambda}}{P_{\tilde{x}^{t}_{n}}\left(u\right)}\int {z{P_{\bar{x}^{t}_{n}|\hat{h}_{n}}}\left(u|\hat{h}_{n} = z\right)} P_{\hat{h}_{n}}(z)dz \notag \\
	&= \frac{{\epsilon\lambda{P_{\bar{x}_{n}^{t}}}\left(u\right)}}{P_{\tilde{x}^{t}_{n}}\left(u\right)}\int {z{P_{\hat{h}_{n}|{\bar{x}_{n}^{t}}}}\left(z|{\bar{x}_{n}^{t}=u}\right)} dz = \frac{{\epsilon\lambda{P_{\bar{x}_{n}^{t}}}\left(u\right)}}{P_{\tilde{x}^{t}_{n}}\left(u\right)}E_{\hat{h}_{n}|\bar{x}_{n}^{t} = u}\left[\hat{h}_{n}\right],
	\label{denoiser}
	\end{align}
	where $\hat{h}_{n}$ denotes the random variable to characterize the channel distribution of user $n$ experienced by the receiver and $\bar{x}_{n}^{t}=\hat{h}_{n}+\tau_{t}v$.
	The equality $(a)$ is based on the distribution of $x_{n}$ in Eq. \eqref{distri_x}.
	It can be seen from Eq. \eqref{denoiser} that the MMSE denoiser depends on the distribution of $\bar{x}_{n}^{t}$, the distribution of $\tilde{x}^{t}_{n}$, and the posterior mean of $\hat{h}_{n}$ given $\bar{x}_{n}^{t}$.
	The distribution of $\bar{x}_{n}^{t}$ has been given by Lemma \ref{lemma_dis_x_bar}, then we derive the distribution of $\tilde{x}^{t}_{n}$ and the posterior mean of $\hat{h}_{n}$ given $\bar{x}_{n}^{t}$ in the following.
	
	According to Eq. \eqref{GaussianAppro}, i.e., $\tilde{x}^{t}_{n}=x_{n}+\tau_{t}v$, and the distribution of $x_{n}$ in Eq. \eqref{distri_x}, we have
	\begin{align}
	P_{\tilde{x}^{t}_{n}}\left(u\right)\overset{(a)}{=} &\epsilon \lambda \frac{{{\mathrm{Q}_1}\left(\frac{|\mu_{t}u|}{{\sqrt {{{\sigma_t^2} \mathord{\left/
								{\vphantom {{\sigma_t^2} 2}} \right.
								\kern-\nulldelimiterspace} 2}} }},\frac{\varsigma }{{\sqrt {{{\sigma_t^2} \mathord{\left/
								{\vphantom {{\sigma_t^2} 2}} \right.
								\kern-\nulldelimiterspace} 2}} }}\right)}}{{\lambda \pi \left(\beta  + \tau_{t}^2\right)}}\exp \left( - \frac{{|u|^2}}{{\beta  + \tau_{t}^2}}\right) \notag \\
							&+\left(1-\epsilon \lambda\right)\frac{1}{\pi \tau_{t}^2}\exp\left(-\frac{|u|^2}{\tau_{t}^2}\right),
	\label{dis_hat_x}
	\end{align}
	where $\mu_{t}=\frac{\beta}{\beta  + \tau_{t}^2}$, $\sigma_t^2=\frac{\beta\tau_{t}^2}{\beta+\tau_{t}^2}$, and the equality $(a)$ is obtained from the distribution of $\bar{x}_{n}^{t}$ in Lemma \ref{lemma_dis_x_bar}.
	
	Now, considering $\bar{x}_{n}^{t}=\hat{h}_{n}+\tau_{t}v$, the posterior mean of $\hat{h}_{n}$ given $\bar{x}_{n}^{t}$ is derived as
	\begin{align}
	&E_{\hat{h}_{n}|\bar{x}_{n}^{t} = u}\left[\hat{h}_{n}\right]=\int z {P_{\hat{h}_{n}|\bar{x}_{n}^{t}}}\left(z|\bar{x}_{n}^{t} = u\right) dz \notag \\
	&=\frac{1}{{P_{\bar{x}_{n}^{t}}}\left(u\right)} \int z P_{\bar{x}_{n}^{t}|\hat{h}_{n}}\left(u|\hat{h}_{n} = z\right)P_{\hat{h}_{n}}(z) dz \notag \\
	&=\frac{1}{{P_{\bar{x}_{n}^{t}}}\left(u\right)} \frac{1}{{\lambda \pi \left(\beta  + \tau_{t}^2\right)}}\exp \left( - \frac{{|u|^2}}{{\beta  + \tau_{t}^2}}\right) \int \frac{z}{{\pi \sigma_{t}^{2}}}\exp \left( - \frac{{|z - \mu_{t}u|^2}}{\sigma_{t}^{2}}\right)c\left(z\right)dz \notag \\
	&\overset{(a)}{=}{{\mathrm{Q}_1}^{-1}\left( {\frac{{\frac{\beta }{{\beta  + \tau_{t}^2}}|u|}}{{\sqrt {{{\sigma_{t}^2} \mathord{\left/
								{\vphantom {{\sigma_{t}^2} 2}} \right.
								\kern-\nulldelimiterspace} 2}} }},\frac{\varsigma }{{\sqrt {{{\sigma_{t}^2} \mathord{\left/
								{\vphantom {{\sigma_{t}^2} 2}} \right.
								\kern-\nulldelimiterspace} 2}} }}} \right)}\int \frac{z}{{\pi \sigma_{t}^{2}}}\exp \left( - \frac{{|z - \mu_{t}u|^2}}{\sigma_{t}^{2}}\right)c\left(z\right)dz \notag \\
	&\overset{(b)}{=}{{\mathrm{Q}_1}^{ - 1}\left( {\frac{{\frac{\beta }{{\beta  + \tau_{t}^2}}|u|}}{{\sqrt {{{\sigma_{t}^2} \mathord{\left/
								{\vphantom {{\sigma_{t}^2} 2}} \right.
								\kern-\nulldelimiterspace} 2}} }},\frac{\varsigma }{{\sqrt {{{\sigma_{t}^2} \mathord{\left/
								{\vphantom {{\sigma_{t}^2} 2}} \right.
								\kern-\nulldelimiterspace} 2}} }}} \right)}\int_{\varsigma}^{\infty} \int_{0}^{2\pi} \frac{r^2}{{\pi \sigma_{t}^{2}}}e^{j\theta}\exp \left( - \frac{{|re^{j\theta} - ke^{j\varphi}|^2}}{\sigma_{t}^{2}}\right)d\theta dr \notag \\
	&={{\mathrm{Q}_1}^{ - 1}\left( {\frac{{\frac{\beta }{{\beta  + \tau_{t}^2}}|u|}}{{\sqrt {{{\sigma_{t}^2} \mathord{\left/
								{\vphantom {{\sigma_{t}^2} 2}} \right.
								\kern-\nulldelimiterspace} 2}} }},\frac{\varsigma }{{\sqrt {{{\sigma_{t}^2} \mathord{\left/
								{\vphantom {{\sigma_{t}^2} 2}} \right.
								\kern-\nulldelimiterspace} 2}} }}} \right)}e^{j\varphi}\int_{\varsigma}^{\infty} \frac{2r^2}{\sigma_{t}^{2}}\exp\left(-\frac{r^2+k^2}{\sigma_{t}^{2}}\right)\mathbf{I}_{1}\left(\frac{2rk}{\sigma_{t}^{2}}\right)dr \notag \\
	%&= \frac{\beta u}{\beta  + \tau_{t}^2}{{{\mathrm{Q}_1}^{ - 1}\left( {\frac{{\frac{\beta }{{\beta  + \tau_{t}^2}}|u|}}{{\sqrt {{{\sigma_{t}^2} \mathord{\left/
	% {\vphantom {{\sigma_{t}^2} 2}} \right.
	% \kern-\nulldelimiterspace} 2}} }},\frac{\varsigma }{{\sqrt {{{\sigma_{t}^2} \mathord{\left/
	% {\vphantom {{\sigma_{t}^2} 2}} \right.
	% \kern-\nulldelimiterspace} 2}} }}} \right)}}{\mathrm{Q}_2}\left( {\frac{{\frac{\beta }{{\beta  + \tau_{t}^2}}|u|}}{{\sqrt {{{\sigma_{t}^2} \mathord{\left/
	% {\vphantom {{\sigma_{t}^2} 2}} \right.
	% \kern-\nulldelimiterspace} 2}} }},\frac{\varsigma }{{\sqrt {{{\sigma_{t}^2} \mathord{\left/
	% {\vphantom {{\sigma_{t}^2} 2}} \right.
	% \kern-\nulldelimiterspace} 2}} }}} \right) \notag \\
	&=\mu_{t}u{{{\mathrm{Q}_1}^{ - 1}\left( {\frac{|\mu_{t}u|}{{\sqrt {{{\sigma_{t}^2} \mathord{\left/
									{\vphantom {{\sigma_{t}^2} 2}} \right.
									\kern-\nulldelimiterspace} 2}} }},\frac{\varsigma }{{\sqrt {{{\sigma_{t}^2} \mathord{\left/
									{\vphantom {{\sigma_{t}^2} 2}} \right.
									\kern-\nulldelimiterspace} 2}} }}} \right)}}{\mathrm{Q}_2}\left( {\frac{|\mu_{t}u|}{{\sqrt {{{\sigma_{t}^2} \mathord{\left/
							{\vphantom {{\sigma_{t}^2} 2}} \right.
							\kern-\nulldelimiterspace} 2}} }},\frac{\varsigma }{{\sqrt {{{\sigma_{t}^2} \mathord{\left/
							{\vphantom {{\sigma_{t}^2} 2}} \right.
							\kern-\nulldelimiterspace} 2}} }}} \right),
	\label{post_mean_h}
	\end{align}
	where $\mathbf{I}_{1}(\cdot)$ is the modified Bessel function of the first kind and $c\left(z\right)$ is the transmission control function defined in Eq. \eqref{step_cntrl_function}.
	The equality $(a)$ is obtained from the distribution of $\bar{x}_{n}^{t}$ in Lemma \ref{lemma_dis_x_bar} and equality $(b)$ is obtained by transforming the complex integral to the integral in polar coordinates\cite{Donoughue12}, i.e., $z=re^{j\theta}$ and $\mu_{t}u=ke^{j\varphi}$.
	
	Substituting Eqs. \eqref{dis_x_bar}, \eqref{dis_hat_x}, and \eqref{post_mean_h} into \eqref{denoiser}, we can obtain the MMSE denoiser function of AMP algorithm for the proposed transmission control scheme in Eq. \eqref{denoiser_tx_cntrl}.
	
	\subsection{Proof of Theorem \ref{theo_state_evolo_tx_contrl}} \label{AppendixC3}
	According to Eq. \eqref{state_evolo}, we know that the state variable $\tau_{t+1}$ is determined by the MSE of each entry of the estimate $\hat{\mathbf{x}}^{t+1}$ in the $(t+1)$-th iteration. Then, based on the proposed MMSE denoiser, we first derive the MSE of each entry of the estimate $\hat{{x}}^{t+1}_{n} = \eta \left(\tilde{x}_{n}^{t},\beta,t\right)$, which is given by
	\begin{align}
	&{E_{\tilde{x}_{n}^{t},x_{n}}}\left[ {{{\left| {\eta\left(\tilde{x}_{n}^{t},\beta,t\right) - x_{n}} \right|}^2}} \right] \overset{(a)}{=}{E_{\tilde{x}_{n}^{t},x_{n}}}\left[ {{{\left| {E_{x_{n}|\tilde{x}_{n}^{t}}\left[x_{n}\right] - x_{n}} \right|}^2}} \right] \notag \\
	&\overset{(b)}{=}E_{\tilde{x}_{n}^{t}} \left[E_{x_{n}|\tilde{x}_{n}^{t}}\left[ {{\left( {E_{x_{n}|\tilde{x}_{n}^{t}}\left[x_{n}\right] - x_{n}} \right)\left( {E_{x_{n}|\tilde{x}_{n}^{t}}\left[x_{n}\right] - x_{n}} \right)^{\ast}}} \right] \right] \notag \\
	&=E_{\tilde{x}_{n}^{t}}\left[E_{x_{n}|\tilde{x}_{n}^{t}}\left[{x_{n}x_{n}^{\ast}}\right]\right]-E_{\tilde{x}_{n}^{t}}
	\left[E_{x_{n}|\tilde{x}_{n}^{t}}\left[x_{n}\right]\left(E_{x_{n}|\tilde{x}_{n}^{t}}\left[x_{n}\right]\right)^{\ast} \right] \notag \\
	&\overset{(c)}{=}E_{\tilde{x}_{n}^{t}}\left[\phi \left(\tilde{x}_{n}^{t}\right)E_{\hat{h}_{n}|\bar{x}_{n}^{t}=\tilde{x}_{n}^{t}}\left[\hat{h}_{n}\hat{h}_{n}^{\ast}\right]\right]-
	E_{\tilde{x}_{n}^{t}}\left[\phi^2\left(\tilde{x}_{n}^{t}\right)\left|E_{\hat{h}_{n}|\bar{x}_{n}^{t}=\tilde{x}_{n}^{t}}\left[\hat{h}_{n}\right]\right|^2\right],
	\label{mse}
	\end{align}
	where $\phi\left(u\right)=\frac{{\epsilon\lambda{P_{{\bar{x}_{n}^{t}}}}\left(u\right)}}{{{P_{\tilde{x}^{t}_{n}}}\left(u\right)}}$ and the random variables $\hat{h}_{n}$ and $\bar{x}_{n}^{t}$ are defined in Lemma \ref{lemma_dis_x_bar}.
	Note that, the equality $(a)$ is obtained with $\eta\left(\tilde{x}_{n}^{t},\beta,t\right) = E_{x_{n}|\tilde{x}_{n}^{t}}\left[x_{n}\right]$ from Eq. \eqref{denoiser} in the proof of Theorem \ref{theo_denoiser_tx_cntrl}.
	Besides, the equality $(b)$ is obtained via using the Bayes theorem.
	In addition, the equality $E_{x_{n}|\tilde{x}_{n}^{t}}\left[{x_{n}}\right]=\phi\left(\tilde{x}_{n}^{t}\right)E_{\hat{h}_{n}|\bar{x}_{n}^{t}=\tilde{x}_{n}^{t}}\left[\hat{h}_{n}\right]$ is obtained from Eq. \eqref{denoiser} in the proof of Theorem \ref{theo_denoiser_tx_cntrl}.
	Similar to the derivation in Eq. \eqref{denoiser}, we obtain $E_{x_{n}|\tilde{x}_{n}^{t}}\left[{x_{n}x_{n}^{\ast}}\right]=\phi\left(\tilde{x}_{n}^{t}\right)E_{\hat{h}_{n}|\bar{x}_{n}^{t}=\tilde{x}_{n}^{t}}\left[\hat{h}_{n}\hat{h}_{n}^{\ast}\right]$.
	The two equalities lead to the equality $(c)$ in Eq. \eqref{mse}.
	
	From the posterior distribution of $\hat{h}_{n}$ conditioned on $\bar{x}_{n}^{t} = u$ in Eq. \eqref{post_dis_hn}, we have
	\begin{align}
	&E_{\hat{h}_{n}|\bar{x}_{n}^{t}=u}\left[\hat{h}_{n}\hat{h}_{n}^{\ast}\right] =\frac{1}{{P_{\bar{x}_{n}^{t}}}\left(u\right)} \frac{\exp \left( - \frac{{|u|^2}}{{\beta  + \tau_{t}^2}}\right)}{\lambda \pi \left(\beta  + \tau_{t}^2\right)}\int zz^{\ast}\frac{1}{{\pi \sigma_{t}^{2}}}\exp \left( - \frac{{|z - \mu_{t}u|^2}}{\sigma_{t}^{2}}\right)c(z)dz \notag \\
	&\overset{(a)}{=}{{\mathrm{Q}_1}^{ - 1}\left( {\frac{|\mu_{t}u|}{{\sqrt {{{\sigma_{t}^2} \mathord{\left/
								{\vphantom {{\sigma_{t}^2} 2}} \right.
								\kern-\nulldelimiterspace} 2}} }},\frac{\varsigma }{{\sqrt {{{\sigma_{t}^2} \mathord{\left/
								{\vphantom {{\sigma_{t}^2} 2}} \right.
								\kern-\nulldelimiterspace} 2}} }}} \right)}\!\! \int_{\varsigma}^{\infty}\!\!\!\! \int_{0}^{2\pi} \!\!\!\frac{r^3}{{\pi \sigma_{t}^{2}}}\exp \left(-\frac{r^2+k^2}{\sigma_{t}^{2}}+\frac{2rk}{\sigma_{t}^{2}}\cos \left(\theta-\phi\right)\right)d\theta dr \notag \\
	&={{\mathrm{Q}_1}^{ - 1}\left( {\frac{|\mu_{t}u|}{{\sqrt {{{\sigma_{t}^2} \mathord{\left/
								{\vphantom {{\sigma_{t}^2} 2}} \right.
								\kern-\nulldelimiterspace} 2}} }},\frac{\varsigma }{{\sqrt {{{\sigma_{t}^2} \mathord{\left/
								{\vphantom {{\sigma_{t}^2} 2}} \right.
								\kern-\nulldelimiterspace} 2}} }}} \right)}\int_{\varsigma}^{\infty} \frac{2r^3}{\sigma_{t}^{2}}\exp\left(-\frac{r^2+k^2}{\sigma_{t}^{2}}\right)\mathbf{I}_{0}\left(\frac{2rk}{\sigma_{t}^{2}}\right)dr \notag \\
	&={{\mathrm{Q}_1}^{ - 1}\left( {\frac{|\mu_{t}u|}{{\sqrt {{{\sigma_{t}^2} \mathord{\left/
								{\vphantom {{\sigma_{t}^2} 2}} \right.
								\kern-\nulldelimiterspace} 2}} }},\frac{\varsigma }{{\sqrt {{{\sigma_{t}^2} \mathord{\left/
								{\vphantom {{\sigma_{t}^2} 2}} \right.
								\kern-\nulldelimiterspace} 2}} }}} \right)} \times \notag \\
							&\int_{\varsigma}^{\infty} \frac{2r^3}{\sigma_{t}^{2}}\exp\left(-\frac{r^2+k^2}{\sigma_{t}^{2}}\right)\left(\mathbf{I}_{2}\left(\frac{2rk}{\sigma_{t}^{2}}\right)+\frac{\sigma_{t}^{2}}{rk}\mathbf{I}_{1}\left(\frac{2rk}{\sigma_{t}^{2}}\right)\right)dr \notag \\
	%&={{\mathrm{Q}_1}^{-1}\left( {\frac{|\mu_{t}u|}{{\sqrt {{{\sigma_{t}^2} \mathord{\left/
	% {\vphantom {{\sigma_{t}^2} 2}} \right.
	% \kern-\nulldelimiterspace} 2}} }},\frac{\varsigma }{{\sqrt {{{\sigma_{t}^2} \mathord{\left/
	% {\vphantom {{\sigma_{t}^2} 2}} \right.
	% \kern-\nulldelimiterspace} 2}} }}} \right)}\left(v^2{{\mathrm{Q}_3}\left( {\frac{|\mu_{t}u|}{{\sqrt {{{\sigma_{t}^2} \mathord{\left/
	% {\vphantom {{\sigma_{t}^2} 2}} \right.
	% \kern-\nulldelimiterspace} 2}} }},\frac{\varsigma }{{\sqrt {{{\sigma_{t}^2} \mathord{\left/
	% {\vphantom {{\sigma_{t}^2} 2}} \right.
	% \kern-\nulldelimiterspace} 2}} }}} \right)}+\sigma_{t}^2{{\mathrm{Q}_2}\left( {\frac{|\mu_{t}u|}{{\sqrt {{{\sigma_{t}^2} \mathord{\left/
	% {\vphantom {{\sigma_{t}^2} 2}} \right.
	% \kern-\nulldelimiterspace} 2}} }},\frac{\varsigma }{{\sqrt {{{\sigma_{t}^2} \mathord{\left/
	% {\vphantom {{\sigma_{t}^2} 2}} \right.
	% \kern-\nulldelimiterspace} 2}} }}} \right)}\right) \notag \\
	&=\frac{|\mu_{t}u|^2}{\alpha_{3}\left(u,t\right)}+\frac{\sigma_{t}^2}{\alpha_{1}\left(u,t\right)},
	\label{mean_squared_h}
	\end{align}
	where $\alpha_{1}\left(u,t\right)$ is defined in Theorem 1 and $\alpha_{3}\left(u,t\right)$ is given by
\begin{align}
	\alpha_{3} \left(u,t\right)=\frac{{{\mathrm{Q}_1}\left(\frac{|\mu_{t}u|}{{\sqrt {{{\sigma_{t}^2} \mathord{\left/
								{\vphantom {{\sigma_t^2} 2}} \right.
								\kern-\nulldelimiterspace} 2}} }},\frac{\varsigma }{{\sqrt {{{\sigma_{t}^2} \mathord{\left/
								{\vphantom {{\sigma_{t}^2} 2}} \right.
								\kern-\nulldelimiterspace} 2}} }}\right)}}{{{\mathrm{Q}_3}\left(\frac{|\mu_{t}u|}{{\sqrt {{{\sigma_{t}^2} \mathord{\left/
								{\vphantom {{\sigma_t^2} 2}} \right.
								\kern-\nulldelimiterspace} 2}} }},\frac{\varsigma }{{\sqrt {{{\sigma_{t}^2} \mathord{\left/
								{\vphantom {{\sigma_{t}^2} 2}} \right.
								\kern-\nulldelimiterspace} 2}} }}\right)}}.
\end{align}							
Equality $(a)$ is obtained by transforming the complex integral to the integral in polar coordinates, i.e., $z=re^{j\theta}$ and $\mu_{t}u=ke^{j\varphi}$. According to Eqs. \eqref{state_evolo}, \eqref{post_mean_h}, \eqref{mse}, and \eqref{mean_squared_h}, we obtain  \eqref{state_evolo_tx_contrl}.
	\vspace{-3mm}
	\subsection{Proof of Theorem \ref{theo_pf_pm_tx_contrl}} \label{AppendixD3}
	According to the proposed activity detection scheme and the definition of false alarm probability, we derive the false alarm probability of user $n$ as
	\begin{align}
	P_{f,n}&=\int_{|u|>l_{n}}\!\!\!P_{\tilde{x}_{n}^{\infty}|x_{n}}\left(u|x_{n}=0\right)du\overset{(a)}{=}\int_{|u|>l_{n}}{\frac{1}{{\pi \tau_{\infty}^2}}\exp \left( { - \frac{{{{\left| u \right|}^2}}}{{\tau_{\infty}^2}}} \right)du} \notag \\
	& = \exp \left( { - \frac{{l_n^2}}{{\tau_{\infty}^2}}} \right), \forall n.
	\label{pf_t_iteration_proof}
	\end{align}
	Equality $(a)$ is based on Eq. \eqref{GaussianAppro} and $\tilde{x}_{n}^{\infty}\sim\mathcal{CN}(0,\tau_{\infty}^2)$ is conditioned on ${x}_{n}=0$.
	
	We obtain the missed detection probability of user $n$ as
	\begin{align}
	P_{m,n}&=\int_{|u|<l_{n}}P_{\tilde{x}_{n}^{\infty}|x_{n}}\left(u|x_{n}\neq 0\right)dx \notag \\ &\overset{(a)}{=}\int_{\left|u\right| < {l_n}} {\frac{{{\mathrm{Q}_1}\left( {\frac{{\frac{\beta }{{\beta  + \tau _{\infty}^2}}\left|u\right|}}{{\sqrt {{{\sigma_\infty^2} \mathord{\left/
										{\vphantom {{\sigma_\infty^2} 2}} \right.
										\kern-\nulldelimiterspace} 2}} }},\frac{\varsigma }{{\sqrt {{{\sigma_\infty^2} \mathord{\left/
										{\vphantom {{\sigma_\infty^2} 2}} \right.
										\kern-\nulldelimiterspace} 2}} }}} \right)}}{{\lambda \pi (\beta  + \tau _{\infty}^2)}}\exp \left( { - \frac{{\left|u\right|^2}}{{\beta  + \tau_{\infty}^2}}} \right)} du \notag \\
	%&=\int_{r < {l_n}} {\frac{2r{{\mathrm{Q}_1}\left( {\frac{{\frac{\beta }{{\beta  + \tau _{\infty}^2}}r}}{{\sqrt {{{\sigma_t^2} \mathord{\left/
	% {\vphantom {{\sigma_t^2} 2}} \right.
	% \kern-\nulldelimiterspace} 2}} }},\frac{\varsigma }{{\sqrt {{{\sigma_t^2} \mathord{\left/
	% {\vphantom {{\sigma_t^2} 2}} \right.
	% \kern-\nulldelimiterspace} 2}} }}} \right)}}{{\lambda (\beta  + \tau _{\infty}^2)}}\exp \left( { - \frac{{r^2}}{{\beta  + \tau_{\infty}^2}}} \right)} dr \notag \\
	%&\overset{(b)}{=}\frac{1}{{\lambda}}\left[ {\exp \left( { - \frac{{(\beta  + \tau _{\infty}^2){\varsigma ^2}}}{{{\beta ^2} + \sigma _\infty^2(\beta  + \tau _{\infty}^2)}}} \right){{\mathop{\rm Q}\nolimits} _1}\left( {\frac{{{l_n}\sqrt {2({\beta ^2} + \sigma _\infty^2(\beta  + \tau _{\infty}^2))} }}{{{\sigma _\infty}(\beta  + \tau _{\infty}^2)}},\frac{{2\beta \varsigma }}{{{\sigma _\infty}\sqrt {2({\beta ^2} + \sigma _\infty^2(\beta  + \tau _{\infty}^2))} }}} \right)} \right. \notag \\
	%&\left. { - \exp \left( { - \frac{{l_n^2}}{{\beta  + \tau _{\infty}^2}}} \right){{\mathop{\rm Q}\nolimits} _1}\left( {\frac{{\beta {l_n}}}{{(\beta  + \tau _{\infty}^2)\sqrt {{{\sigma _\infty^2} \mathord{\left/
	% {\vphantom {{\sigma _\infty^2} 2}} \right.
	% \kern-\nulldelimiterspace} 2}} }},\frac{\varsigma }{{\sqrt {{{\sigma _\infty^2} \mathord{\left/
	% {\vphantom {{\sigma _\infty^2} 2}} \right.
	% \kern-\nulldelimiterspace} 2}} }}} \right)} \right], \notag \\
	&\overset{(b)}{=}\frac{1}{{\lambda}}\left[ {\exp \left( { - \frac{\varsigma ^2}{\mu_{\infty} \beta  + \sigma _{\infty}^2}} \right){{\mathop{\rm Q}\nolimits} _1}\left({l_n} \sqrt{2 \left(\frac{\mu_{\infty}^2}{\sigma_{\infty}^2}+\frac{\mu_{\infty}}{\beta}\right)}, \frac{\sqrt{2\beta \mu_{\infty}}\varsigma}{\sigma_{\infty}\sqrt{\beta\mu_{\infty}+\sigma_{\infty}^2}}\right)\notag }\right.\\
	&\left.{-\exp \left( { - \frac{{l_n^2}}{{\beta  + \tau_{\infty}^2}}}\right){{\mathop{\rm Q}\nolimits} _1}\left( {\frac{\mu_{\infty} {l_n}}{\sqrt {{{\sigma _{\infty}^2} \mathord{\left/
							{\vphantom {{\sigma _{\infty}^2} 2}} \right.
							\kern-\nulldelimiterspace} 2}}},\frac{\varsigma }{{\sqrt {{{\sigma _{\infty}^2} \mathord{\left/
								{\vphantom {{\sigma _{\infty}^2} 2}} \right.
								\kern-\nulldelimiterspace} 2}} }}} \right)}\right],
	\forall n,
	\label{pm_t_iteration_proof}
	\end{align}
	where $\mu_{\infty}=\frac{\beta}{\beta+\tau_{\infty}^2}$ and $\sigma^2_{\infty}=\frac{\beta\tau_{\infty}^2}{\beta+\tau_{\infty}^2}$.
	The equality $(a)$ is based on the fact $\tilde{x}_{n}^{\infty} = \bar{x}_{n}^{t}$ with $t \to \infty$ conditioned on $x_{n}\neq 0$, i.e., $a_n = 1$, and the distribution of $\bar{x}_{n}^{t}$ in Lemma \ref{lemma_dis_x_bar}.
	The equality $(b)$ is based on the integral involving Marcum Q-functions in \cite{Nuttall75}.

\chapter{Design and Analysis of Coded Slotted ALOHA for mMTC under Erasure Channels}\label{C4:chapter4}

\section{Introduction}
In the previous chapter, we proposed a joint user activity identification and channel estimation to enhance the accuracy of user identification and channel estimation in grant-free random access systems for mMTC.
With the accurate user activity identification and channel estimation, we propose and design a random access system, i.e., the CSA system, over erasure channels to enhance the network throughput in this chapter.

We consider two types of erasure channels in this work, i.e., packet erasure channels and slot erasure channels, in order to capture the effects of practical channel fading and external noise on the transmission scheme design, respectively.
For these two kinds of erasure channels, we characterize the impact of channel erasure on the bipartite graph for the proposed CSA scheme, where no capture effect is considered. Based on this, we derive the EXIT functions for CSA schemes over packet erasure channels and slot erasure channels, which allow an asymptotic analysis of user recovery process to design the code distributions.
Note that, a user may lose all packets over erasure channels.
As a result, the traffic load threshold defined by \cite{Liva2011} is strictly zero for erasure channels. To quantify the performance of CSA schemes over erasure channels, we introduce a recovery ratio $\alpha$ and we define the expected traffic load for the CSA schemes over erasure channels. In particular, the expected traffic load is the largest traffic load such that at least $\alpha$ percent of active users can be successfully recovered. Based on this, we design the probability distributions of repetition codes to maximize the expected traffic load for the CSA scheme over various erasure channels. Comparing with the repetition code distributions presented in \cite{Paolini2011} that are obtained to maximize the traffic load threshold of CSA schemes for collision channels, we demonstrate that our designed code distributions can improve the traffic load threshold and the throughput of CSA schemes over erasure channels.

In addition, we employ the MDS codes for the CSA scheme. According to the bounded distance strategy of MDS decoding \cite{Singleton1964,Wicker1999}, we derive the EXIT functions for MDS codes over packet erasure channels and slot erasure channels, respectively. Based on that, we design the probability distributions of MDS codes for the CSA scheme over erasure channels, to maximize the expected traffic load. Furthermore, we analyze the effect of different design parameters on the performance of CSA schemes with MDS codes for packet erasure channels.

In this work, we also derive the asymptotic throughput of CSA schemes with infinite frame lengths for erasure channels. It is shown that our asymptotic throughput can give a good approximation to the simulated throughput of CSA schemes for erasure channels, where more than two replicas are employed by each user. This is consistent with the claim for the asymptotic analysis of CRDSA schemes over non-erasure channels in \cite{DelRioHerrero2014}, which shows that if more than two replicas are employed by each user, the asymptotic throughput holds well for practical frame lengths.

\section{System Model and Problem Formulation}
\begin{figure*}[!t]
	\par
	\begin{center}
		{\includegraphics[width=1\textwidth]{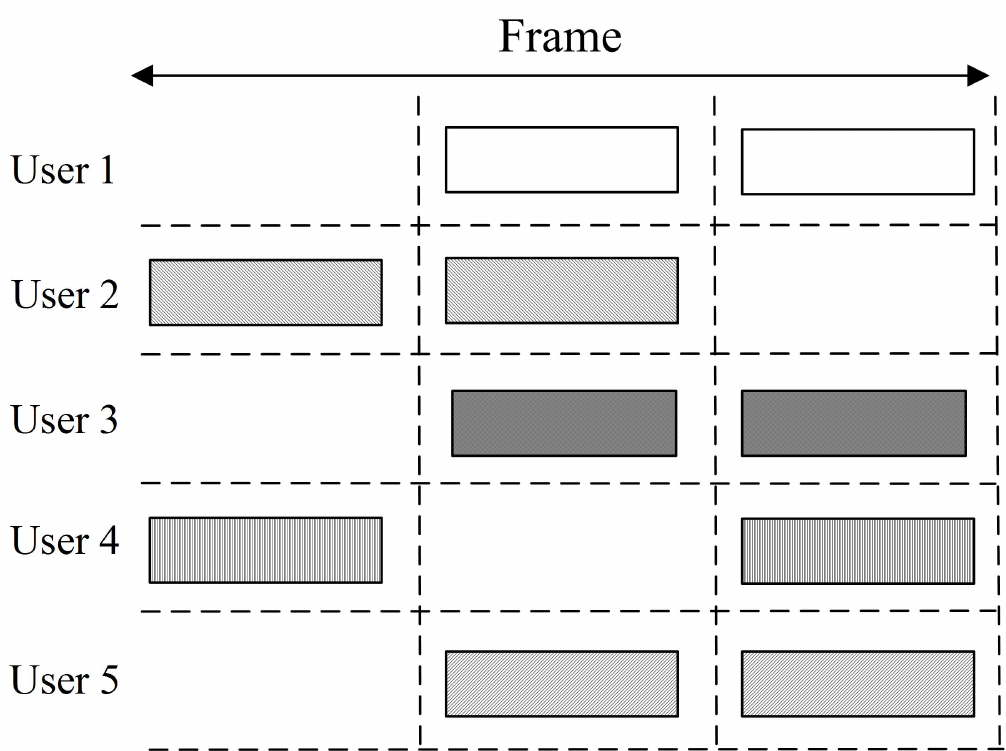}}
	\end{center}
	\caption{CSA system model with erasure channels.}
	\label{fig_system_model_erasure_v4}
\end{figure*}

The CSA system model with erasure rate $\epsilon$ is shown in Fig. \ref{fig_system_model_erasure_v4}. Consider a random access scheme where $M$ active users attempt to transmit data to a common receiver via a shared channel. The MAC frame is a basic data transmission unit that consists of $N$ time slots with identical duration. Each user transmits one data block per MAC frame, denoted by $Z_{m}$ for $m\in\{1,2,\cdots ,M\}$, and all transmissions are slot synchronous.

We assume that the CSA scheme employs a code set ${\mathbf{C}}=\{{\mathbf{c}}_{1},{\mathbf{c}}_{2},\cdots,{\mathbf{c}}_{\theta}\}$, where ${\mathbf{c}}_{h}$ is the $h$-th code with length $n_{h}$ and dimension $k$ for $h\in\{1,2,\cdots ,\theta\}$. For each transmission, the user chooses a code ${\mathbf{c}}_{h}$ from the given code set $\mathbf{C}$ according to a designed probability distribution ${\bm{\Lambda}}=\{\Lambda_{h}\}_{h=1}^{\theta}$, where $\Lambda_{h}$ represents the probability of a code ${\mathbf{c}}_{h}$ being chosen and the average code length is $\bar{n}=\sum_{h=1}^{\theta}\Lambda_{h}n_{h}$. Before the transmission, the data block from each user is divided into $k$ information packets, where $k$ is identical for all $M$ users and the packet length is equal to the time slot duration. The $k$ information packets of each user are then encoded into $n_{h}$ coded packets via its own chosen code ${\mathbf{c}}_{h}$. After encoding, a preamble $P_{m}$ and a pointer are attached to each of the $n_{h}$ coded packets, where the preamble is unique \cite{CoccoVTC11,CoccoICC14,CoccoSatejournal14,Goseling15} and the pointer indicates the location of all other $(n_{h}-1)$ coded packets for each user \cite{Casini2007}. Moreover, each of $n_{h}$ coded packets is equipped with the information about the code chosen by the user. Then, the $n_{h}$ coded packets, denoted by $W_{m}$ for $m\in\{1,2,\cdots ,M\}$ in Fig. \ref{fig_system_model_erasure_v4}, are randomly spread to $n_{h}$ time slots and transmitted to the receiver. The spreading process follows an uniform distribution and the transmitted sequence of each user is denoted as $X_{m}$ for $m\in\{1,2,\cdots ,M\}$ in Fig. \ref{fig_system_model_erasure_v4}. In the CSA scheme, the \emph{offered traffic load} is defined as
\begin{align}
G \triangleq \frac{kM}{N},
\label{def_traffic_load}
\end{align}
which represents the average number of transmitted information packets per time slot.

The channel is a multiple access erasure channel, where $M$ noncooperative users communicate their data blocks to one receiver and the channel from each user to the receiver is the erasure channel with an erasure rate $\epsilon$. In particular, we consider both the packet erasure channel, where each transmitted packet may be erased with probability $\epsilon$, and the slot erasure channel, where all the packets transmitted in a slot may be erased with probability $\epsilon$.
Note that, the packet erasure channel is adopted as a simplification of fading channels, and the slot erasure channel is adopted to model the effect of strong external interference on the received packets in a particular time slot.
In Fig. \ref{fig_system_model_erasure_v4}, the erased packets are represented by the dashed-line squares.
In addition, we assume that all the erased packets are negligible for the packet recovery process at the receiver side.

Based on the received signal sequence within a frame, the receiver can identify the set of active users in each slot and remove the idle slots where no signal is received by using the cross-correlation on preambles.
Here, we assume that each user has a unique preamble and all the collided users in each slot can be identified\footnote[1]{Note that the number of collided users that can be identified in each slot is limited in practice when employing the preambles to identify the collided users.
	In addition, it is non-trivial to determine the maximal number of collided users that can be identified, which is beyond the scope of this work.}.
Then, the packet recovery process begins by detecting the signal in each collision free slot.
Once the signal is detected, its corresponding packet can be recovered and the information about the user is extracted.
The information includes the code ${\mathbf{c}}_{h}$ adopted by the user and the pointer that identifies the slots where other associated coded packets were sent.
For each active user whose information is extracted, the local decoding of the linear block code ${\mathbf{c}}_{h}$ is performed to recover the other associated packets of the user.
Combining the use of preamble and pointer, the receiver can determine the slots where the transmitted packets are erased for the user. By regenerating the signals of recovered packets in the unerased time slots \cite{Liva2011,Paolini2014}, their interference can be removed from the collisions. For a slot with $l$ users' signals, when the interference from the other $(l-1)$ signals has been removed, the residual signal is detected and its corresponding packet is recovered. This process is known as the SIC \cite{Casini2007} and it continues until the packets from all the users are recovered or no more packets can be recovered.

The recovered data block of each user is denoted by $\hat{Z}_{m}$ for $m\in\{1,2,\cdots ,M\}$. If $\hat{Z}_{m}$ is equal to the original data block $Z_{m}$, user $m$ is recovered. For a CSA scheme over the erasure channel, given the code set $\mathbf{C}$, probability distribution $\bm{\Lambda}$, erasure rate $\epsilon$, and traffic load $G$, the recovery probability of users is a random variable depending on the random spreading process and the random erasure of packets. When the frame length of the CSA scheme tends to infinity, the number of packets transmitted in a slot follows a poisson distribution with parameter $\frac{G\bar{n}}{k}$, which characterizes the random spreading process. The number of erased packets can be modeled by a binomial random variable with parameters $\epsilon$ and $\bar{n}M$, which characterizes the random erasure process. In this work, by averaging the recovery probability of users over the two random distributions, the average probability that a user can be recovered is obtained. The average probability depends on the parameter set $\{G, \mathbf{C}, \bm{\Lambda}, \epsilon \}$, denoted by $P_{u}(G, \mathbf{C}, \bm{\Lambda}, \epsilon)$. Then, the average throughput $T$ of the CSA scheme is defined as
\begin{align}
T(G,\mathbf{C},\bm{\Lambda},\epsilon)=GP_{u}(G,\mathbf{C},\bm{\Lambda},\epsilon).
\end{align}
It is noted that in terms of the average throughput, the system performance is only determined by the parameter set $\{G, \mathbf{C}, \bm{\Lambda}, \epsilon \}$ but neither by where the packets are spread randomly nor by which packets are erased randomly.

The packet loss rate (PLR) of the CSA scheme is given by
\begin{align}
\Upsilon(G,\mathbf{C},\bm{\Lambda},\epsilon)=1-\frac{T(G,\mathbf{C},\bm{\Lambda},\epsilon)}{G},
\end{align}
which represents the ratio between the number of unrecovered users' packets and the total number of all active users' packets.

For the CSA scheme over non-erasure collision channels, all users' packets can be recovered by performing the SIC process successfully, if the users' code probability distributions are carefully designed, as in \cite{Liva2011}.
Then, the traffic load threshold $G^{\ast}$ is used to measure the performance of the designed code probability distributions for CSA schemes, which is the largest traffic load such that all users can be recovered with probability close to one \cite{Casini2007,Liva2011}.
However, for the erasure channels that we consider in this work, some users may loss all their packets during the transmission.
In this case, these users' packets cannot be recovered by the SIC process, even the users' code probability distributions are carefully designed.
Therefore, for the erasure channels, it cannot be guaranteed that all users are recovered by the SIC process for any offered traffic load.
This implies that the PLR exhibits an error floor, and therefore, the traffic load threshold $G^{\ast}$ is zero for the erasure channels, as stated in \cite{IvanovBAP2015}.
Since the CSA scheme cannot recover all users' transmitted packets for the erasure channels, we introduce a term, called the user recovery ratio $\alpha$ for $0 < \alpha < 1$, to measure the ratio between the number of successfully recovered users and the number of all active users.
Accordingly, we define the expected traffic load with the user recovery ratio $\alpha$, denoted by $G_{\alpha}^{\ast}$.
In particular, $G_{\alpha}^{\ast}$ is the largest traffic load such that at least $\alpha$ percent of active users can be successfully recovered for the CSA scheme over erasure channels, that is
\begin{align}
G_{\alpha}^{\ast}(\mathbf{C},\bm{\Lambda},\epsilon)\overset{\Delta}{=}\max \left\{G \Big | \frac{T({G},\mathbf{C},\bm{\Lambda},\epsilon)}{G} \geq \alpha \right\}.
\label{def_opt_traffic_load}
\end{align}
Thus, given $\mathbf{C}$, $\epsilon$, and $\alpha$, we can maximize $G_{\alpha}^{\ast}$ by designing the code probability distribution
\begin{align}
{\bf{\Lambda^{\ast}}}\overset{\Delta}{=}\arg \max_{\bm{\Lambda}}G_{\alpha}^{\ast}(\mathbf{C},{\bm{\Lambda}},\epsilon).
\label{def_opt_dis}
\end{align}

It is noteworthy that the user recovery ratio $\alpha$ is directly related to the PLR $\Upsilon(G,\mathbf{C},\bm{\Lambda},\epsilon)$, that is $\Upsilon(G,\mathbf{C},\bm{\Lambda},\epsilon)=1-\frac{T(G,\mathbf{C},\bm{\Lambda},\epsilon)}{G}\leq 1- \alpha$ for any $G \leq G_{\alpha}^{\ast}$. In other words, over the erasure channels, the expected traffic load $G_{\alpha}^{\ast}$ is the largest traffic load which is achieved with the PLR no more than $(1-\alpha)$.

\section{Bipartite Graph and EXIT Chart Analysis}
In this section, we provide the bipartite graph and the EXIT chart analysis for CSA schemes, which are essential analytical tools to characterize the asymptotic performance.
\subsection{Bipartite Graph}
The CSA scheme can be represented by a bipartite graph $g=(U,S,E)$ in \cite{Liva2011}. Here, the set $U=\{u_1,\cdots,u_M\}$ represents $M$ user nodes (UN), $S=\{s_1,\cdots,s_N\}$ represents $N$ time slot nodes (SN) and $E=\{e_{mn}\}$ indicates the $n$-th SN is attempted by the $m$-th UN.

The number of edges connected to a UN or SN is the node degree \cite{Luby2001}. Then, in the CSA scheme, a user employing the code ${\mathbf{c}}_{h}$ is represented as a degree-$n_{h}$ UN, and a slot with the $l$-packets collision is a degree-$l$ SN in graph $g$. Thus, the code probability distribution of the CSA scheme is equivalent to the UN degree distribution of graph $g$. In the following, we refer to the code probability distribution as the UN degree distribution for simplification, and it is given by $\bm{\Lambda}$. Let ${\bf{\Psi}}=\{\Psi_{l}\}_{l=1}^{M}$ be the SN degree distribution of graph $g$. In the asymptotic regime where both $M$ and $N$ tend to infinity, the probability $\Psi_{l}$ that a SN has degree $l$ (i.e., the probability of an $l$-packets collision event within one SN) is given by \cite{Paolini2011}
\begin{align}
\label{def_thpt_infinite}
\lim_{N\rightarrow \infty, M\rightarrow \infty}\Psi_{l}=\frac{e^{\frac{-G}{R}}(\frac{G}{R})^{l}}{l!},
\end{align}
where $R=\frac{k}{\bar{n}}$ is the average transmission rate for the given code set $\mathbf{C}$ and UN degree distribution $\bm{\Lambda}$. In addition, define $\lambda(x)$ and $\rho(x)$ as the edge degree distribution polynomial of UN and SN of graph $g$ in  \cite{LubyMichaelMitzenmacherMAminShokrollahiDanielASpielmanVolkerStemann}, respectively. They can be represented as
\begin{align}
\lambda(x)\overset{\Delta}{=}\sum_{h=1}^{\theta}\lambda_{h}x^{n_{h}-1}
\hspace{2em}
\text{and}
\hspace{2em}
\rho(x)\overset{\Delta}{=}\sum_{l=1}^{M}\rho_{l}x^{l-1},
\label{def_edge_dis}
\end{align}
where $\lambda_{h}$ and $\rho_{l}$ are the fraction of edges emanating from degree-$n_{h}$ UNs and degree-$l$ SNs in graph $g$, respectively. They can be obtained from $\Lambda_{h}$ and $\Psi_{l}$, given by
\begin{align}
\lambda_{h}=\frac{\Lambda _{h}n_{h}}{\sum_{h=1}^{\theta}\Lambda _{h}n_{h}}
\hspace{2.5em}
\text{and}
\hspace{2.5em}
\rho _{l}=\frac{\Psi _{l}l}{\sum_{l=1}^{M}\Psi _{l}l},
\label{relation_nd_eg_dis}
\end{align}
respectively. Using Eq. \eqref{relation_nd_eg_dis}, $\rho(x)$ in Eq. \eqref{def_edge_dis} for $M, N\rightarrow \infty$ can be expressed as \cite{Paolini2011}
\begin{align}
\rho(x)=\exp\left(-\frac{G}{R}(1-x)\right).
%&=\sum\nolimits_{l=1}^{M}\frac{\Psi _{l}l}{\sum\nolimits_{l=1}^{M}\Psi _{l}l}x^{l-1}
\label{sn_eg_dis_colli_channel}
\end{align}
It is obvious that $\rho(x)$ depends on the traffic load $G$, given a transmission rate $R$.

\subsection{EXIT Chart Analysis}
Based on the analogy between the packet recovery process of CSA scheme and the iterative belief-propagation erasure-decoding, an EXIT chart \cite{Brink2001,Clark1999,Ashikhmin2004} can be used to visualize the evolution of this recovery process in \cite{Liva2011,Paolini2011,Paolini2014}. At the $i$-th iteration, let $p_{i}$ be the average probability that a packet cannot be recovered after the SIC. Let $q_{i}$ be the average probability that a packet cannot be recovered after the local decoding. For a degree-$l$ SN in the bipartite graph $g$, the edge connected to it can be recovered by SIC, if and only if $(l-1)$ edges of this SN are recovered successfully by the local decoding. This is given by $p_{i}^{(l)}=1-(1-q_{i})^{l-1}$, where $p_{i}^{(l)}$ is the average probability that a packet cannot be recovered from this SN at the $i$-th iteration. By averaging this function over the SN edge degree distribution $\rho(x)$ of the graph $g$, the EXIT function $f_{s}(q_{i})$ of SN decoder is obtained as \cite{Liva2011}
\begin{align}
f_{s}(q_{i})=\sum_{l=1}^{M}\rho _{l}(1-(1-q_{i})^{l-1})=1-\rho (1-q_{i}).
\label{exit_func_sn_colli_channel}
\end{align}
Note that, since $p_{i}$ is equal to $f_{s}(q_{i})$, $p_{i}=1-\rho (1-q_{i})$ is obtained.

As for the EXIT function of UN decoder, consider a degree-$n_{h}$ UN in graph $g$. By the local decoding, the probability $q_{i}$ outgoing from UNs can be computed from the priori probability $p_{i-1}$ from SNs. Then, the EXIT function of a degree-$n_{h}$ UN is obtained, denoted by $q_{i}^{(h)}=f_{u}^{(h)}(p_{i-1})$. By averaging this function over the UN edge degree distribution $\lambda(x)$, the EXIT function $f_{u}(p_{i-1})$ of UN decoder is expressed as \cite{Liva2011}
\begin{align}
f_{u}(p_{i-1})=\sum_{h=1}^{\theta}\lambda_{h}f_{u}^{(h)}(p_{i-1}).
\label{exit_func_un_colli_channel}
\end{align}
Due to $q_{i}=f_{u}(p_{i-1})$, Eq. \eqref{exit_func_un_colli_channel} can be written as $q_{i}=\sum_{h=1}^{\theta}\lambda_{h}f_{u}^{(h)}(p_{i-1})$.

Based on the obtained two EXIT functions, the iterative recovery process can be described by an EXIT chart, where the horizontal axis and vertical axis represent $p$ and $q$ in \cite{Liva2011}, respectively. The iteration proceeds as long as $p_{i+1}<p_{i}$, and stops if $p_{i+1}=p_{i}$. With $p_{i+1}=f_{s}(f_{u}(p_{i}))$, the stopping condition of iteration can be formulated as $f_{u}(p_{i})=f_{s}^{-1}(p_{i})$, which corresponds to an intersection of $f_{u}(p_{i})$ and $f_{s}^{-1}(p_{i})$ in the EXIT chart. Note that, the recursion of Eq. \eqref{exit_func_sn_colli_channel} and Eq. \eqref{exit_func_un_colli_channel} holds for the asymptotic setting where $N,M\rightarrow\infty$. Thus, in the following, the design of code distributions is performed in this asymptotic setting, but its effectiveness for a finite frame length will be verified in the section of numerical results.

\section{Design of Coded Slotted ALOHA for Erasure Channels}
In this section, the erasure channel is introduced as a practical channel model, and the design of UN degree distribution $\bf{\Lambda^{\ast}}$ for CSA scheme over erasure channels is proposed to maximize the expected traffic load $G_{\alpha}^{\ast}$. In particular, both the packet erasure channel and the slot erasure channel are considered. For both cases, while the iterative recovery process can be represented by a bipartite graph, only the subgraph induced by the erasure is involved in the recovery process. In other words, the subgraph consists of the unerased edges and it is exploited for the recovery process.
Recall the bipartite graph in Section III, it is known that the code probability distribution is equivalent to the UN degree distribution of the entire bipartite graph.
Thus, by analyzing the relationship of degree distributions for the subgraph and the entire bipartite graph, we derive the EXIT functions w.r.t. the degree distributions of the entire bipartite graph. Based on that, we can design the UN degree distribution of the entire bipartite graph to maximize the expected traffic load $G_{\alpha}^{\ast}$, which is equivalent to the design of code probability distributions. Here, the repetition codes and MDS codes are employed to illustrate the design of code distributions for erasure channels.

\subsection{Design of CSA Schemes for Packet Erasure Channels}
For wireless communications, when a transmitted packet experiences a deep fading, the corresponding received signal is too weak to be recovered and its impact on the other received signals in the same time slot is very limited to be ignorable. Then, in order to analyze the effect of deep fading on the design of code distributions, we consider the packet erasure channel in this section. In particular, when a transmitted packet experiences a deep fading and its fading gain is less than a given threshold $\delta$, i.e., $|h|<\delta$, the packet is assumed to be erased and the corresponding erasure probability is $\epsilon=\mathrm{Pr}(|h| < \delta)$.

Denote by $g=(U,S,E)$ and $\tilde{g}=(\tilde{U},\tilde{S},\tilde{E})$ as the entire bipartite graph and the subgraph induced by the erasure, respectively. Let $\lambda(x)$ and $\rho(x)$ be the UN and SN edge degree distribution polynomials of $g$, respectively. The corresponding polynomials for $\tilde{g}$ are given by $\tilde{\lambda}(x)$ and $\tilde{\rho}(x)$. For packet erasure channels, $\tilde{\lambda}(x)$ and $\tilde{\rho}(x)$ are obtained from $\lambda(x)$ and $\rho(x)$ by a random erasure of edges with probability $\epsilon$. Based on that, we can derive the EXIT functions for the CSA schemes, which are associated with $\lambda(x)$ and $\rho(x)$ in the graph $g$. They are presented in the following Lemma \ref{exit_sn_era} and Lemma \ref{exit_un_era}.

\begin{lemm} Let $N, M\rightarrow \infty$ for a constant traffic load $G$. Let $p_{i}$ and $q_{i}$ be the average probability of a packet being unrecovered by the SN decoder and UN decoder at the $i$-th iteration of the recovery process, respectively. Over the packet erasure channel with an erasure rate $\epsilon$, the EXIT function of the SN decoder for the CSA scheme is given by
\vspace{-5mm}
	\begin{equation}
	p_{i}=1-\exp\left(-\frac{G}{R}(1-\epsilon)q_{i}\right),
	\label{exit_func_sn_pct_ers_channel}\vspace{-5mm}
	\end{equation}
	where $R$ is the average transmission rate.
	\label{exit_sn_era}
\end{lemm}

\emph{\quad Proof: }Please refer to Appendix \ref{AppendixA}. \QEDA

Denote by $f'_{s}(q_{i})$ as the EXIT function of SN decoder for packet erasure channels, and we have
\begin{align}
f'_{s}(q_{i})=1-\exp\left(-\frac{G}{R}(1-\epsilon)q_{i}\right),
\label{exit_func_sn_pct_ers_channel_fs}
\end{align}
which represents the updating process from $q_{i}$ to $p_{i}$ for the SN decoder or the SIC process. According to Lemma \ref{exit_sn_era}, it is obvious that the process is determined by the specific traffic load $G$, transmission rate $R$, and erasure rate $\epsilon$. Then, the updating process from $p_{i-1}$ to $q_{i}$ for the UN decoder is obtained as follows.

\begin{lemm} The EXIT function $f'_{u}(p_{i-1})$ of the UN decoder over packet erasure channels is
	\begin{align}
	\!\!\!\!f'_{u}(p_{i-1})=\sum_{h=1}^{\theta}\lambda_{h}\sum_{j=1}^{n_{h}}\dbinom{n_{h}-1}{j-1}\left(
	1-\epsilon \right) ^{j-1}\epsilon^{n_{h}-j}f'^{(j)}_{u}(p_{i-1}),
	\label{exit_func_un_pct_ers_channel}
	\end{align}
	where $\lambda_{h}$ is the probability of an edge emanating from a degree-$n_{h}$ UN in the graph $g$ and $f'^{(j)}_{u}(p_{i-1})$ is the EXIT function of a degree-$j$ UN in the subgraph $\tilde{g}$.
	\label{exit_un_era}
\end{lemm}

\emph{\quad Proof: }Let $\tilde{\lambda}_{j}$ represent the probability that an edge is connected to a degree-$j$ UN in the subgraph $\tilde{g}$. Similarly to Eq. \eqref{relation_sn_eg_nd_dis_poly}, we have
	\begin{align}
	\tilde{\lambda}_{j}&=\sum_{h=1}^{\theta}\lambda_{h}\dbinom{n_{h}-1}{j-1}\left(
	1-\epsilon \right)^{j-1}\epsilon^{n_{h}-j}.
	\label{relation_un_eg_nd_dis}
	\end{align}
	%
	%\begin{align}
	%\tilde{\lambda}_{j}=\sum_{h\in\{h|n_{h}\geq j\}}\lambda_{h}\dbinom{n_{h}-1}{j-1}\left(
	%1-\epsilon \right) ^{j-1}\epsilon^{n_{h}-j}
	%\label{relation_un_eg_nd_dis}.
	%\end{align}
	By substituting Eq. \eqref{relation_un_eg_nd_dis} into Eq. \eqref{exit_func_un_colli_channel}, we have Eq. \eqref{exit_func_un_pct_ers_channel}. The proof completes. \QEDA

In Lemma \ref{exit_un_era}, we show the update from $p_{i-1}$ to $q_{i}$ for the UN decoder. Note that, the function $f'^{(j)}_{u}(p_{i-1})$ depends on the type of codes employed by the users. Now, we consider repetition codes and MDS codes, and derive the corresponding EXIT functions in the following corollaries.

\begin{coro} Let each component code $\textbf{c}_{h}$ of the code set $\mathbf{C}$ be a length-$(h+1)$ repetition code, for $h\in\{1,2,\cdots\theta\}$. With this repetition code set $\mathbf{C}$, the EXIT function of the UN decoder for the CSA scheme over packet erasure channels is given by
	\begin{align}
	f'_{ru}(p_{i-1})=\sum_{h=1}^{\theta}\frac{(h+1)\Lambda_{h}}{\bar{n}}\left((1-\epsilon )p_{i-1}+\epsilon\right)^{h}.
	\label{exit_func_un_pct_ers_channel_rc}
	\end{align}
	\label{exit_un_rep_ers}
\end{coro}
\emph{\quad Proof: }For repetition codes, an edge emanating from a degree-$j$ UN in the subgraph $\tilde{g}$ cannot be recovered only if all other $(j-1)$ edges of this UN have not been recovered by the SN decoder. Then, the EXIT function $f'^{(j)}_{u}(p_{i-1})$ of a degree-$j$ UN is
	\begin{align}
	f'^{(j)}_{u}(p_{i-1})=p_{i-1}^{j-1},
	\label{exit_func_un_pct_ers_channel_rc_jth_un}
	\end{align}
	which leads to Eq. \eqref{exit_func_un_pct_ers_channel_rc} by substituting Eq. \eqref{exit_func_un_pct_ers_channel_rc_jth_un} into Eq. \eqref{exit_func_un_pct_ers_channel}. The proof completes. \QEDA

\begin{coro} Let each component code $\textbf{c}_{h}$ of the code set $\mathbf{C}$ be a $(k+h,k)$ MDS code, for a fixed $k$ and $h\in\{1,2,\cdots\theta\}$. With this MDS code set $\mathbf{C}$, the EXIT function of the UN decoder for the CSA scheme over packet erasure channels is given by
	\begin{align}
	\label{exit_func_un_pct_ers_channel_mdsc}
	&f'_{mu}(p_{i-1})=\sum\limits_{d=k+1}^{k+\theta}\left[\sum\limits_{j=1}^{k}\dbinom{d-1}{j-1}%
	(1-\epsilon)^{j-1}\epsilon ^{d-j}+\sum\limits_{j=k+1}^{d}\dbinom{d-1}{j-1} \right. \notag\\
	&\left. {(1-\epsilon)^{j-1}\epsilon ^{d-j}\left(\sum\limits_{w=0}^{k-1}\dbinom{j-1}{w}%
		\left(1-p_{i-1}\right)^{w}p_{i-1}^{j-1-w}\right)}\right]\frac{d\Lambda_{d-k}}{\bar{n}}.
	\end{align}\vspace{-1cm}
	\label{exit_un_mds_ers}
\end{coro}

\emph{\quad Proof: }Please refer to Appendix \ref{AppendixB}. \QEDA

From Corollary \ref{exit_un_rep_ers} and Corollary \ref{exit_un_mds_ers}, it can be seen that the updating process from $p_{i-1}$ to $q_{i}$ for the UN decoder is associated with UN degree distribution $\bm{\Lambda}$ of the entire graph $g$. Then, we can design $\bm{\Lambda}$ to optimize the updating process, so that the expected traffic load is maximized.
It is noteworthy that since the repetition codes are the special case of the MDS codes with $k=1$, Corollary \ref{exit_un_mds_ers} is a generalization of Corollary \ref{exit_un_rep_ers}. In particular, Eq. \eqref{exit_func_un_pct_ers_channel_mdsc} in Corollary \ref{exit_un_mds_ers} with $k=1$ is identical to \eqref{exit_func_un_pct_ers_channel_rc} in Corollary \ref{exit_un_rep_ers}.

\begin{theo} Let $N, M\rightarrow\infty$ for a constant traffic load $G$. The recursion of probability $p_{i}$ over packet erasure channels with erasure rate $\epsilon$ is given by
	\begin{align}
	\!\!\!p_{i}=&1-\exp\left(-\frac{G}{R}(1-\epsilon)\sum_{h=1}^{\theta}\lambda_{h}\sum_{j=1}^{n_{h}}\dbinom{n_{h}-1}{j-1}\left(
	1-\epsilon \right) ^{j-1} \epsilon^{n_{h}-j}f'^{(j)}_{u}(p_{i-1})\right),\!
	\label{recus_pi_pkt_erasure_channel}
	\end{align}
	where $R$, $\lambda_{h}$, and $f'^{(j)}_{u}(p_{i-1})$ are defined in Lemma \ref{exit_sn_era} and Lemma \ref{exit_un_era}.
\end{theo}

\emph{\quad Proof: }The theorem follows from Lemma \ref{exit_sn_era} and Lemma \ref{exit_un_era} by considering $p_{i}=f'_{s}(f'_{u}(p_{i-1}))$, where $f'_{u}(\cdot)$ and $f'_{s}(\cdot)$ are given by Eq. \eqref{exit_func_un_pct_ers_channel} and Eq. \eqref{exit_func_sn_pct_ers_channel_fs}, respectively. \QEDA

\begin{figure}[!t]
	\centering
		{\includegraphics[height=2.9in,width=3.45in]{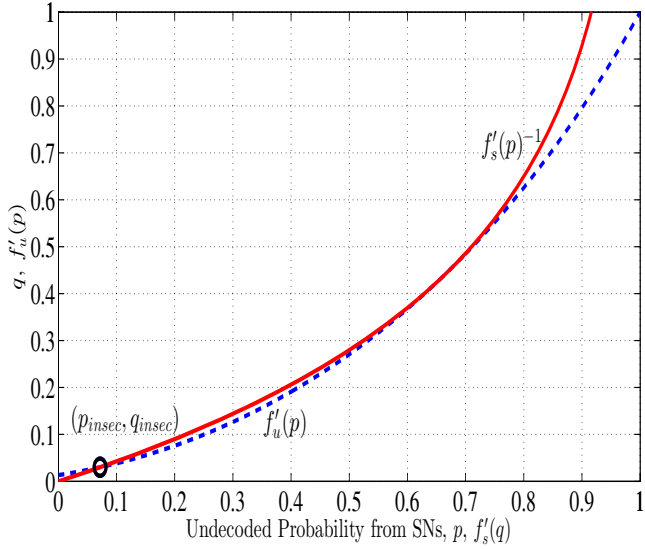}\label{fig_repe}}\vspace{-2mm}
		\caption{EXIT chart for repetition codes with distribution $\Lambda_{\mathrm{r2}}(x)=0.0915x^{2}+0.8113x^{3}+0.0972x^{6}$, packet erasure probability $\epsilon=0.1$, and $G=0.8604$.}
		\label{exit_rep_packet}
\end{figure}
\begin{figure}
	\centering
		{\includegraphics[height=2.9in,width=3.45in]{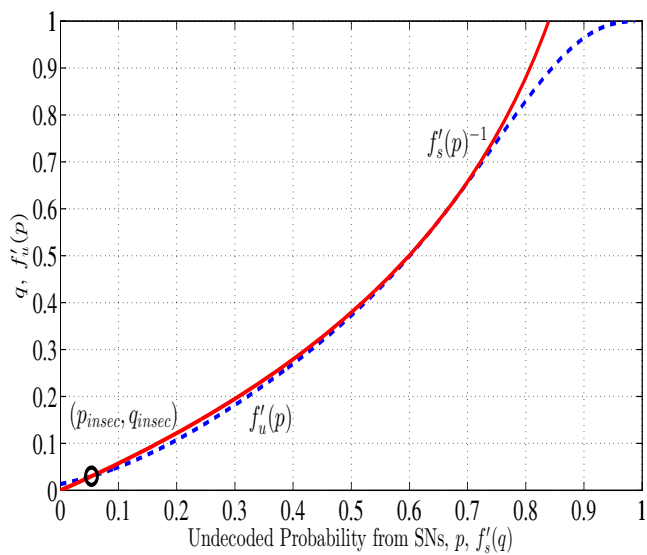}\label{fig_mds}}\vspace{-3mm}
		\caption{EXIT chart for MDS codes with distribution $\Lambda_{\mathrm{m2}}(x)=0.3081x^{5}+0.3904x^{6}+0.3015x^{12}$, packet erasure probability $\epsilon=0.1$, and $G=0.8131$.}
		\label{exit_mds_packet}
\end{figure}
The iterative process between the SIC and local decoding can be characterized by the recursion in Eq. \eqref{recus_pi_pkt_erasure_channel} and visualized by an EXIT chart, where the horizontal axis and vertical axis represent $p$ and $q$, respectively. The EXIT charts for repetition codes and MDS codes are shown in Fig. \ref{exit_rep_packet} and Fig. \ref{exit_mds_packet}, respectively. From the Section III, it is known that the iteration proceeds until $f'_{u}(p_{i})$ and $f'_{s}(p_{i})^{-1}$ intersect in the EXIT chart. From the Fig. \ref{exit_rep_packet} and Fig. \ref{exit_mds_packet}, it can be seen that the intersection $(p_{\mathrm{insec}},q_{\mathrm{insec}})$ is not at the origin $(0,0)$, which is caused by the non-zero erasure rate. Then, the recovery process with a higher user recovery probability can be achieved by obtaining an intersection closer to the origin $(0,0)$. From Lemma \ref{exit_sn_era} and Lemma \ref{exit_un_era}, it is obvious that the intersection of two functions depends on the traffic load $G$ and the UN degree distribution $\bm{\Lambda}$ of graph $g$ for the given $R$ and $\epsilon$. Thus, provided by the intersection satisfies that at least $\alpha$ percent of active users can be successfully recovered, the degree distributions $\bm{\Lambda}^{\ast}$ can be designed to maximize the expected traffic load $G_{\alpha}^{\ast}$ for the given $R$ and $\epsilon$.

%\begin{figure}[!t]
%{\ }
%\par
%\begin{center}
%{\includegraphics[height=2.6in,width=4in]{Graphes/ir_pq_erasure_packet_v1.pdf}}
%\end{center}
%\caption{The curves of probability pair ($p$, $q$) for repetition codes in packet
%erasure channel with $\epsilon=0.1$, $\bar{n}=3.2$}
%\label{fig_ir_pq_erasure_packet_v1}
%\end{figure}
%\begin{figure}[!t]
%\par
%\begin{center}
%{\includegraphics[height=2.6in, width=4in]{Graphes/pq_mds_erasure.pdf}}
%\end{center}
%\caption{The curves of probability pair ($p$, $q$) for MDS codes in packet
%erasure channel with $\epsilon=0.1$, $k=3$, $R=0.4$}
%\label{fig_pq_mds_erasure}
%\end{figure}

For the CSA scheme with repetition codes, we consider a code set with five repetition codes, i.e., $\theta=5$, and the maximum code length is $n_{\max}=6$. For the erasure rate $\epsilon=0.1$ and recovery ratio $\alpha=0.97$, the UN degree distributions $\bm{\Lambda}^{\ast}$ are obtained by linear programming, as shown in Table \ref{table_opdd}. In the table, the polynomial representation $\Lambda^{\ast}(x)$ of $\bm{\Lambda}^{\ast}$ is used. It can be seen that as $\bar{n}$ increases, the expected traffic load $G_{\alpha}^{\ast}$ increases for the same erasure rate. Note that, $\alpha=0.97$ is used as a design example to illustrate the obtained UN degree distributions in the work and other values of $\alpha$ can also be chosen. In addition, we would like to point out that the UN degree distributions and the performance of PLR depend on the value of $\alpha$, which will be shown in detail in the following.

\begin{table}
	\centering
	\caption{The UN Degree Distributions For Repetition Codes With Packet Erasure $\epsilon=0.1$, $\alpha=0.97$, $n_\mathrm{max}=6$, $\theta=5$.}\vspace{5mm}
	\begin{tabular}{ccc}
		\hline
		$\bar{n}$& $\Lambda^{\ast}(x)$ & $G_{\alpha}^{\ast}$  \\ \hline
		$3$& $\Lambda_{\mathrm{r1}}(x)=0.0695x^{2}+0.8957x^{3}+0.0348x^{5}$ & $0.835$  \\
		$3.2$& $\Lambda_{\mathrm{r2}}(x)=0.0915x^{2}+0.8113x^{3}+0.0972x^{6}$ & $0.8604$  \\
		$3.6$& $\Lambda_{\mathrm{r3}}(x)=0.1529x^{2}+0.5963x^{3}+0.2508x^{6}$ & $0.8877$   \\ \hline
	\end{tabular}
	\label{table_opdd}
\end{table}
\begin{table}
	\centering
	\caption{The UN Degree Distributions For MDS Codes With Packet Erasure $\epsilon=0.1$, $\alpha=0.97$.}\vspace{5mm}
	\begin{tabular}{cccccc}
		\hline
		$k$ & $R$ & $n_\mathrm{max}$ & $\theta$ & $\Lambda^{\ast}(x)$ & $G_{\alpha}^{\ast}$  \\ \hline
		$2$ & $0.4$ & $8$ & $6$ & $\Lambda_{\mathrm{m1}}(x)=0.5943x^{4}\hspace{-0.5mm}+\hspace{-0.5mm}0.2075x^{5}\hspace{-0.5mm}+\hspace{-0.5mm}0.1982x^{8}$ & \hspace{-0.5mm}$0.7875$  \\ \hline
		$3$ & $0.4$ & $12$ & $9$ & $\Lambda_{\mathrm{m2}}(x)=0.3081x^{5}\hspace{-0.5mm}+\hspace{-0.5mm}0.3904x^{6}\hspace{-0.5mm}+\hspace{-0.5mm}0.3015x^{12}$ & \hspace{-0.5mm}$0.8131$  \\
		& $0.5$ & $13$ & $10$ & $\Lambda_{\mathrm{m3}}(x)=0.2924x^{5}\hspace{-0.5mm}+\hspace{-0.5mm}0.4152x^{6}\hspace{-0.5mm}+\hspace{-0.5mm}0.2924x^{7}$ & \hspace{-0.5mm}$0.661$  \\ \hline
		$4$ & $0.5$ & $10$ & $6$ & $\Lambda_{\mathrm{m4}}(x)=0.169x^{6}\hspace{-0.5mm}+\hspace{-0.5mm}0.4414x^{7}\hspace{-0.5mm}+\hspace{-0.5mm}0.3896x^{10}$ & \hspace{-0.5mm}$0.6708$  \\
		& $0.55$ & \hspace{-1mm}$10$ & $6$ & \hspace{-4mm}$\Lambda_{\mathrm{m5}}(x)=0.1341x^{6}\hspace{-0.5mm}+\hspace{-0.5mm}0.4896x^{7}\hspace{-1mm}+\hspace{-0.5mm}0.3454x^{8}\hspace{-0.5mm}+\hspace{-0.5mm}0.0309x^{9}$ & \hspace{-5mm}$0.5945$\hspace{-4.5mm}  \\ \hline
		%3.6 & $0.1529x^{2}+0.5963x^{3}+0.2508x^{6}$ & 0.8877   \\ \hline
	\end{tabular}
	\label{table_opddmds}
\end{table}
For the CSA scheme with $(k+h,k)$ MDS codes, the code sets with different number of component codes $(\theta)$ and different maximum code length $n_{\max}$ are adopted for different $k$ and $R$. For $\epsilon=0.1$ and $\alpha=0.97$, the UN degree distributions $\bm{\Lambda}^{\ast}$ are designed and shown in Table \ref{table_opddmds}. In Table \ref{table_opddmds}, the exponent of $x$ represents the value of $(k+h)$ for a $(k+h,k)$ MDS code and $\Lambda^{\ast}(x)$ is the polynomial representation of $\bm{\Lambda}^{\ast}$. From the table, it can be seen that for the same rate $R$, a higher expected traffic load $G_{\alpha}^{\ast}$ can be achieved by using a higher code dimension $k$, at the expense of a higher decoding complexity. For example, with the average rate $R=0.4$, $G_{\alpha}^{\ast}=0.8131$ can be obtained for $k=3$, whereas $G_{\alpha}^{\ast}$ only reaches $0.7875$ for $k=2$.

\subsection{Design of CSA Schemes for Slot Erasure Channels}
In this section, we adopt the slot erasure channel to model the case that a strong external interference or noise may overwhelm the superposition of all the transmitted signals in a particular time slot and the corresponding packets cannot be recovered.
In particular, when the ratio of the sum of received powers to the power of external interference plus noise in a time slot, given by $I$, is less than a given threshold $\Delta$, i.e., $I < \Delta$, we assume that all the packets transmitted in this slot are erased and the erasure probability is $\epsilon=\mathrm{Pr}(I < \Delta)$.

\begin{lemm} Let $N, M\rightarrow \infty$ for a constant traffic load $G$. Let $p_{i}$ and $q_{i}$ be the average probability of a packet being unrecovered by the SN decoder and UN decoder at the $i$-th iteration of the recovery process, respectively. Over the slot erasure channel with an erasure rate $\epsilon$, the EXIT function $F'_{s}(q_{i})$ of the SN decoder for the CSA scheme is
	\begin{align}
	F'_{s}(q_{i})=1-\exp\left(-\frac{G}{R}q_{i}\right),
	\label{exit_func_sn_slt_ers_channel}
	\end{align}
	where $R$ is the average transmission rate.
	\label{exit_sn_slot_era}
\end{lemm}
%\text{ \ \ \ \ \ }
\emph{\quad Proof: }Please refer to Appendix \ref{AppendixC}. \QEDA

Using the EXIT function of the SN decoder, Lemma \ref{exit_sn_slot_era} indicates the updating process from $q_{i}$ to $p_{i}$ by the SN decoder for slot erasure channels. The EXIT function of the UN decoder is derived in the following lemma, which represents the update from $p_{i-1}$ to $q_{i}$ by the UN decoder.

\begin{figure}[!t]
	\centering
		{\includegraphics[height=2.8in,width=3.45in]{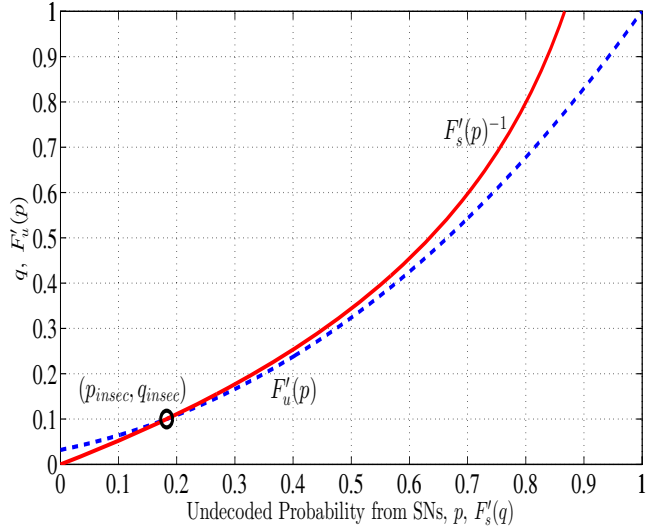}\label{fig_repe_slot}}\vspace{-3mm}
\caption{EXIT chart for repetition codes with distribution $\Lambda_{\mathrm{r5}}(x)=0.0915x^{2}+0.8111x^{3}+0.0974x^{6}$, short erasure probability $\epsilon=0.1$, and $G=0.7744$.}
\label{exit_rep_slot}
\end{figure}
\begin{lemm}
	The EXIT function $F'_{u}(p_{i-1})$ of the UN decoder over slot erasure channels is
	\begin{align}
F'_{u}(p_{i-1})=\sum_{h=1}^{\theta}\lambda_{h}\sum_{j=1}^{n_{h}}\dbinom{n_{h}-1}{j-1}\left(
	1-\epsilon \right)^{j-1}\epsilon^{n_{h}-j}F'^{(j)}_{u}(p_{i-1}),
	\label{exit_func_un_slt_ers_channel}
	\end{align}
	where $\lambda_{h}$ is the probability of an edge emanating from a degree-$n_{h}$ UN in the graph $g$ and $F'^{(j)}_{u}(p_{i-1})$ is the EXIT function of a degree-$j$ UN in the subgraph $\tilde{g}$.
	\label{exit_un_slot_era}
\end{lemm}

\emph{\quad Proof: }Consider a slot erasure channel with an erasure rate $\epsilon$. Let $\delta$ be the packet erasure rate in this channel. Then, we have
	\begin{align}
	\delta=\frac{\epsilon Ne}{Ne}=\epsilon,
	\label{slt_pct_ers_rt_slt_channel}
	\end{align}
	where $e$ is the average number of received packets per slot. Based on Eq. \eqref{slt_pct_ers_rt_slt_channel}, Eq. \eqref{exit_func_un_slt_ers_channel} follows from Eq. \eqref{exit_func_un_pct_ers_channel}. The proof completes. \QEDA
%Equation (\ref{exit_func_un_slt_ers_channel}) follows from (\ref{exit_func_un_pct_ers_channel}) by using $\delta=\epsilon$, where $\delta$ and $\epsilon$ represent the packet erasure rate and slot erasure rate for slot erasure channels, respectively. As all slots may be erased independently with the same probability $\epsilon$, the equation $\delta=\epsilon$ can be obtained.

From Lemma \ref{exit_un_era} and Lemma \ref{exit_un_slot_era}, it is obvious that the EXIT function of the UN decoder is the same for both packet erasure channels and slot erasure channels. Then, both Corollary \ref{exit_un_rep_ers} and Corollary \ref{exit_un_mds_ers} are also valid for slot erasure channels.

\begin{figure}[!t]
	\centering
	{\includegraphics[height=2.85in,width=3.45in]{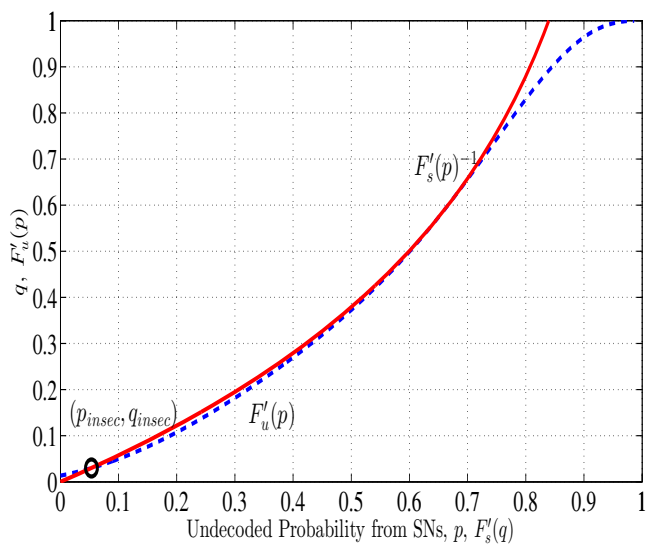}\label{fig_mds_slot}}\vspace{-3mm}
	\caption{EXIT chart for MDS codes with distribution $\Lambda_{\mathrm{m7}}(x)=0.3081x^{5}+0.3904x^{6}+0.3015x^{12}$, short erasure probability $\epsilon=0.1$, and $G=0.7318$.}
	\label{exit_mds_slot}
\end{figure}
\begin{theo} Let $N, M\rightarrow\infty$ for a constant traffic load $G$. The recursion of probability $p_{i}$ over slot erasure channels with erasure rate $\epsilon$ is given by
	\begin{align}
	p_{i}=1-\exp\left(-\frac{G}{R}\sum_{h=1}^{\theta}\lambda_{h}\sum_{j=1}^{n_{h}}\dbinom{n_{h}-1}{j-1} \left(1-\epsilon \right)^{j-1}\epsilon^{n_{h}-j}F'^{(j)}_{u}(p_{i-1})\right),
	\label{recus_pi_slot_erasure_channel}
	\end{align}
	where $R$, $\lambda_{h}$, and $F'^{(j)}_{u}(p_{i-1})$ are defined in Lemma \ref{exit_sn_slot_era} and Lemma \ref{exit_un_slot_era}.
\end{theo}
\emph{\quad Proof: }The theorem follows from Lemma \ref{exit_sn_slot_era} and Lemma \ref{exit_un_slot_era} by considering $p_{i}=F'_{s}(F'_{u}(p_{i-1}))$, where $F'_{u}(\cdot)$ and $F'_{s}(\cdot)$ are given by Eq. \eqref{exit_func_un_slt_ers_channel} and Eq. \eqref{exit_func_sn_slt_ers_channel}, respectively. The proof completes. \QEDA

For the slot erasure channel, the EXIT chart representations of the iterative recovery process for CSA scheme with repetition codes and MDS codes are shown in Fig. \ref{exit_rep_slot} and Fig. \ref{exit_mds_slot}, respectively. Similarly to the packet erasure channel, due to the non-zero erasure rate, the intersection of $F'_{u}(p_{i})$ and $F'_{s}(p_{i})^{-1}$ is $(p_{\mathrm{insec}},q_{\mathrm{insec}})$ instead of the origin $(0,0)$. Thus, provided by the intersection satisfies that at least $\alpha$ percent of active users can be successfully recovered, the degree distributions $\bm{\Lambda}^{\ast}$ can be obtained to maximize $G_{\alpha}^{\ast}$ for the given $R$ and $\epsilon$.

\begin{table}[t]
	\center
	\caption{The UN Degree Distributions For Repetition Codes With Slot Erasure $\epsilon=0.1$, $\alpha=0.97$, $n_\mathrm{max}=6$, $\theta=5$.}\vspace{5mm}
	\begin{tabular}{ccc}
		\hline
		$\bar{n}$ & $\Lambda^{\ast}(x)$ & $G_{\alpha}^{\ast}$  \\ \hline
		$3$ & $\Lambda_{\mathrm{r4}}(x)=0.0695x^{2}+0.8958x^{3}+0.0347x^{5}$ & $0.7515$  \\
		$3.2$ & $\Lambda_{\mathrm{r5}}(x)=0.0915x^{2}+0.8111x^{3}+0.0974x^{6}$ & $0.7744$  \\
		$3.6$ & $\Lambda_{\mathrm{r6}}(x)=0.1532x^{2}+0.5964x^{3}+0.2504x^{6}$ & $0.7989$   \\ \hline
	\end{tabular}
	\label{table_opddslot}
\end{table}
With the same code sets and the same assumptions as packet erasure channels $(\epsilon=0.1, \alpha=0.97)$, the UN degree distributions of the CSA scheme over slot erasure channels are shown in Table \ref{table_opddslot} and Table \ref{table_opddmdsslot}, which correspond to the repetition codes and MDS codes, respectively. From Table \ref{table_opddslot} and Table \ref{table_opddmdsslot}, it can be seen that while the UN degree distributions are almost same as those of packet erasure channels under the same assumptions, the expected traffic load $G^{\ast}_{\alpha}$ for slot erasure channels is always lower than the value for packet erasure channels. For example, when $\bar{n}$ is set as $3.2$ for repetition codes, $G^{\ast}_{\alpha}=0.8604$ can be achieved for packet erasure channels, but $G^{\ast}_{\alpha}$ is only $0.7744$ for slot erasure channels.

\begin{table}[t]
	\center
	\caption{The UN Degree Distributions for MDS Codes with Slot Erasure $\epsilon=0.1$, $\alpha=0.97$.}\vspace{5mm}
	%\resizebox{\columnwidth}{!}{
	\begin{tabular}{cccccc}
		\hline
		$k$ & $R$ & $n_\mathrm{max}$ & $\theta$ & $\Lambda^{\ast}(x)$ & $G_{\alpha}^{\ast}$ \\ \hline
		$2$ & $0.4$ & $8$ & $6$ & \hspace{-2mm}$\Lambda_{\mathrm{m6}}(x)=0.5943x^{4}\hspace{-0.5mm}+\hspace{-0.5mm}0.2077x^{5}\hspace{-0.5mm}+\hspace{-0.5mm}0.198x^{8}$ & \hspace{-2mm}$0.7087$  \\ \hline
		$3$ & $0.4$ & $12$ & $9$ & \hspace{-2mm}$\Lambda_{\mathrm{m7}}(x)=0.3081x^{5}\hspace{-0.5mm}+\hspace{-0.5mm}0.3904x^{6}\hspace{-0.5mm}+\hspace{-0.5mm}0.3015x^{12}$ & \hspace{-2mm}$0.7318$  \\
		& $0.5$ & $13$ & $10$ & \hspace{-2mm}$\Lambda_{\mathrm{m8}}(x)=0.2924x^{5}\hspace{-0.5mm}+\hspace{-0.5mm}0.4152x^{6}\hspace{-0.5mm}+\hspace{-0.5mm}0.2924x^{7}$ & \hspace{-2mm}$0.5949$  \\ \hline
		$4$ & $0.5$ & $10$ & $6$ & \hspace{-2mm}$\Lambda_{\mathrm{m9}}(x)=0.169x^{6}\hspace{-0.5mm}+\hspace{-0.5mm}0.4415x^{7}\hspace{-0.5mm}+\hspace{-0.5mm}0.3895x^{10}$ & \hspace{-2mm}$0.6037$ \\
		& $0.55$ & $10$ & $6$ & \hspace{-2mm}$\Lambda_{\mathrm{m10}}(x)=0.1344x^{6}\hspace{-0.5mm}+\hspace{-0.5mm}0.488x^{7}\hspace{-1mm}+\hspace{-0.5mm}0.3482x^{8}\hspace{-0.5mm}+\hspace{-0.5mm}0.0294x^{9}$ & \hspace{-3mm}$0.535$\hspace{-1.5mm}  \\ \hline
		%      \hline
		%$k$ & $R$ & $n_\mathrm{max}$ & $\theta$ & $\Lambda^{\ast}(x)$ & $G_{\alpha}^{\ast}$  \\ \hline
		%$k=2$ & $0.4$ & $8$ & $6$ & $\Lambda_{\mathrm{m6}}(x)=0.5943x^{4}+0.2077x^{5}+0.198x^{8}$ & $0.7087$  \\ \hline
		%$k=3$ & $0.4$ & $12$ & $9$ & $\Lambda_{\mathrm{m7}}(x)=0.3081x^{5}+0.3904x^{6}+0.3015x^{12}$ & $0.7318$  \\
		%      & $0.5$ & $13$ & $10$ & $\Lambda_{\mathrm{m8}}(x)=0.2924x^{5}+0.4152x^{6}+0.2924x^{7}$ & $0.5949$  \\ \hline
		%$k=4$ & $0.5$ & $10$ & $6$ & $\Lambda_{\mathrm{m9}}(x)=0.169x^{6}+0.4415x^{7}+0.3895x^{10}$ & $0.6037$  \\
		%      & $0.55$ & $10$ & $6$ & $\Lambda_{\mathrm{m10}}(x)=0.1344x^{6}+0.488x^{7}+0.3482x^{8}+0.0294x^{9}$ & $0.535$  \\ \hline
	\end{tabular}
	%}
	\label{table_opddmdsslot}
\end{table}
\begin{table}[h]
	\center
	\caption{The UN Degree Distributions For Repetition Codes With Packet Erasure $\epsilon=0.1$, $\bar{n}=3$, $n_\mathrm{max}=6$, $\theta=5$.}\vspace{5mm}
	\begin{tabular}{cc}
		\hline
		$\alpha$ & $\Lambda^{\ast}(x)$ \\ \hline
		0.99 & $\Lambda_{r7}(x)=0.9996x^{3}+0.0004x^{4}$ \\
		0.98 & $\Lambda_{r8}(x)=0.0093x^{2}+0.9845x^{3}+0.0062x^{5}$ \\
		0.95 & $\Lambda_{r9}(x)=0.1757x^{2}+0.6485x^{3}+0.1758x^{4}$  \\
		0.9 & $\Lambda_{r10}(x)=0.3138x^{2}+0.3721x^{3}+0.3141x^{4}$ \\ \hline
	\end{tabular}
	\label{tb_rpt_diff_alpha_pak}
\end{table}
\begin{table}[h]
	\center
	\caption{The UN Degree Distributions For MDS Codes With Packet Erasure $\epsilon=0.1$, $k=3$, $R=0.4$, $n_\mathrm{max}=12$, $\theta=9$.}\vspace{5mm}
	\begin{tabular}{cc}
		\hline
		$\alpha$ & $\Lambda^{\ast}(x)$ \\ \hline
		0.99 & $\Lambda_{m11}(x)=0.0665x^{5}+0.6725x^{6}+0.261x^{12}$ \\
		0.98 & $\Lambda_{m12}(x)=0.2172x^{5}+0.4965x^{6}+0.2863x^{12}$  \\
		0.95 & $\Lambda_{m13}(x)=0.4003x^{5}+0.2830x^{6}+0.3167x^{12}$  \\
		0.9 & $\Lambda_{m14}(x)=0.0177x+0.4108x^{5}+0.2471x^{6}+0.3244x^{12}$  \\ \hline
	\end{tabular}
	\label{tb_mds_diff_alpha_pak}
\end{table}
\begin{table}[h]
	\center
	\caption{The UN Degree Distributions For Repetition Codes With Slot Erasure $\epsilon=0.1$, $\bar{n}=3$, $n_\mathrm{max}=6$, $\theta=5$.}\vspace{5mm}
	\begin{tabular}{cc}
		\hline
		$\alpha$& $\Lambda^{\ast}(x)$  \\ \hline
		0.99 & $\Lambda_{r11}(x)=0.9996x^{3}+0.0004x^{4}$ \\
		0.98 & $\Lambda_{r12}(x)=0.01x^{2}+0.9863x^{3}+0.0047x^{5}$ \\
		0.95 & $\Lambda_{r13}(x)=0.1757x^{2}+0.6486x^{3}+0.1757x^{4}$  \\
		0.9 & $\Lambda_{r14}(x)=0.3137x^{2}+0.3723x^{3}+0.314x^{4}$ \\ \hline
	\end{tabular}
	\label{tb_rpt_diff_alpha_slot}
\end{table}
\begin{table}[h]
	\center
	\caption{The UN Degree Distributions For MDS Codes With Slot Erasure $\epsilon=0.1$, $k=3$, $R=0.4$, $n_\mathrm{max}=12$, $\theta=9$.}\vspace{5mm}
	\begin{tabular}{cccc}
		\hline
		$\alpha$ & $\Lambda^{\ast}(x)$  \\ \hline
		0.99 & $\Lambda_{m15}(x)=0.0666x^{5}+0.6722x^{6}+0.2612x^{12}$ \\
		0.98 & $\Lambda_{m16}(x)=0.2172x^{5}+0.4964x^{6}+0.2864x^{12}$  \\
		0.95 & $\Lambda_{m17}(x)=0.4003x^{5}+0.283x^{6}+0.3167x^{12}$  \\
		0.9 & $\Lambda_{m18}(x)=0.0182x+0.4095x^{5}+0.2482x^{6}+0.3241x^{12}$  \\ \hline
	\end{tabular}
	\label{tb_mds_diff_alpha_slot}
\end{table}

In addition, the designed UN degree distributions of the repetition codes and MDS codes with different $\alpha$ over the packet erasure channels and slot erasure channels are presented in Table \ref{tb_rpt_diff_alpha_pak} to Table \ref{tb_mds_diff_alpha_slot}.
From Table \ref{tb_rpt_diff_alpha_pak} and Table \ref{tb_rpt_diff_alpha_slot}, it can be seen that for both the packet erasure channels and slot erasure channels, the designed UN distributions of the repetition codes tend to the monomial $x^{3}$ when increasing $\alpha$ from $0.9$ to $0.99$.
The reason why this monomial is $x^{3}$ is due to the constraint on the average code length $\bar{n}=3$.

Furthermore, the similar observation can be obtained for the MDS codes from Table \ref{tb_mds_diff_alpha_pak} and Table \ref{tb_mds_diff_alpha_slot}.
Recall that the expected traffic load $G_{\alpha}^{\ast}$ is the largest traffic load achieved with the PLR no more than $(1-\alpha)$.
Thus, as $\alpha$ approaches one, the PLR will approach zero for any $G \leq G_{\alpha}^{\ast}$.
Since the PLR for the CSA scheme over the erasure channels exhibits an error floor, increasing the value of $\alpha$ will decrease the PLR and its corresponding error floor.
%
%The PLR approaching zero implies a lower error floor of PLR.
%
A lower error floor of the PLR can be achieved by using the UN degree distribution with a monomial polynomial or a regular UN degree distribution.
This is analogous to the characteristic that regular LDPC codes can achieve a lower error floor than the irregular LDPC codes in \cite{Luby01}.
Therefore, for the CSA scheme over erasure channels, the UN degree distribution tends to be monomial or regular as $\alpha$ approaches to one.

\section{Performance Analysis of CSA Schemes Over Erasure Channels}
In this section, we analyze the effect of design parameters on the expected traffic load $G^{\ast}_{\alpha}$ for MDS codes. Furthermore, we derive the asymptotic throughput of CSA schemes with the infinite frame lengths for erasure channels.

\subsection{Analysis of MDS Codes for Erasure Channels}
\begin{figure}[t]
	\par
	\begin{center}
		{\includegraphics[height=2.85in,width=3.45in]{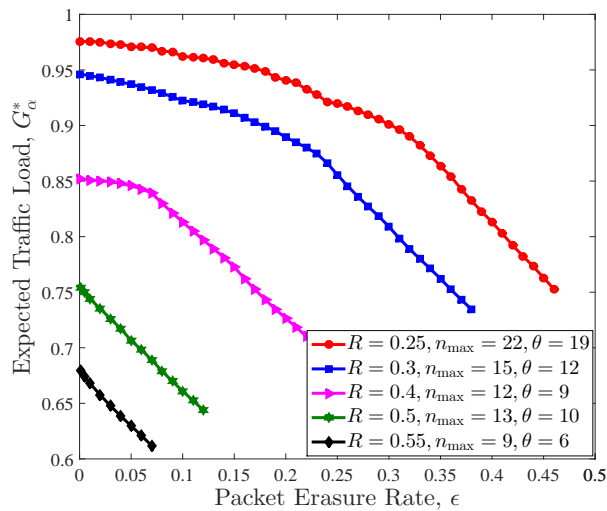}}
	\end{center}\vspace{-8mm}
	\caption{Expected traffic load $G_{\alpha}^{\ast}$ versus packet erasure rate for MDS codes with $k=3$ and different $R$.}
	\label{fig_mds_multirate_packeterasure_gstar}
\end{figure}

As a generalization of repetition codes, the MDS codes are employed with more design parameters (e.g. the number of information packets $k$). Thus, we analyze the effect of different parameters on the performance of MDS codes, including the transmission rate $R$, the number of information packets $k$, and the maximum code length $n_{\max}$. Here, we employ packet erasure channels with $\epsilon=0.1$ as an example to illustrate the analysis. In Fig. \ref{fig_mds_multirate_packeterasure_gstar}, we show the expected traffic load $G^{\ast}_{\alpha}$ for various transmission rates $R$, where $\alpha=0.97$ and $G^{\ast}_{\alpha}$ is obtained analytically. From the figure, it can be seen that for each transmission rate $R$, there exists a threshold of the erasure rate, denoted by $\epsilon_{\mathrm{th}}$. When the channel erasure rate $\epsilon\geq\epsilon_{\mathrm{th}}$, the expected traffic load $G^{\ast}_{\alpha}$ decreases almost linearly with increasing erasure rate $\epsilon$. However, for the channel erasure rate $\epsilon<\epsilon_{\mathrm{th}}$, the expected traffic load $G^{\ast}_{\alpha}$ does not change significantly. In addition, the threshold $\epsilon_{\mathrm{th}}$ is lower for a higher transmission rate $R$. For example, the threshold value can reach $0.25$ for $R=0.25$, while it is only $0.05$ for $R=0.4$.

%when the erasure rates $\epsilon$ are small, the maximum traffic load $G^{\ast}_{\alpha}$ changes slightly by increasing $\epsilon$. This is because that less packets are erased by small erasure rates, so that the iterative decoding process is affected less obviously. In addition, with increasing the erasure rate $\epsilon$, the obvious decrease of $G^{\ast}_{\alpha}$ occurs later for smaller transmission rates $R$. For example, the obvious decrease of maximum traffic load $G_{\alpha}^{\ast}$ begins from $\epsilon=0.25$ for $R=0.25$, while for $R=0.4$ the %starting point is $\epsilon=0.05$.

%the effect of different rates on maximum traffic load $G_{\alpha}^{\ast}$ is firstly analyzed, where the number of information packets $k$ is fixed for simplification. When less packets are erased for small erasure rates, the effect of erasure on the iterative decoding process is less obvious. Then, for small erasure rates, the maximum traffic load $G_{\alpha}^{\ast}$ changes slightly by increasing erasure rates, as shown in Fig.\ref{fig_mds_multimaxdegree_packeterasure_gstar_v4}. Moreover, more coded packets can be generated and transmitted for smaller transmission rates. Thus, with increasing the erasure rate $\epsilon$, the obvious decrease of $G^{\ast}_{\alpha}$ occurs later for smaller transmission rates.

\begin{figure}[t]
	\par
	\begin{center}
		{\includegraphics[height=2.9in,width=3.45in]{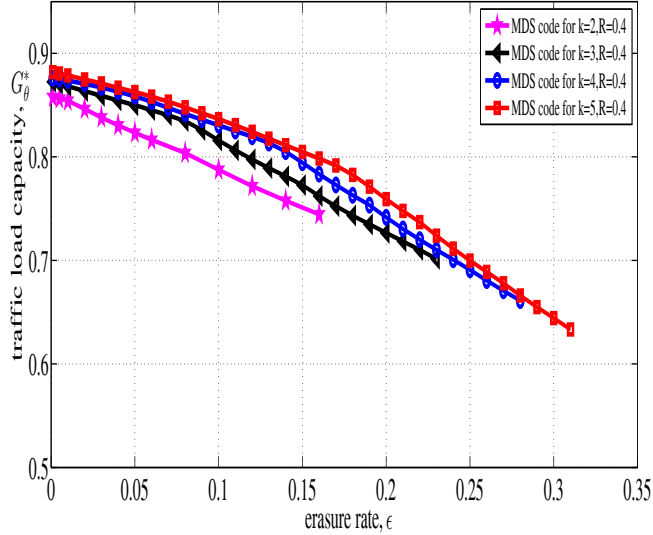}}
	\end{center}\vspace{-8mm}
	\caption{Expected traffic load $G_{\alpha}^{\ast}$ versus packet erasure rate for MDS codes with $R=0.4$ and different $k$.}
	\label{fig_mds_multik_packeterasure_gstar}
\end{figure}

Furthermore, we illustrate the expected traffic load $G_{\alpha}^{\ast}$ for different $k$ but a fixed transmission rate $R$, as shown in Fig. \ref{fig_mds_multik_packeterasure_gstar}. It is obvious that a higher $G_{\alpha}^{\ast}$ can be achieved by using larger $k$. For example, $G_{\alpha}^{\ast}=0.84$ of $k=5$ can be achieved for $\epsilon=0.1$, while $G_{\alpha}^{\ast}$ is only $0.7875$ for $k=2$.

%the different numbers of information packets are then employed for a fixed transmission rate $R$. From Fig.\ref{fig_mds_multik_packeterasure_gstar}, it is obvious that the maximum traffic load $G_{\alpha}^{\ast}$ can be improved by using higher values of $k$. For example, the $G_{\alpha}^{\ast}=0.84$ of $k=5$ can be reached, while $G_{\alpha}^{\ast}$ is only 0.7875 for $k=2$.
\begin{figure}[t]
	\par
	\begin{center}
		{\includegraphics[height=2.85in,width=3.45in]{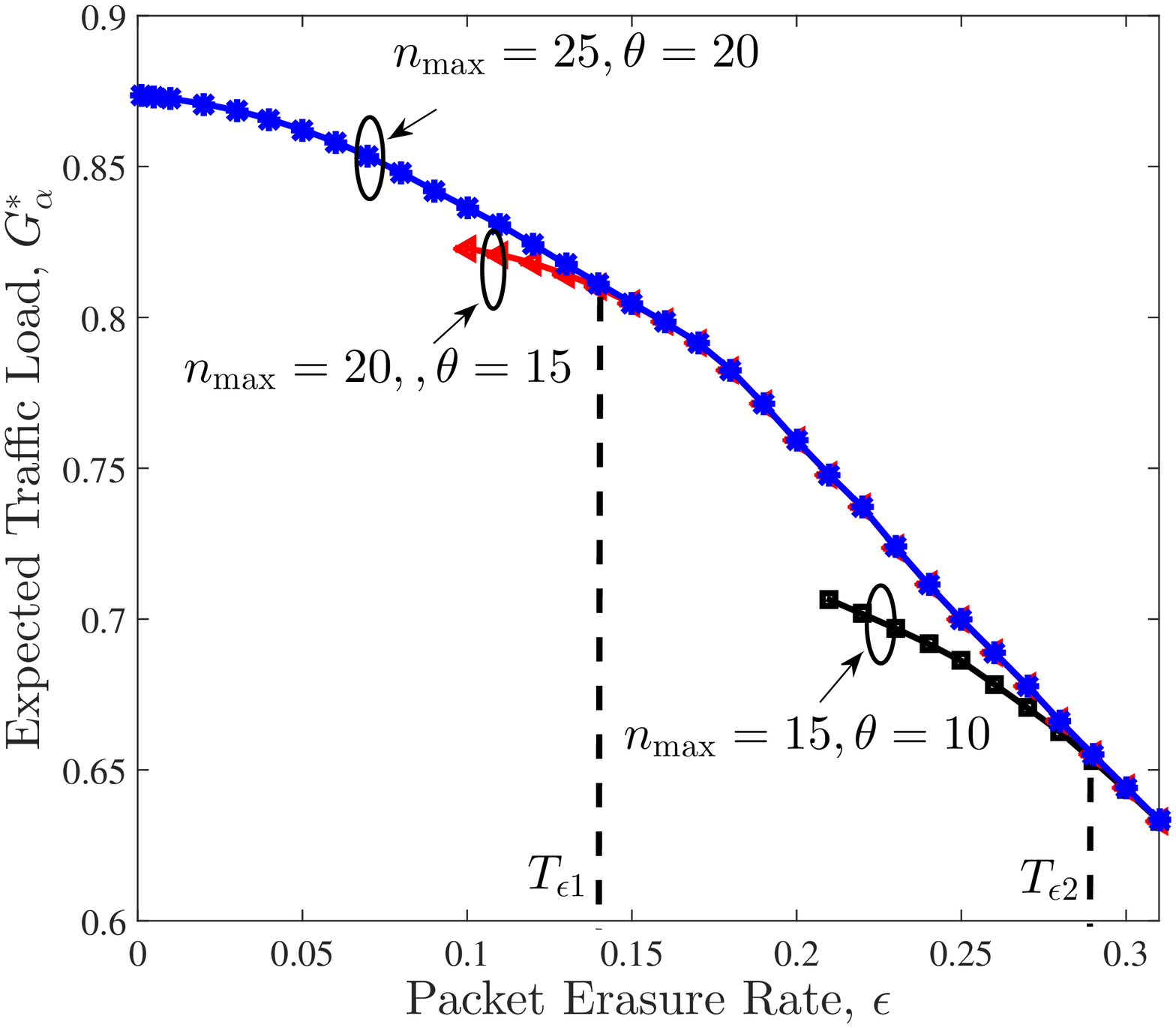}}
	\end{center}\vspace{-8mm}
	\caption{Expected traffic load $G_{\alpha}^{\ast}$ versus packet erasure rate for MDS codes with $k=5$, $R=0.4$, and different $n_{\max}$.}
	\label{fig_mds_multimaxdegree_packeterasure_gstar}
\end{figure}

Finally, we analyze the effect of the maximum code length $n_{\max}$ on the expected traffic load $G_{\alpha}^{\ast}$. From Fig. \ref{fig_mds_multimaxdegree_packeterasure_gstar}, it can be seen that when the erasure rate $\epsilon$ is higher than $T_{\epsilon1}$, the MDS codes with $n_{\max}=20$ and $n_{\max}=25$ can achieve the same expected traffic load $G_{\alpha}^{\ast}$. In addition, when the erasure rate $\epsilon$ is higher than $T_{\epsilon2}$, the MDS codes with $n_{\max}=15$ and $n_{\max}=20$ can achieve the same expected traffic load $G_{\alpha}^{\ast}$.

\subsection{Asymptotic Throughput Analysis}
In this part, we derive the asymptotic throughput of the CSA scheme for erasure channels, where the frame length $N$ and the number of users $M$ tend to infinity. Here, we employ repetition codes to illustrate the derivation of the asymptotic throughput for packet erasure channels.

\begin{lemm}
	For the packet erasure channel with an erasure rate $\epsilon$, the probability that an interfering packet can be removed is given by
	\begin{align}
	P_{\mathrm{pck\_cl}}(N_{\mathrm{iter}})=1-\sum_{h=1}^{\theta}\frac{n_{h}\Lambda _{h}}{\bar{n}}\left(1+\left(\epsilon-1\right)P_{p}(N_{\mathrm{iter}}-1)\right)^{n_{h}-1},
	\label{p_rmv_interference}
	\end{align}
	where $N_{\mathrm{iter}}$ is the number of iteration, $P_{p}(N_{\mathrm{iter}}-1)$ represents the probability that one packet can be recovered at the $(N_{\mathrm{iter}}-1)$-th iteration, and $n_{h}$ is the number of transmitted packets for users choosing the code $(n_{h},k)$.
	\label{pck_rm}
\end{lemm}

\emph{\quad Proof: }Please refer to Appendix \ref{AppendixD}. \QEDA

From Lemma \ref{pck_rm}, it can be seen that $P_{\mathrm{pck\_cl}}(N_{\mathrm{iter}})$ is associated with $P_{p}(N_{\mathrm{iter}}-1)$, i.e., the packet recovery probability at the $(N_{\mathrm{iter}}-1)$-th iteration. In addition, at the $N_{\mathrm{iter}}$-th iteration, the packet recovery probability $P_{p}(N_{\mathrm{iter}})$ depends on $P_{\mathrm{pck\_cl}}(N_{\mathrm{iter}})$. Thus, a recursive function of the packet recovery probability is obtained, which characterizes the iterative recovery process. This recursive function is summarized in the following lemma.

\begin{lemm}
	For the packet erasure channel with an erasure rate $\epsilon$, the packet recovery probability at the $N_{\mathrm{iter}}$-th iteration is given by
	\begin{align}
	\label{def_p_ps}
	P_{p}(N_{\mathrm{iter}})=\mathrm{exp}\left(-G(1-\epsilon)\sum_{h=1}^{\theta}n_{h}\Lambda _{h}\left(1+(\epsilon-1)P_{p}(N_{\mathrm{iter}}-1)\right)^{n_{h}-1}\right).
	\end{align}
	\label{p_ps}
\end{lemm}\vspace{-1cm}

\emph{\quad Proof: }Please refer to Appendix \ref{AppendixE}. \QEDA

According to Eq. \eqref{def_p_ps}, the recursive calculation of $P_{p}$ is initialized with $P_{p}(0)=0$. With $N_{\mathrm{iter}}^{\max}$ iterations, we can obtain $P_{p}(N_{\mathrm{iter}}^{\max})$ as the average packet recovery probability, given by $P_{p}$. Based on $P_{p}$ from Lemma \ref{p_ps}, the asymptotic throughput can be obtained, which is stated in the following theorem.

\begin{theo}
	For the CSA scheme over the packet erasure channel with an erasure rate $\epsilon$, the asymptotic throughput with $N,M \rightarrow \infty$ is given by
	\begin{align}
	T=G-\sum_{h=1}^{\theta}G{\Lambda}_{h}\left(1+(\epsilon-1)P_{p}\right)^{n_{h}}.
	\label{def_thpt}
	\end{align}
	\label{anal_thpt_csa}
\end{theo}\vspace{-1cm}

%\begin{IEEEproof}
\emph{\quad Proof: }Please refer to Appendix \ref{AppendixF}. \QEDA

Note that for the practical frame lengths, the replicas of collided packets in a slot may also collide in other slots, which is referred to as the event of stopping sets \cite{DelRioHerrero2014,IvanovBAP2015}. While the asymptotic throughput does not consider the impact of the event of stopping sets on the CSA performance, we would point out that the derived asymptotic throughput can give a good approximation to the throughput of the CSA scheme over erasure channels, which will be demonstrated by the simulation. This is consistent with the claim for the asymptotic analysis of CRDSA schemes over non-erasure channels in \cite{DelRioHerrero2014}, which shows that if more than two replicas are employed by each user, the asymptotic throughput holds well for the practical frame lengths\footnote[2]{Here, the practical frame length refers to $N=1000$ time slots.}.

\section{Numerical Results}
Simulation results are presented in this section. The results include the throughput versus traffic load, the PLR versus traffic load, and the expected traffic load $G_{\alpha}^{\ast}$ versus erasure rate for the CSA schemes. For the packet erasure channels and slot erasure channels, the corresponding proposed distributions $\Lambda_{\mathrm{r2}}(x)$ and $\Lambda_{\mathrm{r5}}(x)$ from Table \ref{table_opdd} and Table \ref{table_opddslot} are used in the following simulations. Comparing with the code distributions presented in \cite{Paolini2011} that are obtained to maximize the traffic load threshold for non-erasure collision channels, we demonstrate that our designed code distributions can improve the expected traffic load and the throughput of CSA schemes over erasure channels. In the simulations, we assume a fixed frame size of $N=1000$ slots and a given code set ${\mathbf{C}}=\{{\mathbf{c}}_{1},{\mathbf{c}}_{2},\cdots,{\mathbf{c}}_{5}\}$ to be chosen by the users, where ${\mathbf{c}}_{h}$ is a repetition code with $(h+1)$-length for $h\in\{1,2,\cdots ,5\}$.

\begin{figure}[!t]
	\par
	\begin{center}
		{\includegraphics[height=2.85in,width=3.45in]{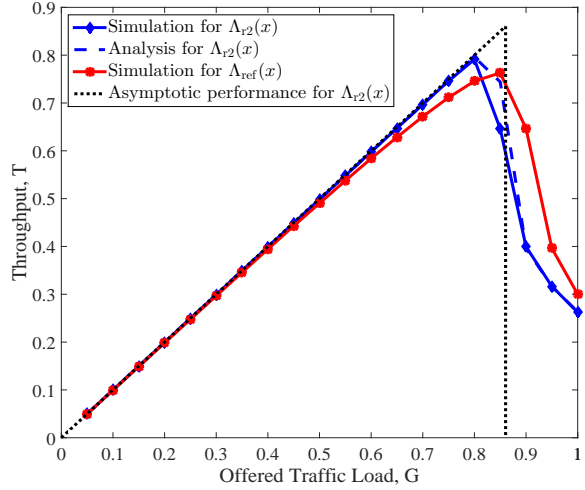}}\vspace{-10mm}
	\end{center}
	\caption{Throughput versus traffic load for $\Lambda_{\mathrm{r2}}(x)=0.0915x^{2}+0.8113x^{3}+0.0972x^{6}$, $\Lambda_{\mathrm{ref}}(x)=0.5631x^{2}+0.0436x^{3}+0.3933x^{5}$, $N=1000$, and packet erasure rate $\epsilon=0.1$.}
	\label{fig_thpt_propsd_anal_simu_bechmak}
\end{figure}

For the packet erasure channel with an erasure rate $\epsilon=0.1$, the throughput for the degree distribution $\Lambda_{\mathrm{r2}}(x)$ from Table \ref{table_opdd} is shown in Fig. \ref{fig_thpt_propsd_anal_simu_bechmak}. The throughput for the distribution $\Lambda_{\mathrm{ref}}(x)$ in \cite{Paolini2011} is also provided as a reference, which is designed to maximize the traffic load threshold for non-erasure collision channels.
From Fig. \ref{fig_thpt_propsd_anal_simu_bechmak}, it can be seen that the CSA scheme with our designed distribution $\Lambda_{\mathrm{r2}}(x)$ achieves a peak throughput close to $0.8$, while the peak throughput of CSA scheme with the distribution $\Lambda_{\mathrm{ref}}(x)$ is $0.75$.
More importantly, we also show the asymptotic performance of the CSA scheme in a dotted line in Fig. \ref{fig_thpt_propsd_anal_simu_bechmak}, when both $M$ and $N$ tend to infinity. It is obvious that the throughput of CSA scheme increases linearly with the traffic load up to $G=0.75$ for the degree distribution $\Lambda_{\mathrm{r2}}(x)$, while for the degree distribution $\Lambda_{\mathrm{ref}}(x)$ the linear relationship only holds up to $G=0.4$. According to Eq. \eqref{def_opt_traffic_load}, the expected traffic load $G_{\alpha}^{\ast}=0.8$ of degree distribution $\Lambda_{\mathrm{r2}}(x)$ can be achieved for $\alpha=0.97$, but $G_{\alpha}^{\ast}$ of the degree distribution $\Lambda_{\mathrm{ref}}(x)$ is only $0.6$. In other words, for any $G\leq G_{\alpha}^{\ast}$, at least $97\%$ of all active users can be successfully recovered. Thus, for packet erasure channels, our proposed degree distribution $\Lambda_{\mathrm{r2}}(x)$ allows to improve the expected traffic load $G_{\alpha}^{\ast}$ about $33\%$ w.r.t. the degree distribution $\Lambda_{\mathrm{ref}}(x)$ presented in \cite{Paolini2011}.

Moreover, for the distribution $\Lambda_{\mathrm{r2}}(x)$, the simulated throughput is compared with the analytical throughput in Fig. \ref{fig_thpt_propsd_anal_simu_bechmak}. The analytical results are obtained from Eq. \eqref{def_thpt}. It is shown that our obtained asymptotic throughput can give a good approximation to the simulated throughput of CSA schemes over erasure channels, where the frame length is $N=1000$ and the average number of replicas per user is $\bar{n}=3.2$. In fact, the asymptotic throughput can hold well for the practical frame lengths, if more than two replicas are employed by each user averagely, i.e., $\bar{n}>2$ \cite{DelRioHerrero2014}.

%upper bound can match the simulated throughput tightly in particular for traffic load $G\leq0.8$. Due to the increasing occurrence probability of stopping sets, although the gap between simulation results and the upper bound is more obvious for the traffic load approaching to one, the upper bound can predict the trend of simulated throughput.

%Fig. \ref{fig_thpt_propsd_anal_simu_bechmak} shows that the asymptotic analysis obtained from Eq. \eqref{def_thpt} can give a good approximation to the simulated throughput of CSA schemes over erasure channels.
%%
%This is consistent with the claim for the asymptotic analysis of CRDSA schemes over non-erasure channels in \cite{DelRioHerrero2014}, which shows that if more then two replicas are employed by each user, the asymptotic throughput holds well for the practical frame lengths.
%%
%As pointed by \cite{herrero09}, $3$ or $4$ replicas can achieve the best performance of CRDSA, in terms of the system throughput.
%%
%Thus, our design of probability distributions for the CSA schemes over erasure channels is based on the average number of replicas between $3$ and $4$, and the asymptotic analysis is well justified in practice.

\begin{figure}[!t]
	\par
	\begin{center}
		{\includegraphics[height=2.85in,width=3.45in]{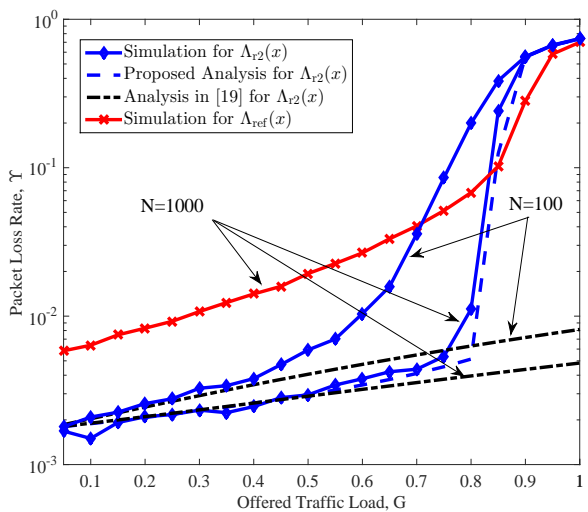}}\vspace{-10mm}
	\end{center}
	\caption{Packet loss rate versus traffic load for $\Lambda_{\mathrm{r2}}(x)=0.0915x^{2}+0.8113x^{3}+0.0972x^{6}$, $\Lambda_{\mathrm{ref}}(x)=0.5631x^{2}+0.0436x^{3}+0.3933x^{5}$, and packet erasure rate $\epsilon=0.1$.}
	\label{fig_plr_propsd_anal_simu_bchmakv2}
\end{figure}
In Fig. \ref{fig_plr_propsd_anal_simu_bchmakv2}, for the two distributions, i.e., $\Lambda_{\mathrm{r2}}(x)$ and $\Lambda_{\mathrm{ref}}(x)$, the PLRs of CSA schemes over the packet erasure channel are compared, where the erasure rate is $\epsilon=0.1$.
It can be seen that compared to $\Lambda_{\mathrm{ref}}(x)$, the PLR for $\Lambda_{\mathrm{r2}}(x)$ exhibits a much lower error floor, while it has a slight performance loss at the high PLR.
In particular, a $\mathrm{PLR}=10^{-2}$ is achieved with $G=0.28$ for the CSA scheme with $\Lambda_{\mathrm{ref}}(x)$, while the CSA scheme with $\Lambda_{\mathrm{r2}}(x)$ allows achieving this PLR at $G=0.8$.
In addition, we illustrate our derived asymptotic PLR and the error floor analysis of PLR in \cite{IvanovBAP2015} for the CSA scheme with the distribution $\Lambda_{\mathrm{r2}}(x)$.
Here, two frame lengths of $N=100$ and $N=1000$ are considered and the average number of replicas is $\bar{n}=3.2$.
The proposed asymptotic PLR is obtained by calculating a recursive function of the packet recovery probability iteratively, which characterizes the iterative recovery process for CSA schemes with asymptotically long frames.
From Fig. \ref{fig_plr_propsd_anal_simu_bchmakv2}, it can be seen that the proposed asymptotic PLR gives a good approximation to the simulation result for $N=1000$ in the whole region of offered traffic loads, i.e., $0.05 \leq G \leq 1$.
However, the asymptotic analysis does not provide an accurate estimation for the PLR when $N=100$.
Furthermore, it can also be seen from the figure that for both the small and large frame lengths, i.e., $N=100$ and $N=1000$, the error floor analysis of PLR in \cite{IvanovBAP2015} is accurate for the low to medium traffic loads.
In other words, the approach in \cite{IvanovBAP2015} is valid in the error floor region for all values of $N$, but not for high traffic loads.

\begin{figure}[!t]
	\par
	\begin{center}
		{\includegraphics[height=2.85in,width=3.45in]{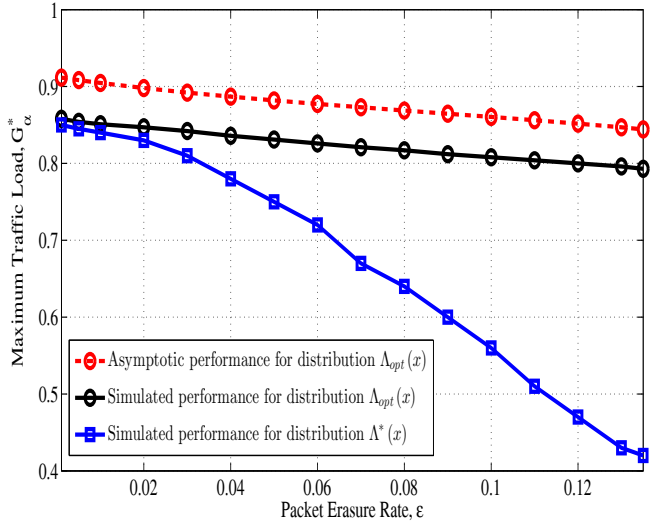}}\vspace{-8mm}
	\end{center}
	\caption{Simulated and asymptotic $G_{\alpha}^{\ast}$ for packet erasure
		channels with various $\epsilon$.}
	\label{fig_g_star_new_old_asym_erasure_packet_v1}
\end{figure}
Furthermore, with different erasure rates $\epsilon$ ranging from $0.001$ to $0.135$, we show the relationship between $\epsilon$ and $G_{\alpha}^{\ast}$ in Fig. \ref{fig_g_star_new_old_asym_erasure_packet_v1}. For different $\epsilon$, the corresponding degree distributions $\Lambda^{\ast}(x)$ are obtained by using the proposed method. From Fig. \ref{fig_g_star_new_old_asym_erasure_packet_v1}, it can be seen that the gap between the analytical results obtained from the EXIT chart analysis and the simulation results for distribution $\Lambda^{\ast}(x)$ is constant. Compared to the distribution $\Lambda_{\mathrm{ref}}(x)$ in \cite{Paolini2011} that is designed to maximize the traffic load threshold for non-erasure collision channels, our designed degree distribution can significantly improve the expected traffic load of CSA schemes over erasure channels, especially for the high erasure rates. With an erasure rate up to $0.135$, $G_{\alpha}^{\ast}$ of $\Lambda^{\ast}(x)$ can reach $0.79$, while $G_{\alpha}^{\ast}$ of $\Lambda_{\mathrm{ref}}(x)$ is only $0.42$. In this case, the expected traffic load of CSA scheme is increased by nearly $86\%$.

\begin{figure}[!t]
	\par
	\begin{center}
		{\includegraphics[height=2.8in,width=3.45in]{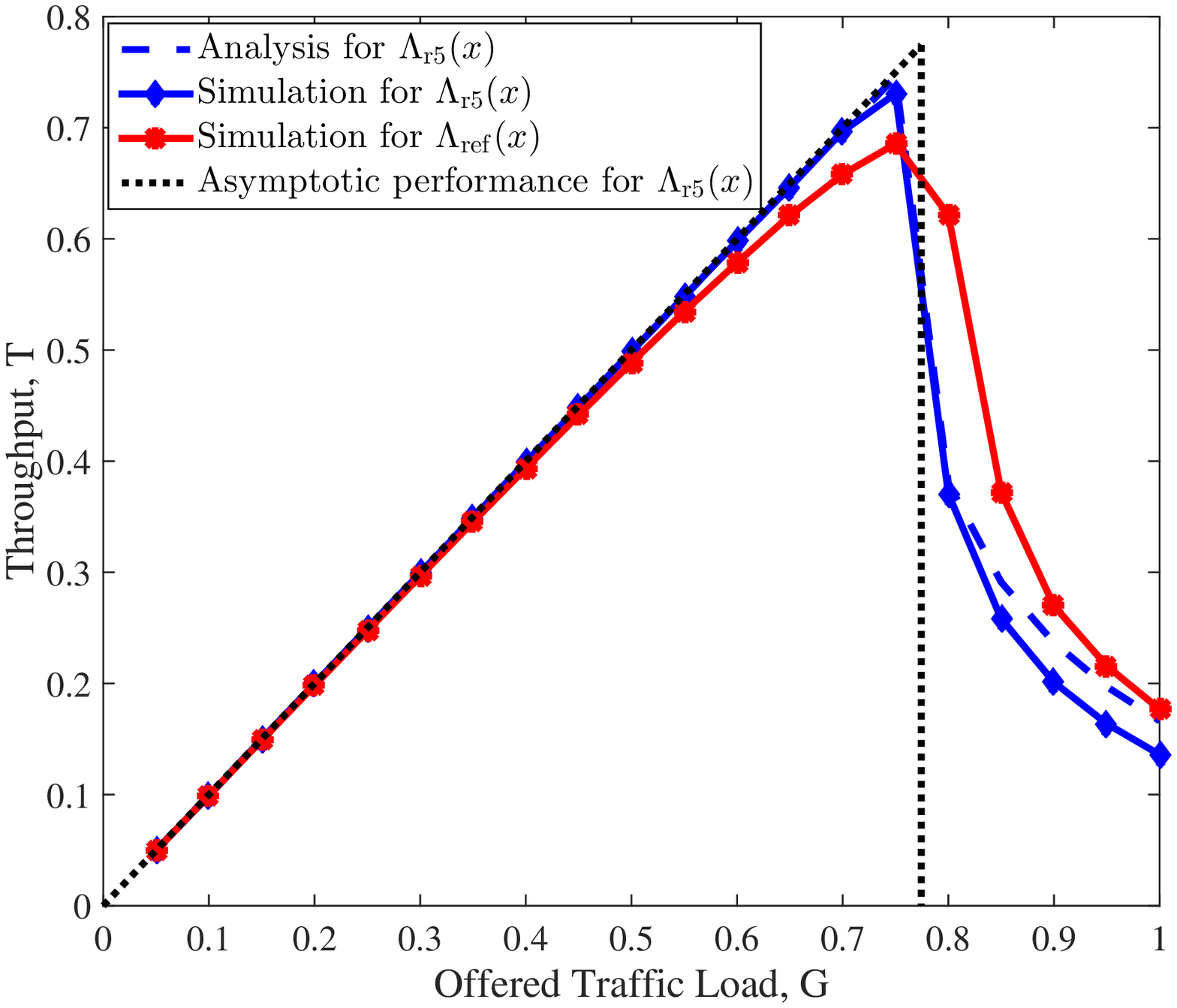}}
	\end{center}\vspace{-8mm}
	\caption{Throughput versus traffic load for $\Lambda_{\mathrm{r5}}(x)=0.0915x^{2}+0.8111x^{3}+0.0974x^{6}$, $\Lambda_{\mathrm{ref}}(x)=0.5631x^{2}+0.0436x^{3}+0.3933x^{5}$, $N=1000$, and slot erasure rate $\epsilon=0.1$.}\vspace{-20mm}
	\label{fig_rep_slot_erasure_tg_v3}
\end{figure}
For the slot erasure channel with an erasure rate $\epsilon=0.1$, the performance of the proposed code distribution $\Lambda_{\mathrm{r5}}(x)$ in Table \ref{table_opddslot} and the distribution $\Lambda_{\mathrm{ref}}(x)$ are compared, in terms of the throughput, as shown in Fig. \ref{fig_rep_slot_erasure_tg_v3}. From the figure, it can be seen that the peak throughput of $\Lambda_{\mathrm{r5}}(x)$ is higher than that of $\Lambda_{\mathrm{ref}}(x)$. Moreover, the asymptotic throughput of distribution $\Lambda_{\mathrm{r5}}(x)$ is provided. It is shown that the asymptotic throughput can give a good approximation to the simulation results well in particular for $G\leq 0.8$.

\begin{figure}[!t]
	\par
	\begin{center}
		{\includegraphics[height=2.85in,width=3.45in]{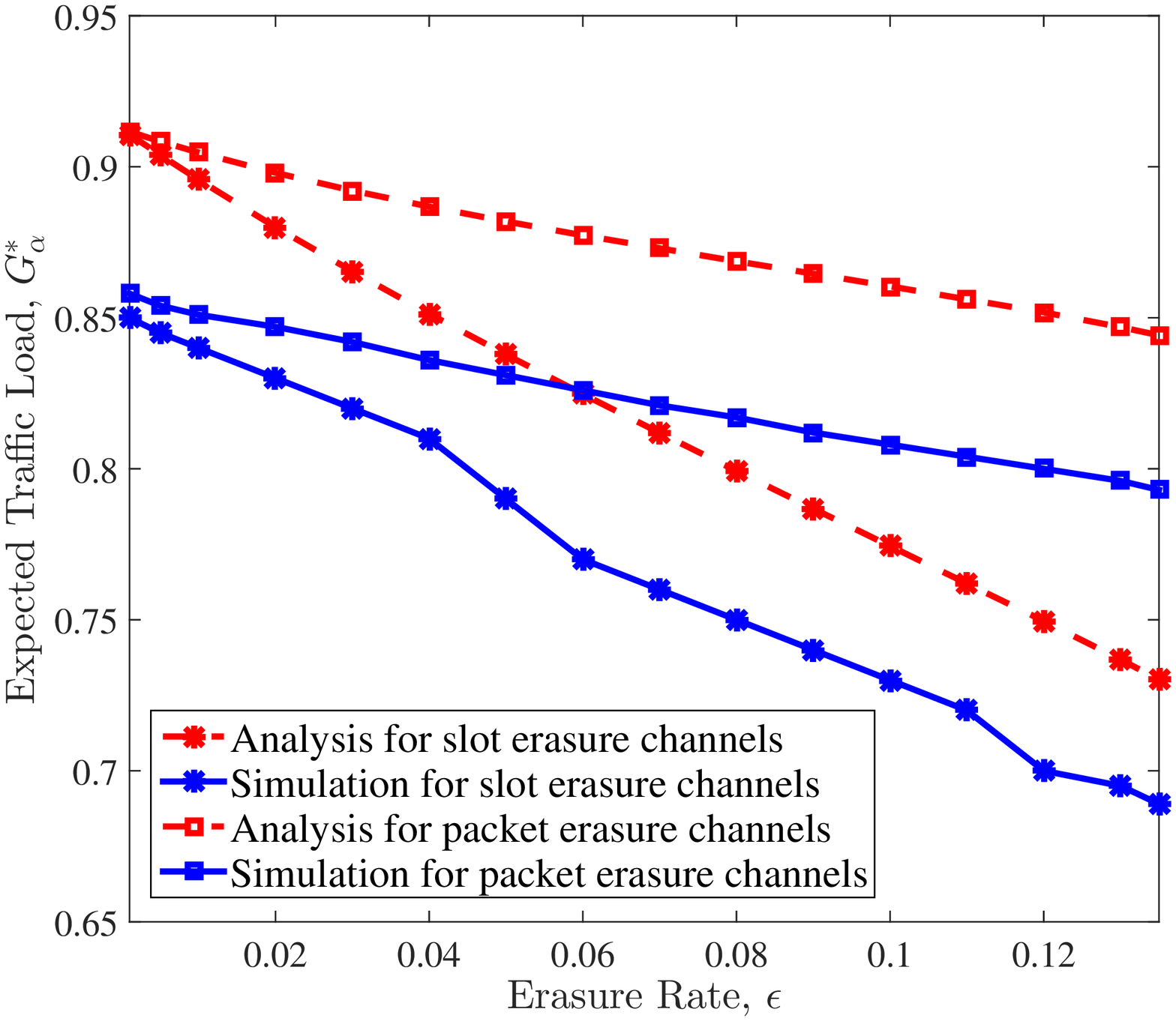}}\vspace{-10mm}
	\end{center}
	\caption{Simulated and asymptotic $G_{\alpha}^{\ast}$ for erasure channels with various $\epsilon$.}
	\label{fig_g_star_asym_erasure_packet_slot}
\end{figure}

The expected traffic loads $G_{\alpha}^{\ast}$ for packet erasure channels and slot erasure channels are compared in both the simulation and asymptotic settings, as shown in Fig. \ref{fig_g_star_asym_erasure_packet_slot}. It can be seen that $G_{\alpha}^{\ast}$ for the slot erasure channel is always less than that of the packet erasure channel. The trend is more obvious for higher erasure rates, which can be explained as follows. For packet erasure channels, the erased packets reduce the degree of SNs, which is beneficial for the SIC process to resolve collisions. However, for slot erasure channels, the degree of unerased slots remains the same, which does not improve the probability for the SIC process to resolve collisions.

\section{Chapter Summary}
In this chapter, we designed the code probability distributions for the CSA schemes over erasure channels to maximize the expected traffic load. For the erasure channels, the EXIT functions of the CSA scheme were derived as functions of the degree distributions of the entire bipartite graph, although only a subgraph consisting of unerased edges was involved in the recovery process. Based on that, we designed the probability distributions for the CSA scheme with repetition codes and MDS codes over both packet erasure channels and slot erasure channels, to maximize the expected traffic load. It was shown that the proposed code distributions could maximize the expected traffic load and thus achieve an improved throughput for the erasure channels, compared to the code distributions designed for non-erasure collision channels in \cite{Paolini2011}. In addition, the asymptotic CSA throughput was derived to predict the performance of the CSA schemes in the presence of erasure. The asymptotic throughput was demonstrated to give a good approximation to the simulation results, when more than two replicas are employed by each user.

\section{Appendix}
\subsection{Proof of Lemma \ref{exit_sn_era}} \label{AppendixA}
Consider a packet with $(l-1)$ interfering packets in a time slot. Each transmitted packet is erased with probability $\epsilon$. Then, the probability that the desired packet is received with $(v-1)$ interfering packets is $\dbinom{l-1}{v-1}\left(1-\epsilon \right)^{v-1}\epsilon^{l-v}$. Averaging this probability over the SN edge degree distribution $\rho(x)$ of the graph $g$ leads to the SN edge degree distribution $\tilde{\rho}(x)$ of the subgraph $\tilde{g}$, given by
\begin{align}
\tilde{\rho}(x)&=\sum_{l=1}^{M}\rho_{l}\sum_{v=1}^{l}\dbinom{l-1}{v-1}\left(
1-\epsilon \right)^{v-1}\epsilon^{l-v}x^{v-1} \notag \\
&=\rho\left((1-\epsilon)x+\epsilon\right).
\label{relation_sn_eg_nd_dis_poly}
\end{align}
Recall Eq. \eqref{exit_func_sn_colli_channel}, we have
\begin{align}
p_{i}=1-\tilde{\rho}(1-q_{i}).
\label{exit_sn_pckers}
\end{align}
Lemma \ref{exit_sn_era} can thus be obtained by substituting Eq. \eqref{sn_eg_dis_colli_channel} and \eqref{relation_sn_eg_nd_dis_poly} into Eq. \eqref{exit_sn_pckers}.

\subsection{Proof of Corollary \ref{exit_un_mds_ers}} \label{AppendixB}
For the UN associated with a $(k+h,k)$ MDS code, only when at least $k$ edges connected to the UN have been recovered, all its edges can be recovered \cite{Singleton1964}. Then, an edge connected to the UN cannot be recovered by the UN decoder, when one of the following cases happens:
\\
\indent Case 1: There are less than $k$ edges connected to this UN in the subgraph $\tilde{g}$;
\\
\indent Case 2: While the degree of the UN is more than $k$ in the subgraph $\tilde{g}$, less than $k$ edges are recovered. \\
Then, the probability $q_{i}$ from the UN decoder is the sum probability of two cases, given by
\begin{align}
\label{exit_func_un_pct_ers_channel_mdsc_hth_un}
q_{i}=\sum\limits_{j=1}^{k}\tilde{\lambda}_{j}+\sum\limits_{j=k+1}^{k+\theta}\left[\sum\limits_{w=0}^{k-1}%
\dbinom{j-1}{w}\left( 1-p_{i-1}\right) ^{w}p_{i-1}^{j-1-w}\right]\tilde{\lambda}_{j}.
\end{align}
Eq. \eqref{exit_func_un_pct_ers_channel_mdsc} follows from Eq. \eqref{exit_func_un_pct_ers_channel_mdsc_hth_un}.

\subsection{Proof of Lemma \ref{exit_sn_slot_era}} \label{AppendixC}
Consider a slot where $v$ packets are received. In the slot erasure channel, if $v \neq 0$ this slot is not erased and all the transmitted packets are received. The SN degree distribution of the subgraph $\tilde{g}$ is
\begin{align}
\tilde{\Psi}_{v}=
\begin{cases}
\epsilon +(1-\epsilon )\exp (-\frac{G}{R})\hspace*{8mm} \mathrm{if}\hspace*{1mm}v=0, \\
(1-\epsilon )\frac{e^{-\frac{G}{R}}}{v!}(\frac{G}{R})^{v}\text{ \ \ \ \ \ \ \ \ \
	\ }\mathrm{if}\hspace*{1mm}v\neq 0.
\end{cases}
\label{sn_nd_dis_slt_channel_rx_p1}
\end{align}
The SN edge degree distribution of the subgraph $\tilde{g}$ is
\begin{align}
\tilde{\rho}(x)=\sum_{v=1}^{M}\frac{(\frac{G%
	}{R})^{v-1}}{(v-1)!}e^{-\frac{G}{R}}x^{v-1}.
\label{sn_eg_dis_slt_channel_rx_p2}
\end{align}
Substituting Eq. \eqref{sn_eg_dis_slt_channel_rx_p2} into Eq. \eqref{exit_sn_pckers}, Lemma \ref{exit_sn_slot_era} follows.

\subsection{Proof of Lemma \ref{pck_rm}} \label{AppendixD}
Consider a packet with $(j-1)$ received replicas. If any of its $(j-1)$ replicas has been recovered at the $(N_{\mathrm{iter}}-1)$-th iteration, the packet is recovered and its interference can be removed at the $N_{\mathrm{iter}}$-th iteration. This event happens with the probability $P_{\mathrm{pck\_cl}}^{j}(N_{\mathrm{iter}})=1-(1-P_{p}(N_{\mathrm{iter}}-1))^{j-1}$. The probability that an interfering packet can be removed at the $N_{\mathrm{iter}}$-th iteration is
\begin{align}
P_{\mathrm{pck\_cl}}(N_{\mathrm{iter}})&=\sum_{j=1}^{n_{\max}}\tilde{\lambda}_{j}(1-\left(1-P_{p}(N_{\mathrm{iter}}-1)\right)^{j-1}) \notag \\
%&=1-\sum_{j=1}^{n_{\max}}\tilde{\lambda}_{j}(1-P_{\mathrm{ps}}(N_{\mathrm{iter}}-1|G,{\bf\Lambda}))^{j-1} \notag \\
&=1-\tilde{\lambda}\left(1-P_{p}(N_{\mathrm{iter}}-1)\right),
\label{p_rmv_interference_p1}
\end{align}
where $\tilde{\lambda}_{j}$ is the probability that a packet has $(j-1)$ received replicas and $n_{\max}$ is the maximum code length in the given code set. Based on Eq. \eqref{relation_un_eg_nd_dis}, we have
\begin{eqnarray}
%\label{p_rmv_interference_p2_a}
P_{\mathrm{pck\_cl}}(N_{\mathrm{iter}})&=&1-\lambda\left(1+(\epsilon-1)P_{p}(N_{\mathrm{iter}}-1)\right) \notag \\
%\label{p_rmv_interference_p2_b}
&=&1-\sum_{h=1}^{\theta}\frac{n_{h}\Lambda _{h}}{\bar{n}}\left(1+(\epsilon-1)P_{p}(N_{\mathrm{iter}}-1)\right)^{n_{h}-1},
\label{p_rmv_interference_p2}
\end{eqnarray}
where the second equation is obtained from the definition of $\lambda_{h}$ in Eq. \eqref{relation_nd_eg_dis}.

\vspace{-5mm}
\subsection{Proof of Lemma \ref{p_ps}} \label{AppendixE}
A packet can be recovered only when there is no interference in its corresponding slot, i.e., no collision happens or all interfering packets have been removed. This event happens with probability
\begin{align}
P_{p}(N_{\mathrm{iter}})&=\tilde{\rho}_{1}+\sum_{v=2}^{M}\tilde{\rho}_{v}\left(P_{\mathrm{pck\_cl}}(N_{\mathrm{iter}})\right)^{v-1} =\tilde{\rho}\left(P_{\mathrm{pck\_cl}}(N_{\mathrm{iter}})\right),
\label{def_p_ps_p1}
\end{align}
where $\tilde{\rho}_{v}$ is the probability that a packet is received with $(v-1)$ interfering packets. Based on Lemma \ref{pck_rm}, the recursive function of $P_{p}$ can be obtained as
\begin{align}
\label{def_p_ps_p2}
P_{p}(N_{\mathrm{iter}})=\tilde{\rho}\left(1-\sum_{h=1}^{\theta}\frac{n_{h}\Lambda _{h}}{\bar{n}}(1+(\epsilon-1)P_{p}(N_{\mathrm{iter}}-1))^{n_{h}-1}\right).
\end{align}
In an asymptotic setting where $N,M \rightarrow \infty$, we have $\tilde{\rho}(x)=\mathrm{exp}\left(-\frac{G}{R}(1-\epsilon)(1-x)\right)$. By substituting $\tilde{\rho}(x)$ into Eq. \eqref{def_p_ps_p2}, Eq. \eqref{def_p_ps} is obtained.

\subsection{Proof of Theorem \ref{anal_thpt_csa}} \label{AppendixF}
The user with $j$ received packets can be recovered when at least one of the $j$ packets has been recovered. This event happens with the probability $P_{u}=1-(1-P_{p})^{j}$. The probability that a user can be recovered is given by
\begin{align}
P_{u}&=\sum_{j=1}^{n_{\max}}\tilde{\Lambda}_{j}(1-(1-P_{p})^{j})=1-\tilde{\Lambda}(1-P_{p}),
\label{def_thpt_p1}
\end{align}
where $\tilde{\Lambda}_{j}$ is the probability that a user has $j$ received packets. Noting that $\tilde{\Lambda}(x)=\Lambda\left((1-\epsilon)x+\epsilon\right)$, we have
\begin{align}
\label{def_thpt_p1}
P_{u}=1-\sum_{h=1}^{\theta}\Lambda_{h}(1+(\epsilon-1)P_{p})^{n_{h}}.
\end{align}
Eq. \eqref{def_thpt} follows by considering $T=GP_{u}$. The proof completes.

\chapter{Physical-layer Network Coding based Decoding Scheme in Random Access Systems for mMTC}\label{C5:chapter5}

%\nomenclature{$a$}{The number of angels per unit area}%
%\nomenclature{$N$}{The number of angels per needle point}%
%\nomenclature{$A$}{The area of the needle point}%
%
%\ifpdf
%    \graphicspath{{1_introduction/figures/PNG/}{1_introduction/figures/PDF/}{1_introduction/figures/}}
%\else
%    \graphicspath{{1_introduction/figures/EPS/}{1_introduction/figures/}}
%\fi
\section{Introduction}
In the previous chapter, we designed the CSA system under erasure channels to enhance the network throughput.
The design in Chapter 4 mainly focused on the transmitter side.
With the framework of CSA system, we propose an efficient data decoding scheme for the receiver to further improve the network throughput in this chapter.
%To address the above issues, we proposed a low-complexity PNC decoding scheme for RA schemes in \cite{Zhuo16}.
%
The proposed decoding scheme consists of a low-complexity physical-layer network coding (PNC)-based decoding algorithm and an enhanced message-level SIC algorithm.
In particular, we exploit the PNC-based decoding scheme to obtain multiple linear combinations of users' packets in each time slot of a MAC frame.
%
%we first select the optimized coefficients of linear combinations in terms of minimizing their decoding error probability, and then only decode the linear combinations associated with these optimized coefficients.
%%
%This avoids the exhaustive decoding of all possible linear combinations and therefore reduces the computation complexity of the random access system.
%
Then, we propose an enhanced message-level SIC algorithm to jointly exploit NC messages across multiple time slots to achieve an improved throughput of the CSA system.
We also propose an analytical framework for the proposed decoding scheme and optimize the design of CSA scheme based on the analysis. For the ease of presentation, only BPSK is considered in this work.
%We also propose an analytical framework for the PNC-based decoding scheme and optimize the proposed scheme based on the analysis in this paper.
%
The main contributions of the work in this chapter are summarized below.

1. We propose an efficient data decoding scheme to improve the throughput of CSA systems designed in Chapter 4.
We first present a low-complexity PNC-based decoding scheme to obtain multiple linear combinations of users' packets in each time slot of a MAC frame, where a linear combination of users' packets is called a network-coded (NC) message.
In the low-complexity PNC-based decoding scheme, we first select the optimized coefficients of NC messages in terms of minimizing their decoding error probability, and then only decode the NC messages associated with these optimized coefficients.
This avoids the exhaustive decoding of all possible NC messages and therefore reduces the computation complexity of the random access system.
In addition, we propose a simple yet effective message-level SIC decoding algorithm to jointly exploit the NC messages from the low-complexity PNC-based decoding scheme, in order to improve the throughput of CSA systems.
In the proposed message-level SIC algorithm, the degree-$2$ NC messages, i.e., the linear combinations of two users' packets, will be exploited to enhance the message-level SIC process.
In particular, when no more clean packets exist in the message-level SIC process, degree-$2$ NC messages will be canceled from other NC messages in which they are involved.
In this way, new clean packets might be created and the message-level SIC can proceed further.
This allows more users' packets to be recovered and therefore improves the throughput of CSA systems.
By representing the relationship between the NC messages and the users' packets in a frame with a graph, the degree-$2$ NC messages that are exploited in the decoding process can be found by leveraging the method of searching length-$4$ cycles of the Tanner graph.
It is shown that the proposed enhanced message-level SIC decoding algorithm can achieve a considerably improved system throughput, compared to the conventional message-level SIC decoding algorithm.
%
%Note that, the designed NC coefficient matrix also ensures that NC messages are decoded in ascending of decoding error probability, which can decreases the number of decoded NC messages to obtain a given number of NC messages

%Note that, in contrast to the signal-level SIC process adopted in \cite{Casini2007,Herrero09,Liva2011,Paolini2014a}, the proposed packet-level SIC decoding algorithm avoids the re-encoding and re-modulation operations, and therefore reduces the complexity of SIC process.

%%
%2. We introduce the Tanner graph representation of the frame-based decoding, by exploiting the analogy between the decoded NC messages from slot-based decoding and the Luby transform (LT) codes.
%%
%Based on that, we propose an improved \textit{packet-level} SIC decoding algorithm for the frame-based decoding, which utilizes the correlation among the NC messages across different time slots.
%%
%In particular, both the single-user messages and the NC messages associated with multiple users's messages are exploited to recover the users' messages in the improved packet-level SIC decoding algorithm.
%%
%This is different from the conventional packet-level SIC decoding algorithm, in which only the single-user messages are exploited.

2. We propose an analytical framework for the PNC-based decoding scheme and derive the throughput for the proposed scheme.
%
%Due to the randomness of users' transmissions, we first obtain the cumulative distribution function (CDF) and the probability density function (PDF) of minimum decoding distance of each NC message in a time slot.
%
In the proposed analytical framework, we first analyze the average number of successfully decoded NC messages per time slot.
For the time slot with two collided users, we derive a closed-form expression of the average number of successfully decoded NC messages.
Then, we analytically obtain the number of independent NC messages in a MAC frame, by taking into account all decoded NC messages and the redundancy between these NC messages introduced by multiple replicas.
Based on these results, we obtain a tight analytical approximation of the system throughput for the proposed scheme.
It is demonstrated that the analytical results match well with the simulations, in terms of both the average number of decoded NC messages per time slot and the throughput.

3. We optimize the number of replicas transmitted by each user to further improve the system throughput and energy efficiency.
Interestingly, we find that the optimized scheme is a generalization of the CRDSA and CRDSA++ schemes.
In particular, at the low-to-medium offered load regime, a user's packet should be repeated more than twice in the optimized scheme to achieve a high throughput, which coincides with the transmission scheme in CRDSA++.
At the high offered load regime, a user's packet should be repeated only twice in the optimized scheme, which is the same as that in the CRDSA scheme.
We also find that for all offered loads, the user's packet should be repeated twice, in order to optimize the system energy efficiency.
%

%Note that, in our conference version \cite{Zhuo16}, we only present an effective approach for designing the NC coefficient matrix, to achieve a low-complexity decoding scheme for NC messages.
%%
%Compared to our previous work in \cite{Zhuo16}, we further propose an improved packet-level SIC scheme to significantly improve the system throughput in this paper.
%%
%Moreover, in this paper, we present the analysis framework for the PNC based decoding scheme, derive the analytical throughput for the proposed scheme, and optimize the scheme to maximize the throughput.\color{black}
\section{System Model and Problem Formulation}
Consider a random access system where $M$ users attempt to communicate with a common receiver via a shared channel\footnote{Note that, we consider a non-relay setting in this work, where the receiver serves as a termination point and needs to exploit the PNC scheme to recover all these transmitted packets explicitly rather than their network-coded messages.}.
%\color{red}e.g. the uplink of a wireless sensor network with $M$ sensors or a satellite return link from $M$ terminals to a satellite.
%
We assume that each user has one packet to be transmitted to the receiver in a MAC frame, where a MAC frame consists of $N$ equal-duration time slots.
The duration of each time slot equals the length of a user packet, denoted by $S$.
The offered traffic load is defined as $G=\frac{M}{N}$, which represents the average number of transmitted users' packets per time slot.
We assume that all the transmissions are slot synchronous\footnote{The synchronization can be achieved by obtaining the relative delay to the receiver for each user in the initial association phase \cite{Roberts1975}.}, i.e., the attempts of packet transmission are time aligned with the slots.
%Even though the relative delays are usually obtained only up to some precise in practice, the effect of the less precise delays on the synchronization can be mitigated by including the guard time within each slot.
%%
%The issue is common among all approaches to the slotted random access.
%%
%Moreover, the problem of synchronization for PNC decoding has also been discussed in \cite{Cocco12,CoccoSatejournal14}, and the methods to compensate for the effect of non-perfect synchronization are proposed.
%%
%Therefore, we assume the transmissions are slot synchronous in this paper.
%While the synchronization may be less accurate in practice, there are many effective methods to compensate for the effect of less accurate synchronization on the detection \cite{Cocco12}.
%%
%In addition, the symbol duration is relatively large for the M2M communications, since the users transmit messages with low data rates.
%%
%It causes the relative offset among users be within a fraction of the symbol duration.
%%
%Then, the asynchronism among users is relatively small, and it can be addressed for the PNC decoder in \cite{lu2012,Cocco12,rossetto2009}, which is beyond the scope of this paper.
%

For the ease of presentation, we first present a real-valued system model, i.e., the transmitted signals and channel coefficients are real-valued.
Then, we will show that a complex-valued model can be transformed into a real-valued model.
Now, for simplicity, we focus on an un-coded system to present our proposed decoding scheme\footnote{The simulation results for the coded systems will be presented to illustrate the performance of the proposed decoding scheme in practical applications.}.

Consider an un-channel-coded real-valued random access model.
Let $\mathbf{u}_{m}\in\{0,1\}^{1\times S}$ be the length-$S$ binary packet of user $m$, $m\in \{1,2,\ldots,M\}$.
The packet $\mathbf{u}_{m}$ is mapped to the modulated signal $\mathbf{x}_{m}\in\{-1,1\}^{1\times S}$ by using the BPSK modulation.
%
%Here, $\gamma$ is used to normalize the symbol energy, i.e., $E(x_{m}^2)=1$.
%
The modulated signal $\mathbf{x}_{m}$ is repeated $r$ times and transmitted over $r$ randomly selected time slots within a MAC frame, which is widely adopted in the CSA based schemes, such as CSA, CRDSA, and IRSA \cite{Casini2007,Herrero09,Paolini2014}.
%
%The repetition of transmitting one signal is commonly adopted by the CSA scheme.
%In this case, the number of transmissions per time slot follows a Poisson distribution for a large MAC frame in \cite{Paolini2011}, i.e., $N\rightarrow \infty$.
%
%with mean $\frac{rG}{M}$, where $G=\frac{M}{N}$ is the normalized traffic load to the network.

%
Assume that $K$ users transmit packets in the $n$-th time slot simultaneously, where $0 \leq K \leq M$ and $1 \leq n \leq N$.
In this case, we say that the collision size of this time slot is $K$ or simply call this time slot as a \textit{$K$-collision time slot}.
Let $I(n)=(I(n,1),I(n,2),...,I(n,K))$ be the user indices associated with the entire user set $\{1,2,\ldots,M\}$ for the $K$ active users in the $n$-th time slot, satisfying $1\leq I(n,k)\leq M$ and $I(n,k-1)<I(n,k)$, $k\in\{1,2,\ldots,K\}$.
In other words, for the $k$-th active user in time slot $n$, its index in the set $\{1,2,\ldots,M\}$ is $I(n,k)$.
%Let $I(n)=(I(n,1),I(n,2),...,I(n,K))$ be the indices of the $K$ active users in the $n$-th time slot, where $I(n,k)$ represents the index of the $k$-th user, satisfying $1\leq I(n,k)\leq M$ and $I(n,k-1)<I(n,k)$, $k\in\{1,2,\ldots,K\}$.
%
Then, the transmitted signal matrix is given by $\mathbf{X}[n]=\left[\mathbf{x}_{I(n,1)}^{T},\ldots,\mathbf{x}_{I(n,K)}^{T}\right]^{T} \in \{-1,1\}^{K\times S}$.
The fading channel coefficient vector from the $K$ users to the receiver is denoted by $\mathbf{h}[n]=\left[h_{I(n,1)},\ldots,h_{I(n,K)}\right]^{T} \in \mathbb{R}^{K \times 1}$, whose the $k$-th element is the channel coefficient from user $I(n,k)$ to the receiver, $k\in\{1,\ldots,K\}$.
The received signal in the $n$-th time slot is given by
\vspace{-1mm}
\begin{align}
\mathbf{y}[n]=\sqrt{E_{s}}\mathbf{h}^{T}[n]\mathbf{X}[n]+\mathbf{z}[n],
\label{sys_m1}
\end{align}
\par
\vspace{-3mm}
\noindent
where $E_{s}$ is the average transmit power per symbol and each element of $\mathbf{z}[n] \in \mathbb{R}^{1 \times S}$ is the AWGN at the receiver with zero mean and variance $\sigma_{z}^{2}$.
The transmit SNR is defined as $\gamma=\frac{E_s}{\sigma_{z}^2}$, which is represented as SNR by default.
%The signal-to-noise ratio (SNR) is defined as $\gamma=\frac{E_s}{\sigma_{z}^2}$.
%
In this work, we consider an equal transmit energy for all users\footnote{Note that, by incorporating the transmit power per symbol in users' channel gains, the proposed scheme can be generalized to the case where different users employ different transmit symbol energy levels. Moreover, the proposed scheme can be applied to the MIMO case, by extending the NC message decoding method to MIMO systems via a similar approach in \cite{Guo15}.}.
Besides, each user experiences an independent block fading channel, i.e., the users' channels are static within a MAC frame but vary independently across the different users.
%Besides, each user experiences an independent block fading channel, i.e., the channel for a user is static within a MAC frame but varies among the MAC frames independently.
%
Also, the receiver has perfect CSI of active users for coherent signal detection, which can be achieved by exploiting the orthogonal pilots.

In practical systems, both the transmitted signals and channel coefficients are complex-valued.
We note that with considering the aligned carrier phases of all active users, a complex-valued random access model can be equivalently represented by a real-valued RA model.
%We note that a complex-valued random access model can be equivalently represented by a real-valued RA model.
%
To this end, we denote $\tilde{\mathbf{X}}[n]$, $\tilde{\mathbf{h}}[n]$, $\tilde{\mathbf{z}}[n]$, and $\tilde{\mathbf{y}}[n]$ as the complex-valued transmitted signal matrix, channel coefficient vector, noise, and received signal in the $n$-th time slot, respectively.
Then, the complex-valued model can be represented as
\vspace{-2mm}
\begin{align}
\tilde{\mathbf{y}}[n]=\sqrt{E_{s}}\tilde{\mathbf{h}}^T[n]\tilde{\mathbf{X}}[n]+\tilde{\mathbf{z}}[n].
\label{sys_m1_complx}
\end{align}
\par
\vspace{-2mm}
\noindent
In particular, when the carrier phases of $K$ active users are aligned, the $1\times K$ complex-valued channel model is equivalent to the $2\times 2K$ real-valued channel model, given by
%In particular, the $1\times K$ complex-valued channel model is equivalent to the $2\times 2K$ real-valued channel model, given by
\begin{align}
\hspace{-3mm}\left[\hspace{-1mm}{\begin{array}{*{20}{c}}
	{{\mathop{\rm Re}\nolimits} (\tilde{\mathbf{y}}[n])}\\
	{{\mathop{\rm Im}\nolimits} (\tilde{\mathbf{y}}[n])}
	\end{array}}\hspace{-1mm} \right] \hspace{-1mm}= &\sqrt {{E_s}}\left[\!\hspace{-2mm} {\begin{array}{*{20}{c}}
	{{\mathop{\rm Re}\nolimits} (\widetilde {\bf{h}}^T[n])}&{- {\mathop{\rm Im}\nolimits} (\widetilde {\bf{h}}^T[n])}\\
	{{\mathop{\rm Im}\nolimits} (\widetilde {\bf{h}}^T[n])}&{{\mathop{\rm Re}\nolimits} (\widetilde {\bf{h}}^T[n])}
	\end{array}}\hspace{-1mm} \!\right]\!\!\left[\hspace{-2mm} {\begin{array}{*{20}{c}}
	{{\mathop{\rm Re}\nolimits} (\widetilde {\bf{X}}[n])}\\
	{{\mathop{\rm Im}\nolimits} (\widetilde {\bf{X}}[n])}
	\end{array}}\hspace{-1mm} \right] + \left[\hspace{-1mm} {\begin{array}{*{20}{c}}
	{{\mathop{\rm Re}\nolimits} (\tilde{\mathbf{z}}[n])}\\
	{{\mathop{\rm Im}\nolimits} (\tilde{\mathbf{z}}[n])}
	\end{array}} \hspace{-1mm}\right].
\label{sys_m1_real2complx}
\end{align}
\par
\noindent
Therefore, in the sequel, we will adopt the real-valued RA model in Eq. \eqref{sys_m1} for presenting the design and analysis of our proposed decoding scheme.

Based on the received superimposed signal $\mathbf{y}[n]\in \mathbb{R}^{1 \times S}$, the receiver attempts to decode \textit{multiple} linear combinations of the $K$ active users' packets, which are referred to as NC messages, by exploiting the PNC decoding.
%
%The linear combinations of users' messages are called as network coded (NC) messages.
%
The PNC decoding is conducted in each time slot independently.
When the PNC decoding process is performed for the $N$ time slots in a MAC frame, the receiver proceeds to a message-level SIC decoding across all the time slots to recover the users' packets.
%
%If a user's message is successfully recovered by the SIC decoding, then the user's transmission is successful.
%%
%Otherwise, the user's transmission is failed.
%
Denote $R$ as the number of recovered users' packets, where $0 \leq R \leq M$.
The throughput of the RA scheme is defined as
%the user's transmission is successful and the user is called the successful user. Then, the throughput is given by
\vspace{-2mm}
\begin{align}
T=\frac{R}{N},
%T(G,\textbf{C},{\bf{\Lambda}})\triangleq kS/N
\label{def_thp}
\end{align}
\par
\vspace{-2mm}
\noindent
which represents the average number of recovered users' packets per time slot \cite{Casini2007}.
Obviously, a larger $T$ means a higher spectral efficiency.
%The throughput can be employed to characterize the performance of a RMA scheme.
%
In this work, we aim to propose a decoding scheme for the PNC-based collision recovery system to efficiently recover more users' packets and thereby improve the system throughput.
%
%In particular, the proposed RMA scheme consists of the slot based NC message decoding and the frame based user message decoding, which will be presented in details in the following.

\section{Physical-layer Networking Coding based Decoding Scheme}
In this section, we present the PNC-based decoding scheme for the RA system.
It consists of the NC message decoding in a time slot and the recovery of users' packets in a frame.
For the NC message decoding, an approach for designing the optimized coefficients of NC messages and an NC message decoding algorithm with the designed NC coefficients are presented.
For the recovery of users' packets, we propose an enhanced message-level SIC algorithm to improve the system throughput.
In the following, we will present the two parts in details.
Note that, we employ the one-symbol packet for each user, i.e., $S=1$, to facilitate the presentation of our proposed decoding scheme.
\subsection{NC Message Decoding}
Consider a $K$-collision time slot $n$.
The receiver is assumed to decode $L$, $L\leq K$, NC messages\footnote{Note that, the number of NC messages to be decoded in a time slot, i.e., $L$, is determined by the collision size of this time slot and the computation complexity constraint of practice systems.
	In this work, we assume that the computation resources are sufficient and $L$ equals the collision size of the time slot, i.e., $L=K$.} from its received signal $y[n]$ in the time slot $n$.
The $l$-th NC message, $l\in\{1,2,\ldots,L\}$, is
\begin{align}
w_{l}[n]=\mathbf{g}_{l}^{T}[n]\otimes \mathbf{u}[n],
\label{Eq_NC_messages}
\end{align}
\par
\noindent
where $\mathbf{g}_{l}[n]\in \{0,1\}^{K\times 1}$ is the coefficient vector of the $l$-th NC message $w_{l}[n]$, namely the NC coefficient vector, and ${\bf{u}}[n]=[u_{I(n,1)},\ldots,u_{I(n,K)}]^{T}\in \{0,1\}^{K\times 1}$ is the vector of $K$ users' packets.
%Denote by ${\bf{u}}[n]=[u_{I(n,1)},u_{I(n,2)},\ldots,u_{I(n,K)}]^{T}$ the vector of message symbols from $K$ users.
%%
%The $l$-th NC message can be given by
%
%%
%Note that, the entries of $\mathbf{g}_{l}[n]$ take values in $\mathrm{GF}(q)$, i.e., $\mathbf{g}_{l}[n]\in \mathrm{GF}(q)$.
%
The $L$ NC messages in this time slot are then given by
\begin{align}
\mathbf{w}[n]=\mathbf{G}^{T}[n]\otimes \mathbf{u}[n],
\end{align}
\par
\noindent
where vector $\mathbf{w}[n]=\left[w_{1}[n],\ldots,w_{L}[n]\right]^{T} \in \{0,1\}^{L\times 1}$ is an NC message vector and matrix $\mathbf{G}[n]=\left[\mathbf{g}_{1}[n],\ldots,\mathbf{g}_{L}[n]\right] \in \{0,1\}^{K\times L}$, namely the NC coefficient matrix, collects the coefficient vectors of all decoded NC messages, i.e., $\mathbf{w}[n]$.

As discussed in \cite{Lei2016}, the probability that an NC message can be successfully decoded is determined by the effective minimum decoding distance w.r.t. its NC coefficient vector.
%
%The minimum NC set-distance is determined by the given coefficient vector.
%
Then, we can optimize the NC coefficient matrix $\mathbf{G}[n]$ to maximize the effective minimum decoding distance and thus minimize the decoding error probability for each NC message, which will be presented below.
%
%As a result, it can obtain as many NC messages as possible by decoding the $L$ NC messages.
%%
%The optimization of the coefficient matrix $\mathbf{G}[n]$ is summarized below.
\subsubsection{Optimal Design of NC coefficient Matrix}\label{opt_design}
Denote $\mathbf{x}[n]$ and $\hat{\mathbf{x}}[n]$ as two different transmitted BPSK signal vectors of $K$ users in the $n$-th time slot.
Define
\begin{align}
\boldsymbol{\delta}[n]\triangleq \frac{1}{2}\left(\mathbf{x}[n]-\hat{\mathbf{x}}[n]\right) \label{Eq_def_delta}
\end{align}%
\par
\noindent
as the difference vector (DV) of the signal vector pair $\left( \mathbf{x}[n]%
\text{, }\hat{\mathbf{x}}[n]\right)$, where $\boldsymbol{\delta}[n] \in \left\{-1, 0, 1\right\}^{K\times 1}$.
Given the CSI of $K$ users, i.e., ${\mathbf{h}}[n]$, the Euclidean distance between the superimposed signals corresponding to the two transmitted signal vectors is given by
\begin{align}
\sqrt{E_{s}}\left|\mathbf{h}^{T}[n]\left( \mathbf{x}[n]-\hat{\mathbf{x}}[n]%
\right)\right|=\sqrt{E_{s}}\left|\mathbf{h}^{T}[n]%
\boldsymbol{\delta}[n]\right|.
\end{align}
\par
\noindent
Then, the minimum Euclidean distance for all signal vector pairs $\left( \mathbf{x}[n]%
\text{, }\hat{\mathbf{x}}[n]\right)$ is
\begin{align}
d_{1}[n]=\underset{_{_{\boldsymbol{\delta}[n]\in \left\{-1, 0, 1\right\}^{K\times 1},\text{ }\left\Vert{\boldsymbol{\delta}[n]}\right\Vert
			\neq 0}}}{\text{min}}\sqrt{E_{s}}\left|\mathbf{h}^{T}[n]%
{\boldsymbol{\delta}}[n]\right|.  \label{Eq_def_DV1}
\end{align}
\par
\noindent
It is noteworthy that $d_1[n]$ is the effective minimum decoding distance for the conventional complete decoding scheme, where the users' packets $\mathbf{u}[n]$ are decoded directly from the received signal.
%
%The DV that corresponds to the minimum Euclidean distance $d_{1}[n]$ is thus given by
The DV of two transmitted signal vectors, which can achieve the minimum Euclidean distance $d_{1}[n]$ between the received superimposed signals, is selected and denoted by $\Delta_1[n]$, given by
\begin{align}
\mathbf{\Delta }_{1}[n]=\underset{_{_{\boldsymbol{\delta}[n]\in \left\{-1, 0, 1\right\}^{K \times 1},\text{ }\left\Vert{\boldsymbol{\delta}[n]}\right\Vert
			\neq 0}}}{\text{arg min}}\sqrt{E_{s}}\left|\mathbf{h}^{T}[n]%
{\boldsymbol{\delta}}[n]\right|,  \label{Eq_def_DV1}
\end{align}%
\par
\noindent
where the vector of all zeros is not considered for the selection.
The second minimum Euclidean distance over all signal vector pairs $\left( \mathbf{x}[n]%
\text{, }\hat{\mathbf{x}}[n]\right)$ is denoted by $d_{2}[n]$, and its corresponding DV is $\mathbf{\Delta }_{2}[n]$, subject to
\begin{align}
\text{Rank}\left(\bmod\left(\left[ \mathbf{\Delta }_{1}[n],\mathbf{\Delta }_{2}[n]\right],2\right)\right)=2.
\label{4_rank}
\end{align}
\par
\noindent
Note that, Eq. \eqref{4_rank} ensures that the decoded NC messages in a time slot are linearly independent.
%Here, $\bmod\left(\mathbf{X},q\right)$ represents the component-wise modulo-$q$ operation of matrix $\mathbf{X}$, and $\mathrm{Rank}(\mathbf{X})$ obtains the rank of matrix $\mathbf{X}$ over $\mathrm{GF}(q)$.
%
Similarly, let $d_{j}[n]$ be the $j$-th minimum Euclidean distance for all signal vector pairs $\left( \mathbf{x}[n]\text{, }\hat{\mathbf{x}}[n]\right)$ and $\mathbf{\Delta }_{j}[n]$ be the corresponding DV, subject to
\begin{align}
\hspace{-2mm}\text{Rank}\left(\bmod\!\left(\left[ \mathbf{\Delta }_{1}[n],\ldots,\mathbf{\Delta }_{j}[n]\right]\!,2\right)\right)\!=\!j,\text{for }j\!=\!3,4,\cdots.\hspace{-1mm}
\end{align}
\par
\noindent
From $\mathbf{\Delta }_{j}[n]$, we obtain $\mathbf{v}_{j}[n]=\bmod\left(\mathbf{\Delta }_{j}[n],2\right) \in \{0,1\}^{K \times 1} $, $j\in \{1,2,\ldots,K\}$. Matrix $\mathbf{V}[n]=\left[ \mathbf{v}_{K}[n],\mathbf{v}_{K-1}[n],\cdots ,\mathbf{v}_{1}[n]\right] \in \{0,1\}^{K \times K}$ collects all the vectors $\mathbf{v}_{j}[n]$.
Based on matrix $\mathbf{V}[n]$, we introduce the following theorem which reveals the structure of the optimal NC coefficient matrix to make the superimposed signals with smaller Euclidean distance have the same underlying NC message.
Thus, the minimum distance of superimposed signals with different underlying NC messages is maximized and the decoding error probability of the associated NC messages is minimized.

\begin{theo}
	Consider a $K$-collision time slot $n$. Assume that the receiver can decode $L$, $L \leq K$, NC messages in the time slot. In the asymptotic regime where the SNR is sufficiently large, i.e., $\gamma \to \infty$, ${\bf{G}}^{*}[n]$ is an optimal NC coefficient matrix, if and only if it satisfies
	\begin{align}
	(\mathbf{G}^{\ast}[n])^{T}\otimes\mathbf{V}[n]=\mathbf{T}[n], \label{Eq_Theorem2}
	\end{align}
	\par
	\noindent
	where $\mathbf{T}[n] \in \{0,1\}^{L \times K}$ is a lower-triangular matrix with non-zero diagonal entries, i.e.,
	\begin{align}
	\mathbf{T}[n] = \left[
	\begin{matrix}
	a_{1,1} & 0 & \ldots & 0 & \ldots & 0\\
	a_{2,1} & a_{2,2} & \ldots & 0 & \ldots & 0\\
	\vdots & \vdots & \ddots & \vdots & & \vdots\\
	a_{L,1} & a_{L,2} & \ldots & a_{L,L} & \ldots & 0%
	\end{matrix}%
	\right],
	\label{Eq_Tmatrix}
	\end{align}
	\par
	\noindent
	with $a_{l,l}\neq 0$ for $l\in\{1,\cdots ,L\}$.
	\label{Theo_optmzed_g}
\end{theo}

\emph{\quad Proof: }The proof process is similar to that in \cite{Lei2016} and is omitted here. \QEDA
%Due to the page limitation, we omit the proof here and the interested readers can follow a similar approach as that in  to show the result.

%
It can be seen from Theorem \ref{Theo_optmzed_g} that for the $l$-th NC message with the optimal coefficients given by Eq. \eqref{Eq_Theorem2} and Eq. \eqref{Eq_Tmatrix}, its effective minimum distance is equal to the $(K-l+1)$-th minimum Euclidean distance $d_{K-l+1}[n]$. This is explained as follows.
Denote $\left( \mathbf{x}_{j}[n]\text{, }\hat{\mathbf{x}}_{j}[n]\right)$ as the signal vector pair which corresponds to the $j$-th minimum Euclidean distance $d_{j}[n]$, $1 \leq j \leq K$.
Then the DV of $\left( \mathbf{x}_{j}[n]\text{, }\hat{\mathbf{x}}_{j}[n]\right)$ is $\mathbf{\Delta }_{j}[n]$, which corresponds to the $(K-j+1)$-th column of matrix $\mathbf{V}[n]$.
From Eq. \eqref{Eq_Theorem2}, it can be seen that the element $a_{l,(K-j+1)}$ of $\mathbf{T}[n]$ represents the difference of underlying NC messages of $\mathbf{x}_{j}[n]$ and $\hat{\mathbf{x}}_{j}[n]$ w.r.t. the $l$-th NC coefficient vector $\mathbf{g}_{l}[n]$.
If $a_{l,(K-j+1)}=0$, $\mathbf{x}_{j}[n]$ and $\hat{\mathbf{x}}_{j}[n]$ have an identical underlying NC message w.r.t. $\mathbf{g}_{l}[n]$.
It implies that $\mathbf{x}_{j}[n]$ and $\hat{\mathbf{x}}_{j}[n]$ do not need to be distinguished and  the $j$-th minimum Euclidean distance $d_{j}[n]$ is irrelevant to the decoding of the $l$-th NC message.
Otherwise, $\mathbf{x}_{j}[n]$ and $\hat{\mathbf{x}}_{j}[n]$ have different underlying NC messages w.r.t. $\mathbf{g}_{l}[n]$, and the effective minimum distance for the $l$-th NC message is not larger than $d_{j}[n]$.
By defining $\mathbf{T}[n]$ as a lower-triangular matrix with non-zero diagonal entries, given by Eq. \eqref{Eq_Tmatrix}, we can ensure that given the $l$-th NC coefficient vector $\mathbf{g}_{l}[n]$, $1 \leq l \leq L$, the $j$-th minimum Euclidean distance $d_{j}[n]$ is irrelevant to the decoding of the $l$-th NC message for $1 \leq j \leq (K-l)$.
Therefore, the effective minimum decoding distance for the $l$-th NC message is increased to $d_{K-l+1}[n]$.
%According to Eq. \eqref{Eq_Theorem2} and \eqref{Eq_Tmatrix}, it can be seen that the first NC coefficient vector $\mathbf{g}_{1}[n]$, i.e., the first column of $\mathbf{G}^{\ast}[n]$, is set in such a way that the NC message is identical for all symbol vector pairs that correspond to the $k$-th minimum distance, $k=1,2,\cdots ,K-1$.
%%
%It implies that the effective minimum distance w.r.t. the first NC message $w_{1}[n]$ is increased to the $K$-th smallest Euclidean distance of transmitted symbol vectors.
%%
%Similarly, the second NC coefficient vector $\mathbf{g}_{2}[n]$ is selected, so that the effective minimum distance w.r.t. the second NC message $w_{2}[n]$ is increased to the $(K-1)$-th smallest Euclidean distance of transmitted symbol vectors.
%
%Furthermore, due to Eq. \eqref{indp_dv}, $\mathbf{g}_{2}[n]$ is independent of $\mathbf{g}_{1}[n]$ over $\mathrm{GF}(q)$.
%
As the decoding error probability of an NC message is determined by the effective minimum decoding distance at the medium-to-high SNR regime, the optimized NC coefficients lead to a lower error probability of NC messages and thereby more decoded NC messages.

\vspace{-3mm}
\subsubsection{Decoding of NC Messages}
\vspace{-3mm}Given the optimized NC coefficient matrix $\mathbf{G}^{\ast}[n]$, we decode each NC message separately by using a component-wise MAP algorithm \cite{Robertson95}.
%
%This component-wise MAP decoding algorithm has a much lower decoding complexity than the joint MAP decoding algorithm which tries to decode all the NC messages jointly.
%
In particular, based on the received signal $y[n]$ and the optimized NC coefficient vector $\mathbf{g}^{\ast}_{l}[n]$ for the $n$-th time slot, the $l$-th NC message, $l\in\{1,2,\ldots,L\}$, can be decoded by\vspace{-3mm}
\begin{align}
\widehat{w}_{l}[n]& =\underset{w_{l}[n]}{\text{arg max}}\text{ }p\left(
w_{l}[n]\big|y[n],\mathbf{g}^{\ast}_{l}[n]\right) \notag \\
& =\underset{w_{l}[n]}{\text{arg
		max}}\hspace{-3mm}\sum\limits_{\mathbf{u}[n]:(\mathbf{g}^{\ast}_{l}[n])^{T}\otimes\mathbf{u}[n]=w_{l}[n]}\hspace{-8mm}p\left( \mathbf{u}[n]\big|%
y[n],\mathbf{g}^{\ast}_{l}[n]\right) \notag\\
%& =\underset{w_{l}[n]}{\text{arg max}}%
%\sum\limits_{\mathbf{u}[n]:\mathbf{g}^{T}_{l}[n]\otimes\mathbf{u}[n]=w_{l}[n]}p\left(y[n]|%
%\mathbf{u}[n]\right)  \notag \\
& =\underset{w_{l}[n]}{\text{arg max}}\hspace{-3mm}\sum\limits_{\mathbf{u}[n]:(\mathbf{g}^{\ast}_{l}[n])^{T}\otimes\mathbf{%
		u}[n]=w_{l}[n]}\!\!\!\!\!\!\!\!\!\!\!\exp \left( \frac{-\left|y[n]\!-\!\mathbf{h}^{T}[n] 2\sqrt{E_{s}}\left( \mathbf{u}[n]\!-\!\frac{1}{2}\mathbf{1}\right)%
	\right|^{2}}{2\sigma ^{2}}\right),  \label{Eq_MAP_Relay}
\end{align}
where $\mathbf{1}$ is the all-ones vector with the same length as $\mathbf{u}[n]$.
From Eq. \eqref{Eq_MAP_Relay}, it can be seen that given the NC coefficient vector $\mathbf{g}^{\ast}_{l}[n]$, the posteriori probability of NC message $w_{l}[n]$ is calculated as the sum of the likelihood functions of all the user packet vectors $\mathbf{u}[n]$ corresponding $w_{l}[n]$.
Then, the NC message with the maximum posteriori probability is decided as the $l$-th decoded NC message $\widehat{w}_{l}[n]$.

Let $\widehat{\mathbf{w}}[n]=\left[\widehat{w}_{1}[n],\widehat{w}_{2}[n],\ldots,\widehat{w}_{L}[n]\right]^{T}$ collect the $L$ decoded NC messages in the $n$-th time slot and $\widetilde{\mathbf{w}}[n]$ collect the correctly decoded NC messages\footnote{The correctness of decoded NC messages can be identified by performing the channel decoding or the cyclic redundancy check of packets.}, given by
%\footnote[3]{In order to determine whether the decoding of an NC message is correct, the decoding error probability of this NC message is employed. If the decoding error probability is below a certain threshold, e.g. $10^{-6}$, we believe that the decoding of this NC message is correct. Otherwise, the decoding of this NC message is not correct.}
\vspace{-4mm}
\begin{equation}
\widetilde{\mathbf{w}}[n]=(\widetilde{\mathbf{G}}^{\ast}[n])^{T}\otimes \mathbf{u}[n],
\label{ct_nc_ms}\vspace{-3mm}
\end{equation}
\par
\noindent
where $\widetilde{\mathbf{G}}^{\ast}[n]$ is the NC coefficient matrix associated with $\widetilde{\mathbf{w}}[n]$ and it is a sub-matrix of $\mathbf{G}^{\ast}[n]$.
%
%In the following, the NC messages that are correctly decoded are called as the decoded NC messages for simplify.
%
After performing the NC messages decoding in all time slots of a frame, the correctly decoded NC messages within a frame, i.e., $\widetilde{\mathbf{w}}[n]$, $1 \leq n \leq N$, are jointly exploited to recover users' packets.
	The presented NC message decoding has a lower decoding complexity than the exhaustive seek and decode scheme in \cite{Cocco14}.
	For the decoding scheme in \cite{Cocco14}, NC messages are continuously decoded until $K$ NC messages are successfully decoded in a $K$-collision time slot.
	Then, $(2^{K}-1)$ NC messages are required to be decoded in the worst case.
	For the presented NC message decoding scheme, at most $K$ NC messages are decoded, and they can be successfully decoded with a high probability since their coefficients are optimized to minimize their decoding error probabilities.
	Obviously, for $K \geq 2$, less NC messages are required to be decoded by the presented scheme.
	%
	%and the performance can be approached for the proposed scheme, in terms of the number of successfully decoded NC messages.
	%%
	%Moreover, the decoding complexity for one NC message is high, particularly considering the channel decoding.
	%%
	%Therefore, the proposed scheme has a much smaller computation complexity compared to that of the exhaustive decoding scheme in a time slot.
	%
	We note that the presented scheme introduces some overheads on the complexity for selecting the coefficients of NC messages.
	However, this additional operation is done only once per time slot, and the associated complexity is very small compared to that of channel decoding.
	Therefore, the extra complexity is negligible.
In addition, the presented NC message decoding scheme can obtain both the individual native packets and the network-coded packets, which correspond to the results obtained by the MUD and the PNC decoders, respectively. As a result, the proposed scheme has a lower implementation complexity than the scheme proposed in \cite{Lu2013,You2015}, where both MUD and PNC decoders are required.
%	In addition, the presented NC message decoding scheme contains the MUD and PNC decoding as special cases.
%	%
%	As a result, the proposed scheme has a lower implementation complexity than the scheme proposed in \cite{Lu2013,You2015}.
%	%
%	In other words, only a single PNC decoder is required in the proposed scheme, while both MUD decoder and PNC decoder are required in \cite{Lu2013,You2015}.
	
\subsection{Enhanced Message-level SIC Algorithm}
While the NC message decoding is independently performed in each time slot, the multiple replicas of a user packet in a MAC frame introduce a correlation among the NC messages across several time slots.
By exploiting this correlation and the structure of decoded NC messages, we aim to propose an enhanced message-level SIC decoding algorithm to achieve a higher throughput in this part.
%
%The frame-based decoding aims to exploit this correlation to recover more users' messages, which in turn leads to an improved throughput.
%%
%Traditionally, only the message of a single user can be used to perform interference cancellation in the frame-based decoding, which is termed as the conventional message-level SIC algorithm.
%%
%However, this may limit the recovery of more users' messages, in particular for the high traffic load.
%%
%Therefore, we propose an improved message-level SIC decoding algorithm, which exploits the structure of the decoded NC messages to achieve a higher throughput.
%

To facilitate the discussion of our proposed message-level SIC algorithm, we first formulate the problem of recovering users' packets from the decoded NC messages.
We then introduce a graph representation for the recovering process of users' packets.
The enhanced message-level SIC algorithm is presented at last.

Let $\widetilde{\mathbf{w}}=\left[\widetilde{\mathbf{w}}^{T}[1],\widetilde{\mathbf{w}}^{T}[2],\ldots,\widetilde{\mathbf{w}}^{T}[N]\right]^{T}$ and $\widetilde{\mathbf{G}}^{\ast}=\left[\widetilde{\mathbf{G}}^{\ast}[1],\widetilde{\mathbf{G}}^{\ast}[2],\ldots,\widetilde{\mathbf{G}}^{\ast}[N]\right]$ collect all the decoded NC messages in a MAC frame and their associated NC coefficients, respectively.
Besides, we denote vector $\mathbf{u}$ as a collection of all users' packets in a MAC frame.
We have
\vspace{-3mm}
\begin{equation}
\widetilde{\mathbf{w}}=(\widetilde{\mathbf{G}}^{\ast})^{T} \otimes \mathbf{u}.
\label{frame_probm_formul}\vspace{-5mm}
\end{equation}
\par
\noindent
The recovery of users' packets is to obtain $\mathbf{u}$ from Eq. \eqref{frame_probm_formul}, given $\widetilde{\mathbf{w}}$ and $\widetilde{\mathbf{G}}^{\ast}$.

The relationship between decoded NC messages and users' packets can be characterized by a Tanner graph, which includes two disjoint node sets and multiple edges connecting the two node sets \cite{Tanner81}.
%The relationship between decoded NC messages and users' messages can be characterized by a graph representation.
%
In particular, each element in $\mathbf{u}$ is represented as a variable node and each element in $\widetilde{\mathbf{w}}$ is a constraint node.
The connections between variable nodes and constraint nodes are specified by $\widetilde{\mathbf{G}}^{\ast}$. For example, if the element $\widetilde{g}^{\ast}_{k,l}=1$ in the matrix $\widetilde{\mathbf{G}}^{\ast}$, there is an edge between the variable node $k$ and the constraint node $l$.
If $\widetilde{g}^{\ast}_{k,l}=0$, there is no edge between variable node $k$ and constraint node $l$.

\begin{figure}[t]
	\centering
	\subfigure[Graph representation.]
	{\label{fig_frame based decoding2a} %% label for first subfigure
		\includegraphics[width=0.4\textwidth]{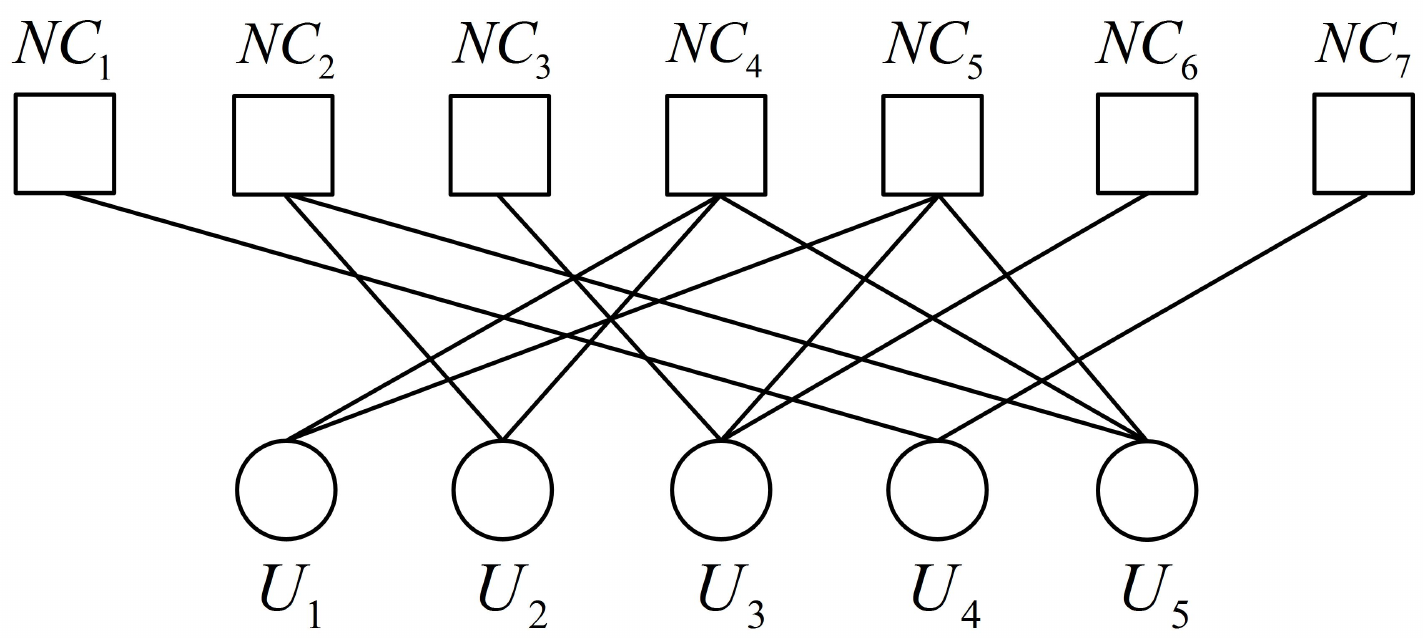}}\hspace*{3mm}
	\subfigure[SIC iteration 1.]
	{\label{fig_frame based decoding2b} %% label for second subfigure
		\includegraphics[width=0.4\textwidth]{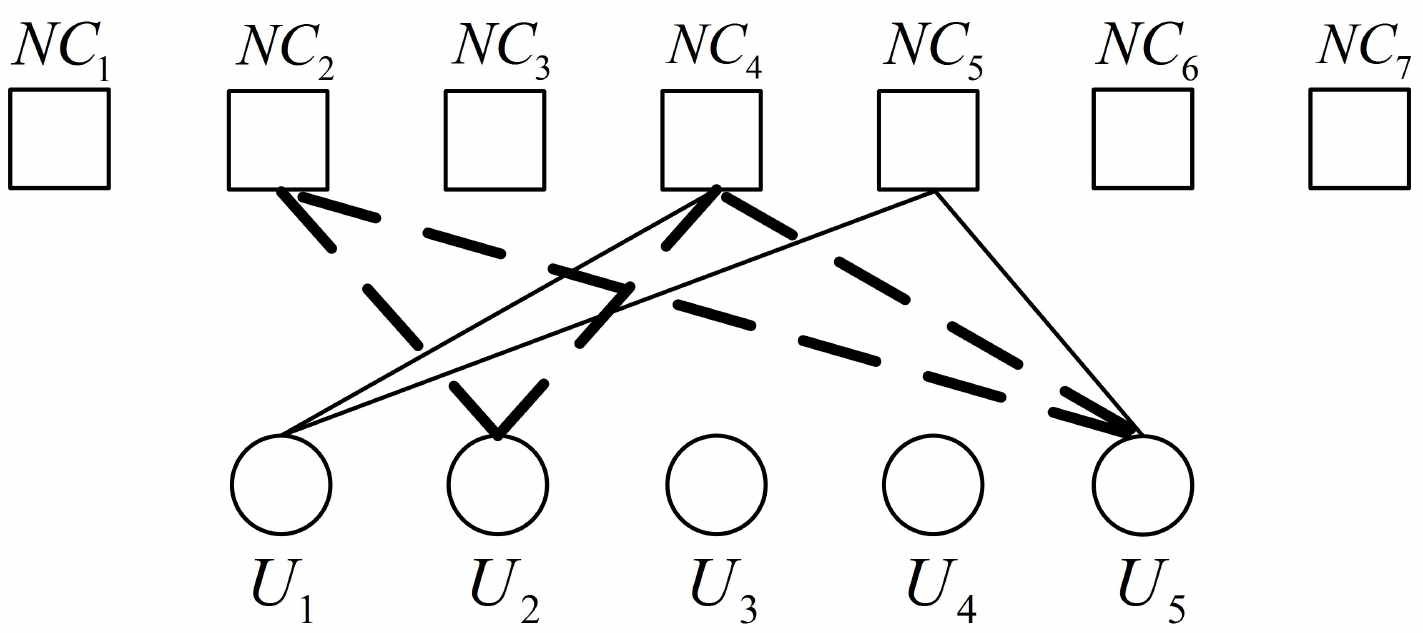}}
	\hspace{10mm}
	\subfigure[SIC iteration 2.]
	{\label{fig_frame based decoding2c} %% label for second subfigure
		\includegraphics[width=0.4\textwidth]{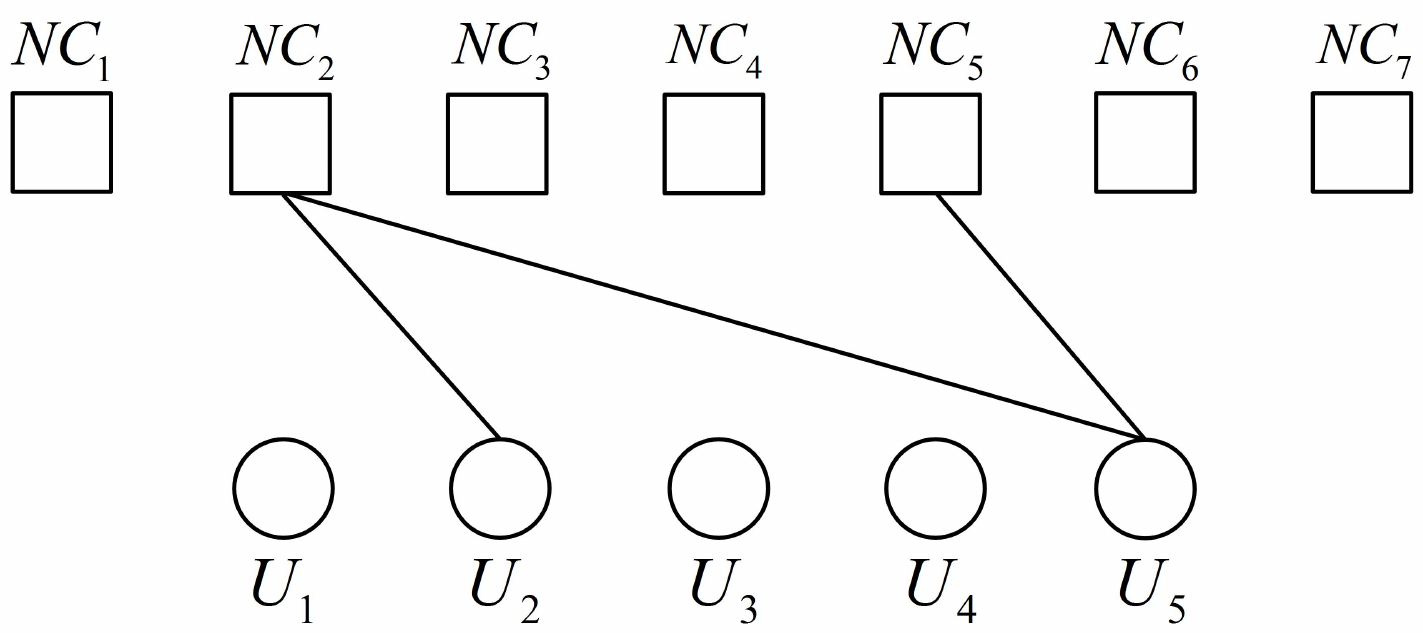}}\hspace*{3mm}
	\subfigure[SIC iteration 3.]
	{\label{fig_frame based decoding2d} %% label for second subfigure
		\includegraphics[width=0.4\textwidth]{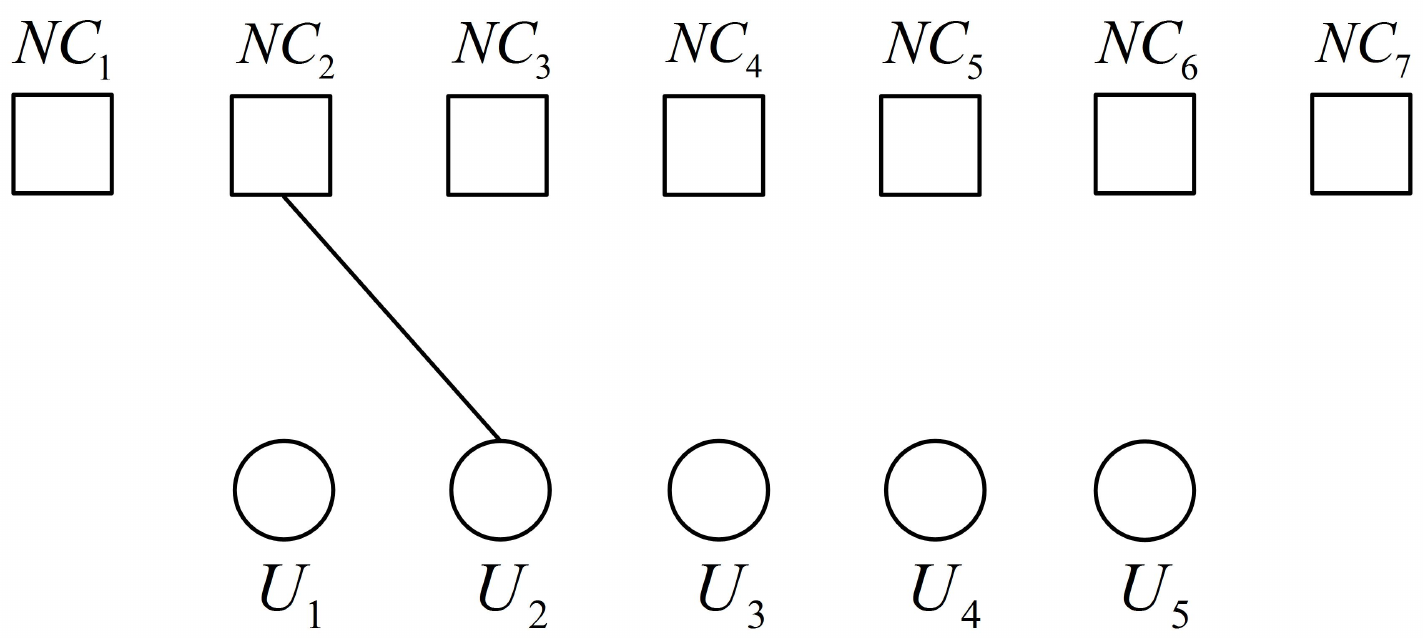}}\vspace{-3mm}
	\subfigure
	{\label{fig_frame based decoding_legend}%% label for first subfigure
		\includegraphics[width=0.5\textwidth]{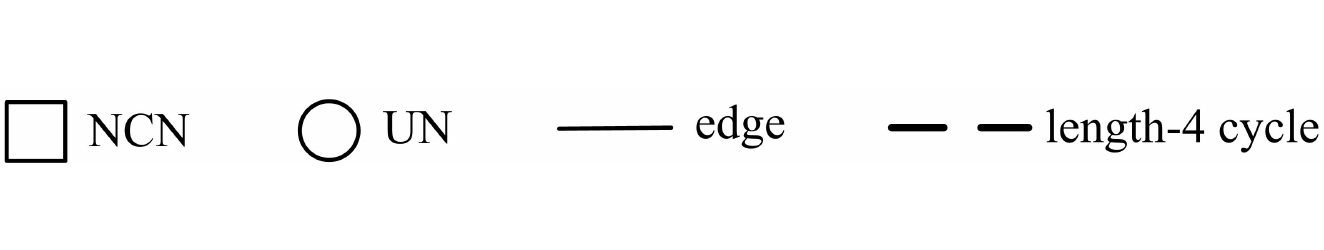}}\hspace*{3mm}\vspace{-5mm}
	\caption{A graph representation of the enhanced message-level SIC algorithm.}\vspace{-4mm}
	\label{fig_frame_based_decoding_3}
\end{figure}

%\begin{figure}[!t]
%{\ }
%\par
%\begin{center}
%{\includegraphics[width=6in]{Graphes/frame_based_decoding5.pdf}}
%\end{center}\vspace{-10mm}
%\caption{Graph Representation of the proposed message-level SIC algorithm}
%\label{fig_frame_based_decoding_3}
%\end{figure}
%
An example of the graph representation associated with
\vspace{-3mm}
\begin{align}
\widetilde{\mathbf{G}}^{\ast}= \left[
\begin{matrix}
0 & 0 & 0 & 1 & 1 & 0 & 0 \\
0 & 1 & 0 & 1 & 0 & 0 & 0 \\
0 & 0 & 1 & 0 & 1 & 1 & 0 \\
1 & 0 & 0 & 0 & 0 & 0 & 1 \\
0 & 1 & 0 & 1 & 1 & 0 & 0%
\end{matrix}%
\right]
\label{expl_op_mx}
\end{align}\vspace{-5mm}
\par
\noindent
is shown in Fig. \ref{fig_frame based decoding2a}.
In the figure, each decoded NC message is represented by a square, namely an NC node (NCN), and each user is represented by a circle, namely a user node (UN).
An edge connects an NCN to a UN only if the decoded NC message contains this user's packet.
The degree of the $l$-th NCN is the number of users that are associated with it or the weight of the $l$-th column of $\widetilde{\mathbf{G}}^{\ast}$.
The degree of the $k$-th UN is the number of decoded NC messages that are associated with it or the weight of the $k$-th row of $\widetilde{\mathbf{G}}^{\ast}$.

From the graph representation, it can be seen that the performance for recovering users' packets is affected by the decoding algorithm, given the NC coefficient matrix $\widetilde{\mathbf{G}}^{\ast}$.
For the conventional message-level SIC decoding algorithm proposed by \cite{Lu2013}, the degree-$1$ NCN (or termed as the native packet in \cite{Lu2013}) is successively exploited to cancel its interference from all its associated NCNs at the packet level.
However, for the considered system the structure of $\widetilde{\mathbf{G}}^{\ast}$ (or the degree distribution of variable nodes) is determined by the random channel realizations.
Then, it is difficult to guarantee a high probability of degree-$1$ NCNs during the decoding process, resulting in a degraded performance of the conventional message-level SIC decoding algorithm.
Based on this consideration, we propose an enhanced message-level SIC decoding algorithm.
In particular, we not only cancel the degree-$1$ NCNs in the graph, but further cancel the degree-$2$ NCNs when there is no degree-$1$ NCN.
This cancellation may generate new degree-$1$ NCNs and enable the message-level SIC decoding to continue.
We summarize the proposed message-level SIC decoding algorithm in \textbf{Algorithm} \ref{alg2}.
It is noteworthy that when multiple length-$4$ cycles with overlapping lines are available in the graph, we randomly choose one length-$4$ cycle which includes one degree-$2$ NCN and one degree-$3$ NCN, and cancel the degree-$2$ NCN from the degree-$3$ NCN within this length-$4$ cycle in this work.
\begin{table}[t]
	%\vspace{-2mm}
	\begin{algorithm} [H]                    % enter the algorithm environment
		\caption{Enhanced Message-Level SIC Decoding Algorithm}     % give the algorithm a caption
		\label{alg2}                             % and a label for \ref{} commands later in the document
		\begin{algorithmic} [1]
			%\footnotesize          % enter the algorithmic environment
			\REPEAT
			\REPEAT
			\STATE Find and release all degree-$1$ NCNs to recover their associated packets
			\STATE Cancel the recovered packets' replicas from the NCNs that contain them as the associated packets
			\UNTIL
			No degree-$1$ NCN is found
			\IF{All users' packets are recovered}
			\STATE Break
			\ENDIF
			%\STATE break
			\STATE Find the degree-$2$ NCNs that are involved in the length-$4$ cycles of the Tanner graph
			\STATE Cancel the degree-$2$ NCNs from the other NCNs in their corresponding length-$4$ cycles by performing the exclusive-or operation
			\UNTIL
			No degree-$1$ and degree-$2$ NCNs are found
		\end{algorithmic}
	\end{algorithm}
\end{table}

\begin{figure}[t]
	\begin{center} {\includegraphics[width=3.45in]{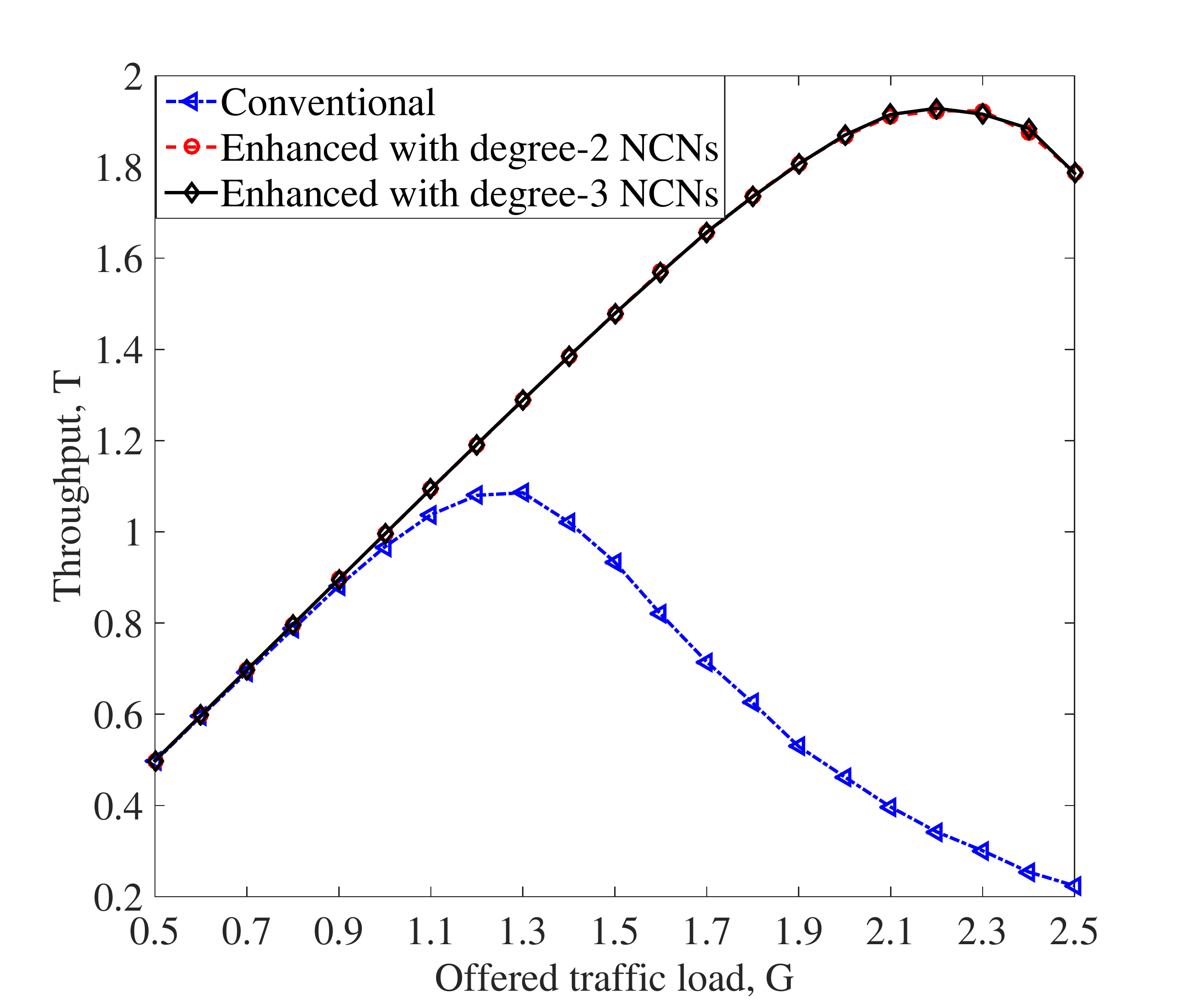}}\vspace{-5mm}
	\end{center}
	\caption{Throughput of the enhanced message-level SIC algorithm by canceling the NCNs with different degrees for $r=2$.}\vspace{-3mm}
	\label{fig_improved_sic_diff_degree}
\end{figure}

We note that the proposed message-level SIC algorithm only exploits the benefits of canceling the degree-$2$ NCNs and will not proceed to higher degree NCNs by considering the implementation complexity.
By borrowing the algorithms in the parity-check matrix of a low-density parity-check (LDPC) code in \cite{Fossorier04}, length-$4$ cycles in the graph can be effectively found.
%The cancellation process of degree-$2$ NCNs can be implemented effectively by borrowing the algorithms which are used to find length-$4$ cycles in the parity-check matrix of a low-density parity-check (LDPC) code in \cite{Fossorier04}.
%
Moreover, the cancellation process of degree-$2$ NCNs can be implemented effectively by performing the exclusive-or of the binary values of degree-$2$ NCNs and the binary values of other NCNs in the associated length-$4$ cycles.
%The cancellation process of degree-$2$ NCNs can be implemented effectively by borrowing the algorithms which are used to find length-$4$ cycles in the parity-check matrix of a low-density parity-check (LDPC) code in \cite{Fossorier04}.
%
Therefore, the increased implementation complexity for the proposed message-level SIC decoding is small.
More importantly, the enhanced message-level SIC decoding algorithm can considerably improve the throughput for the proposed decoding scheme.
In fact, the improvement of canceling NCNs, whose degrees are larger than two, is marginal.
To see this, we compare the throughput of the proposed decoding scheme obtained by canceling NCNs of different degrees in Fig. \ref{fig_improved_sic_diff_degree}.
It can be seen that compared to the conventional message-level SIC process \cite{Lu2013}, the throughput is significantly enhanced by using the enhanced message-level SIC process with canceling the degree-$2$ NCNs, particularly at the high traffic load regime.
This is because the number of degree-$1$ NCNs is smaller for the higher traffic load and the less degree-$1$ NCNs causes the conventional message-level SIC process cannot perform effectively to recover more user packets.
However, there is no noticeable throughput improvement achieved by canceling the degree-3 NCNs, compared to our proposed message-level SIC decoding algorithm.

\subsection{Example}
\begin{figure}[t]
	\begin{center}
		{\includegraphics[width=3in]{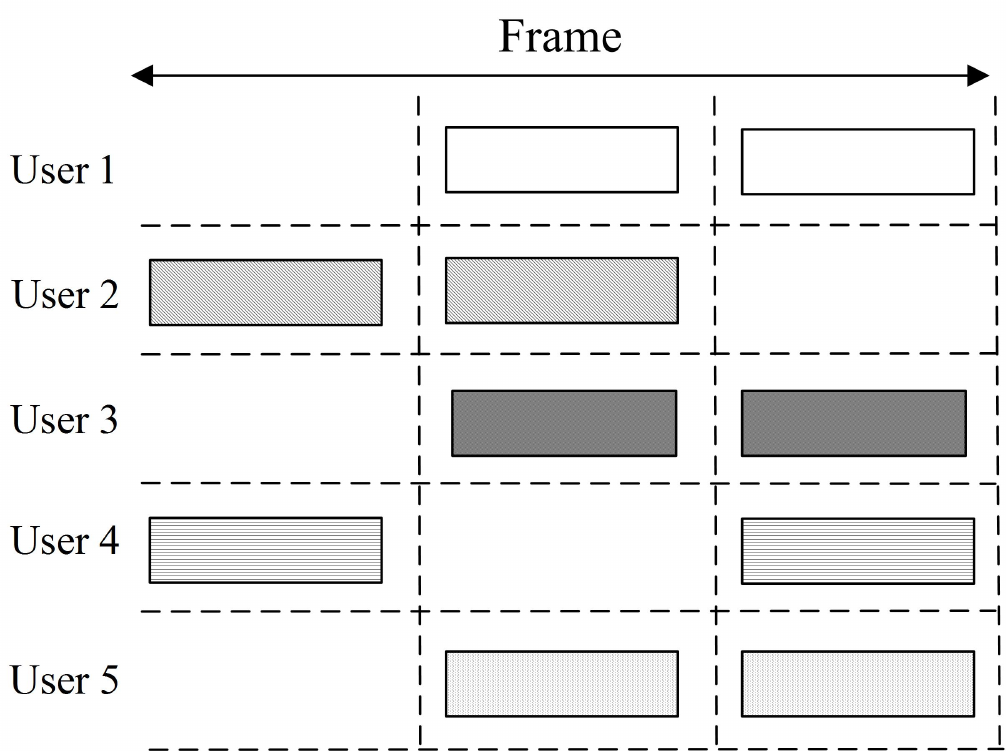}}\vspace{-5mm}
	\end{center}
	\caption{An example of the proposed decoding algorithm with $M=5$ users and $N=3$ time slots.}
	\label{fig_system_model_erasure_v5}\vspace{-5mm}
\end{figure}

In this part, we provide an example to further illustrate the proposed decoding scheme.
Consider a MAC frame with $N=3$ time slots and $M=5$ users.
Each user repeats its packet $u_{m}\in \{0,1\}$ twice ($r=2$), $m\in\{1,2,\ldots,5\}$, and transmits them in two randomly selected time slots.
Assume the packets are transmitted as shown in Fig. \ref{fig_system_model_erasure_v5} and the channel realization of the five users is given by $\mathbf{h}=[1.1809,0.4066,1.0353,0.375,0.782]^{T}$.
%
%Here, we consider the real-valued channel gains for simplicity.

For the first time slot, two users are active and their channel coefficients are $0.4066$ and $0.375$, respectively.
Therefore, the matrix $\mathbf{V}[1] \in \{0,1\}^{2 \times 2}$ can be calculated by using Eqs. \eqref{Eq_def_delta}, \eqref{Eq_def_DV1}, and \eqref{4_rank}, given as
\begin{align}
\mathbf{V}[1]=\left[
\begin{matrix}
1 & 1 \\
1 & 0 \\%
\end{matrix}%
\right].
\label{expl_op_mx}
\end{align}
With $\mathbf{V}[1]$, the optimal NC coefficient matrix ${\bf{G}}^{*}[1]$ can be designed according to Theorem \ref{Theo_optmzed_g}.
The optimal NC coefficient matrix is
\begin{align}
\mathbf{G}^{\ast}[1]\!=\!\left[
\begin{matrix}
0 & 1 \\
1 & 0 %
\end{matrix}%
\right]\!\!.
\end{align}
Then, the two NC messages associated with $\mathbf{G}^{\ast}[1]$ are decoded by using the MAP algorithm in Eq. \eqref{Eq_MAP_Relay}.
Among them, $\widetilde{\mathbf{w}}[1]=1$ is correctly decoded and the associated coefficient vector is $\mathbf{g}_{1}^{\ast}[1]=\left[0,1\right]^{T}$.
The corresponding coefficient vector w.r.t. all users is obtained as $\widetilde{\mathbf{G}}^{\ast}[1]=\left[0,0,0,1,0\right]^{T}$.
For the second and the third time slots, the similar process is conducted and the correctly decoded NC messages are obtained as
\vspace{-4mm}
\begin{equation}
\widetilde{\mathbf{w}}[2]=[1,1,0]^{T}, \widetilde{\mathbf{w}}[3]=[0,1,1]^{T},
\label{expl_nc_seperate}\vspace{-4mm}
\end{equation}
\par
\noindent
respectively. The associated coefficient matrices w.r.t. all users are written as
\vspace{-4.5mm}
\begin{equation}
\widetilde{\mathbf{G}}^{\ast}[2]\!=\!\left[
\begin{matrix}
0 & 0 & 1 \\
1 & 0 & 1 \\
0 & 1 & 0 \\
0 & 0 & 0 \\
1 & 0 & 1 %
\end{matrix}%
\right]\!\!,
\widetilde{\mathbf{G}}^{\ast}[3]\!=\!\left[
\begin{matrix}
1 & 0 & 0 \\
0 & 0 & 0 \\
1 & 1 & 0 \\
0 & 0 & 1 \\
1 & 0 & 0 %
\end{matrix}%
\right].\!\!
\label{expl_op_mx_seperate}\vspace{-4.5mm}
\end{equation}
\par
\noindent
%We first design the NC coefficients and decode the associated NC messages in each time slot by using the proposed NC message decoding in Section III-A.
%%
%For the three time slots, the correctly decoded NC messages are given by
%\begin{align}
%\widetilde{\mathbf{w}}[1]=1, \widetilde{\mathbf{w}}[2]=[1,1,0]^{T}, \widetilde{\mathbf{w}}[3]=[0,1,1]^{T},
%\label{expl_nc_seperate}
%\end{align}
%\par
%\noindent
%respectively. The associated coefficient matrices are written as
%\begin{align}
%\widetilde{\mathbf{G}}^{\ast}[1]\!=\!\left[
%\begin{matrix}
%0 \\
%0 \\
%0 \\
%1 \\
%0%
%\end{matrix}%
%\right]\!\!,
%\widetilde{\mathbf{G}}^{\ast}[2]\!=\!\left[
%\begin{matrix}
%0 & 0 & 1 \\
%1 & 0 & 1 \\
%0 & 1 & 0 \\
%0 & 0 & 0 \\
%1 & 0 & 1 %
%\end{matrix}%
%\right]\!\!,
%\widetilde{\mathbf{G}}^{\ast}[3]\!=\!\left[
%\begin{matrix}
%1 & 0 & 0 \\
%0 & 0 & 0 \\
%1 & 1 & 0 \\
%0 & 0 & 1 \\
%1 & 0 & 0 %
%\end{matrix}%
%\right].\!\!
%\label{expl_op_mx_seperate}
%\end{align}
%\par
%\noindent
%

By collecting all the decoded NC messages and the NC coefficient matrices in the three time slots, the decoded NC messages and the NC coefficient matrix for recovering users' packets are written as $\widetilde{\mathbf{w}}=\left[1,1,1,0,0,1,1\right]^{T}$ and
\vspace{-3mm}
\begin{equation}
\widetilde{\mathbf{G}}^{\ast}=\left[
\begin{matrix}
0 & 0 & 0 & 1 & 1 & 0 & 0 \\
0 & 1 & 0 & 1 & 0 & 0 & 0 \\
0 & 0 & 1 & 0 & 1 & 1 & 0 \\
1 & 0 & 0 & 0 & 0 & 0 & 1 \\
0 & 1 & 0 & 1 & 1 & 0 & 0%
\end{matrix}%
\right],
\label{expl_op_mx}\vspace{-3mm}
\end{equation}
\par
\noindent
respectively.
Therefore, the users' packets are recovered under a constraint
\vspace{-3mm}
\begin{equation}
\left[1,1,1,0,0,1,1\right]^{T}=\left[
\begin{matrix}
0 & 0 & 0 & 1 & 1 & 0 & 0 \\
0 & 1 & 0 & 1 & 0 & 0 & 0 \\
0 & 0 & 1 & 0 & 1 & 1 & 0 \\
1 & 0 & 0 & 0 & 0 & 0 & 1 \\
0 & 1 & 0 & 1 & 1 & 0 & 0%
\end{matrix}%
\right]^{T}\hspace{-2mm}\otimes \mathbf{u},
\label{expl_usr_msg}\vspace{-3mm}
\end{equation}
\par
\noindent
where $\mathbf{u}$ collects the five users' packets.
Eq. \eqref{expl_nc_seperate} shows that seven NC messages are decoded within the three time slots.
It implies that the NC message decoding can effectively extract information from the physical signals, which is essential for the throughput improvement of the proposed decoding scheme.
Based on Eq. \eqref{expl_usr_msg}, the recovering process of users' packets is performed, which is shown in Fig. \ref{fig_frame_based_decoding_3}.
The graph representation for recovering users' packets is presented in Fig. \ref{fig_frame based decoding2a}.
%, where the edge is labeled as '1' if its corresponding message can be recovered.
%%
%Otherwise, the edge is labeled as '0'.
%
It can be seen that there are four degree-$1$ NCNs, i.e., $NC_{1}$, $NC_{3}$, $NC_{6}$, and $NC_{7}$.
Then, the packets of user $3$ and user $4$ are recovered by releasing the four degree-$1$ NCNs.
The recovered packets' replicas are canceled from the corresponding NCNs, i.e., the packet replica of user 3 is canceled from $NC_{5}$, as shown in Fig. \ref{fig_frame based decoding2b}.
%
%In Fig. \ref{fig_frame_based_decoding_3}. b, the recovery process begins by releasing the NC messages with only one associated message, e.g. the NC messages $NC_{1}$ and $NC_{3}$.
%%
%Then, the messages of user 3 and user 4 can be recovered and their replicas are removed from the corresponding NC messages, e.g. the message replica of user 3 is removed from $NC_{5}$.
%
However, this cancellation does not generate new degree-$1$ NCNs and there are no more degree-$1$ NCNs in the Tanner graph.
In this case, the degree-$2$ NCNs that are included in the length-$4$ cycles are exploited.
Here, we exploit $NC_{2}$ and its associated length-$4$ cycle, which is highlighted with the dashed lines in the figure.
By performing the exclusive-or of $NC_{2}$ and $NC_{4}$, the packet combination of user $2$ and user $5$ can be canceled from $NC_{4}$.
In this way, $NC_{4}$ becomes a degree-$1$ NCN, so that the packet of user $1$ is recovered and its replica is canceled from $NC_{5}$, which is shown in Fig. \ref{fig_frame based decoding2c}.
Now, the degree of $NC_{5}$ becomes one and the packet of user $5$ is recovered.
By canceling the packet replicas of user $5$ from $NC_{2}$ and $NC_{4}$, both remained NCNs become degree-$1$ NCNs and the packet of user $2$ is recovered, as shown in Fig. \ref{fig_frame based decoding2d}.
We can see from this example that our proposed message-level SIC decoding algorithm can recover all the users' packets, while the conventional message-level SIC decoding algorithm can only recover two users' packets.
Hence, the proposed scheme can effectively improve the system throughput.

\section{Performance Analysis}
In this section, we analyze the throughput performance for the proposed decoding scheme.
To this end, we first derive the average number of decoded NC messages in a $K$-collision time slot for our proposed NC message decoding.
When $K=2$, we obtain a closed-form expression of the average number of decoded NC messages.
Then, the total number of decoded NC messages in a MAC frame is calculated.
As a user's packet might be included in several NC messages from different time slots and these NC messages might be dependent, this introduces some redundancy among decoded NC messages in different time slots.
Hence, we characterize the redundancy among decoded NC messages and the number of independent NC messages for recovering users' packets.
Then, we derive a tight approximation of throughput for the proposed decoding scheme.
Furthermore, based on the throughput approximation, we optimize the number of replicas to improve the system throughput and energy efficiency.
Note that, in order to facilitate the presentation of our analysis, the one-symbol packet for each user, i.e., $S=1$, is assumed and the corresponding symbol-wise decoding error probability is analyzed in this section.
We emphasize that the analysis can predict the performance of the general case with multi-symbols packet blocks, since the block-wise decoding error probability is determined by the symbol-wise decoding error probability for a given channel code.

\subsection{Number of NC Messages in a $K$-Collision Time Slot}
Consider the proposed NC message decoding, the decoding error probability of an NC message is determined by the effective minimum decoding distance w.r.t. its NC coefficient vector.
Next, we derive the minimum decoding distance for each NC message in a $K$-collision time slot.
Due to the random channel realizations, the minimum decoding distance w.r.t. an NC coefficient vector is random.
Therefore, we first derive its CDF and PDF.

\vspace{-0.5mm}
To start with, we first consider a $2$-collision time slot $n$, where the receiver attempts to decode $L=2$ NC messages.
Recall \textit{Theorem} $1$ and its discussion, the minimum decoding distance for the $l$-th NC message is equal to the $(K-l+1)$-th minimum Euclidean distance $d_{K-l+1}[n]$.
Then, for a $2$-collision time slot, i.e., $K=2$, the minimum decoding distance for the first NC message, i.e., $l=1$, is equal to the second minimum Euclidean distance $d_{2}[n]$.
The minimum decoding distance for the second NC message is equal to the minimum Euclidean distance $d_{1}[n]$.
%
%As the decoding error probability of an NC message is determined by the effective minimum distance w.r.t. its coefficient vector, the decoding error probability of first NC message can be obtained from $d_{2}[n]$.
%%
%However, due to the random access of users to the time slot, $d_{2}[n]$ is a random variable.
%
Thus, the CDF and PDF of the minimum decoding distances for the first and the second NC messages can be obtained by analyzing the CDF and PDF of $d_{2}[n]$ and $d_{1}[n]$, respectively.

\vspace{-0.5mm}
Denote the channel coefficients of two active users in the $2$-collision time slot $n$ as $h_{1}$ and $h_{2}$, respectively.
We assume that $h_{1}$ and $h_{2}$ are two i.i.d. Gaussian variables following the distribution\footnote{For the ease of exposition, we focus on the case of symmetric users with identical channel distribution for the performance analysis. Note that, the proposed performance analysis can be easily generalized to the scenarios with independent but different channel distributions among users.} $g_{h}(x)$, where $g_{h}(x)=\frac{1}{\sqrt{2 \pi}\sigma_{h}}\exp \left(-\frac{x^2}{2\sigma_{h}^2}\right)$.
Due to the symmetry of $g_{h}(x)$, the analysis for the case of $|h_{1}|>|h_{2}|$ is same as that for $|h_{1}|<|h_{2}|$.
Therefore, we only analyze the case of $|h_{1}|>|h_{2}|$ in the following.
When $|h_{1}|>|h_{2}|$, the two smallest Euclidean distance in the constellation of the received signals are $2\sqrt{E_{s}}|h_{2}|$ and $2\sqrt{E_{s}}(|h_{1}| - |h_{2}|)$.
In this case, the probability that the second minimum distance $d_{2}[n]$ is no more than $D$ is
\vspace{-3mm}
\begin{align}
&P\left(d_{2}[n]\leq D \big| |h_{1}|>|h_{2}| \right)  \notag \\
&=P\left(\max\{2|h_{2}|, 2(|h_{1}| - |h_{2}|)\}\leq D \big| |h_{1}|>|h_{2}| \right) \notag \\
%&=P\left(0 < 2|h_{2}|\leq D, 0 < 2(|h_{1}| - |h_{2}|) \leq D\right) \notag \\
&\overset{(a)}{=}4\int_{0}^{\frac{D}{2\sqrt{E_{s}}}}g(h_{2})\int_{h_{2}}^{\frac{D}{2\sqrt{E_{s}}}+h_{2}}g(h_{1})dh_{1}dh_{2}, %\notag \\
%&\overset{(b)}{=}4\int_{0}^{\frac{D}{2\sqrt{E_{s}}}}\frac{1}{\sigma\sqrt{2 \pi}}\exp \left(-\frac{h_{2}^2}{2\sigma^2}\right)\int_{h_{2}}^{\frac{D}{2\sqrt{E_{s}}}+h_{2}}\frac{1}{\sigma\sqrt{2 \pi}}\exp \left(-\frac{h_{1}^2}{2\sigma^2}\right)dh_{1}dh_{2},
\label{example_d2_cdf_case1}
\end{align}
\par
\vspace{-3mm}
\noindent
where the equality (a) is obtained from the fact $g_{h}(|x|)=2g_{h}(x)$.
As Eq. \eqref{example_d2_cdf_case1} is symmetric w.r.t. the two users, the CDF of the second minimum
Euclidean distance $d_{2}[n]$, i.e., the probability that $d_{2}[n]$ is no more than $D$, is written as
\vspace{-3mm}
\begin{align}
&F_{d_{2}}(D)=P\left(d_{2}[n]\leq D \right)=2 P\left(d_{2}[n]\leq D \big| |h_{1}|>|h_{2}| \right) \notag \\
%&=8\int_{0}^{\frac{D}{2\sqrt{E_{s}}}}g(h_{2})\int_{h_{2}}^{\frac{D}{2\sqrt{E_{s}}}+h_{2}}g(h_{1})dh_{1}dh_{2} \notag \\
&\overset{(a)}{=}8\int_{0}^{\frac{D}{2\sqrt{E_{s}}}}\frac{1}{\sigma\sqrt{2 \pi}}\exp \left(-\frac{h_{2}^2}{2\sigma^2}\right)\int_{h_{2}}^{\frac{D}{2\sqrt{E_{s}}}+h_{2}}\frac{1}{\sigma\sqrt{2 \pi}}\exp \left(-\frac{h_{1}^2}{2\sigma^2}\right)dh_{1}dh_{2}.
\label{example_d2_cdf}
\end{align}
\par
\vspace{-6mm}
\noindent

By differentiating the CDF of $d_2[n]$ w.r.t. $D$, the PDF of $d_{2}[n]$ is obtained by
\vspace{-9mm}
\begin{align}
&\hspace{-2.5mm}f_{d_{2}}(D)=\frac{d}{dD}F_{d_{2}}(D) \notag \\
&\hspace{-2.5mm}=\frac{\sqrt{2}}{\sqrt{\pi E_{r}}}\exp\!\left(-\frac{D^2}{8E_{r}}\right)\!\!\left(\mathrm{erf}\!\left(\!\frac{D}{\sqrt{2E_{r}}}\!\right)\!-\!\mathrm{erf}\!\left(\!\frac{D}{2\sqrt{2E_{r}}}\!\right)\right)\notag \\
&\hspace{-2.5mm}-\!\!\frac{1}{\sqrt{\pi E_{r}}}\exp\!\left(-\frac{D^2}{16E_{r}}\right)\!\!\left(\mathrm{erf}\!\left(\!\frac{D}{4\sqrt{E_{r}}}\!\right)\!-\!\mathrm{erf}\!\left(\!\frac{3D}{4\sqrt{E_{r}}}\!\right)\right),
\label{pdf_d2}
\end{align}
\par
\begin{figure}[t]
	\centering
	{ %% label for first subfigure
		\includegraphics[height=2.85in,width=3.45in]{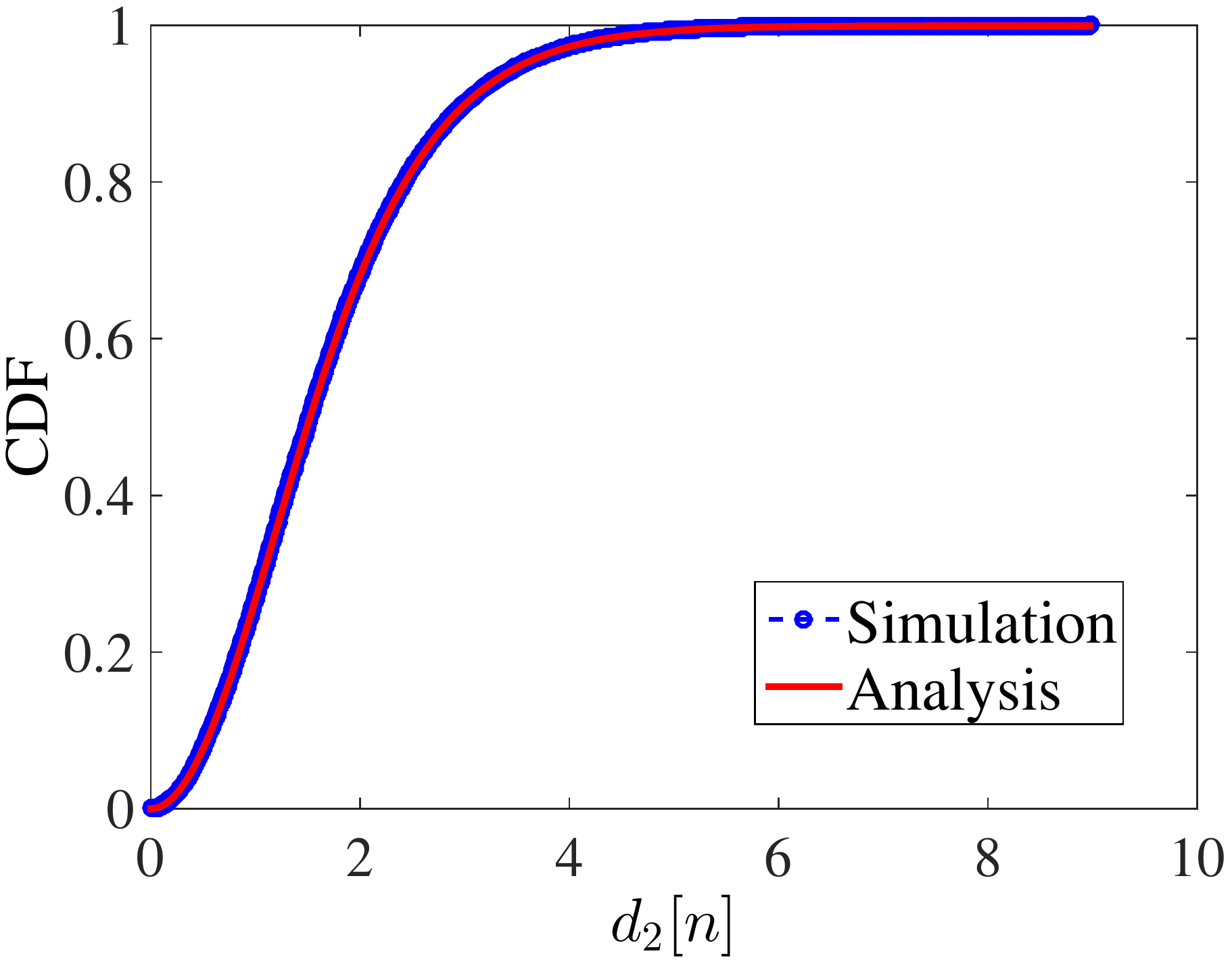}}
	\caption{CDF of ${d_{2}[n]}$ for $K=2$ and $E_{s}=0$ dB.}
	\label{fig_cdf_d2}
\end{figure}

\vspace{-5mm}
\noindent
which is equal to the PDF of the minimum decoding distance for the first NC message $w_{1}$.
Function $\mathrm{erf}(\cdot)$ denotes the error function, and $E_{r}$ denotes the received power per symbol, given by $E_{r}=E_{s}\sigma_{h}^2$.
It can be seen from Eq. \eqref{pdf_d2} that the PDF of $d_{2}[n]$ is associated with the received power per symbol $E_{r}$.
We show the derived CDF and PDF of $d_{2}[n]$ in Fig. \ref{fig_cdf_d2} and Fig. \ref{fig_pdf_d2}, respectively, and compare them with the corresponding Monte Carlo simulation results.
\begin{figure}
	\centering
	{%% label for second subfigure
		\includegraphics[height=2.85in,width=3.45in]{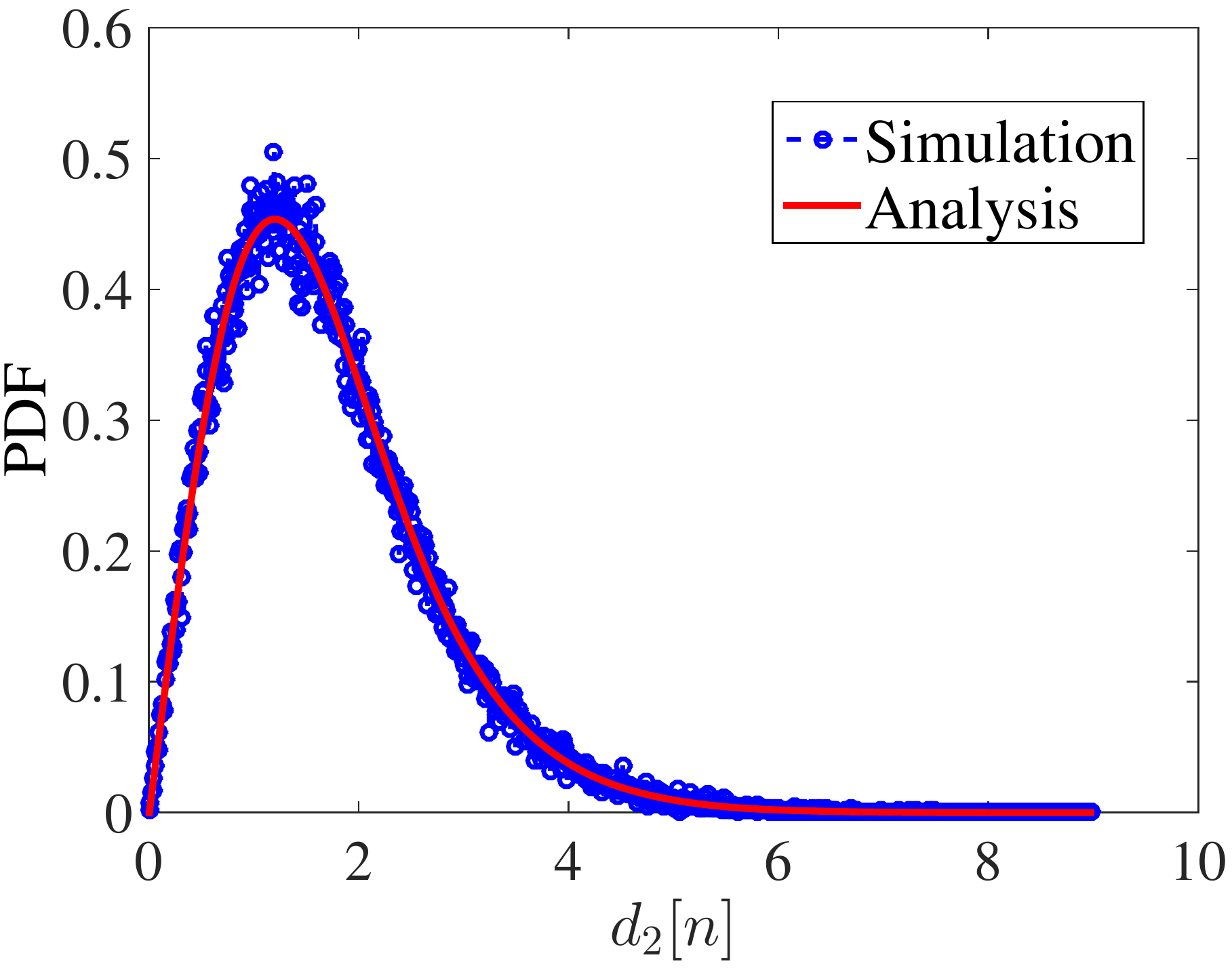}}\vspace{-1mm}
	\caption{PDF of ${d_{2}[n]}$ for $K=2$ and $E_{s}=0$ dB.}
	\label{fig_pdf_d2}
\end{figure}
It can be seen that analytical results well match with simulation results.

%\begin{figure}[!t]
%\centering
%\begin{subfigure}
%{\includegraphics[height=2in,width=2.6in]{Graphes/compare_simu_anal_num_nc_k2_2nd_dis_cdf_complex.pdf}\label{fig_compare_simu_anal_num_nc_k2_2nd_dis_cdf}}
%\end{subfigure}
%\hspace{-3mm}
%\begin{subfigure}
%{\includegraphics[height=2in,width=2.6in]{Graphes/compare_simu_anal_num_nc_k2_2nd_dis_pdf_complex.pdf}\label{fig_compare_simu_anal_num_nc_k2_2nd_dis_pdf}}
%\end{subfigure}
%\caption{The CDF and PDF of $d_{2}[n]$ for $K=2$ and $E_{s}=0 dB$. (a) Simulated and analytical CDF of $d_{2}[n]$. (b) Simulated and analytical PDF of $d_{2}[n]$.}
%\end{figure}

With the PDF of $d_{2}[n]$, i.e., $f_{d_{2}}(D)$, the error probability of the first NC message $w_{1}[n]$ at the medium-to-high SNR regime is given by
\vspace{-3mm}
\begin{align}
&P_{e}(w_{1})\approx \int_{0}^{\infty}Q\left(\frac{D}{2\sigma_{z}}\right)f_{d_{2}}(D)dD \notag \\
&=\frac{1}{2}+\frac{2}{\pi}\left(\mathrm{arctan}\left(\sqrt{\frac{\gamma\sigma_{h}^2}{\gamma \sigma_{h}^2+2}}\right)+\mathrm{arctan}\left(\sqrt{\frac{\gamma\sigma_{h}^2}{\gamma \sigma_{h}^2+1}}\right)\right.\notag \\
&\left.
-\mathrm{arctan}\left(\sqrt{\frac{4\gamma\sigma_{h}^2}{\gamma \sigma_{h}^2+5}}\right)-\mathrm{arctan}\left(\sqrt{\frac{9\gamma\sigma_{h}^2}{\gamma \sigma_{h}^2+5}}\right)\right).
%&=\frac{1}{2}+\frac{2}{\pi}\left(\mathrm{arctan}\left(\frac{\sqrt{\gamma^2\sigma_{h}^4+\gamma\sigma_{h}^2}+\sqrt{\gamma^2\sigma_{h}^4+2\gamma\sigma_{h}^2}}{\sqrt{\gamma^2\sigma_{h}^4+3\gamma\sigma_{h}^2+2}-\gamma\sigma_{h}^2}\right)-\mathrm{arctan}\left(\sqrt{\frac{4\gamma\sigma_{h}^2}{\gamma \sigma_{h}^2+5}}\right)-\mathrm{arctan}\left(\sqrt{\frac{9\gamma\sigma_{h}^2}{\gamma \sigma_{h}^2+5}}\right)\right)
\label{pe_w1}
\end{align}
\par
\vspace{-3mm}
\noindent
From Eq. \eqref{pe_w1}, we can observe that the error probability of $w_{1}[n]$ is a function of the receive SNR, i.e., $\gamma \sigma_{h}^2$.
In addition, it can be verified from the differential of $P_{e}(w_{1})$ w.r.t. $\gamma \sigma_{h}^2$ that $P_{e}(w_{1})$ is a decreasing function of $\gamma \sigma_{h}^2$.
Correspondingly, the probability that $w_{1}[n]$ can be decoded is given by $P_{d}(w_{1})=1-P_{e}(w_{1})$, which implies that the decoding probability $P_{d}(w_{1})$ increases with increasing the receive SNR $\gamma \sigma_{h}^2$.

By the same token, the CDF of the minimum Euclidean distance $d_{1}[n]$ can be written as
\vspace{-3mm}
\begin{align}
&F_{d_{1}}(D)=P\left(d_{1}[n]\leq D\right)=1-2P\left(d_{1}[n]> D \big| |h_{1}|>|h_{2}|\right) \notag \\
&=1\!-\!\frac{8}{2\pi \sigma^2}\int_{\!\frac{D}{2\sqrt{E_{s}}}}^{\infty}\exp \left(-\frac{h_{2}^2}{2\sigma^2}\right)\int_{\frac{D}{2\sqrt{E_{s}}}+h_{2}}^{\infty}\exp \left(-\frac{h_{1}^2}{2\sigma^2}\right)dh_{1}dh_{2}.
\label{cdf_d2}
\end{align}
\par
\vspace{-3mm}
\noindent
%
%For the second NC message $w_{2}$, it corresponds to the second column vector of coefficient matrix $\mathbf{G}^{\ast}[n]$.
%%
%Recall Eq. \eqref{Eq_Theorem2}, the effective minimum distance w.r.t. the second column vector of $\mathbf{G}^{\ast}[n]$ is the minimum distance of received symbol vectors, i.e., $d_{1}$.
%%
%Thus, in order to derive the pair-wise error probability of decoding $w_{2}$, the CDF and PDF of $d_{1}$ should be obtained.
%%
%%
%Given the fading coefficients of two transmitting users in a $K=2$ time slot, i.e., the two i.i.d. Rayleigh variables $h_{1}$ and $h_{2}$, the CDF of $d_{1}$ can be given by
%\begin{align}
%F_{d_{1}}(D)=&P(d_{1}[n]\leq D) \notag \\
%=&1-P(h_{1}> {\frac{D}{2\sqrt{E_{s}}}})P(h_{2}-h_{1} > {\frac{D}{2\sqrt{E_{s}}}})-P(h_{2}> {\frac{D}{2\sqrt{E_{s}}}})P(h_{1}-h_{2} > {\frac{D}{2\sqrt{E_{s}}}}) \notag \\
%=&1-2\int_{{\frac{D}{2\sqrt{E_{s}}}}}^{\infty}g(h_{2})\int_{{\frac{D}{2\sqrt{E_{s}}}}+h_{2}}^{\infty}g(h_{1})dh_{1}dh_{2}.
%\label{cdf_d1}
%\end{align}
%The minimum Euclidean distance $d_{1}[n]$ can be two times one of the two users' received energy or two times their difference, i.e., $2\sqrt{E_{s}}h_{1}$ or $2\sqrt{E_{s}}h_{2}$ or $2\sqrt{E_{s}}(h_{1}-h_{2})$ ($2\sqrt{E_{s}}(h_{2}-h_{1})$), where it is assumed the energy of transmitted symbol is one.
%
%Then, the complementary CDF of $d_{1}[n]$, i.e., $1-F_{d_{1}}(D)$, is obtained when both $2\sqrt{E_{s}}h_{1}$ ($2\sqrt{E_{s}}h_{2}$) and $2\sqrt{E_{s}}(h_{2}-h_{1})$ ($2\sqrt{E_{s}}(h_{1}-h_{2})$) are larger than $D$, which correspond to the two parts in Eq. \eqref{cdf_d1}, respectively.
%
By differentiating the CDF of $d_{1}[n]$ w.r.t. $D$, the PDF of $d_{1}[n]$ is given by
\vspace{-3mm}
\begin{align}
&f_{d_{1}}(D)=\frac{d}{dD}F_{d_{1}}(D) \notag \\
&=\frac{\exp\!\left(\frac{-D^2}{16E_{r}}\right)}{\sqrt{\pi E_{r}}}\!\!\left[1\!-\!\mathrm{erf}\left(\frac{3D}{4\sqrt{E_{r}}}\right)\!\!+\!\!\sqrt{2}\exp\!\left(\frac{-D^2}{16E_{r}}\right)\!\!\left(1-\mathrm{erf}\left(\frac{D}{\sqrt{2E_{r}}}\right)\right)\right],
\label{pdf_d1}
\end{align}
\par
\begin{figure}[t]
	\centering
	{ %% label for first subfigure
		\includegraphics[height=2.85in,width=3.45in]{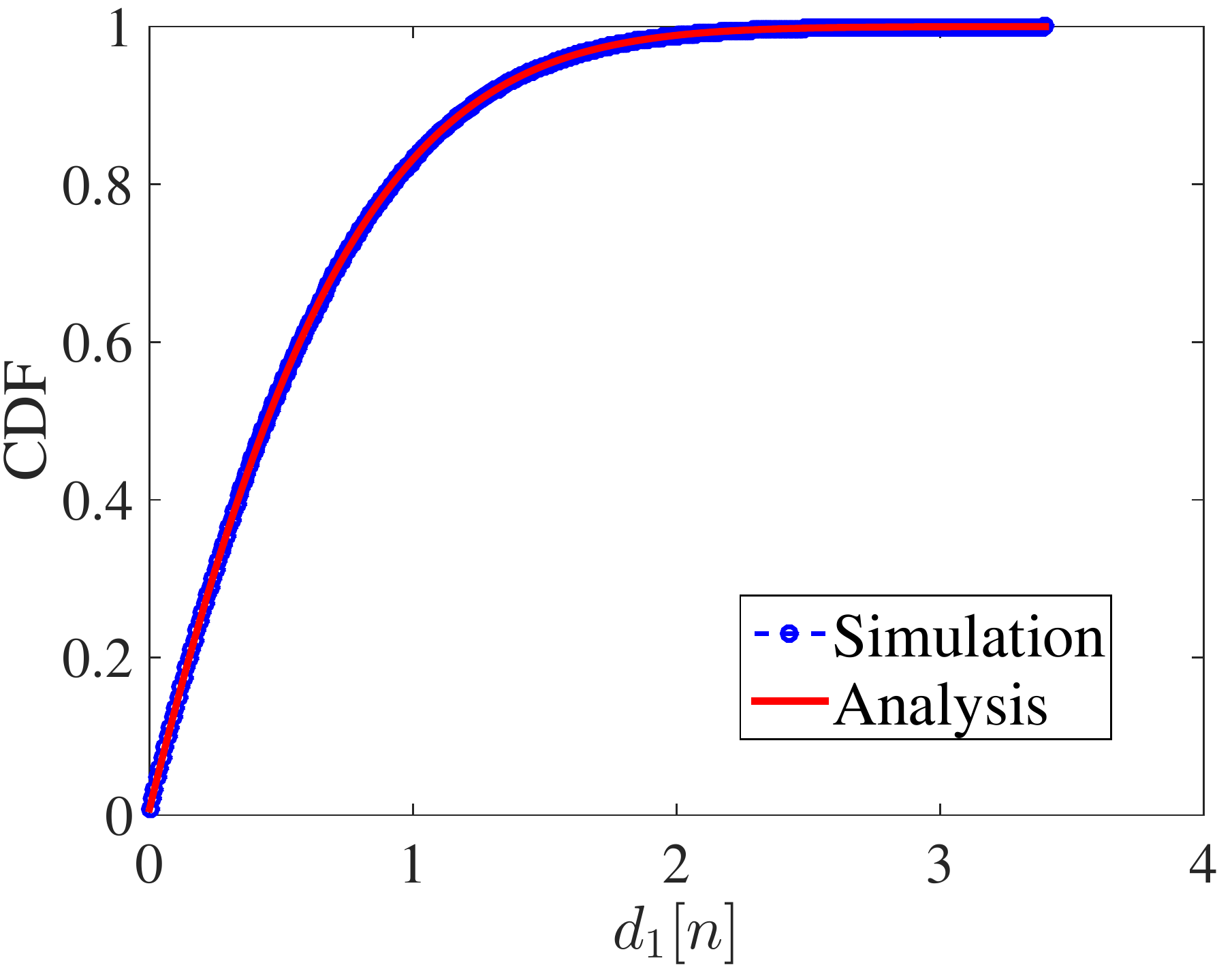}}
\caption{CDF of ${d_{1}[n]}$ for $K=2$ and $E_{s}=0$ dB.}
\label{fig_cdf_d1}
\end{figure}
\begin{figure}
\centering
	{%% label for second subfigure
		\includegraphics[height=2.85in,width=3.45in]{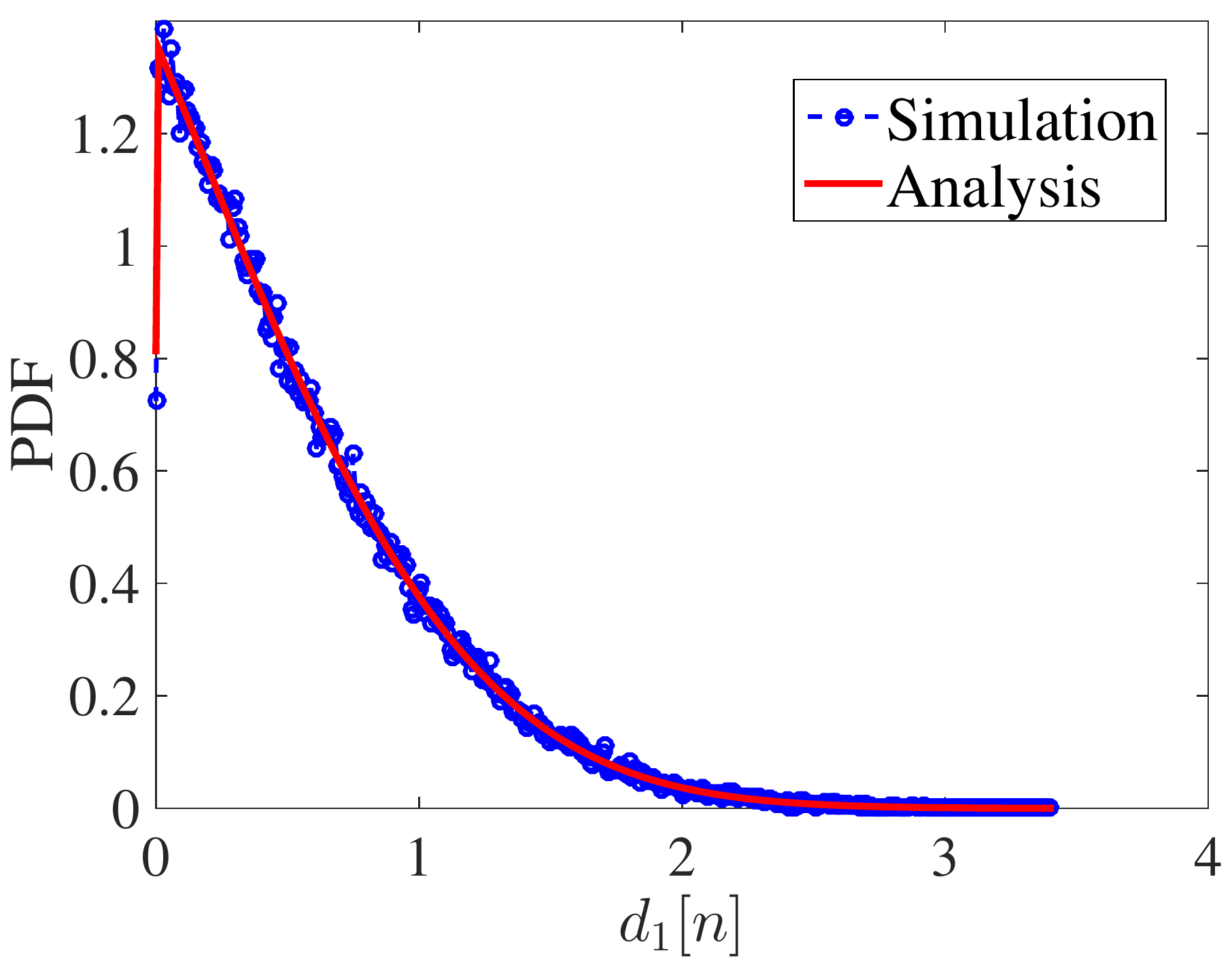}}\vspace{-1mm}
	\caption{PDF of ${d_{1}[n]}$ for $K=2$ and $E_{s}=0$ dB.}\vspace{-2mm}
	\label{fig_pdf_d1}
\end{figure}

\vspace{-3mm}
\noindent
which is equal to the PDF of the minimum decoding distance for the second NC message $w_{2}$.
Similar to $d_{2}[n]$, the PDF of $d_{1}[n]$ also depends on the received symbol energy $E_{r}$, as shown in Eq. \eqref{pdf_d1}.
The derived CDF and PDF of $d_{1}[n]$ are compared with the corresponding Monte Carlo simulation results in Fig. \ref{fig_cdf_d1} and Fig. \ref{fig_pdf_d1}, respectively.
Again, the analytical results match perfectly with the simulation results.


%\begin{figure}[t]
%	\centering
%	\subfigure[CDF of ${d_{1}[n]}$.]
%	{\label{fig_cdf_d1} %% label for first subfigure
%		\includegraphics[width=1.67in]{Graphes/compare_simu_anal_num_nc_k2_1st_dis_cdf_complex.pdf}}\vspace{-1mm}
%	\subfigure[PDF of ${d_{1}[n]}$.]
%	{\label{fig_pdf_d1} %% label for second subfigure
%		\includegraphics[width=1.67in]{Graphes/compare_simu_anal_num_nc_k2_1st_dis_pdf_complex.pdf}}\vspace{-1mm}
%	\caption{The CDF and PDF of ${d_{1}[n]}$ for $K=2$ and $E_{s}=0$ dB.}
%	\label{cdf_pdf_d1}\vspace{-5mm}
%\end{figure}

Based on the PDF of $d_{1}[n]$, i.e., $f_{d_{1}}(D)$, the error probability of the second NC message $w_{2}[n]$ at the medium-to-high SNR regime is given by
\begin{align}
&P_{e}(w_{2})\approx \int_{0}^{\infty}Q\left(\frac{D}{2\sigma_{z}}\right)f_{d_{1}}(D)dD \notag \\
&=-\frac{3}{2}+\frac{2}{\pi}\left(\mathrm{arctan}\left(\sqrt{\frac{1}{2\gamma \sigma_{h}^2}}\right)+\mathrm{arctan}\left(\sqrt{\frac{1}{\gamma \sigma_{h}^2}}\right)\right.\notag \\
&\left.
+\mathrm{arctan}\left(\sqrt{\frac{4\gamma\sigma_{h}^2}{\gamma \sigma_{h}^2+5}}\right)+\mathrm{arctan}\left(\sqrt{\frac{9\gamma\sigma_{h}^2}{\gamma \sigma_{h}^2+5}}\right)\right).
\label{pe_w2}
\end{align}
\par
\noindent
We can observe from Eq. \eqref{pe_w2} that the error probability of $w_{2}[n]$ only depends on the receive SNR, i.e., $\gamma \sigma_{h}^2$ and it can be verified from the differential of $P_{e}(w_{2})$ w.r.t. $\gamma \sigma_{h}^2$ that $P_{e}(w_{2})$ is a decreasing function of $\gamma \sigma_{h}^2$.
The probability that $w_{2}[n]$ can be decoded is $P_{d}(w_{2})=1-P_{e}(w_{2})$.

%\begin{figure}[t]
%\footnotesize
%\centering
%\vspace*{-6mm}
%\begin{minipage}{.47\textwidth}
%\centering
%\includegraphics[width=\textwidth]{Graphes/compare_simu_anal_num_nc_k2_1st_dis_cdf_complex.pdf}\vspace*{-3mm}
%a) Simulated and analytical CDF of $d_{1}[n]$.
%\label{cdf_d1}
%\end{minipage}
%\hspace*{1.5mm}
%\begin{minipage}{.47\textwidth}
%\centering
%\includegraphics[width=\textwidth]{Graphes/compare_simu_anal_num_nc_k2_1st_dis_pdf_complex.pdf}\vspace*{-3mm}
%b) Simulated and analytical PDF of $d_{1}[n]$.
%\label{pdf_d1}
%\end{minipage}
%\vspace*{-2mm}
%\caption{The CDF and PDF of $d_{1}[n]$ for $K=2$ and $E_{s}=0 dB$.}\vspace*{-10mm}
%\label{cdf_pdf_d1}%
%\end{figure}

%\begin{figure}[!t]
%\centering
%\begin{subfigure}
%{\includegraphics[height=2in,width=2.6in]{Graphes/compare_simu_anal_num_nc_k2_1st_dis_cdf_complex.pdf}\label{fig_compare_simu_anal_num_nc_k2_1st_dis_cdf}}
%\end{subfigure}
%\hspace{-3mm}
%\begin{subfigure}
%{\includegraphics[height=2in,width=2.6in]{Graphes/compare_simu_anal_num_nc_k2_1st_dis_pdf_complex.pdf}\label{fig_compare_simu_anal_num_nc_k2_1st_dis_pdf}}
%\end{subfigure}
%\caption{The CDF and PDF of $d_{1}[n]$ for $K=2$ and $E_{s}=0 dB$. (a) Simulated and analytical CDF of $d_{1}[n]$. (b) Simulated and analytical PDF of $d_{1}[n]$.}
%\end{figure}
With $P_{d}(w_{1})$ and $P_{d}(w_{2})$, we have the closed-form expression of the average number of decoded NC messages for a $2$-collision time slot, which is written as
%Based on the probability that the NC message $w_{l}[n]$ can be correctly decoded for $l=1,2$, the number of decoded NC messages for the $2$-collision time slot can be obtained by
\vspace{-5mm}
\begin{align}
\eta_{2}=P_{d}(w_{1})+P_{d}(w_{2})=2-P_{e}(w_{1})-P_{e}(w_{2}).
\label{num_nc_k2}
\end{align}
\par
\vspace{-6mm}
\noindent
%where $P_{e}(w_{1})$ and $P_{e}(w_{2})$ are the decoding error probability of NC message $w_{1}[n]$ and $w_{2}[n]$, given by Eq. \eqref{pe_w1} and \eqref{pe_w2}, respectively.

%
%It is noteworthy that since the PDF of $d_{2}[n]$ and $d_{1}[n]$, i.e., $f_{d_{2}}(D)$ and $f_{d_{1}}(D)$, decrease with increasing $E_{s}$ and $Q\left(\frac{D}{2\sigma}\right)$ decreases with $\sigma$, both $P_{e}(w_{1})$ and $P_{e}(w_{2})$ decrease with increasing $\mathrm{SNR}$, defined as $\rho=\frac{E_{s}}{\sigma^2}$.
%%
%Then, according to Eq. \eqref{num_nc_k2}, $\eta_{2}$ increases with $\rho$.
%%
%When $\rho$ becomes enough high, $\eta_{2}$ may approach two.
%%
%It implies that for the high $\mathrm{SNR}$, it is a high probability that both the two NC messages can be correctly decoded in the $2$-collision time slot.
%%
%As the two decoded NC messages are linearly independent, the two users' messages can be successfully recovered with a high probability.

Now, we consider the general collision size $K$.
Similar to the case of $K=2$, we need to derive the PDF of minimum decoding distance for each NC message.
Then, the decoding error probability of each NC message is obtained and the average number of decoded NC messages in a time slot can be estimated.
However, obtaining the PDF of minimum decoding distance for each NC message is a tedious task for general $K$, if it is not impossible.
Instead, it will be obtained numerically to reveal the system performance.
Denote the PDF of minimum decoding distance for the $l$-th NC message $w_{l}[n]$ by $f_{d_{K-l+1}}(D)$, $l\in\{1,2,\ldots,L\}$, and the error probability of $w_{l}[n]$ is given by
\vspace{-4mm}
\begin{align}
P_{e}(w_{l})\approx \int_{0}^{\infty}Q\left(\frac{D}{2\sigma_{z}}\right)f_{d_{K-l+1}}(D)dD.
\label{pe_wk}
\end{align}
\par
\vspace{-5mm}
\noindent
%where $l=1,2,\cdots,K$.
%
The probability that $w_{l}[n]$ can be decoded is $P_{d}(w_{l})=1-P_{e}(w_{l})$. Therefore, the average number of decoded NC messages in a $K$-collision time slot is given by
\vspace{-5mm}
\begin{align}
\eta_{K}=\sum_{l=1}^{L}P_{d}(w_{l})=L-\sum_{l=1}^{L}P_{e}(w_{l}).
\label{num_nc_gene_k}
\end{align}
\par
\begin{figure}[t]\vspace{-1mm}
	\par
	\begin{center}
		{\includegraphics[height=2.85in,width=3.45in]{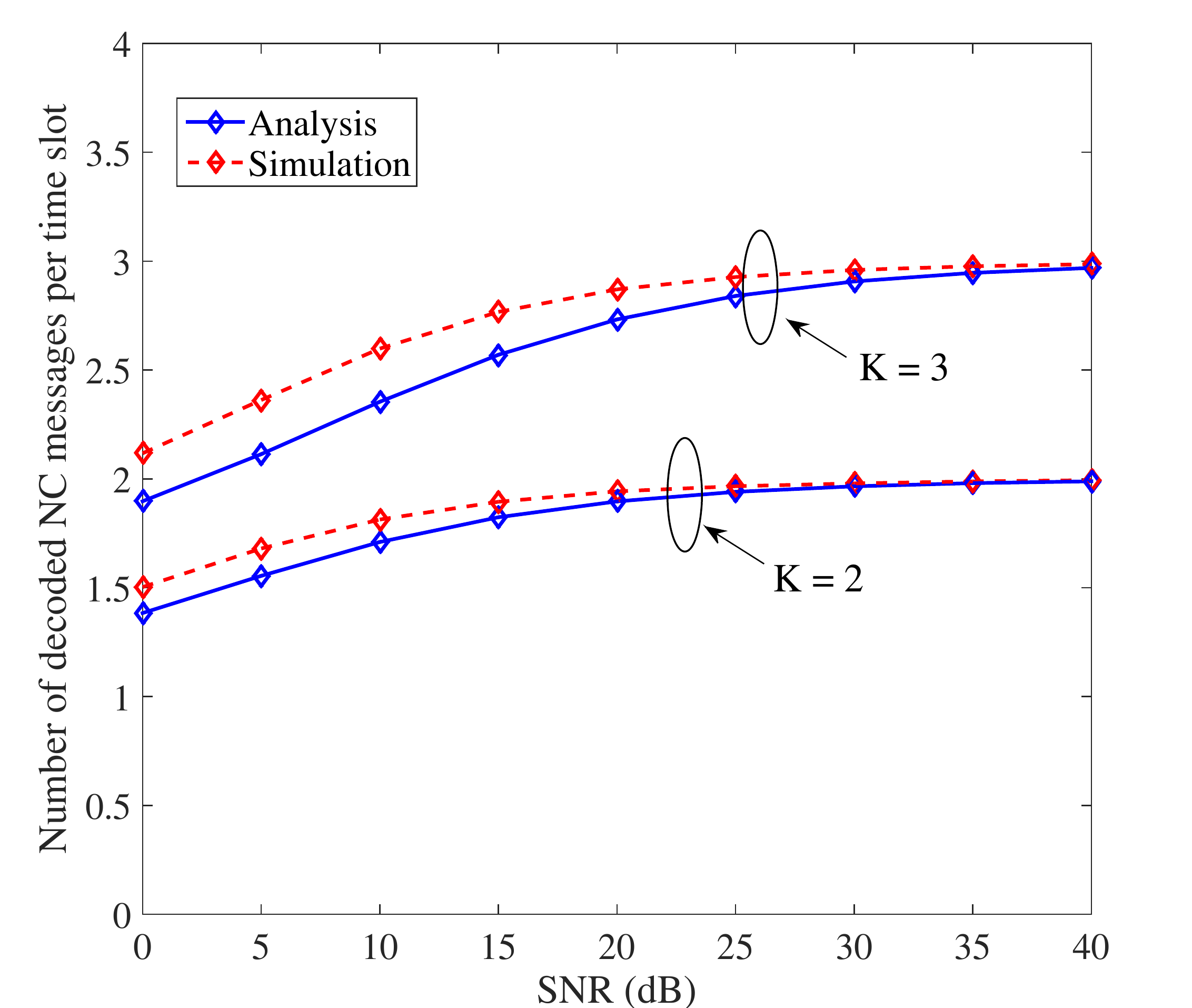}}
	\end{center}\vspace{-5mm}
	\caption{The simulated and analytical number of obtained NC messages per time slot for $M=150$ users.}
	\label{fig_compare_k23_num_nc_pslot_anal_simu_v2}\vspace{-5mm}
\end{figure}

\vspace{-5mm}
\noindent
%
%For the general $K$-collision time slot, it is also a high probability that all the $K$ NC messages can be correctly decoded at the high $\mathrm{SNR}$ region, which is similar to the case of $K=2$.
%%
%As the $K$ decoded NC messages are linearly independent, all the $K$ users' messages in the time slot can be recovered with a high probability.
The derived number of decoded NC messages in a time slot is compared with the Monte Carlo simulation result in Fig. \ref{fig_compare_k23_num_nc_pslot_anal_simu_v2}, where the collision sizes with $K=2$ and $K=3$ are considered.
%
%Both Monte Carlo simulation results and analysis results are shown there.
%%
It can be seen that the derived analytical results coincide with the simulation results well, particularly at medium-to-high SNRs.
In fact, the adopted pairwise error probability in Eq. \eqref{pe_wk} offers a more accurate approximation of the NC messages' error probability at the medium-to-high SNR regime, compared to the case at the low SNR regime.
%the proposed slot based decoding can decode multiple NC messages in each time slot on average.
%%
%At a medium-to-high SNR regime, the proposed decoding scheme can almost decode $K$ NC messages in a $K$-collision time slot.
%%
%This means the proposed slot based decoding can effectively extract the information from the physical signals.
%%
%In addition,

%
Note that for a large collision size, it is difficult to extract information from the physical signal, i.e., it is hard to decode an NC message.
In addition, the decoding complexity is high in such a time slot.
By considering these, we assume that only the time slot with a collision size no greater than $K_\mathrm{max}$ is decoded\footnote{The value of $K_\mathrm{max}$ is generally determined by the practical systems, e.g. the decoding complexity and system performance requirement.}.
For a time slot with $K > K_\mathrm{max}$, $\eta_{K}=0$.

\vspace{-3mm}

\subsection{Total Number of NC Messages in a Frame}
Having the average number of decoded NC messages in a $K$-collision time slot, we proceed to obtain the total number of decoded NC messages in a MAC frame.
Consider $M$ users transmitting packets in a MAC frame with $N$ time slots.
Each user repeats its packet $r$ times.
Then, the probability that $K$ users transmit packets simultaneously in a time slot, i.e., the time slot is a $K$-collision time slot, is given by
\vspace{-1mm}
\begin{eqnarray}
\Psi_{K}=\dbinom{M}{K}\left(\frac{r}{N}\right)^{K}\left(1-\frac{r}{N}\right)^{M-K}.
\label{node_dis}
\end{eqnarray}
\par
\vspace{-1mm}
\noindent
%where $K=1,\ldots,M$.
%
%Thus, $\{\Psi_{K}\}_{K=1}^{M}$ is obtained as the probability distribution that a time slot has the different collision sizes.
%
Recall Eq. \eqref{num_nc_gene_k}, $\eta_{K}$ is the average number of decoded NC messages in a $K$-collision time slot for $K\in\{1,\ldots,K_\mathrm{max}\}$.
By averaging $\eta_{K}$ over $\Psi_{K}$, the average number of decoded NC messages in a time slot is written as
\vspace{-1mm}
\begin{eqnarray}
N_{s}=\sum_{K=1}^{K_\mathrm{max}}\eta_{K}\Psi_{K}.
\label{num_s}
\end{eqnarray}
\par
\vspace{-1mm}
\noindent
Then, the total number of decoded NC messages in a MAC frame with $N$ time slots is
\vspace{-1mm}
\begin{eqnarray}
N_\mathrm{f}=N N_{s}=N\sum_{K=1}^{K_\mathrm{max}}\eta_{K}\dbinom{M}{K}\left(\frac{r}{N}\right)^{K}\left(1-\frac{r}{N}\right)^{M-K}\hspace{-2mm}.
\label{num_nc}
\end{eqnarray}
\par
\vspace{-1mm}
\noindent

In Eq. \eqref{num_nc}, $N_\mathrm{f}$ is obtained by summing up the numbers of decoded NC messages in all time slots within a MAC frame.
This does not consider the correlation or redundancy between NC messages from multiple time slots.
%
%Though the proposed slot based decoding scheme assures that the decoded NC messages in a time slot are independent, the decoded NC messages in different time slots might be dependent.
%%
%This is because each user's message is repeated by $r$ times in a MAC frame, which introduces the correlation of NC messages from multiple time slots.
%
Next, we characterize this redundancy to obtain the number of \emph{independent} NC messages in a MAC frame.
It will facilitate the analysis of throughput for the proposed decoding scheme.

\subsection{Throughput of the Proposed Scheme}
Consider one replica of a particular user's packet.
The replica is received in a $K$-collision time slot with a probability
\vspace{-5mm}
\begin{align}
\Psi_{K|1}=\dbinom{M-1}{K-1}\left(\frac{r}{N}\right)^{K-1}\left(1-\frac{r}{N}\right)^{M-K}.
\end{align}
\par
\vspace{-1mm}
\noindent
%
%where $K=1,\ldots,M$.
%
%Then, the size probability distribution of collisions where a particular user's message is received is denoted by $\{\Psi_{K|1}\}_{K=1}^{M}$.
%
Recall the NC message decoding, for a $K$-collision time slot the average number of decoded NC messages is $\eta_{K}$ and the average decoding probability of an NC message is $p_{K}=\frac{\eta_{K}}{K}$.
As all the decoded NC messages in a time slot are linearly independent, their NC coefficient matrix has rank $\eta_{K}$.
This means that for a $K$-collision time slot, the average number of users' packets that can be recovered is upper bounded by $\eta_{K}$ and the average recovery probability of a user's packet is upper bounded by $p_{K}$.
By averaging $p_{K}$ over $\Psi_{K|1}$, the upper bound on the average recovery probability of a packet in a time slot is given by
\vspace{-1.5mm}
\begin{eqnarray}
\chi_{1}=\sum_{K=1}^{K_\mathrm{max}}p_{K}\Psi_{K|1}=\sum_{K=1}^{K_\mathrm{max}}p_{K}\dbinom{M-1}{K-1}\left(\frac{r}{N}\right)^{K-1}\!\!\left(1-\frac{r}{N}\right)^{M-K}\hspace{-2mm}.
\label{prb_dcd_nc}
\end{eqnarray}
\par
\vspace{-1.5mm}
\noindent
To facilitate the explanation of our derivation for the number of redundant NC messages, we will adopt the Tanner graph representation, whereby the cause of redundant NC messages is explained and the number of redundant NC messages is derived.
In the Tanner subgraph that corresponds to a time slot, $\chi_{1}$ is the upper bound on the probability that a UN is connected to a degree-1 NCN.
When the UN is connected to more than one degree-$1$ NCN from different time slots within a MAC frame, these degree-$1$ NCNs are redundant for the UN.
For example, both the degree-$1$ $NC_{3}$ and degree-$1$ $NC_{6}$ are connected to $U_{3}$ in Fig. \ref{fig_frame_based_decoding_3}, then either degree-$1$ $NC_{3}$ or degree-$1$ $NC_{6}$ is redundant.
If a UN is connected to $l$ degree-$1$ NCNs in the full Tanner graph associated with the MAC frame for $l\in\{2,\ldots,r\}$, then it results in $(l-1)$ redundant NCNs.
Then the number of redundant NCNs for $M$ UNs is upper bounded by
\vspace{-2mm}
\begin{align}
\varphi_{1}=M \sum_{l=2}^{r}(l-1)\dbinom{r}{l}\chi_{1}^{l}(1-\chi_{1})^{r-l}.
\label{num_rd_s}
\end{align}
\par
\vspace{-2mm}
\noindent
Note that $\varphi_{1}$ is an upper bound on the number of redundant NCNs for $M$ UNs. This is because $\varphi_{1}$ depends on $\chi_{1}$ and $\chi_{1}$ is an upper bound, as shown in Eq. \eqref{num_rd_s}.
In particular, apart from the probability that a UN is connected to a degree-$1$ NCN, i.e., the probability of a packet being recovered, $\chi_{1}$ includes the probability of a UN being connected to a degree-$2$ or higher degree NCN.
If a UN is connected a degree-$2$ or higher-degree NCN, the NCN may not be redundant for the UN.
However, when deriving $\varphi_{1}$, such an NCN is considered as a redundant NCN by its associated UNs.
For example, a degree-2 NCN is considered as a redundant NCN by its two associated UNs, and it contributes to the number of redundant NCNs twice.
%
%If the connected NCN has degree-$1$, it is redundant for the UN.
%%
%Otherwise, the NCN is not redundant for any its associated UNs.
%%
%However, when deriving $\varphi_{1}$, a connected NCN is considered to be redundant by the UN and all its associated UNs.
%%
%For example, a degree-2 NCN is considered as a redundant NCN by its two associated UNs, and it contributes to the number of redundant NCNs $\varphi_{1}$ twice.
%
%\color{blue}the probability $p_{K}$ includes the probability of a UN being connected to a degree-$2$ or higher degree NCN, apart from the probability that the UN is connected to a degree-$1$ NCN.
%%
%\color{black}If a UN is connected a degree-$2$ or higher degree NCN, this NCN may not be redundant for the UN.
%%
%However, when deriving $\varphi_{1}$, such an NCN is considered as a redundant NCN by all its associated UNs.
%%
%For example, a degree-2 NCN is considered as a redundant NCN by its two associated UNs, and it contributes to the number of redundant NCNs $\varphi_{1}$ twice.
%
%However, when a UN is connected to a NCN whose degree is more than one, the NCN may not be redundant for this UN and the gap of upper bound $\varphi_{1}$ is generated.
%%
%Moreover, this NCN is considered as the redundant NCN by all its associated users, when calculating $\varphi_{1}$.
%
%It implies that one NCN is counted as a redundant NCN multiple times.
%
%This is the reason why $\varphi_{1}$ overestimates the number of redundant NCNs for $M$ UNs.
%
Thus, we need to analyze the number of degree-$2$ or higher degree NCNs to compensate the overestimation effect.

We first consider the degree-$2$ NCNs.
The probability of obtaining an NCN in a $K$-collision time slot is $p_{K}$.
As the decoded NCN has a degree of two with probability ${\dbinom{K}{2}}/{(2^{K}-1)}$, where each degree is assumed to be chosen uniformly by an NCN, the probability of obtaining a degree-$2$ NCN in a $K$-collision time slot is given by
\vspace{-1.5mm}
\begin{align}
p_{K|2}=\dbinom{K}{2}\frac{p_{K}}{2^{K}-1}.
\end{align}
\par
\vspace{-1.5mm}
\noindent
For the two UNs connected to a degree-$2$ NCN, the corresponding two users' packets are received in a $K$-collision time slot with a probability
\vspace{-1mm}
\begin{eqnarray}
\Psi_{K|2}=\dbinom{M-2}{K-2}\left(\frac{r}{N}\right)^{K-2}\left(1-\frac{r}{N}\right)^{M-K}.
\label{prb_k_general}
\end{eqnarray}
\par
\vspace{-1mm}
\noindent
%where $2 \leq K \leq M$.
%
By averaging $p_{K|2}$ over $\Psi_{K|2}$, the probability of obtaining a degree-$2$ NCN in a time slot is written as
\vspace{-1mm}
\begin{eqnarray}
\chi_{2}=\sum_{K=2}^{K_\mathrm{max}}p_{K|2}\Psi_{K|2}=\sum_{K=2}^{K_\mathrm{max}}p_{K|2}\dbinom{M-2}{K-2}\left(\frac{r}{N}\right)^{K-2}\!\!\left(1-\frac{r}{N}\right)^{M-K}\hspace{-2mm}.
\label{prb_dcd_nc_2}
\end{eqnarray}
\par
\vspace{-1mm}
\noindent
As there are $\dbinom{M}{2}\left(\frac{r}{N}\right)^2$ possible degree-2 NCNs, the number of degree-2 NCNs can be estimated by
\vspace{-1mm}
\begin{eqnarray}
\varphi_{2}=\dbinom{M}{2}\left(\frac{r}{N}\right)^2\chi_{2}.
\label{num_dcd_nc_2}
\end{eqnarray}
\par
\vspace{-1mm}
\noindent
Each degree-$2$ NCN contributes to the number of redundant NCNs twice when deriving $\varphi_{1}$.
Thus, the effect of degree-$2$ NCNs on the overestimation of $\varphi_{1}$ can be compensated by $\varphi_{1}-\varphi_{2}$.

Similar to the case of degree-$2$ NCNs, the effect of degree-$i$ NCNs on the overestimation of $\varphi_{1}$ for $i\in\{3,4,\ldots,K_\mathrm{max}\}$ can be compensated by estimating the number of degree-$i$ NCNs.
In particular, the probability of obtaining a degree-$i$ NCN in a $K$-collision time slot is given by
\vspace{-1mm}
\begin{align}
p_{K|i}=\dbinom{K}{i}\frac{p_{K}}{2^{K}-1},
\end{align}
\par
\vspace{-1mm}
\noindent
where $3 \leq i \leq K_\mathrm{max}$ and $i \leq K \leq K_\mathrm{max}$.
\begin{figure}[t]\vspace{-5mm}
	\par
	\begin{center}
		{\includegraphics[height=2.85in,width=3.45in]{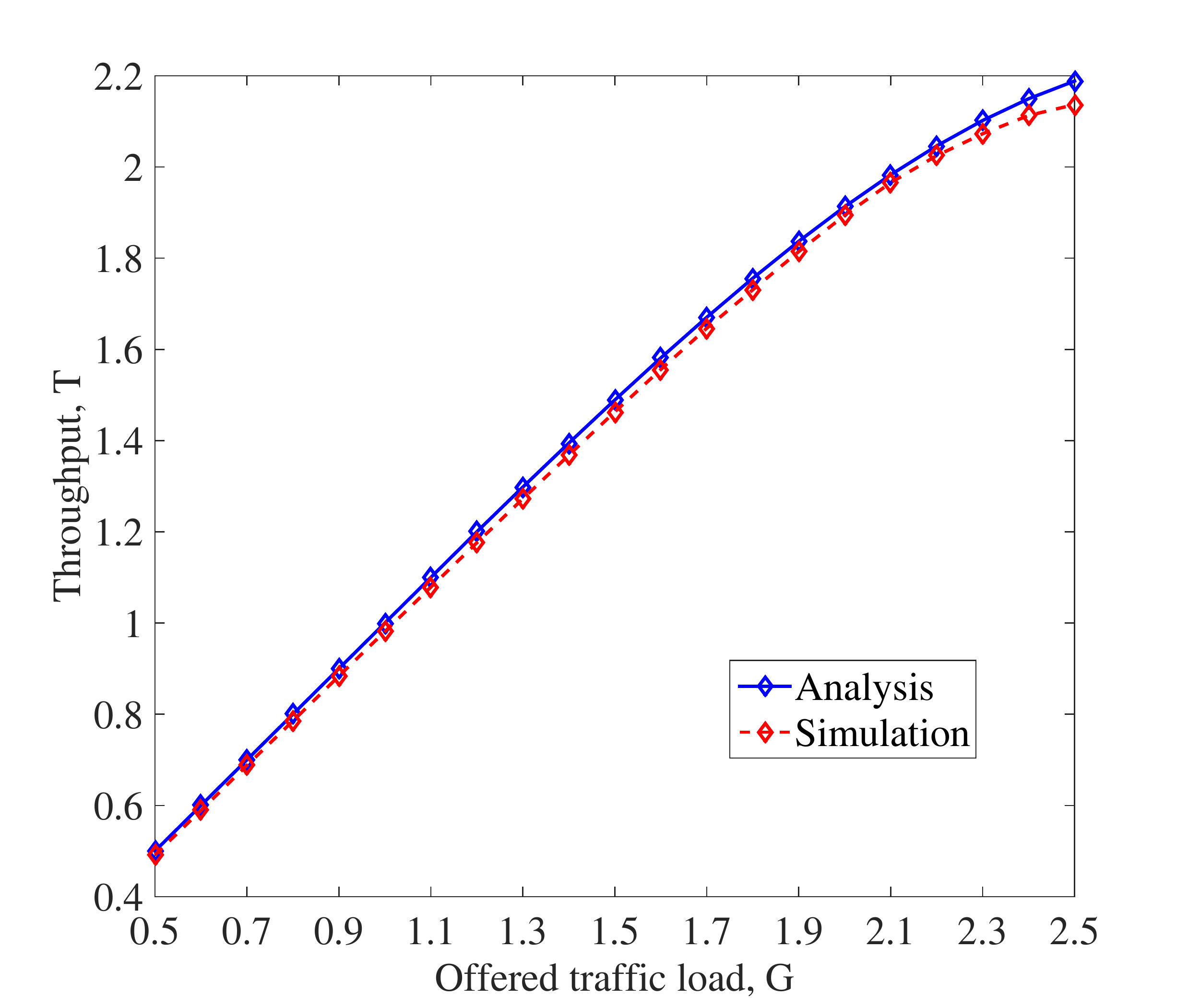}}
	\end{center}\vspace{-5mm}
	\caption{The simulated and analytical throughput for SNR $=15$ dB, $N=100$, $r=2$, and $K_\mathrm{max}=7$.}
	\label{fig_thpt_real_anal_simu_2}
\end{figure}
For the $i$ UNs connected to a degree-$i$ NCNs, the corresponding $i$ users' packets are received in a $K$-collision time slot with a probability
\vspace{-1mm}
\begin{equation}
\Psi_{K|i}=\dbinom{M-i}{K-i}\left(\frac{r}{N}\right)^{K-i}\left(1-\frac{r}{N}\right)^{M-K}.
\label{prb_k_general}
\end{equation}
By averaging $p_{K|i}$ over $\Psi_{K|i}$, the probability of obtaining a degree-$i$ NCN in a time slot is
\vspace{-1mm}
\begin{eqnarray}
\chi_{i}=\sum_{K=i}^{K_\mathrm{max}}p_{K|i}\Psi_{K|i}=\hspace{-2mm}\sum_{K=i}^{K_\mathrm{max}}p_{K|i}\dbinom{M-i}{K-i}\left(\frac{r}{N}\right)^{K-i}\!\!\left(1-\frac{r}{N}\right)^{M-K}\hspace{-2mm}.
\label{prb_dcd_nc_j}
\end{eqnarray}
\par
\noindent
%which is equal to the upper bound on the probability that a NCN is degree-$j$ in the bipartite graph.
%
As there are $\dbinom{M}{i}\left(\frac{r}{N}\right)^i$ possible degree-$i$ NCNs, the number of degree-$i$ NCNs is estimated as
\vspace{-1.5mm}
\begin{equation}
\varphi_{i}=\dbinom{M}{i}\left(\frac{r}{N}\right)^i\chi_{i}.
\label{num_dcd_nc_2}\vspace{-1.5mm}
\end{equation}
Each degree-$i$ NCN contributes to the number of redundant NCNs $i$ times when deriving $\varphi_{1}$.
Then the effect of degree-$i$ NCNs on the overestimation of $\varphi_{1}$ can be compensated by $\varphi_{1}-(i-1)\varphi_{i}$.
By compensating the effects of degree-$i$ NCNs on the overestimation of $\varphi_{1}$, $i\in\{2,\ldots,K_\mathrm{max}\}$, the number of dependent NC messages in a MAC frame is estimated as
\vspace{-1.5mm}
\begin{equation}
N_\mathrm{rd}=\varphi_{1}-\sum_{i=2}^{K_\mathrm{max}}(i-1)\varphi_{i}.
%=M\sum_{l=2}^{r}(l-1)\dbinom{r}{l}\chi_{1}^{l}(1-\chi_{1})^{r-l}-\sum_{i=2}^{K_\mathrm{max}}(i-1)\dbinom{M}{i}\left(\frac{r}{N}\right)^i\chi_{i}
\label{num_rd_general}\vspace{-1.5mm}
\end{equation}

%
%It is noteworthy that while the degree of a NCN is chosen from the channel realizations of users, we assume that each degree is chosen uniformly by a NCN for simplifying the analysis.
%%
%Moreover, we mainly consider the redundancy of NC messages that is caused by transmitting more replicas for each user.
%%
%However, when the same users collide in more than one time slot, the redundancy of NC messages may also be caused.
%%
%This event happens with a relatively small probability, and its effect on the number of redundant NC messages can be negligible.
%

By subtracting the number of redundant NC messages from the total number of decoded NC messages in a MAC frame, the number of independent NC messages that contribute to the recovery of users' packets can be obtained by
\vspace{-1mm}
\begin{equation}
N_{u}=N_\mathrm{f}-N_\mathrm{rd}.
\label{indpd_general}\vspace{-1mm}
\end{equation}
%where $N_\mathrm{nc}$ and $N_\mathrm{rd}$ are given by Eq. \eqref{num_nc} and Eq. \eqref{num_rd_general}, respectively.
%
As all the independent NC messages contribute to the throughput, the system throughput for the proposed decoding scheme can be approximated by
\vspace{-1.5mm}
\begin{equation}
T=\frac{N_{u}}{N}.
\label{tpt_anal_equ}\vspace{-1.5mm}
\end{equation}
The approximated throughput is compared with the Monte Carlo simulation result in Fig. \ref{fig_thpt_real_anal_simu_2}.
%
%Both Monte Carlo simulation results and analysis results are shown there.
%%
It can be seen that the approximation result matches the simulation result tightly.

\subsection{Optimal Number of Replicas}
From the above analysis, it is known that the throughput for the proposed decoding scheme is related to the number of replicas for each user's packet in a MAC frame, i.e., $r$.
In addition, $r$ affects the energy efficiency of RA schemes, which is important for the power-limited MTC devices.
Thus, we will study the optimal $r$ for maximizing the system throughput and energy efficiency.

%Firstly, the system throughput is considered.
%

\begin{figure}[t]\vspace{-5mm}
\begin{center}
{\includegraphics[height=2.85in,width=3.45in]{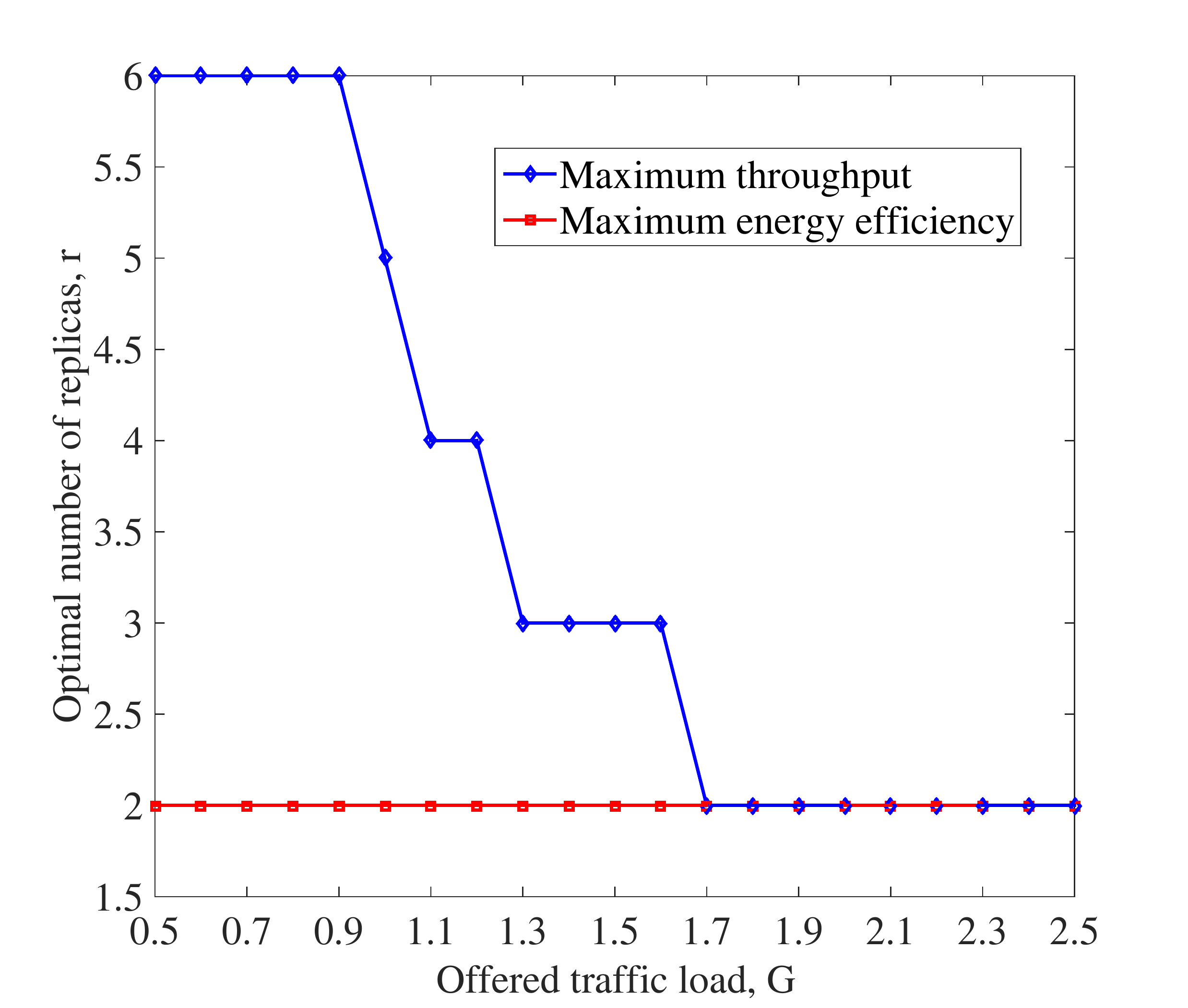}}
\end{center}\vspace{-5mm}
\caption{Optimal number of replicas versus offered traffic load for SNR $=15$ dB, $N=100$, and $K_\mathrm{max}=7$.}
\label{fig_opt_n_max_thpt_max_engy}\vspace{-3mm}
\end{figure}

Firstly, $r$ is optimized to maximize the system throughput.
%
%From the performance analysis in the last section, it is known that the throughput is associated with $r$ via the number of independent NC messages $N_{u}$, as shown in Eq. \eqref{indpd_general} and Eq. \eqref{tpt_anal_equ}.
%%
%From Eq. \eqref{indpd_general}, we can see that both the number of decoded NC messages fed to the frame based decoding $N_{nc}$ and the number of redundant NC messages $N_{rd}$ are associated with $r$.
%%
%Moreover, both $N_{nc}$ and $N_{rd}$ have the similar trend for an increasing $r$.
%%
%In particular, both $N_{nc}$ and $N_{rd}$ increase with $r$ for the low-to-medium offered traffic load, and they decrease with increasing $r$ for the high offered traffic load.
%%
%%When the offered traffic load is high, both $N_{nc}$ and $N_{rd}$ decrease with increasing $r$.
%%
%As a result, the $r$ should be optimized to achieve a balance between the two factors.
%
For a given offered traffic load, we calculate the throughput for all $r$ in a reasonable range, i.e., $2 \leq r \leq 6$, by using Eq. \eqref{indpd_general} and Eq. \eqref{tpt_anal_equ}.
The particular value of $r$ with the highest throughput is determined as the optimal $r$.
In Fig. \ref{fig_opt_n_max_thpt_max_engy}, the optimal values of $r$ are presented for the different offered traffic loads $G$.
%, where $\eta_{2}$ is derived in Eq. \eqref{num_nc_k2} and $\eta_{K}$ for $K=3,4,\ldots,K_\mathrm{max}$ is obtained numerically by using Eq. \eqref{num_nc_gene_k}
It can be seen from Fig. \ref{fig_opt_n_max_thpt_max_engy} that at the low-to-medium offered traffic load regime, the users' packets should be repeated more.
This is because for the low-to-medium traffic load, more NC messages can be obtained by increasing $r$ while the collision size is not significant.
%
%
%load, though both $N_{nc}$ and $N_{rd}$ increase with $r$, the increase of $N_{nc}$ is faster than that of $N_{rd}$.
%%
%Then the throughput increases with $r$ at a low-to-medium offered traffic load regime.
%%
At the high offered load regime, the users' packets should be repeated less.
This is because the large collision size at this regime will degrade the performance of the NC message decoding.
In addition, we optimize $r$ to achieve a higher energy efficiency.
We assume that each packet replica consumes a unit of power.
Then, a user's packet utilizes $r$ units of power, if it is repeated $r$ times.
The energy efficiency is defined as
\vspace{-1.5mm}
\begin{align}
\tau=\frac{T}{r},
\label{energy_eff}
\end{align}
\par
\vspace{-1.5mm}
\noindent
which represents the average number of recovered users' packets per time slot per unit of power.
Based on the obtained throughput for $2 \leq r \leq 6$ by using Eq. \eqref{indpd_general} and Eq. \eqref{tpt_anal_equ}, we calculate the associated energy efficiency.
The value of $r$ with the highest energy efficiency is determined as the optimal $r$.
In Fig. \ref{fig_opt_n_max_thpt_max_engy}, the optimal values of $r$ to achieve a higher energy efficiency are presented for different offered traffic loads $G$.
Interestingly, it can be seen from Fig. \ref{fig_opt_n_max_thpt_max_engy} that the optimal $r$ remains unchanged as two for all the traffic loads.
%
%It is different from the optimal $r$ to maximize the throughput, where the users' messages should be repeated by more than two times at the low-to-medium offered traffic load regime.
%%
This is because for $r=2$ the throughput is already nearly optimal at the low-to-medium offered traffic load regime.
In this case, the increase of $r$ will not significantly increase the throughput but will significantly increase the transmit power.
Therefore, $r=2$ is optimal for the energy efficiency, at the low-to-medium offered traffic load regime.
At the high traffic load regime, i.e., $G \geq 1.7$, $r=2$ is optimal for both energy efficiency and throughput.

\section{Numerical Results}
In this section, we present numerical results for the proposed decoding scheme in coded systems.
%
%We show the number of decoded NC messages per time slot for various collision sizes to demonstrate the slot based decoding performance.
%%
%The system throughput and energy efficiency are also investigated.
%
%the system throughput, which characterizes the performance of the frame based user message decoding.
%
In simulations, we consider the complex-valued Rayleigh channel for all users. The Long-Term Evolution (LTE) Turbo code with code rate $\frac{1}{2}$ and an information block length of 424 is employed.
%
%the coded system, where the $R=\frac{1}{2}$ 3GPP Turbo codes with $424$-length information and the modulation of QPSK are employed.
%
Each frame consists of $N=100$ time slots and each user's packet is repeated twice $(r=2)$.
\begin{figure}[!t]\vspace{-5mm}
	\begin{center}
		{\includegraphics[height=2.85in,width=3.45in]{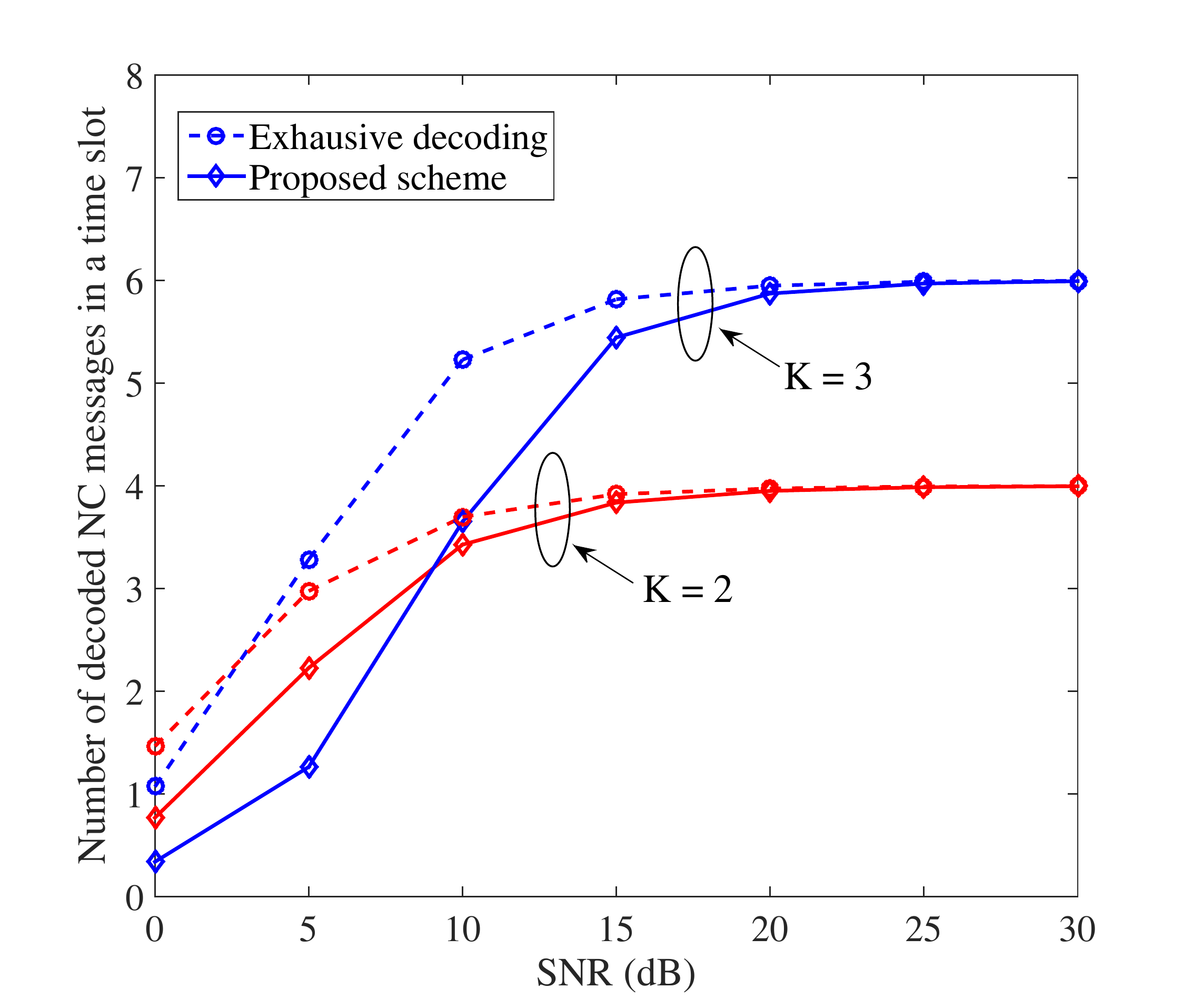}}
	\end{center}\vspace{-5mm}
	\caption{The comparison of the number of obtained NC messages per time slot for $M=100$ users.}
	\label{fig_compare_k23_num_nc_pslot_bnchmark_v5(MUD_exhausive_PNC)}
\end{figure}

We first illustrate the number of decoded NC messages in a time slot for the proposed decoding scheme in Fig. \ref{fig_compare_k23_num_nc_pslot_bnchmark_v5(MUD_exhausive_PNC)}, where QPSK modulation and two collision sizes, i.e., $K=2$ and $K=3$, are considered.
As a QPSK modulated sequence can be viewed as two independent BPSK modulated sequences, a $K$-collision complex-valued time slot can be view as a $2K$-collision real-valued time slot, where $2K$ NC messages will be decoded.
It can be seen from the figure that the proposed NC message decoding can decode multiple NC messages in each time slot on average.
At the medium-to-high SNR regime, it can decode $2K$ NC messages most of the time in a $K$-collision time slot.
In addition, we compare the proposed decoding scheme to the exhaustive decoding scheme used in \cite{Cocco14}, in terms of the number of decoded NC messages in a time slot.
In the exhaustive decoding, all possible NC messages are exhaustively searched until $2K$ linearly independent NC messages are correctly decoded the $K$-collision time slot.
Obviously, the performance achieved by the exhaustive decoding scheme provides an upper bound in terms of the number of decoded NC messages in each time slot.
However, it comes at the expense of a very high decoding complexity.
In contrast, the proposed NC message decoding can achieve this upper bound at the medium-to-high SNR regime with a relatively low complexity.
%It can be seen that our proposed slot based decoding can achieve this upper bound at a medium-to-high SNR regime.
%
%Note that, for the CRDSA++ scheme, the PNC technology is not exploited and no messages are decoded in the $K$-collision time slots, $K\geq 2$.

\begin{figure}[t]
	\begin{center}
		{\includegraphics[height=2.85in,width=3.45in]{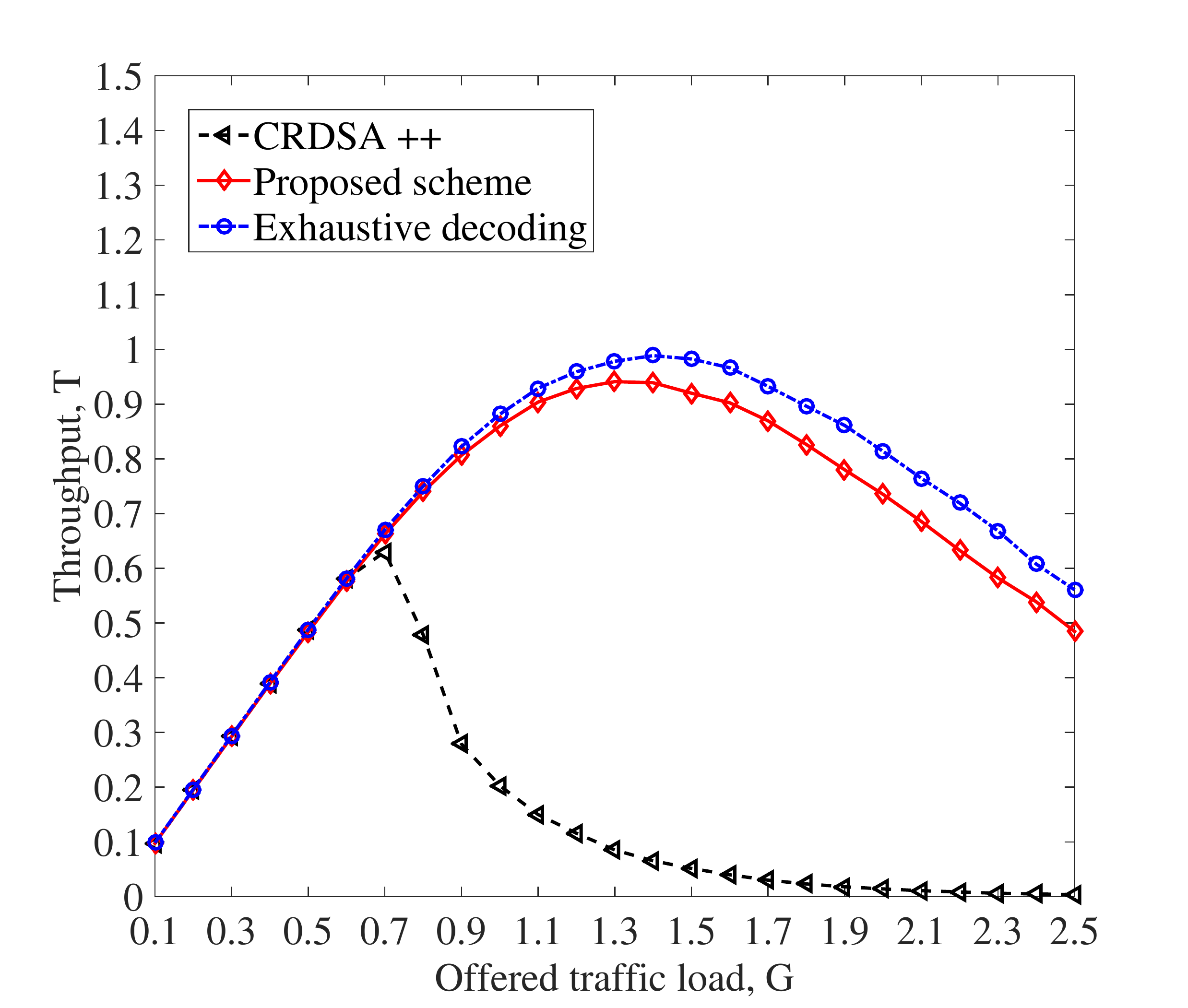}}\vspace{-5mm}
	\end{center}
	\caption{Throughput versus offered traffic load for SNR $=15$ dB, $N=100$, and $K_\mathrm{max}=5$.}
	\label{fig_compare_thpt_bnchmark_v3(mud_exhausive_pnc)}
\end{figure}
Now, we present the throughput of the proposed decoding scheme in Fig. \ref{fig_compare_thpt_bnchmark_v3(mud_exhausive_pnc)}.
The throughput obtained by the exhaustive decoding \cite{Cocco14} and the CRDSA++ scheme with three replicas are also provided for comparison, where the exhaustive decoding scheme refers to the exhaustive NC message decoding combined with the enhanced message-level SIC algorithm.
%The throughput obtained by the exhaustive decoding \cite{Cocco14} and the CRDSA++ scheme with three replicas are also provided for comparison.
%
It can be seen that for the three considered schemes, the throughput first increases and then decreases with increasing the offered traffic load.
The offered traffic load corresponding to the peak throughput is called the traffic load threshold.
It can be observed from Fig. \ref{fig_compare_thpt_bnchmark_v3(mud_exhausive_pnc)} that the performance of our proposed decoding scheme approaches that of the exhaustive decoding scheme, in terms of the traffic load threshold and the peak throughput.
In addition, the throughput gap between our proposed decoding scheme and the exhaustive decoding scheme is marginal for the low-to-medium traffic load.
%
%This is because there are more small-sized collisions than the large-sized collisions for the low-to-medium traffic load, and it has a high probability to successfully decode $2K$ NC messages by using the proposed decoding scheme approaches for the small
%More importantly, since the exhaustive seek and decode is not necessary for the proposed scheme, the proposed scheme has the lower complexity than the benchmark.
%
Compared to the CRDSA++ scheme, our proposed decoding scheme has an excellent performance in a network with much higher offered loads.
In other words, it can support more users in RA networks.

\begin{figure}[t]
	\begin{center}
		\hspace{-2mm}{\includegraphics[height=2.85in,width=3.45in]{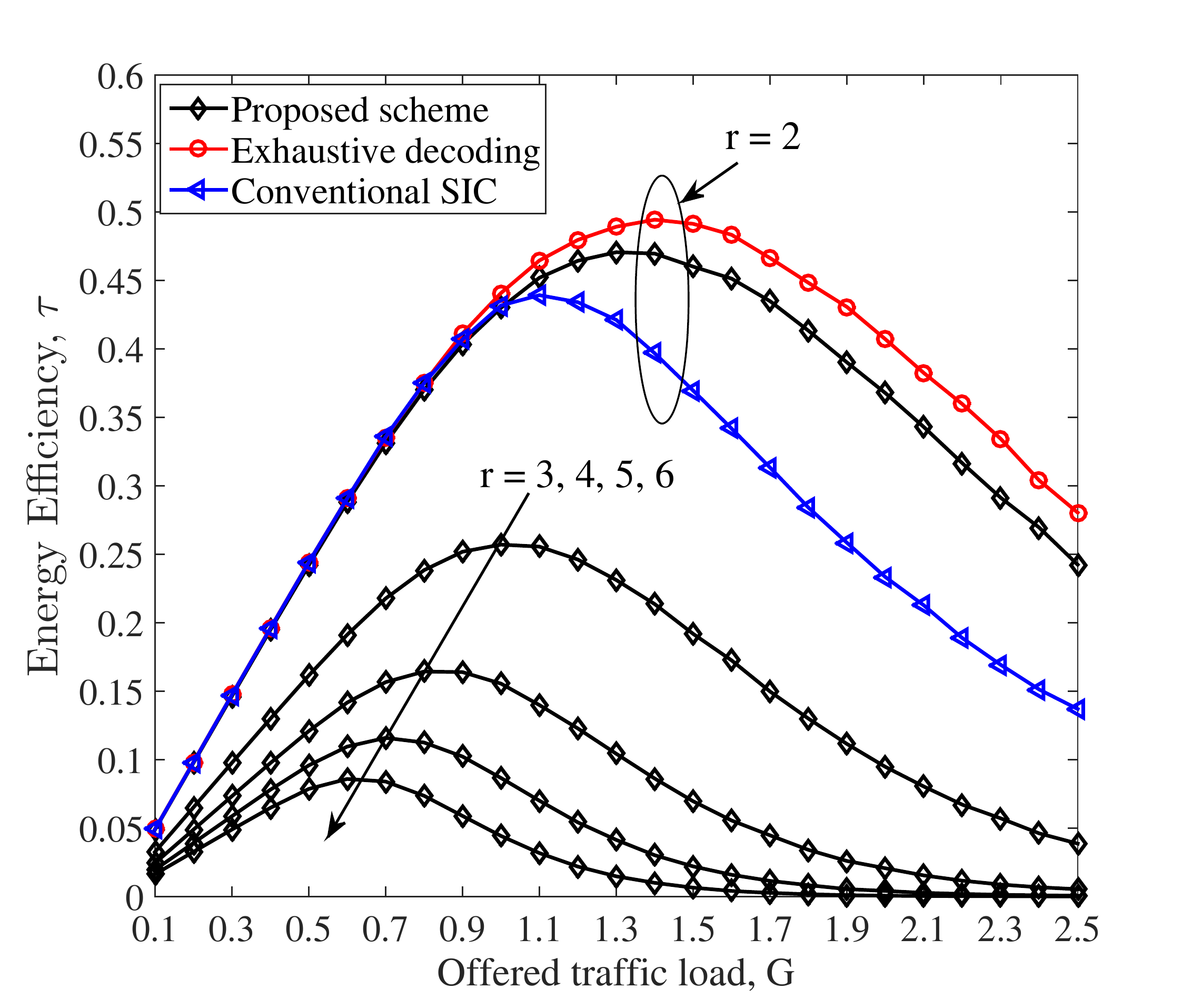}}\vspace{-5mm}
	\end{center}
	\caption{Simulated energy efficiency for SNR $=15$ dB and different $r$.}\vspace{-1cm}
	\label{fig_enegy_efficy_diff_r}
\end{figure}
At last, the simulated energy efficiency for various $r$ is depicted in Fig. \ref{fig_enegy_efficy_diff_r} to verify the obtained optimal number of replicas in Section IV-D.
%At last, the simulated energy efficiency for various $r$ is depicted in Fig. \ref{fig_enegy_efficy_diff_r} to verify the optimized results in Section V.
%
It can be seen that when $r=2$, the RA scheme has the largest energy efficiency for $2 \leq r \leq 6$.
In addition, the energy efficiency decreases with an increased $r$.
The results align with the optimized results in Section IV-D.
Moreover, we compare the energy efficiency of the proposed scheme with the exhaustive decoding and conventional SIC scheme, to illustrate the performance comparison in terms of the NC message decoding and the users' packet recovering, respectively.
From the comparison between the proposed scheme and the exhaustive decoding, where only the NC message decoding scheme is different for the both schemes, we can observe that the proposed scheme can achieve a close-to-optimal energy efficiency achieved by the exhaustive decoding, through utilizing a low-complexity NC message decoding scheme.
It is also seen that compared to the conventional message-level SIC scheme, the proposed scheme can greatly improve the energy efficiency by exploiting the proposed message-level SIC scheme.

\section{Chapter Summary}
In this chapter, we proposed an enhanced low-complexity PNC-based decoding scheme for the CSA system to improve the system throughput.
In the proposed scheme, the linear combinations of collided packets in each time slot were first decoded by exploiting a low-complexity PNC decoding scheme.
Based on the decoded linear combinations within a MAC frame, an enhanced message-level SIC algorithm was proposed to recover more users' packets.
We presented an analytical framework for the PNC-based decoding scheme and derived a tight approximation of the system throughput for the proposed scheme.
Moreover, we optimized the number of replicas transmitted by each user to further improve the system throughput and energy efficiency.
%
%Interestingly, the optimization results show that the optimal number of replicas, in terms of the energy efficiency, is a constant for all offered loads.
%%
%On the other hand, the optimal number of replicas that maximizes the system throughput decreases as the offered load increases.
%
Numerical results showed that the derived analytical results matched well with the simulation results.
Furthermore, the proposed scheme achieved a considerably improved throughput.
%\section{Acknowledgement}
%The authors would like to thank Prof. Marco Chiani from University of Bologna, Italy, for his invaluable comments.\vspace{-3mm}

\chapter{Conclusions and Future Research Topics}\label{C6:chapter6}

%\nomenclature{$a$}{The number of angels per unit area}%
%\nomenclature{$N$}{The number of angels per needle point}%
%\nomenclature{$A$}{The area of the needle point}%
%
\ifpdf
    \graphicspath{{1_introduction/figures/PNG/}{1_introduction/figures/PDF/}{1_introduction/figures/}}
\else
    \graphicspath{{1_introduction/figures/EPS/}{1_introduction/figures/}}
\fi

In this final chapter, we summarize the research work and highlight the contributions of this thesis. We also outline some future research directions arising from this work.

\section{Conclusions}
This thesis has investigated the design of joint user activity identification and channel estimation, the random access, and the data detection schemes for mMTC. In the following, we briefly review the main results of each chapter.

In Chapter 3, we proposed a decentralized transmission control scheme and designed an AMP-based user activity identification and channel estimation algorithm to improve the system performance.
By adopting the state evolution technique, we analyzed the user activity identification performance in terms of the false alarm probability and the missed detection probability.
Additionally, we derived the statistical characteristics of packet delay and the closed-form expression of network throughput.
Based on that, the transmission control scheme was optimized to maximize the network throughput.
Simulation results demonstrated that compared to the conventional scheme without transmission control, the proposed scheme can significantly improve the user identification and channel estimation performance, reduce the packet delay, and enhance the network throughput.

In addition to study the accurate user activity identification and channel estimation, we designed the CSA schemes over erasure channels to improve the network throughput in Chapter 4, where both packet erasure channels and slot erasure channels are considered.
Particularly, by deriving the EXIT functions and optimizing their convergence behavior, we designed the code probability distributions for the CSA schemes with repetition codes and MDS codes.
Furthermore, we analyzed the asymptotic throughput of the CSA schemes over erasure channels by considering an infinite frame length, which has been verified to match well with the throughput for a practical frame length.
Simulation results demonstrated that the designed code probability distributions can improve the throughput of CSA schemes over erasure channels.

For the CSA system, we proposed a low-complexity and efficient data decoding scheme to further improve the system throughput in Chapter 5.
We first presented a low-complexity PNC decoding scheme to obtain NC messages in each time slot of a MAC frame.
Then, we designed an enhanced message-level SIC algorithm to wisely exploit the obtained NC messages across multiple time slots to recover more users' packets.
For the proposed data decoding scheme, an analytical framework was presented and the accurate approximation of system throughput was derived.
By employing the analytical results, we optimized the CSA system to maximize the throughput and the energy efficiency, respectively.

\section{Future Work}
Future wireless communication networks will strive to accommodate such a massive number of connected devices, satisfy the stringent requirement on latency, and at the same time, guarantee diverse QoS requirements \cite{Wong17,Boccardi14}.
This brings about several challenges, including the reduction of signalling overhead for user access and the design of efficient collision resolution methods.
In Chapter 3, we have proposed the joint user activity identification and channel estimation scheme in grant-free random access systems, which can significantly reduce the signalling overhead.
In Chapters 4-5, we have designed the random access scheme and the data decoding algorithm to efficiently resolve packet collisions and improve the system throughput.
However, there are still many research issues to be addressed.
In the following, we propose some future research directions arising from the work presented in this thesis.

\subsection{User Activity Identification and Channel Estimation with Diverse QoS Requirements}
One extension of Chapter 3 is to design the user activity identification and channel estimation scheme with diverse QoS requirements.
MTC has a variety of applications ranging from smart metering to intelligent transport systems \cite{Dawy17, Abbas17}.
This wide range of applications results in very diverse QoS requirements, where QoS requirements include the energy and delay requirements \cite{Dawy17}.
Furthermore, both the energy consumption and delay are impacted by the user activity identification and channel estimation scheme.
Therefore, designing an efficient user activity identification and channel estimation scheme to satisfy the varying QoS requirements is an essential research problem for practical MTC systems.

\subsection{Design of CSA Schemes for Low-Latency MTC}
In Chapter 4 of this thesis, we designed the CSA scheme in the time domain.
In particular, each user encodes its packet segments according to a designed code probability distribution and transmits these encoded packet segments over multiple time slots.
The employment of multiple time slots inevitably results in a large system latency, which may not satisfy the stringent latency requirement in mission-critical applications  \cite{Chen17}.
Therefore, it is essential to introduce other domains, e.g. the frequency domain and the spatial domain, to decrease the system latency and improve the to throughput.
In conjunction with the time domain, introducing the OFDM system for the frequency domain \cite{Wong99} or the MIMO system for the spatial domain \cite{Marzetta10} can lead to the CSA schemes with ultra-low latencies, which will be an interesting subject of future work.

\subsection{Data Decoding Algorithm for CSA Schemes with User Identification and Channel Estimation Errors}
In Chapter 5 of this thesis, we assumed the perfect user activity identification and channel estimation for designing the data decoding algorithm.
However, in practical systems, the user activity identification and channel estimation error is inevitable \cite{Liu18ii} and needs to be considered for designing the data decoding algorithm.
In particular, the channel estimation error may degrade the performance of original PNC-based decoding scheme \cite{Yasami11} in each time slot and the user activity identification error may cause the original message-level SIC algorithm infeasible.
Therefore, with the user activity identification and channel estimation errors, designing the data decoding algorithm for CSA schemes is a challenging but essential issue to be addressed.

\clearpage{\pagestyle{empty}\cleardoublepage}

% references
\renewcommand{\bibname}{Bibliography}
\addcontentsline{toc}{chapter}{\protect\numberline{}{Bibliography}}
\singlespacing
{\bibliographystyle{IEEEtran}
\bibliography{refer_thesis}
}

\end{document}